%% file: PhDThesisv4.tex
\documentclass[final]{inatesis2}
\usepackage{amssymb}
\usepackage{amsmath}
\usepackage{tensor}
\usepackage[chapter]{theorems}
\usepackage{symbols}
\usepackage{url}
\usepackage{wasysym}
\usepackage{graphicx}
\usepackage{epstopdf}
\usepackage[small]{caption} 
\usepackage{threeparttable} 
\usepackage{threeparttablex}
\usepackage{fancyhdr} 
\usepackage{appendix}
\usepackage{longtable}
\usepackage{environ}
\usepackage{nomencl}
\usepackage{bpchem}
\usepackage{doi}
\usepackage{natbib}
\usepackage{aas_macros}
\usepackage[nottoc]{tocbibind}
\usepackage{float}
\usepackage{etoolbox}
\usepackage{lscape}
\usepackage{subfig}
\usepackage{morefloats}
\usepackage{lettrine}
\usepackage{lscape}
\usepackage{rotating}

\usepackage{makeidx}  
\makeindex

\usepackage{epigraph}
\makenomenclature

\newcommand{\ha}{\mathrm{H}\alpha}
\newcommand{\hb}{\mathrm{H}\beta}
\newcommand{\hg}{\mathrm{H}\gamma}
\newcommand{\Oii}{\mathrm{O\ II}}
\newcommand{\Oiii}{\mathrm{O\ III}}
\newcommand{\Nii}{\mathrm{N\ II}}
\newcommand{\Sii}{\mathrm{S\ II}}
\newcommand{\logd}{\log}
\newcommand{\hii}{H\,\textsc{ii}}
\newcommand{\hi}{H\,\textsc{i}}
\newcommand{\lsig}{$L(\mathrm{H}\beta) - \sigma$}

\newcommand{\Msol}{M$_\odot$}
\newcommand{\mincir}{\raise-3.truept\hbox{\rlap{\hbox{$\sim$}}\raise4.truept\hbox{$<$}\ }}
\newcommand{\magcir}{\raise-3.truept\hbox{\rlap{\hbox{$\sim$}}\raise4.truept\hbox{$>$}\ }}
\newcommand{\minmag}{\raise-3.truept\hbox{\rlap{\hbox{$<$}}\raise5.truept\hbox{$<$}\ }}
\newcommand{\be}{\begin{equation}}
\newcommand{\ee}{\end{equation}}
 \newcommand{\ba}{\begin{eqnarray}}
\newcommand{\ea}{\end{eqnarray}}
\newcommand{\brr}{\begin{array}}
\newcommand{\err}{\end{array}}
\newcommand{\bc}{\begin{center}}
\newcommand{\ec}{\end{center}}

\renewcommand{\nomname}{List of Symbols}
\def\mean#1{\left< #1 \right>}

\title{Constraining the Parameter Space of the Dark Energy Equation of State Using Alternative Cosmic Tracers}
\author{Ricardo Ch\'avez}
\degree{DOCTOR OF PHILOSOPHY}
\department{ASTROPHYSICS}
\address{Tonantzintla, Puebla, Mexico}
\advisor[Tenured Researcher INAOE, Mexico]{Dr. Roberto Terlevich}{}
\advisorextra[Tenured Researcher INAOE, Mexico]{Dr. Elena Terlevich}{}
\advisorextraa[Tenured Researcher INAOE, Mexico \& NOA, Greece]{Dr. Manolis Plionis}{}

\begin{document}
\frontmatter
\maketitle     

\cleardoublepage   

\pagestyle{fancy} 

%\begin{thankspageCenter}
%  Esta investigaci\'on fue realizada con apoyo del Consejo de Ciencia y Tecnolog\'ia del Estado de Puebla.
%\end{thankspageCenter}

\begin{thankspage}
  To my parents, for their tireless support.
\end{thankspage}

%\approvalpage  % esto hace la pagina de aprobacion
\onehalfspacing
\begin{Declaration}
	\input{ftmt/declaration}
\end{Declaration}

%\singlespacing
%\small 
\begin{ResumenIngles}
	\input{ftmt/abstract}
\end{ResumenIngles}

\begin{Resumen}
	\input{ftmt/resumen}
\end{Resumen}

\begin{Image}
\begin{figure}
        \centering
	{\includegraphics[width=16cm]{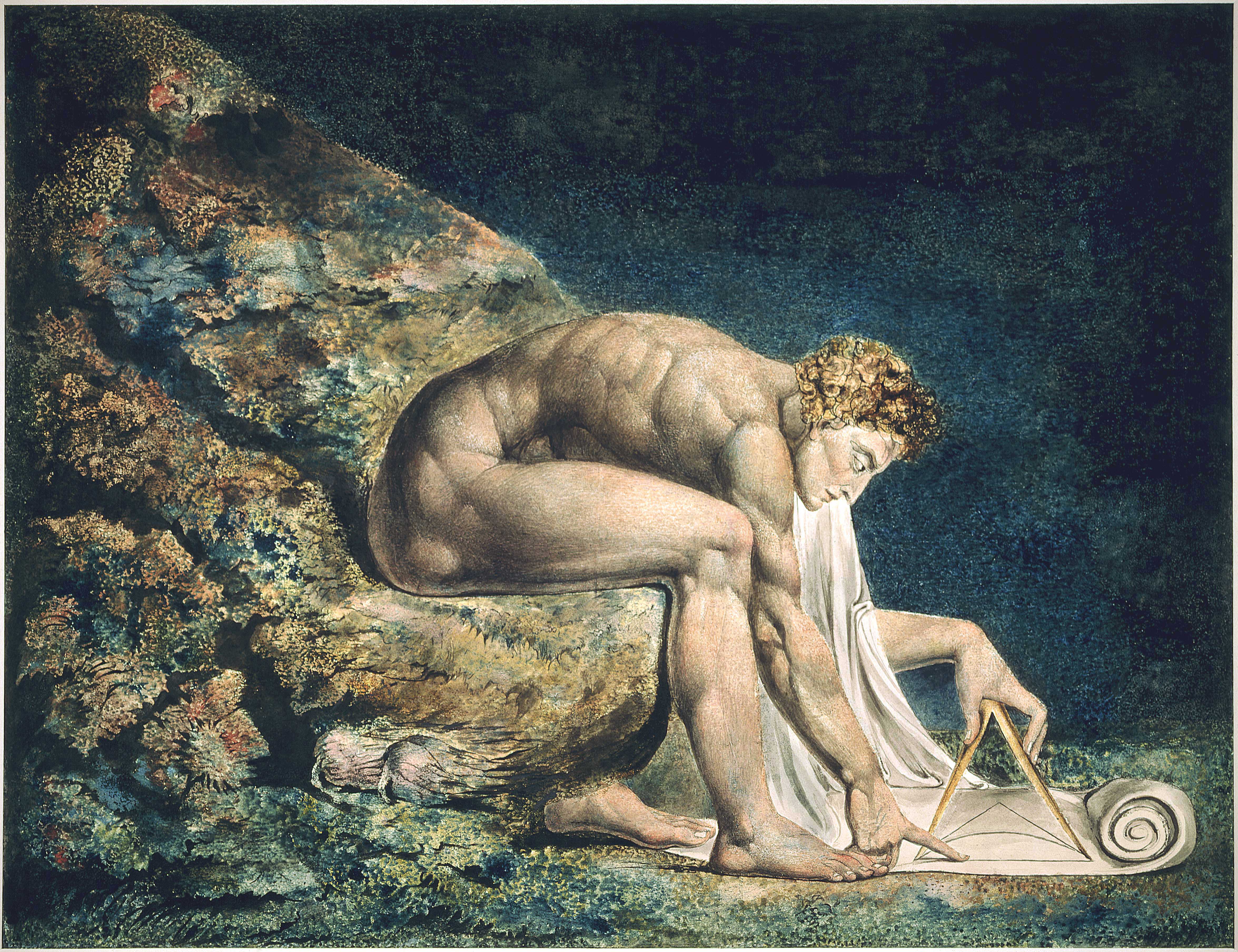}}
	{\caption*{``Newton" by William Blake}}
\end{figure}
\end{Image}

\normalsize 
\singlespacing
\tableofcontents

\onehalfspacing
\begin{Prefacio}
	\input{ftmt/preface}
\end{Prefacio}
%\addcontentsline{toc}{chapter}{Preface}

\begin{Agradecimientos}
	\input{ftmt/acknowledgments}
\end{Agradecimientos}
%\addcontentsline{toc}{chapter}{Acknowledgments}

\addcontentsline{toc}{chapter}{List of Symbols}
\renewcommand{\nompreamble}{We have attempted to keep the basic notation as standard as possible. In general, the notation is defined at its first occurrence in the text. The signature of the metric is assumed to be $(+ , -, -, -)$ and the speed of light, $c$, is taken to be equal to $1$ throughout this work, unless otherwise specified. Throughout this work the Einstein summation convention is assumed.  Note that throughout this thesis the subscript 0 denotes a parameter's present epoch value, unless otherwise specified. }
\cleardoublepage% or \clearpage
\markboth{\nomname}{\nomname}% maybe with \MakeUppercase
\printnomenclature

\mainmatter

\fancyhead[CO]{\nouppercase \rightmark}
\fancyhead[CE]{\nouppercase \leftmark}

%\singlespacing
\onehalfspacing

\input{ch01/ch01} %Intro
\input{ch02/ch02} %Context Distance estrimators cosmology
\input{ch03/ch03} %Physicsl properties fo HII galaxies 
\input{ch04/ch04} %L-Sig Relation of star bursts
\input{ch05/ch05} %Ho from the L-S
\input{ch06/ch06} %The dark energy equation of state
\input{ch07/ch07} %Conclusions 

%apendices
\appendix
\appendixpage
\addappheadtotoc

\input{apa/apa}

\input{apc/apc}
\input{apb/apb}
\input{apd/apd}

\backmatter
\singlespacing
\listoffigures
\listoftables

\renewcommand{\bibname}{References} 
\bibliography{bib/thesis}

%\cleardoublepage
%\phantomsection
%\addcontentsline{toc}{Index}{\indexname}
%\printindex

\end{document}

%% file: ftmt/declaration.tex
This dissertation is my own work and contains nothing which is the outcome of work done in collaboration with others, except as specified in the text and Acknowledgments. 

I hereby declare that my thesis entitled:\\

\emph{Constraining the Parameter Space of the Dark Energy Equation of State Using Alternative Cosmic Tracers} \\

is not substantially the same as any that I have submitted for a degree or diploma or other qualification at any other Research Institute or University.

Some parts of this work have already been published in refereed journals as follows:
\begin{itemize}
	\item Section \ref{sec:3_3}, \emph{\hii\ Galaxies as Cosmological Probes}, has already partially appeared in \citet{Plionis2011}.
	\item Chapter \ref{ch:4}, \emph{The \lsig\ Relation for Massive Bursts of Star Formation}, has already fully appeared in \citet{Chavez2014}.
	\item Chapter \ref{ch:5}, \emph{The Hubble Constant}, has already partially appeared in \citet{Chavez2012}.
	 
\end{itemize}

I further declare that this copy is identical in every respect to the  volume
examined for the Degree, except that any alterations required by the Examiners have been
made.

Date: \today

Signed:

Ricardo Ch\'avez

%% file: ftmt/abstract.tex
We propose to use \hii\ galaxies to trace the redshift-distance relation, by means of their $L(\hb) - \sigma$ correlation, in an attempt to constrain the dark energy equation of state parameter solution space, as an alternative to the cosmological use of type Ia supernovae. 

In order to use effectively high redshift \hii\ galaxies as probes of the dark energy equation of state parameter, we must reassess the $L(\hb) - \sigma$ distance estimator, minimising the observational uncertainties and taking care of the possible associated systematics, such as stellar age, gas metallicity, reddening, environment and morphology.

For a sample of 128 local ($0.02\lesssim z\lesssim 0.2$) compact \hii\ galaxies with high equivalent widths of their Balmer emission lines we obtained ionised gas velocity dispersion from high S/N, high-dispersion spectroscopy (Subaru-HDS and ESO VLT-UVES)  and  integrated H$\beta$ fluxes from low dispersion wide aperture spectrophotometry.

We find that the \lsig\ relation is strong and stable against restrictions in the sample (mostly based on the emission line profiles). The `gaussianity' of the profile is  important for reducing the rms uncertainty of the distance indicator, but at the expense of substantially reducing the sample. By fitting other physical parameters into the  correlation we are able to significantly decrease the scatter without reducing the sample. The size of the starforming region is an important second parameter, while  adding the emission line equivalent width or the continuum colour and metallicity, produces the solution with the smallest rms scatter, $\delta \logd L(\hb) = 0.233$.

We have used the \lsig\ relation from a local sample of \hii\ galaxies and a local calibration or `anchor', given by giant \hii\ regions in nearby galaxies which have accurate distance measurements determined via primary indicators, to obtain a value of $H_0$. Using our best sample of 69 \hii\ galaxies (with $0.01< z <0.16$) and 23 Giant \hii\ regions in 9 galaxies we obtain $H_{0}=74.3 \pm 3.1$ (statistical)$\pm$ 2.9 (systematic) km s$^{-1}$ Mpc$^{-1}$, in excellent agreement with, and independently confirming, the most recent SNa~Ia based results.

Using a local sample (107 sources) and a sample of 21 high redshift \hii\ galaxies, 6 of them with medium-dispersion spectroscopy (ESO VLT-XShooter) and 17 taken from the literature, we have obtained constraints on the planes $H_0 - \Omega_m$, $\Omega_m - w_0$ and $w_0 - w_1$ (CPL model). Results are in line with other recent results although weaker due to the small size of the sample used. We expect to obtain better constraints using a larger high redshift sample. 

%% file: ftmt/resumen.tex
Se propone el uso de la relaci\'on corrimiento al rojo - distancia de las galaxias \hii\, mediada mediante el uso se su correlaci\'on $L(\hb) - \sigma$, con el fin de determinar la funci\'on de Hubble a corrimientos al rojo intermedios y altos, en un intento por restringir el espacio de soluciones de la ecuaci\'on de estado de la energ\'ia oscura, como una alternativa al uso de supernovas del tipo Ia (SNe Ia).

Con el fin de usar eficientemente a las galaxias \hii\ como trazadores cosmol\'ogicos, es necesario calibrar el estimador de distancia $L(\hb) - \sigma$, minimizando las incertidumbres observacionales e identificando los efectos sistem\'aticos asociados, tales como edad estelar, metalicidad del gas, enrojecimiento, medio ambiente y morfolog\'ia. 

Para una muestra local ($0.02\lesssim z\lesssim 0.2$) de 128 galaxias \hii\ compactas con altos anchos equivalentes en sus lineas de emisi\'on de Balmer, se obtuvo la dispersi\'on de velocidades del gas ionizado usando espectroscop\'ia de alta dispersi\'on y alta S/N (Subaru-HDS y ESO VLT-UVES) y flujos integrados de H$\beta$ usando espectro-fotometr\'ia de baja dispersi\'on y apertura ancha. 

Se encontr\'o que la relaci\'on  \lsig\  es resistente y estable ante restricciones en la muestra (basadas mayormente en los perfiles de las l\'ineas de emisi\'on). La `gaussianidad' del perfil es importante para reducir la incertidumbre rms del indicador de distancia, pero a costa de una reducci\'on substancial en el tama\~no de la muestra. Ajustando otros par\'ametros f\'isicos en la correlaci\'on, es posible reducir significativamente la dispersi\'on sin reducir la muestra. El tama\~no de la regi\'on de formaci\'on estelar es un importante segundo par\'ametro, mientras que agregando el ancho equivalente de las l\'ineas de emisi\'on o el color del continuo y la metalicidad, se encuentra la soluci\'on de menor dispersi\'on rms, $\delta \logd L(\hb) = 0.233$. 

Se ha usado el estimador \lsig\ de una muestra local de galaxias \hii\ y una calibraci\'on local o `ancla' de esta correlaci\'on dada por regiones \hii\ gigantes en galaxias cercanas con mediciones de distancias precisas, determinadas mediante indicadores primarios, para obtener un valor de $H_0$. Usando la mejor muestra de 69 galaxias \hii\  (con $0.01< z <0.16$) y 23 regiones \hii\ gigantes en 9 galaxias, se ha obtenido $H_{0}=74.3 \pm 3.1$ (estad\'istico)$\pm$ 2.9 (sistem\'atico) km s$^{-1}$ Mpc$^{-1}$, en excelente acuerdo con, y confirmando independientemente, los resultados mas recientes obtenidos con SNe Ia.

Usando una muestra local de 107 objetos y otra de 21 galaxias \hii\ a alto corrimiento al rojo, 6 de ellas con espectroscop\'ia de mediana dispersi\'on (ESO VLT-XShooter) y 17 tomadas de la literatura, se han obtenido restricciones en los planos $H_0 - \Omega_m$, $\Omega_m - w_0$ y $w_0 - w_1$ (modelo CPL). Los resultados est\'an de acuerdo con otras determinaciones recientes aunque son m\'as d\'ebiles. Se esperan obtener mejores restricciones con una muestra mas grande a alto corrimiento al rojo. 

%% file: ftmt/preface.tex
The terms `cosmological constant', `dark energy' and `modified gravity' have been remainders of our incomplete understanding of the ``physical reality''. Our comprehension has been hampered by incomplete and biased data sets and the consequent theoretical over charged speculation.
 
Knowledge is constructed progressively, harsh and lengthy battles between proud theoretical systems, between judgements,  must be fought before a glimpse of certainty  can be acquired. However, sometimes an apparently tractable \emph{petit} problem has been enough to demolish the noblest system.

The cosmic acceleration, detected at the end of the 1990s, could be one of this class of problems that are the key to a new view of reality. First of all, this problem is related to many  fields in physics, crossing from gravitation to quantum field theory and to the unknown in the embodiment of quantum gravity with its multiple flavours (e.g. string theory, loop quantum gravity, twistor theory, ...). Even more, the quest for a theoretical account of the observed acceleration has given an enormous impetus to the search for alternative theories of gravity.

The theoretical explanations for the cosmic acceleration are many and diverse, first of all we have the cosmological constant as a form of vacuum energy, then we are faced with a multitude of models in which its origin is explained by means of a substance with an exotic equation of state, and finally we encounter explanations based on modifications of the theory of general relativity.

The fact is that the current empirical data is not enough to discriminate between the great number of theoretical models, and therefore if we want to eventually decide on which is the best model we will need more and accurate data.

This work is devoted to explore the possibility of using \hii\ galaxies as probes for the cosmic expansion history. 
Many distinct probes already have been used or proposed, such as type Ia Supernovae, gamma ray bursts, baryon acoustic oscillations, galaxy clusters and weak lensing. From the previously mentioned only type Ia Supernovae, gamma ray bursts and \hii\ galaxies are purely geometrical probes (i.e. related directly to the metric), whereas the others are growth probes (i.e. related to the rate of growth of matter density perturbations) or a combination of both.

The advantage of using \hii\ galaxies over the use of type Ia Supernovae, as probes of cosmic acceleration, is that \hii\ galaxies can be observed easily to higher redshifts. Although, their distance modulus determinations, through their $L(\hb) - \sigma$ relation, have larger uncertainties than those of Supernovae Ia. Nevertheless, \hii\ galaxies would be a valuable complement to the type Ia Supernovae data, especially at high redshifts, and even more they could also be used as an independent confirmation for cosmic acceleration.

\hii\ galaxies are a promising  new avenue for the determination of the cosmic expansion history. Their true value will be seen and assessed during coming years, when a large enough  sample of  intermediate and high redshift objects  have been observed and analysed. 

%% file: ftmt/acknowledgments.tex
I wish to thank my advisors, Roberto Terlevich, Elena Terlevich and Manolis Plionis, for their constant support. I am truly grateful to them for their guidance and patience throughout these years of work. Their clear explanations and examples have many times helped me to make sense of obscure ideas. 

I am also very grateful to Itziar Aretxaga, M{\'o}nica Rodr{\'i}guez, Daniel Rosa-Gonz{\'a}lez, Divakara Mayya, Axel De La Macorra and Rafael Guzm\'an - Llorente  my thesis evaluation committee and examiners, for undertaking the task of reviewing this manuscript and for their useful suggestions.

I am also very grateful to the staff at OAN, OAGH and NAOJ-Subaru for their help during our observing runs and to Elena Terlevich, Fabio Bresolin and Roberto Terlevich for their invaluable advice during our observations.

I also would like to thank to Jorge Melnick, Spyros Basilakos and Elias Koulouridis for their help and many useful observations during our collaborative work.  

I would like to thank the CONACyT (Consejo Nacional de Ciencia y Tecnolog{\'i}a). Without their scholarship (No. 224117), this thesis would not have been possible.

Last, but no least, I would like to thank my family and friends for their love and their help in finding my way through life. If it was not for their continuous support it would have been impossible for me to finish this work.

%% file: ch01/ch01.tex
%chapter ch01.tex ------- 
\chapter{Introduction}

\epigraph{\emph{``Begin at the beginning," the King said gravely, ``and go on till you come to the end: then stop."}}{--- \textup{L. Carroll}, Alice in Wonderland}

\lettrine{O}{ur} current understanding of the cosmological evidence shows that our Universe is homogeneous on the large-scale, spatially flat and in  accelerated expansion; it is composed of baryons, some sort of cold dark matter and a component which acts as having a negative pressure (dubbed 'dark energy' or 'cosmological constant'). 
The Universe underwent an inflationary infancy of  extremely rapid growth, followed by a phase of gentler expansion driven initially by its relativistic and then by its non-relativistic contents but by now its evolution is governed by the dark energy component \citep[e.g.][]{2008PASP..120..235R, Frieman2008}.

The observational evidence for dark energy  was presented in 1998 when two teams studying type Ia supernovae (SNe Ia), the  {\em Supernova Cosmology Project} and the {\em High-$z$
Supernova Search}, found independently that these objects were further away than expected in a Universe without a cosmological constant \citep{Riess1998, Perlmutter1999}. Since then measurements of cosmic microwave background (CMB) anisotropy \citep[e.g.][]{Jaffe2001, Pryke2002, Spergel2007, Ade2013} and of large-scale structure (LSS) \citep[e.g.][]{Tegmark2004, Seljak2005}, in combination with independent Hubble relation measurements \citep{Freedman2001}, have confirmed the accelerated expansion of the Universe.

The accumulated evidence implies that nearly 70\% of the total mass-energy of
the Universe is composed of this mysterious dark energy;  for which 
its nature is still largely unknown.  Possible
candidates of the cause of the accelerated expansion 
are Einstein's cosmological constant, which implies that the dark
energy component is constant in time and uniform in space
\citep{Carroll2001};  or it could be that  the dark
energy is an exotic  form of  matter with a time dependent equation  of state \citep[e.g.][]{Peebles2003, Copeland2006};  or  since 
 the range of validity of General Relativity  (GR) is limited, an extended gravitational theory is needed \citep[e.g.][]{Joyce2014}.

From the previous discussion we can see that understanding the
nature of dark energy is of paramount importance and it could have
deep implications for fundamental physics; it is thus of no surprise that
this problem has been called out prominently in recent policy reports
\citep{Albrecht2006, Peacock2006} where extensive experimental
programs to explore dark energy have been put forward. 

To the present day, the cosmic acceleration has been traced directly only by
means of SNe Ia and at redshifts, $z \sim 1$, a fact which implies
that it is of great
importance to use alternative geometrical probes at higher redshifts
in order to verify the SNe Ia results and to obtain
more stringent constrains in the cosmological parameters solution
space, with the final aim of discriminating among the various theoretical
alternatives that attempt to explain the accelerated expansion of the
Universe \citep[cf.][]{Suyu2012}.

\section{Aims of this Work}
The main objective of this work is to trace the Hubble function using the redshift-distance
relation of \hii\ galaxies, as an alternative to SNe Ia, in an attempt 
to constrain the dark energy equation of state \citep{Plionis2011}.
The main reasons for choosing \hii\ galaxies as alternative tracers of the 
Hubble function, are:
\begin{itemize}
\item \hii\ galaxies can be used as standard candles
  \citep{Melnick2000, Melnick2003, Siegel2005, Plionis2011, Chavez2012} due to the correlation
  between their velocity dispersion and $\hb$-line luminosity
  \citep{Terlevich1981, Melnick1988, Chavez2014}.
\item \hii\ galaxies can be observed to higher redshifts than those
  sampled by current SNe Ia surveys and thus probe a region where
  the Hubble function is more sensitive to the cosmological parameter variations \citep{Melnick2000, Plionis2011}.
\item The use of \hii\ galaxies as alternative high-$z$ tracer will
  enable us, to some extent, to independently verify the SNe Ia based results.
\end{itemize}

In order to use effectively high-$z$ \hii\ galaxies as geometrical
probes, we need to locally re-assess the $L(\hb) - \sigma$ distance
estimator, since this was originally done 30 years ago using non-linear
detectors and without including corrections for effects such as the
galaxy peculiar motions and environmental dependencies. 
Having this objective in mind, we must 
investigate, at low-$z$, all the parameters that can systematically
affect the distance estimator; such as the stellar age, metallicity,
extinction, environment, etc., with the intention to determine
accurately the estimator's zero-point. %MP(26/05/10)

After  recalibration of the $L(\hb) - \sigma$ relation, we can use it to obtain luminosities for \hii\ galaxies to intermediate and high-$z$ and hence luminosity distances. Using a big enough sample it is possible to obtain as good or better constraints to $H_0$ and other cosmological parameters as currently obtained using SNe Ia \citep{Plionis2011}.

\section{Structure of this Work}
Through the second chapter we will be presenting the cosmic
acceleration problem, its observational evidence, its 
implications and possible theoretical explanations and finally the
possible avenues to constrain its parameters solution space in order to
obtain a better understanding of its nature than currently known.

The third chapter explores the fundamental physical properties of \hii\
galaxies as young massive bursts of star formation and justifies their use  as cosmological probe.

The fourth chapter explores the  $L(\hb) - \sigma$ relation for \hii\ galaxies and its 
systematics such as stellar age, metallicity, extinction, environment, etc., with the intention of determining
accurately the estimator's slope.

Through the fifth chapter we apply the \lsig\ relation to the local determination of the Hubble parameter ($H_0$) and explore the related systematics. We analyse the \lsig\ relation for a sample of giant extragalactic \hii\ regions to accurately determine the correlation zero point, then using the already analysed \hii\ galaxies sample we determine the value of $H_0$.

In the sixth chapter we explore the application of the \hii\ galaxies \lsig\ relation to intermediate and high redshifts. We use a sample of 6 high redshift \hii\ galaxies observed using XShooter at the ESO-VLT combined with 17 objects taken from the literature to obtain constraints on the $H_0 - \Omega_m$, $\Omega_m - w_0$ and $w_0 - w_1$ planes.

Finally, in the seventh chapter we summarise  the general conclusion to this work and  the  work to be done to improve the results already obtained.

%% file: ch02/ch02.tex
%chapter ch02.tex 
\chapter{The Expanding Universe}

\epigraph{\emph{We are to admit no more causes of natural things than such as are both true and sufficient to explain their appearances.}}{--- \textup{I. Newton}, Rules of Reasoning in Philosophy : Rule I}

\lettrine{T}{he} current accepted cosmological model explains the history of the Universe as a succession of epochs characterised by their expansion rates. The Universe expansion rate has changed as one of its energy components dominates over all others. Near the beginning of time around $10^{-36}$ seconds after the `big bang', the dominant component is the so called `inflaton' field and the Universe expands exponentially, then after a reheating process, the dominant component is radiation followed by dark matter and at those epochs the expansion of the universe decelerates. Now the Universe is again in a phase of accelerated expansion and we call `dark energy'  the dominant component that causes it.

Our growing comprehension of the expanding Universe picture has advanced as new data has been accumulating through the technological improvements that have revolutionised  astronomy during the past century. The first glimpse of the Universe expansion was obtained using ground based optical spectroscopy \citep{Hubble1929}, now the available data spans essentially all the electromagnetic spectrum, all kinds of astronomical techniques and is obtained through ground as well as space borne observations.

Through time different tracers have been used to measure the expansion rate, initially extragalactic Cepheids, now Supernovae Ia (SNe Ia), the Cosmic Microwave Background (CMB) and also galaxies. %are used as probes. 
Even so, our  current knowledge is insufficient to determine the nature of `dark energy'.

The cause of the cosmic acceleration is one of the most intriguing
problems in all physics. In one form or another it is related to gravitation, high
energy physics, extra dimensions, quantum field theory and even more
exotic areas of physics, as quantum gravity or worm holes. However, we
still know very little regarding the mechanism that drives the
accelerated expansion of the Universe. 

With the recent confirmation of the existence of the Higgs boson, the plot thickens even more. The electroweak phase transition generated by the Higgs potential induces a non vanishing contribution to the vacuum energy at the classical level that is blatantly in discordance with the accepted value for the cosmological constant in the `concordance' ($\Lambda$CDM) cosmological model, thus leaving us with an embarrassingly large fine tuning problem \citep[cf.][]{Sola2013}. 

Due to the lack of a fundamental physical theory explaining the
accelerated expansion, there have been many
theoretical speculations about the nature of dark energy
\citep[e.g.][]{Caldwell2009, Frieman2008}; furthermore and most
importantly, the current observational/experimental data is not
adequate to distinguish between the many adversary theoretical models.

Essentially one can probe dark energy by one or more of the following methods:
\begin{itemize}
\item Geometrical probes of the cosmic expansion, which are directly related to the
  metric like distances and volumes.
\item Growth probes related to the growth rate of the
  matter density perturbations.
\end{itemize}

The existence of dark energy was first inferred from a geometrical
probe, the redshift-distance relation of SNe Ia 
\citep{Riess1998, Perlmutter1999}; this method continues to be the
preeminent way to probe directly the cosmic acceleration. The recent
\emph{Union2.1} \citep{Suzuki2012}, \emph{SNLS3} \citep{Conley2011} and
\emph{PS1} \citep{Rest2013} compilations of SNe Ia data %and other cosmological probes 
are consistent with a cosmological constant,
although the results, within reasonable statistical uncertainty, also agree 
with many dynamical dark-energy models
\citep{Shafer2014}. It is therefore of great importance to trace the Hubble
function to higher redshifts than
currently probed, since at higher redshifts the different models deviate
significantly from each other \citep{Plionis2011}.

In this chapter  we will explore the Cosmic  Acceleration issue; the
first section  is devoted to a general account of  the basics of
theoretical  and observational cosmology, in the second section we 
present a brief
outlook  of  the  observational  evidence which  supports  the  cosmic
acceleration; later we will survey  some of the theoretical
explanations of the accelerating expansion and finally we
will  overview   some probes    used  to test the Universe expansion. 

Finally, a few words of caution regarding the  terminology used. 
Through this chapter we will be using the term \emph{dark energy} as opposed to
\emph{cosmological constant}, in the sense of a time-evolving cause of
the cosmic acceleration. However, in later chapters we will use only the term
\emph{dark energy} since we consider it as the most general model, of
which the \emph{cosmological constant} is (mathematically) a particular case, while it 
effectively reproduces also the phenomenology of some modified gravity
models.

%\section{Brief Historical Sketch}
%Newton, in its cosmological queries, asks about the way to obtain a static universe that is governed by an all pervading attractive force, gravity. That same question leads Einstein to the introduction of a 'cosmological constant' that counteracts the effects of the  gravitational attraction of matter in the universe. 

\section{Cosmology Basics}
The fundamental assumption over which our current understanding of the
Universe is constructed is known as the cosmological principle, which
states that the Universe is homogeneous and isotropic on
large-scales. The evidences that sustain the cosmological principle are
basically the near-uniformity of the CMB
temperature \citep[e.g.][]{2003ApJS..148..175S} and the large-scale distribution of galaxies
\citep[e.g.][]{2005MNRAS.364..601Y}.

Under the assumption of homogeneity and isotropy, the geometrical
properties of space-time are described by the
Friedmann-Robertson-Walker (FRW) metric \citep{1935ApJ....82..284R},
given by
\begin{equation}
 ds^2 = dt^2 - a^2 (t) \left[\frac{dr^2}{1 - kr^2} + r^2 (d \theta^2 + \sin^2 \theta d \phi^2) \right],
\end{equation}
\nomenclature{$a(t)$}{The cosmic scale factor.}%
where $r$, $\theta$, $\phi$ are spatial comoving coordinates
(i.e., where a freely falling particle comes to rest) and $t$ is the
time parameter, whereas $a(t)$ is the cosmic scale factor which at the
present epoch, $t_0$, has a value $a(t_0) = 1$; $k$ is the curvature of the space, such that
$k = 0$ corresponds to a spatially flat Universe, $k = 1$ to a positive
curvature (three-sphere) and $k = -1$ to a negative curvature
(saddle as a 2-D analogue). 
Note that we are using units where the speed of light, $c=1$.

From the FRW metric we can derive the cosmological redshift, i.e. the
amount that a photon's wavelength ($\lambda$) increases due to the
scaling of the photon's energy with $a(t)$, with corresponding definition:
\begin{equation}
  1 + z \equiv \frac{\lambda_0}{\lambda_e} = \frac{a(t_0)}{a(t)} = \frac{1}{a(t)},
\label{eqRS}
\end{equation}
\nomenclature{$z$}{The redshift.}%
where, $z$ is the redshift, $\lambda_0$ is the observer's frame
wavelength and $\lambda_e$ is the emission's frame wavelength. Note
that throughout this thesis the subscript 0 denotes a parameter's
present epoch value.%MP2505 ET

In order to determine the dynamics of the space-time geometry we must
solve the GR field equations for the FRW metric, in the presence of
matter, obtaining the cosmological field equations or
Friedmann-Lema\^{\i}tre equations (for a full derivation see Appendix
\ref{ApA}):
\begin{eqnarray}
  \left(\frac{\dot{a}}{a}\right)^2 &=& \frac{8 \pi G \rho}{3} 
- \frac{k}{a^2} + \frac{\Lambda}{3}, \label{eqF2}\\
  \frac{\ddot{a}}{a} &=& -\frac{4 \pi G}{3} (\rho + 3 p ) + 
\frac{\Lambda}{3}, \label{eqF1}
\end{eqnarray}
\nomenclature{$\rho$}{The mass-energy density.}%
\nomenclature{$p$}{The pressure.}%
\nomenclature{$\Lambda$}{The cosmological constant.}%
where $\rho$ is the total energy density of the Universe, $p$ is the
total pressure and $\Lambda$ is the cosmological constant.%MP2505 ET

In eq.(\ref{eqF2}) we can define the Hubble parameter 
\begin{equation}
  H \equiv \frac{\dot{a}}{a},
\label{eqHP}
\end{equation}
\nomenclature{$H$}{The Hubble parameter.}%
\nomenclature{$h$}{The dimensionless Hubble parameter.}%
of which its present value is conventionally expressed as $H_0 = 100 h\ \mathrm{km\
  s^{-1}\ Mpc^{-1}}$, where $h$ is the dimensionless Hubble parameter and unless otherwise stated we take a value of
 $h = 0.743 \pm 0.043$ \citep{Chavez2012, Freedman2012, Riess2011, Freedman2001, Tegmark2006}.

The time derivative of eq.(\ref{eqF2}) gives:
$$
\ddot{a} = \frac{8 \pi G}{3} \left(\rho a + \frac{\dot{\rho} a^2}{2
    \dot{a}} \right) + \frac{\Lambda a}{3},
$$
and from the above and eq.(\ref{eqF1}) we can eliminate $\ddot{a}$ to obtain
\begin{eqnarray*}
  \frac{- 4 \pi G a}{3} \left[(\rho + 3 p)  + 2 \left(\rho + 
\frac{\dot{\rho}a}{2 \dot{a}} \right) \right] &=& 0\\
  \frac{a}{\dot{a}} \dot{\rho} + 3 (\rho + p) &=& 0,
\end{eqnarray*}
which then gives:
\begin{equation}
 \dot{\rho} + \frac{3 \dot{a}}{a} (\rho + p) = 0,
\label{eqEC}
\end{equation}
which is an expression of energy conservation.%MP2505 ET

Equation (\ref{eqEC}) can be written as
\begin{eqnarray}
 \frac{d(\rho a^3)}{dt} &=& - 3 a^2 \dot{a} p\\
 \frac{d(\rho a^3)}{da} &=& - 3 a^2 p \label{eqpCE},
\end{eqnarray}
and thus:
\begin{equation}
 d(\rho_i a^3) = - p_i da^3,
\label{eqPV}
\end{equation}
where the subscript $i$ runs over all the components of the
Universe. Equation (\ref{eqPV}) is the expanding universe analog
of the first law of thermodynamics, $dE = - pdV$.%MP02505 ET

If we assume that the different components of the cosmological fluid
 have an equation of state of the generic form:
\begin{equation}
  p_i = w_i \rho_i,
\end{equation}
then from eq.(\ref{eqpCE}) we have
\begin{equation}
 \frac{d(\rho_i a^3)}{da} = -3 w_i \rho_i a^2,
\end{equation}
which in the case where the equation of state
parameter depends on time, ie., $w_{i}(a)$, the corresponding density takes the following 
form:
\begin{equation}
  \rho_{i} \propto \exp \left\{ - 3 \int{\frac{da}{a}[1 + w_{i}(a)]}\right\}\;.
\end{equation}
For the particular case where $w_{i}$ is a constant through 
cosmic time, we have
\begin{equation}
 \rho_{i} \propto a^{-3(1 + w_i)},
\end{equation}
where $w_i \equiv p_i/\rho_i$.% is the equation of state parameter. 
\nomenclature{$w$}{The equation of state parameter.}%
These last two equations can be written as a function of redshift, defined
by the eq.(\ref{eqRS}), as:
\begin{align}
\rho_{i} &\propto \exp{\left[3 \int_{0}^{z}{\frac{1 + w_{i}(z')}{1 +
        z'} dz'} \right]}, \label{eqrhow}\\
 \rho_i &\propto (1+z)^{3(1 + w_i)}\;. 
\label{eqER}
\end{align}
For the case of non-relativistic matter (dark matter and baryons), $w_m = 0$ and
$\rho_m \propto (1 + z)^3$, while for relativistic particles (radiation and
neutrinos), $w_r = 1/3$ and $\rho_r \propto (1+z)^4$, while for vacuum
energy (cosmological constant), $w_{\Lambda} = -1$ and for which we have
$p_{\Lambda} = - \rho_{\Lambda} = -\Lambda/8 \pi G$.%MP2505 ET

In general the dark energy equation of state can be parameterized as
\citep[e.g.][]{Plionis2009}
\begin{equation}
 p_w = w(z) \rho_w,
\end{equation}
where 
\begin{equation}
 w(z) = w_0 + w_1 f(z),
\end{equation}
with $w_0 = w(0)$ and $f(z)$ is an increasing function of redshift,
such as $f(z) = z/(1 + z)$ \citep{Chevallier2001, Linder2003, Peebles2003, Dicus2004, Wang2006}.

The so called critical density corresponds to the total energy density of the
Universe. From eq.(\ref{eqF2}),
where we take the cosmological constant as a cosmic fluid, and
the definition of the Hubble parameter,  eq.(\ref{eqHP}), we have that:
\begin{equation}
  \rho_c \equiv \frac{3 H_0^2}{8 \pi G} = 1.88 \times 10^{-29} h^2\ 
\mathrm{g\ cm^{-3}} = 8.10 \times 10^{-47} h^2\ \mathrm{GeV^4}\;.
\end{equation}
\nomenclature{$\rho_{c}$}{The critical density (that required for the Universe to have a flat spatial geometry).}%
This parameter provides a convenient mean to normalize the mass-energy
densities of the different cosmic components, and we can write:
\begin{equation}
 \Omega_i = \frac{\rho_i(t_0)}{\rho_c} \;,
\end{equation}
\nomenclature{$\Omega$}{The mass-energy density nomarlized to the present $\rho_{c}$ value.}%
where the subscript $i$ runs over all the different components of the
cosmological fluid. Using this last definition and eq.(\ref{eqrhow}) 
we can write eq.(\ref{eqF2}) as:
\begin{equation}
  H^2(z) = H_0^2 \left[\Omega_r(1+z)^4 + \Omega_m(1+z)^3 +
    \Omega_k(1+z)^2 + \Omega_w \exp \left( 3 \int_{0}^{z}
{\frac{1 + w(z')}{1 + z'} dz'} \right) \right],
 \label{eqHZ}
\end{equation}
where $\Omega_k$ has been defined as 
$$
 \Omega_k \equiv \frac{- k}{a^2 H_0^2}\;.
$$
By definition we have that $\Omega_r + \Omega_m + \Omega_k + \Omega_w
\equiv 1$, and as a useful parameter we can define $\Omega_0 \equiv \Omega_r
+ \Omega_m + \Omega_w$, such that for a positively curved Universe
$\Omega_0 > 1$ and for a negatively curved Universe $\Omega_0 < 1$.%MP2505 ET

The value of the curvature radius, $R_{curv} \equiv a/\sqrt{| k |}$, is given by
\begin{equation}
 R_{curv} = \frac{H_{0}^{-1}}{\sqrt{|\Omega_{0} - 1|}},
\end{equation}
\nomenclature{$R_{curv}$}{The curvature radius.}%
then its characteristic scale or Hubble radius is given by $H_{0}^{-1} \approx 3000 h^{-1}\ \mathrm{Mpc}$.%MP2505 ET

\subsection{Observational toolkit}
In observational cosmology the fundamental observable is the
redshift, and therefore it is important to express the distance relations
in terms of $z$. The first distance measure to be considered is the
lookback time, i.e. the difference between the age of the Universe at
observation $t_0$ and the age of the Universe, $t$, when the photons were
emitted. From the definitions of redshift, eq.(\ref{eqRS}), and the
Hubble parameter, eq.(\ref{eqHP}), we have:
\begin{eqnarray*}
 \frac{dz}{dt} = -\frac{\dot{a}}{a^2} = -H(z) (1+z)\;,
\end{eqnarray*}
from which we have:
\begin{equation}
 dt = -\frac{dz}{H(z)(1+z)} \;,
\end{equation}
and the lookback time is defined as:
\begin{equation}
 t_0 - t = \int_{t}^{t_0}{dt} = \int_{0}^{z}{\frac{dz'}{H(z')(1+z')}}
 = \frac{1}{H_0} \int_{0}^{z}{\frac{dz'}{(1+z') E(z')}}\;,
\end{equation}
where
\begin{equation} 
  E(z) = \sqrt{\Omega_r(1+z)^4 + \Omega_m(1+z)^3 + \Omega_k(1+z)^2 + 
\Omega_w \exp \left( 3 \int_{0}^{z}{\frac{1 + w(z')}{1 + z'} dz'} \right) }\;.
\end{equation}%MP2505 ET

From the definition of lookback time it is clear that the cosmological
time or the time back to the Big Bang, is given by
\begin{equation}
 t(z) = \int_{z}^{\infty}{\frac{dz'}{(1+z') H(z')}}\;.
\end{equation}%MP2505 ET1006
\nomenclature{$t(z)$}{The cosmological time.}%

In the following discussion it will be useful to have an adequate
parameterization of the FRW metric \citep{Hobson2005} which is given by:
\begin{equation*}
 ds^2 = dt^2 - a^2 (t) \left[d\chi^2 + S^2(\chi) (d \theta^2 + \sin^2 \theta d \phi^2) \right],
\end{equation*}
where the function $r = S(\chi)$ is: 
\begin{equation}
 S(\chi) = \left \{ \begin{array}{lc}
                     \sqrt{k}^{-1} \sin( \chi \sqrt{k})  & \mathrm{if}\ k > 0,\\
                     \chi       & \mathrm{if}\ k =0,\\
                     \sqrt{|k|}^{-1} \sinh( \chi \sqrt{|k|}) & \mathrm{if}\ k < 0,\\
                    \end{array} \right.
\end{equation}
We can see that the comoving distance, i.e., that between two free
falling particles which remains constant with epoch, is defined by:
\begin{equation}
 \chi = \int_{t}^{t_0}{\frac{dt}{a(t)}} = \frac{1}{H_0} \int_{0}^{z}{\frac{dz'}{E(z')}}\;.
\end{equation}
\nomenclature{$\chi$}{The comoving distance.}%
The transverse comoving distance (also called proper distance) is defined as:
\begin{equation}
 D_M(t) = a(t) S(\chi),
\end{equation}
\nomenclature{$D_{M}$}{The proper distance.}%
At the present time and for the case of a flat model we have, $D_M =
a(t_0) \chi = \chi$.%MP2505 ET0610

The angular distance is defined as the ratio of an object's physical
transverse size to its angular size, and can be expressed as:
\begin{equation}
 D_A = \frac{D_M}{1+z} \;.
\end{equation}
\nomenclature{$D_{A}$}{The angular distance.}%
Finally, the luminosity distance is defined by means of the relation
\begin{equation}
 f = \frac{L}{4 \pi D_L^2},
\label{eqDL}
\end{equation}
where $f$ is an observed flux, $L$ is the intrinsic luminosity of the
observed object and $D_L$ is the luminosity distance; from which one obtains:
\begin{equation}
 D_L = (1+z) D_M = (1+z) \int_{0}^{z}{\frac{dz'}{H(z')}}\;.
\label{eqLD}
\end{equation}%MP2505 ET
\nomenclature{$D_{L}$}{The luminosity distance.}%

The distance modulus of a given cosmic object is defined as:
\begin{equation}
 \mu \equiv m - M = 5 \logd(D_L / 10\ \rm{pc})
 \label{eqDM0}
\end{equation}
where $m$ and $M$ are the apparent and absolute magnitude of the
object, respectively. If the distance, $D_L$, is expressed in Mpc then
we have:
\begin{equation}
 \mu = 5 \logd D_L + 25\;.
 \label{eqDM}
\end{equation}
Through this relation and with the use of standard candles, i.e. objects of
fixed absolute magnitude $M$, we can constrain the different
parameters of the cosmological models via the construction of the 
Hubble diagram (the magnitude-redshift relation). 

The following relation is useful when working with fluxes and luminosities instead of magnitudes,
\begin{equation}
 \logd L = \logd f + 0.4\mu + 40.08\;,
 \label{eqLF}
\end{equation}
where $f$ is the observed flux of the object, $L$ the luminosity and $\mu$ the distance modulus as defined in \ref{eqDM}.

The scale factor can be Taylor expanded around its present value:
\begin{eqnarray*}
 a(t) &=& a(t_0) - (t_0 - t) \dot{a}(t_0) + \frac{1}{2} (t_0 -t)^2 \ddot{a}(t_0) - \cdots\\
      &=& a(t_0)[1 - (t_0 -t)H(t_0) - \frac{1}{2} (t_0 -t)^2 q(t_0) H^2(t_0) - \cdots]\\
      &=& 1 + H_0(t - t_0) - \frac{1}{2} q_0 H_0^2 (t-t_0)^2 + \cdots \;, 
\end{eqnarray*}
where the deceleration parameter $q(t)$ is given by
\begin{equation}
 q(t) \equiv -\frac{\ddot{a}(t) a(t)}{\dot{a}^2(t)}\;.
\end{equation}
\nomenclature{$q(t)$}{The deceleration parameter .}%
From the previous definitions we can write an approximation to the distance-redshift relation as
\begin{equation}
 H_0 D_L = z + \frac{1}{2}(1-q_0)z^2 + \cdots \;, 
\label{eqHR}
\end{equation}
where we can recognize that for $z \ll 1$ it can be written as 
\begin{equation}
 H_0 D_L \approx z,
\end{equation}
which is known as the ``Hubble law''.

Finally the comoving volume element, as a function of redshift, can be written as:
\begin{equation}
 \frac{dV}{dz d\Omega} = \frac{S^2(\chi)}{H(z)}\;,
 \label{eqCVE}
\end{equation}
\nomenclature{$dV$}{The comoving volume element.}%
where $\Omega$ is the solid angle.

\subsection{Growth of structure}

The accelerated expansion of the Universe affects 
the evolution of cosmic structures since the
expansion rate influences the growth rate of the density
perturbations.

The basic assumptions regarding the evolution of structure in the Universe
are that the dark matter is composed of non relativistic particles, i.e it
is composed of what is called cold dark matter (CDM), and that the initial spectrum of density
perturbations is nearly scale invariant, $P(k) \sim k^{n_{s}}$, where
the spectral index is $n_{s} \simeq 1$, as it is predicted by
inflation. With this in mind, the linear growth of small amplitude, matter
density perturbations on length scales much smaller than the Hubble
radius is governed by a second order differential equation,
constructed by linearizing the perturbed equations of motions of a
cosmic fluid element and given by:
\begin{equation}
 \ddot{\delta_{k}} + 2H\dot{\delta_{k}} - 4 \pi G \rho_{m} \delta_{k} = 0,
\end{equation}
\nomenclature{$\delta_{k}$}{The perturbations to the mass-energy density decomposed into their Fourier modes. }%
where the perturbations $\delta_k \equiv \delta \rho_{m}(\mathbf{x}, t) /
\bar{\rho}_{m}(t)$ have been decomposed into their Fourier modes of
wave number $k$. The expansion of the Universe enters through the
so-called ``Hubble drag'' term, $2 H \dot{\delta_{k}}$. Note that $\bar{\rho}_{m}$ is
the mean density.

The growing mode solution of the previous differential equation, 
in the standard {\em concordance} cosmological model 
($w_{\Lambda} = -1$) is given by:
\begin{equation}
  \delta_{k} (z) \propto H(z) (5 \Omega_{m} / 2) \int_{z}^{\infty}{\frac{1 + z'}{H^{3}(z')}dz'}\;.
\end{equation}

From the previous equation we obtain that, $\delta_{k}(t)$ is
approximately constant during the radiation dominated epoch, grows as
$a(t)$ during the matter dominated epoch and again is
constant during the cosmic acceleration dominated epoch, in which the
growth of linear perturbations effectively freezes.

\section{Empirical Evidence}

\begin{figure}[tbh]
\centering
\resizebox{0.7\textwidth}{!}{\includegraphics{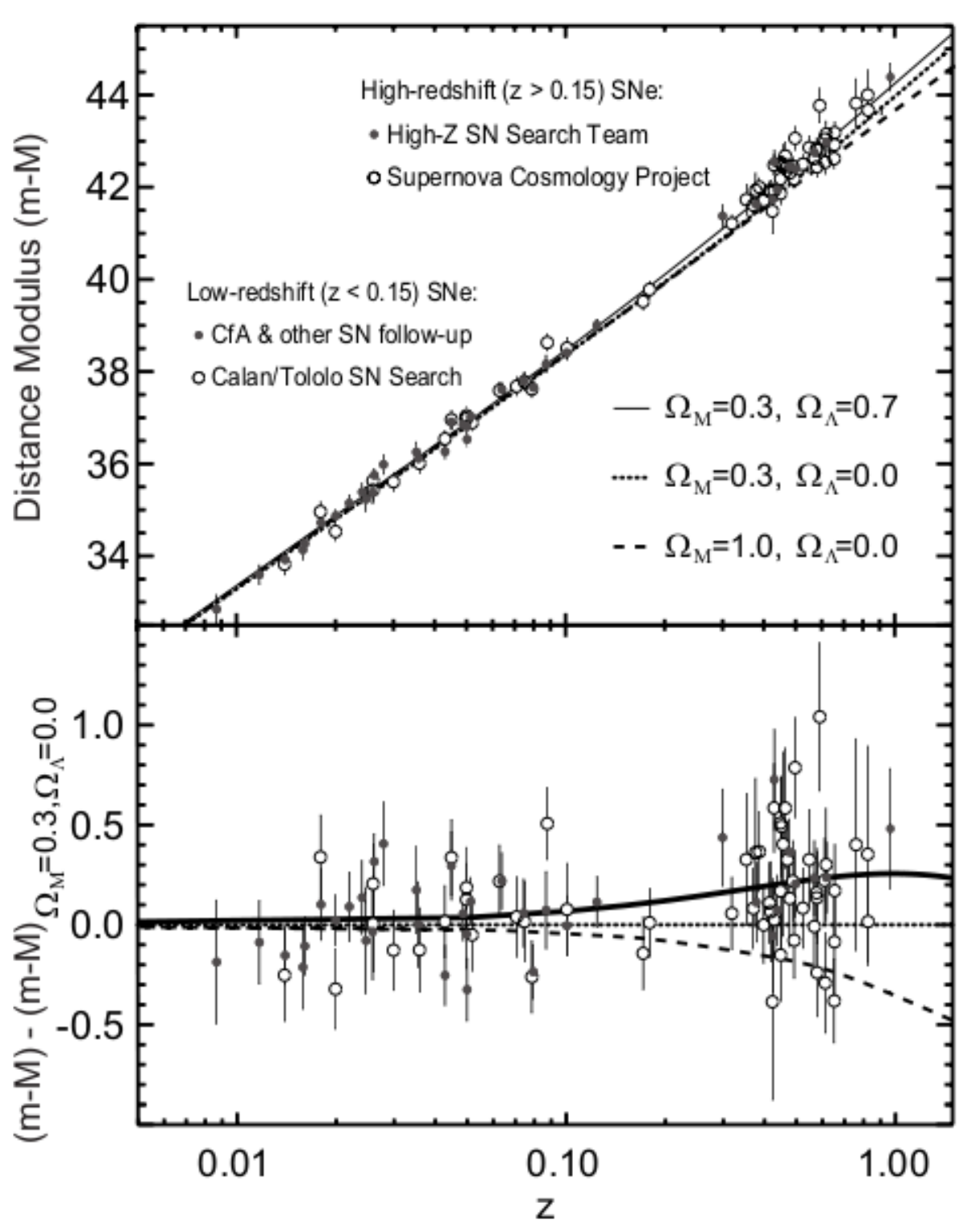}}
\caption[Supernovae Ia Hubble Diagram] {\small \emph{Upper panel:} Hubble
  diagram of type Ia supernovae measured by the Supernova Cosmology
  Project and the High-z Supernova Team. \emph{Lower Panel: }
  Residuals in distance modulus relative to an open Universe with
  $\Omega_{0} = \Omega_{m} = 0.3$. Taken from
  \citet{Perlmutter2003}.}
\label{fig:SnHd}
\end{figure}

The cosmic acceleration was established empirically at the end of the
1990s when two independent teams, the {\em Supernova Cosmology Project} and
the {\em High-z Supernova Search}, succeeded in their attempt to measure the
supernova Hubble diagram up to relatively high redshifts ($z \sim 1$)
\citep{Riess1998, Perlmutter1999}. Surprisingly, both teams found that
the distant supernovae are $\sim 0.25\ \rm{mag}$ dimmer that they would be
in a decelerating universe, indicating that the cosmic expansion has
been accelerating over the past $\sim 7\ \mathrm{Gyr}$ (see Figure
\ref{fig:SnHd}).%MP2505 ET1006

The cosmic acceleration has been verified by many
other probes, and in this section we will briefly review the current
evidence on which this picture of the Universe was constructed.%MP2505 ET

%Cosmic Microwave Background
\subsection{Cosmic microwave background}

%Figure ---- Acoustic temperature spectrum sensitivity
\begin{figure}[tb]
\centering
\resizebox{0.8\textwidth}{!}{\includegraphics{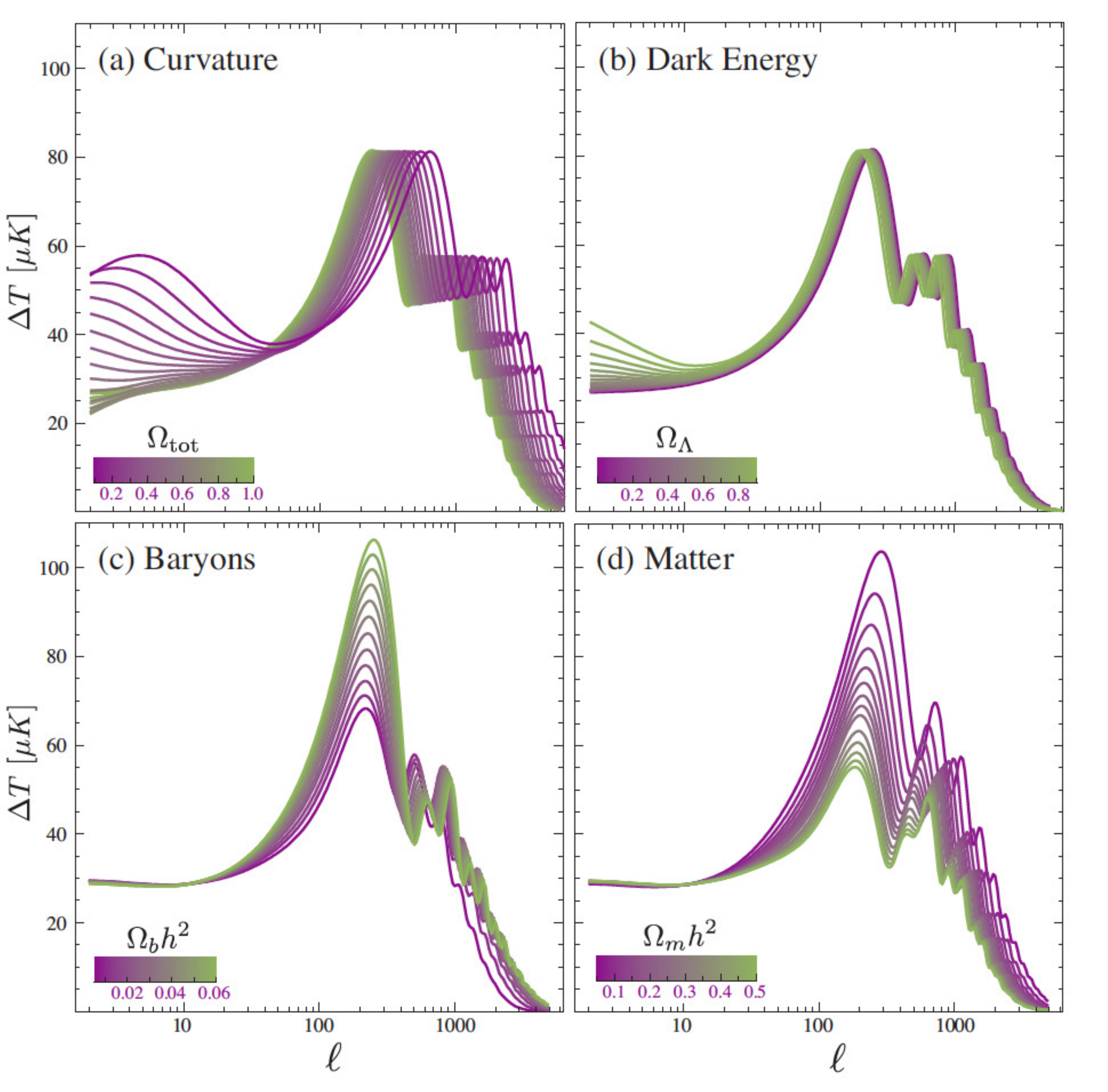}}
\caption[Acoustic temperature spectrum sensitivity] {\small
  Sensitivity of the acoustic temperature spectrum to four fundamental
  cosmological parameters. (a) The curvature as quantified by
  $\Omega_{0}$. (b) The dark energy as quantified by the cosmological
  constant $\Omega_{\Lambda}$ ($w_{\Lambda} = -1$). (c) The 
  baryon density $\Omega_{b}h^{2}$. (d) The matter density
  $\Omega_{m}h^{2}$. All parameters are varied around a fiducial model
  with: $\Omega_{0}= 1, \Omega_{\Lambda} = 0.65, \Omega_{b}h^{2} = 0.02, \Omega_{m}
  h^{2} = 0.147, n =1, z_{ri} = 0, E_{i} = 0$. Taken from
  \citet{Hu2002}.}
\label{fig:AosCMB}
\end{figure}

%Figure ---- Angular power spectrum measurements
\begin{figure}[tb]
\centering
\resizebox{0.9\textwidth}{!}{\includegraphics{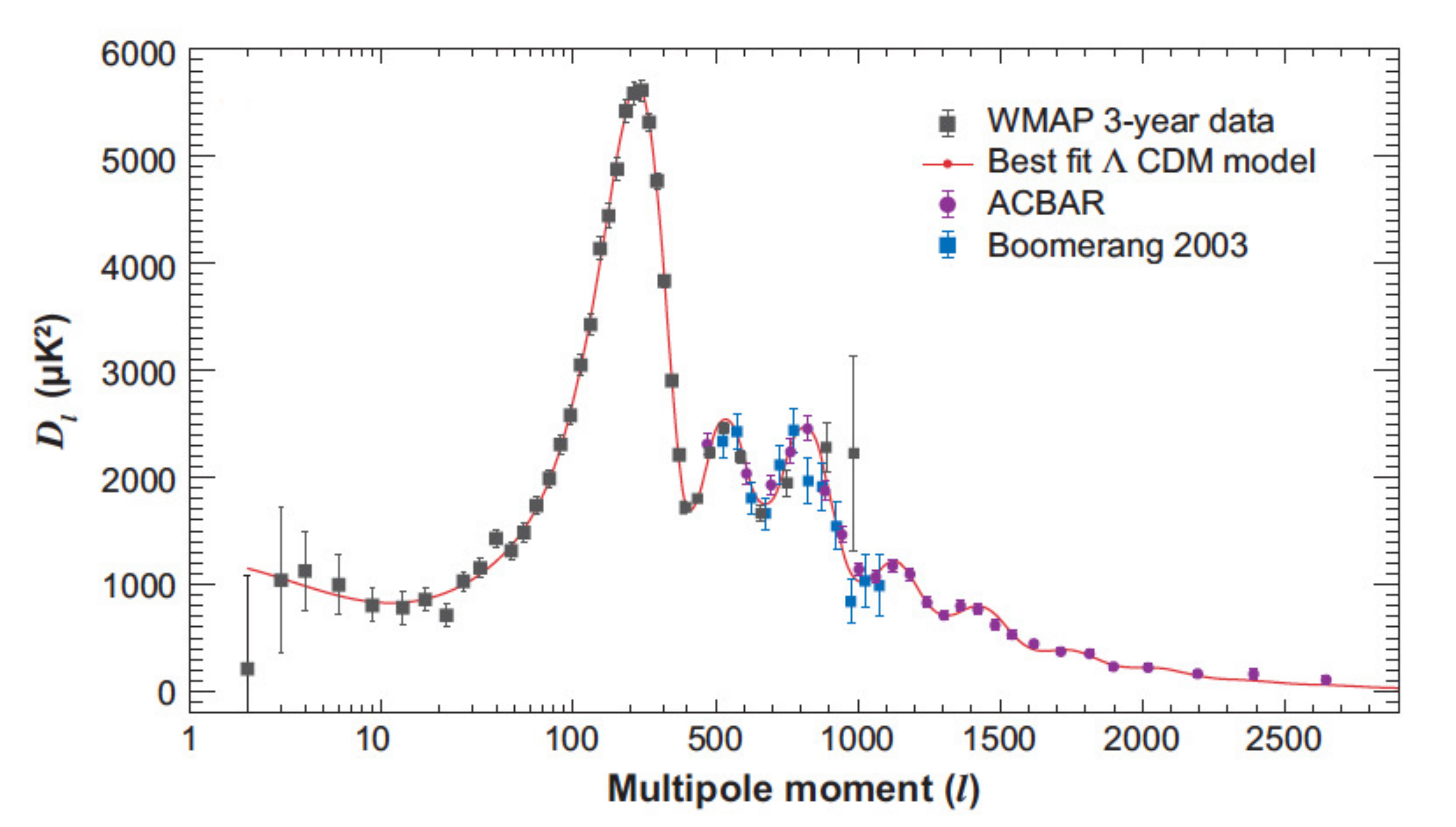}}
\caption[Angular power spectrum measurements] {\small Angular power
  spectrum measurements of the cosmic microwave background temperature
  fluctuations from the Wilkinson Microwave Anisotropy Probe (WMAP),
  Boomerang, and the Arcminute Cosmology Bolometer Array Receiver
  (ACBAR). Taken from \citet{Frieman2008}.}
\label{fig:AosCMB2}
\end{figure}

The measurement of the CMB black body spectrum was one of the most
important tests of the big bang cosmology. The CMB spectrum started being
studied by means of balloon and rocket borne observations and
finally the black body shape of the spectrum was settled in the 1990s by
observations with the FIRAS radiometer at the Cosmic Background
Explorer Satellite (COBE) \citep{Mather1990}, which also showed
that the departures from a pure blackbody were extremely small ($\delta E/ E
\leq 10^{-4}$) \citep{Fixsen1996}.%MP2505 ET

The CMB anisotropies provide a vision of the Universe when photons
decoupled from baryons and before structure developed, about
380000 years after the Big Bang. The angular power spectrum of the CMB
temperature anisotropies is dominated by acoustic peaks that arise
from gravity-driven sound waves in the photon-baryon fluid. The
position and amplitudes of the acoustic peaks indicate it  the
Universe is  spatially flat or not (see Figure \ref{fig:AosCMB}).
Furthermore, in combination with Large Scale Structure (LSS) or independent $H_{0}$
measurements, it shows that the matter contributes only about 25\% of the
critical energy density \citep{Hu2002}.  Clearly, a component of
missing energy is necessary to match both results, a fact which is fully
consistent with the dark energy being an explanation of the accelerated
expansion.%MP2505 ET

Measurements of the angular power spectrum of the CMB have been carried
out in the last ten years by many experiments \citep[e.g.][]{Jaffe2001,
  Pryke2002, Spergel2007, Reichardt2009}]. Figure \ref{fig:AosCMB2}
shows a combination of some recent results where the first acoustic
peak around $l = 200$ is clearly seen, which constrain the spatial curvature of
the universe to be very close to null. 

\begin{figure}[tb]
\centering
\resizebox{0.9\textwidth}{!}{\includegraphics{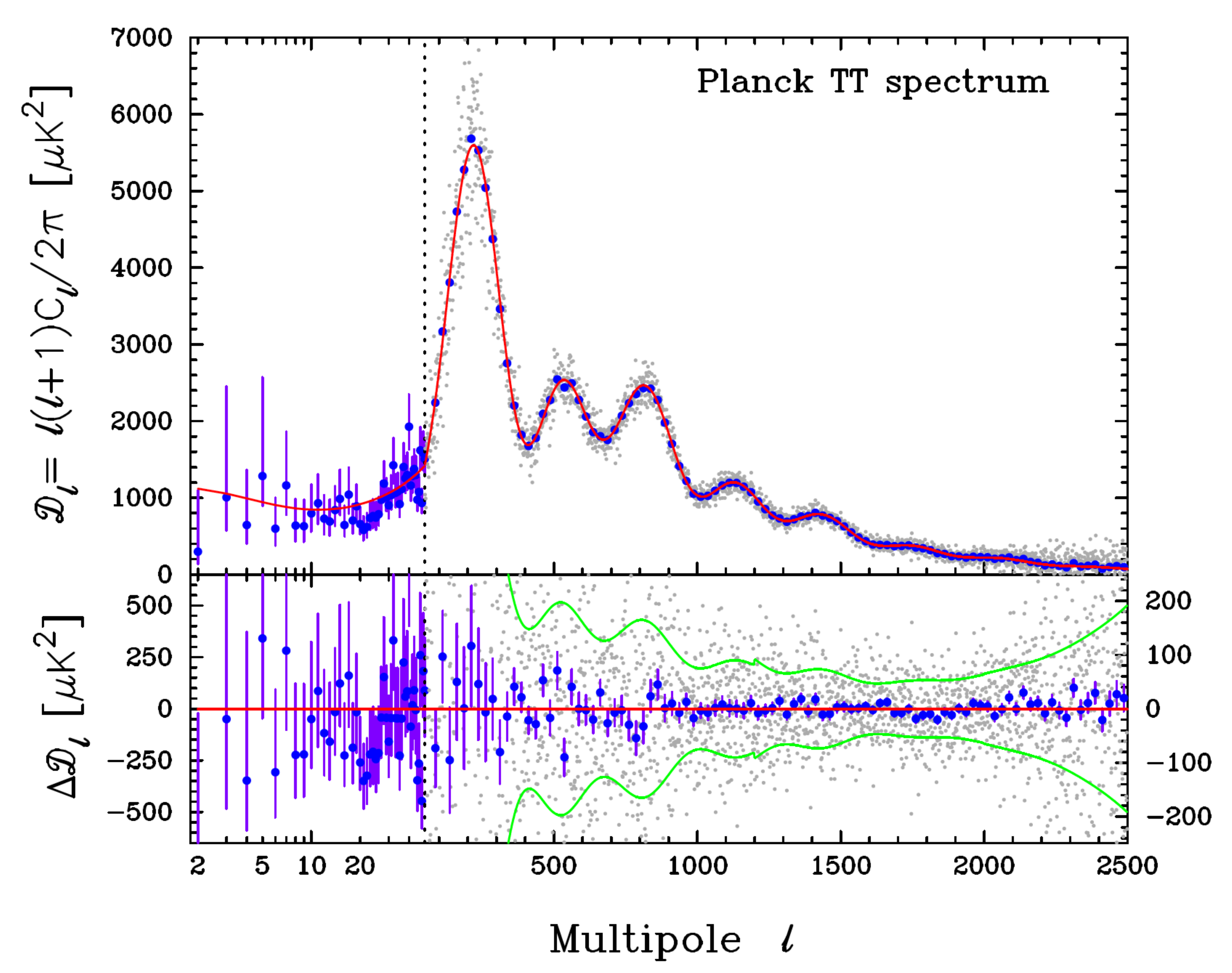}}
\caption[Angular power spectrum from Planck] {\small Angular power
  spectrum measurements of the cosmic microwave background temperature
  fluctuations form Planck. The power spectrum at low multipoles ($l = 2 - 49$) 
  is plotted in a logarithmic multipole scale. Taken from \citet{Ade2013}.}
\label{fig:Planck13}
\end{figure}

The most recent results from the Planck mission \citep{Ade2013} are shown in 
Figure \ref{fig:Planck13}. The Planck mission results are consistent with the standard
spatially-flat six-parameter $\Lambda$CDM cosmology but with a slightly lower value 
for $H_0$ and a higher value for $\Omega_m$ compared with the  SNe Ia results. 
%This discrepancy will be discussed further in section ??. 
When curvature is included, the Planck CMB data is consistent with a flat Universe to percent level precision.

%Figure ---- Angular power spectrum measurements

Although all these results are
consistent with an accelerating expansion of the universe, they alone
are not conclusive; other cosmological data, like the independent measurement of
the Hubble constant, are necessary in order to indicate the cosmic
acceleration. 

%Large-scale structure
\subsection{Large-scale structure}

%Figure ---- BAO peak in clustering of red galaxies 
\begin{figure}[tbp]
\centering
\resizebox{0.9\textwidth}{!}{\includegraphics{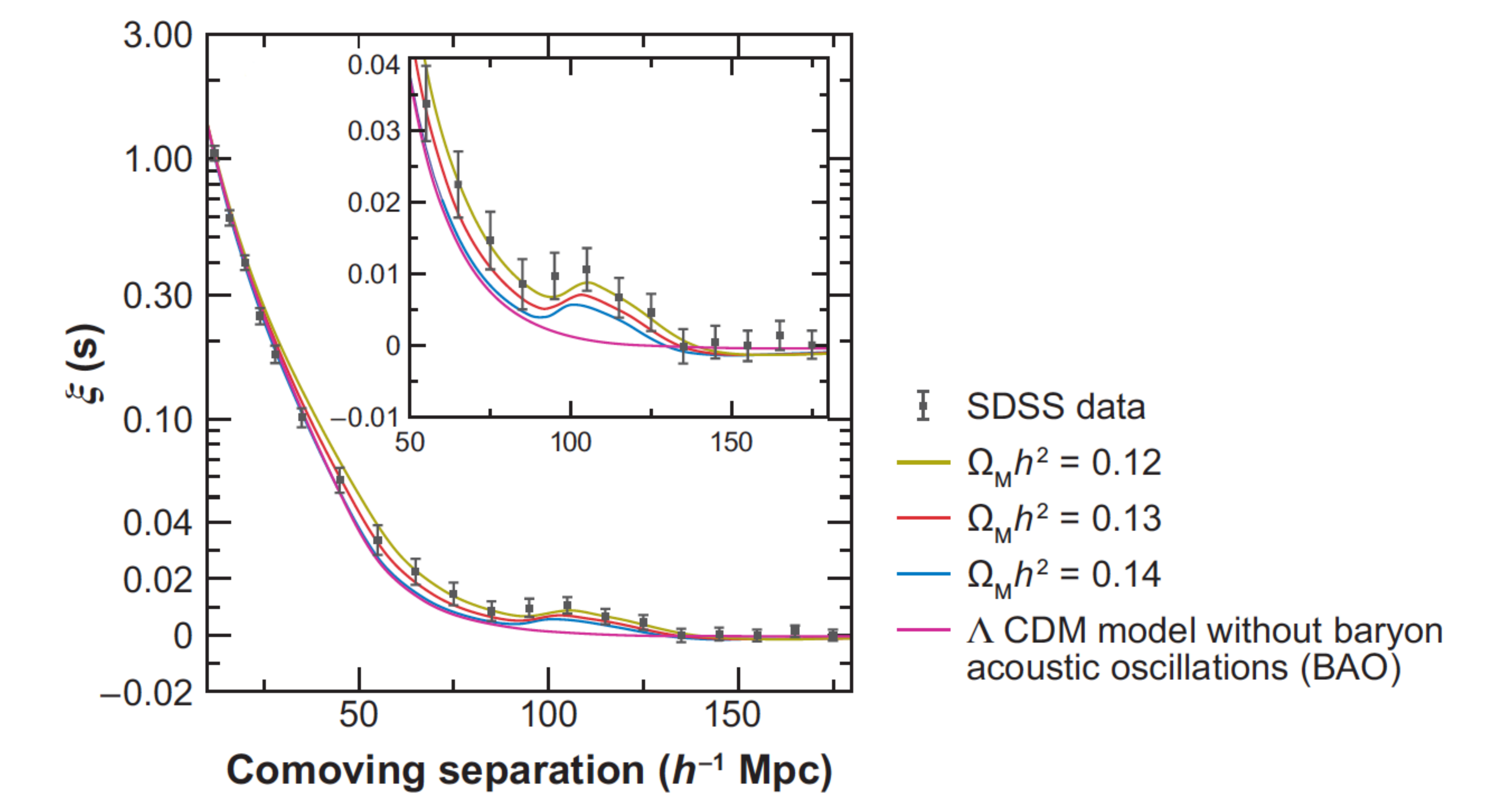}}
\caption[BAO peak in clustering of red galaxies] {\small Detection of
  the baryon acoustic peak in the clustering of luminous red galaxies
  in the Sloan Digital Sky Survey \citep{Eisenstein2005}. The
  two-point galaxy correlation function in redshift space is shown; the inset shows
  an expanded view with a linear vertical axis. Curves correspond to the
  $\rm{\Lambda CDM}$ predictions for $\Omega_{m}h^{2} = 0.12$ (dark
  yellow), $0.13$ (red), and $0.14$ (blue). The magenta curve shows a
  $\rm{\Lambda CDM}$ model without baryonic acoustic oscillations. Taken
  from \citet{Frieman2008}.}
\label{fig:Baop}
\end{figure}

The two-point correlation function of galaxies, as a measure of
distribution of galaxies on large scales, has long been used to
provide constrains on various cosmological parameters.
The measurement of the correlation
function of galaxies from the APM survey excluded, at that time,
the standard cold dark matter (CDM) picture \citep{Maddox1990} and
subsequently argued in favor of a model with a low
density CDM and possibly a cosmological constant \citep{Efstathiou1990}.%MP2505

The baryonic acoustic oscillations (BAO) leave a characteristic
signature in the clustering of galaxies, a bump in the two-point
correlation function at a scale $\sim 100 h^{-1}\ \rm{Mpc}$ that can
be measured today. Measurements of the BAO signature have been carried
out by \citet{Eisenstein2005} for luminous red galaxies of the Sloan
Digital Sky Survey (SDSS). They find results for the value of $\Omega_{m}
h^{2}$ and the acoustic peak at $100 h^{-1}\ \rm{Mpc}$ scale which are
consistent with the outcome of the CMB fluctuation analyses 
(see Figure \ref{fig:Baop}).

The recent work by \citet{Padmanabhan2012} reanalyses the \citet{Eisenstein2005}
sample using an updated algorithm to account for the effects of survey geometry as
well as redshift-space distortions finding similar results, while more recently \citet{Anderson2014}, using the clustering of galaxies from the  SDSS DR11 and in combination with the data from Planck find best fits
of $\Omega_mh^2 = 0.1418 \pm 0.0015$ and $\Omega_m = 0.311 \pm 0.009$.

%Current Supernova Results
\subsection{Current supernovae results}
\label{sec:SNeres}

%Figure ---- SNe Hubble diagram 
\begin{figure}[tb]
\centering
\resizebox{0.7\textwidth}{!}{\includegraphics{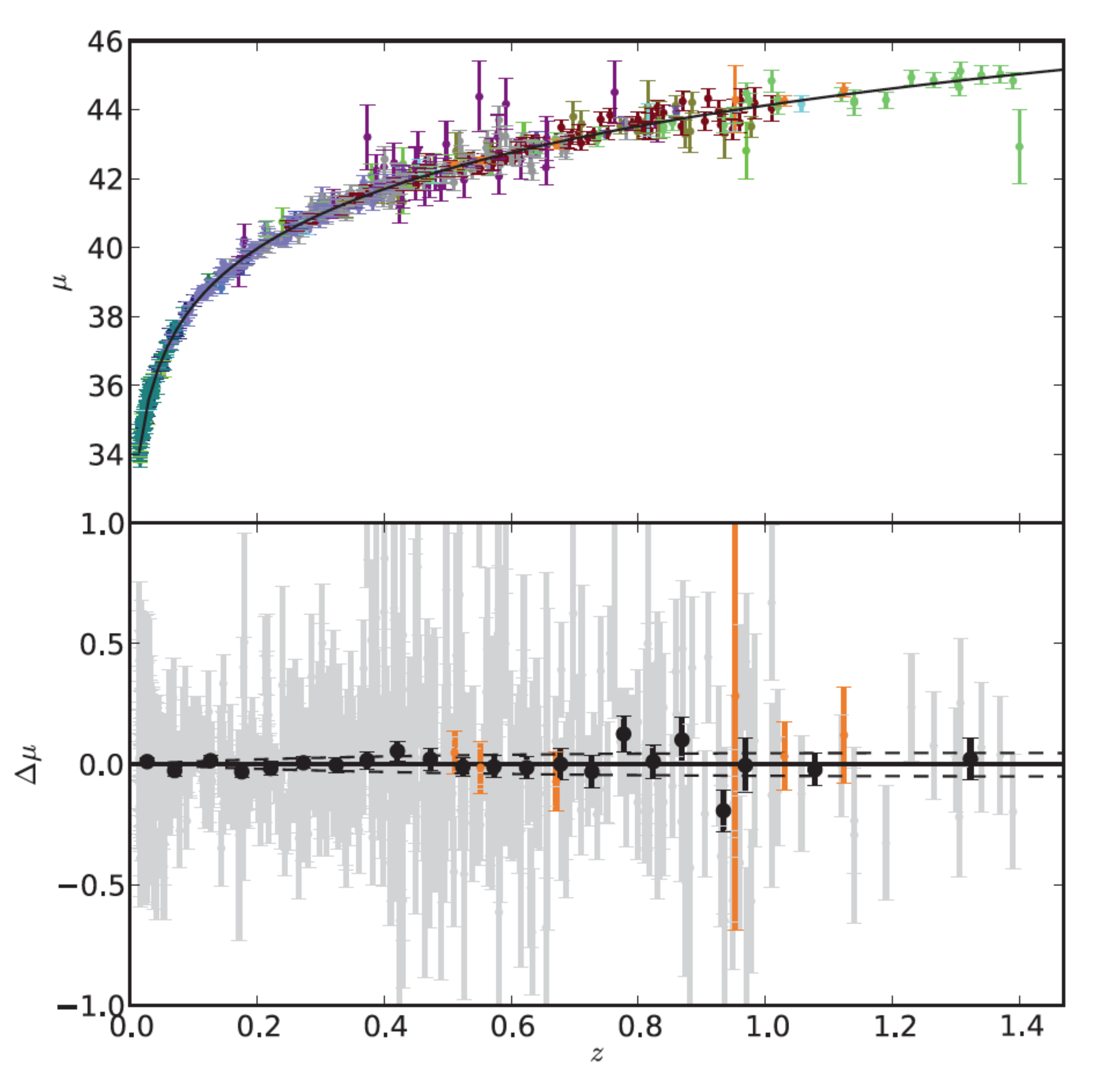}}
\caption[\emph{Union2} compilation Hubble diagram] {\small \emph{Upper panel: } Hubble
  diagram for the \emph{Union2} SNe Ia compilation. The solid line represents
  the best fitted cosmology for a flat Universe including CMB and BAO
  constraints. The different colours indicate the different data. 
  \emph{Lower panel:} Hubble diagram residuals where the
  best fitted cosmology has been subtracted from the light curve shape
  and color corrected peak magnitudes. The grey points show the
  residuals for individual SNa, while the black points show the binned
  values in redshifts bins of $0.05$ for $z < 1.0$ and $0.2$ for $z >
  1.0$. The dashed lines show the expected Hubble diagram residuals
  for cosmological models with $w \pm 0.1$ from the best fitted
  value. Taken from \citet{Amanullah2010}.}
\label{fig:SNeHd}
\end{figure}

%Figure ---- Constraints on cosmological parameters 
\begin{figure}[tb]
\centering
\resizebox{0.86\textwidth}{!}{\includegraphics{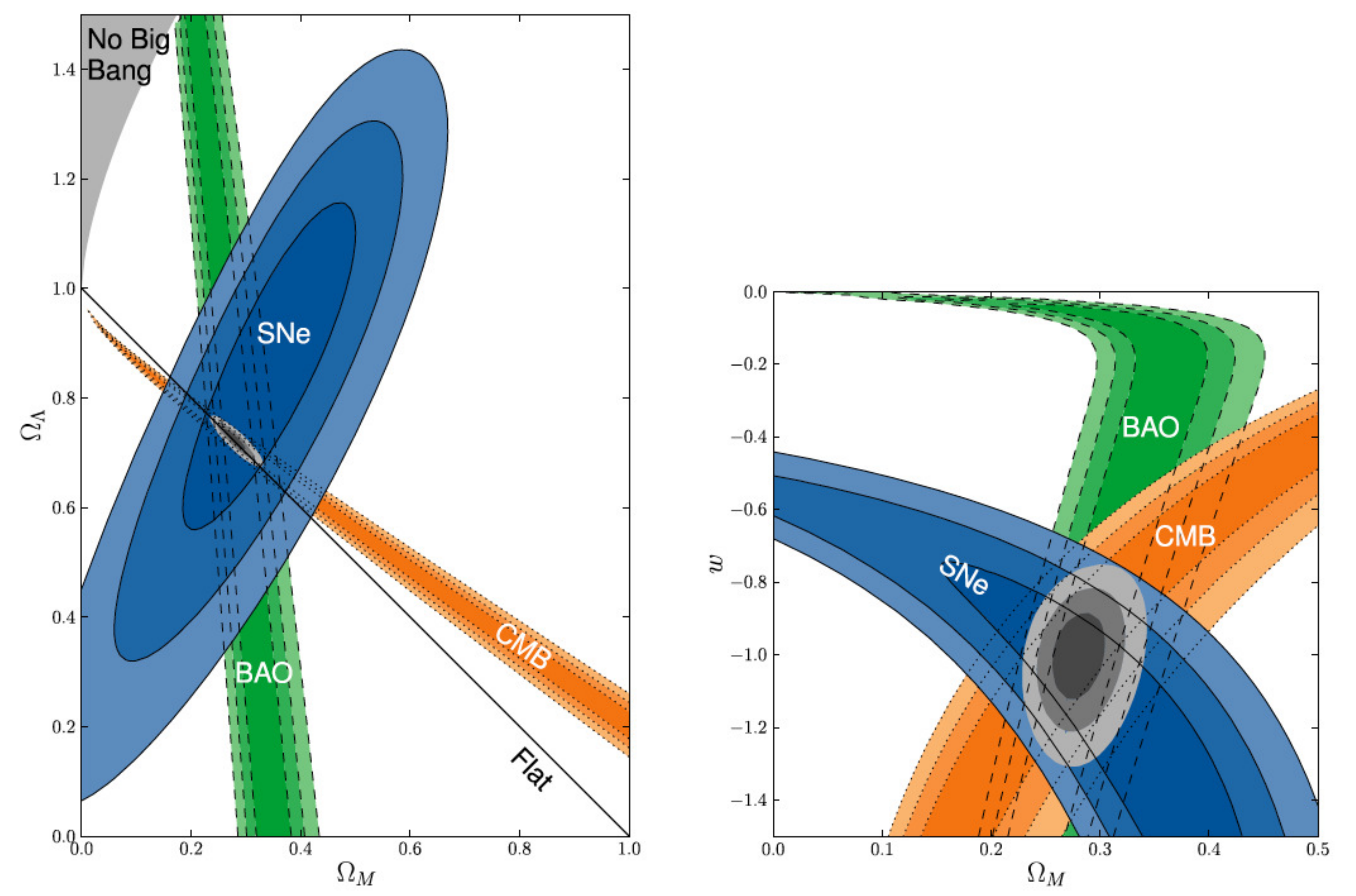}}
\caption[Constraints on cosmological parameters] {\small \emph{Left
    panel:} 68.3 \%, 95.4 \% and 99.7\% confidence regions in the
  ($\Omega_{m}, \Omega_{\Lambda}$) plane from SNe, BAO and CMB with
  systematic errors. Cosmological constant dark energy ($w = -1$) has
  been assumed. \emph{Right panel:} 68.3 \%, 95.4 \% and 99.7\%
  confidence regions in the ($\Omega_{m}, w$) plane from SNe, BAO and
  CMB with systematic errors. Zero curvature and constant $w$ has
  been assumed. Taken from \citet{Amanullah2010}.}
\label{fig:Constraints}
\end{figure}

After the first SNe Ia results were published, concerns were raised
about the possibility that intergalactic extinction or evolutionary
effects could be the cause of the observed distant supernovae dimming
\citep{Aguirre1999, Drell2000}. Since then a number of surveys have
been conducted which have strengthened the evidence for cosmic
acceleration. Observations with the Hubble Space
Telescope (HST), have provided high quality light curves
\citep{Riess2007}, and observations with ground based telescopes, have
permitted the construction of two large surveys,
 based on 4 meter class telescopes, the SNLS
(Supernova Legacy Survey) \citep{Astier2006} and the ESSENCE (Equation
of State: Supernovae Trace Cosmic Expansion) survey
\citep{Miknaitis2007} with spectroscopic follow ups
on larger telescopes.%MP2505 ET

The SNe Ia Hubble diagram has been constantly improved by the addition
of new data, from the above mentioned surveys, mostly at $z <
1.0$. \citet{Amanullah2010} have succeeded in analyzing the current
SNe Ia data (557 objects) homogeneously and have taken care of known systematics,
forming what has been named the \emph{Union2} compilation. Figure
\ref{fig:SNeHd} shows the Hubble diagram based on the \emph{Union2} dataset,
where the solid line represents the best fitted cosmology,
obtained from an iterative $\chi^{2}$-minimization procedure based on:
\begin{equation}
 \chi^{2} = \sum_{\rm{SNe}} {\frac{[\mu_{B}(\alpha, \beta, M_B) - \mu
     (z; \Omega_{m}, \Omega_{w}, w)]^{2}}{\sigma_{\rm{ext}}^{2} + \sigma_{\rm{sys}}^{2} +
\sigma_{\rm{lc}}^{2}}} \;,
\end{equation}
\nomenclature{$\chi^{2}$}{The Chi-square merit function. }%
%where $\Omega_{m}$ and $w$ have been fixed to $0.25$ and $-1$
%respectively and 
where $\sigma_{\rm{lc}}$ is the propagated error of the
covariance matrix of the light curve fit, whereas, $\sigma_{\rm{ext}}$
and $\sigma_{\rm{sys}}$ are the uncertainties associated with the
Galactic extinction correction, host galaxy peculiar velocity and
gravitational lensing, the former, and potential systematic errors the
later. The observed distance modulus is defined as $\mu_B = m_B^{\mathrm{corr}} - M_B$, where $M_B$ is the absolute $B$-band magnitude and $m_B^{\mathrm{corr}} = m_B^{\mathrm{max}} + \alpha x_1 - \beta c$; furthermore $m_B^{\mathrm{max}}$, $x_1$ and $c$ are parameters for each supernova that are weighted by the \emph{nuisance} parameters $\alpha$, $\beta$ and $M_B$ which are fitted simultaneously with the cosmological parameters ($z; \Omega_{m}, \Omega_{w}, w$) which give the model distance modulus $\mu$.  %ET

Combining the data from the three  probes that have been
considered up to now, it is possible to obtain stronger constraints over the
cosmological parameters (see Figure \ref{fig:Constraints}). 

%Table \ref{tab:CPFUnion2}  shows the fitted 
%cosmological parameters from using jointly the SNe Ia, BAO and CMB
%cosmological data. %ET

More recently, \citet{Suzuki2012} have added 23 SNe Ia (10 of which are beyond $z = 1$) to the \emph{Union2} compilation to form the \emph{Union2.1} dataset. Using this improved catalog of SNe Ia jointly with BAO and CMB data they obtain even better constraints to the values of the cosmological parameters as shown in Table \ref{tab:CPFUnion2.1}.
 
%Table --- Cosmological Parameters from fit Union2
\begin{table}[bt]
  \caption[Cosmological parameters fit from \emph{Union2.1}]
  {\small $\chi^2$ minimisation results of cosmological parameters $\Omega_{m}$, $w$ and
    $\Omega_{k}$ and their uncertainties.
% The values of the parameters are followed by the combination of
% statistical and systematic uncertainties. 
Adapted from \citet{Suzuki2012}.}
\begin{center}
\resizebox{0.8\textwidth}{!}{
\begin{tabular} { c c c c }
\hline
\hline
Fit & $\Omega_{m}$ & $\Omega_{k}$ & $w$\\
\hline
 SNe & $0.295^{+0.043}_{-0.040}$ & $0$ (fixed) & $-1$ (fixed)\\
 SNe + BAO + CMB & $0.286^{+0.018}_{-0.017}$ & $-0.004^{+0.006}_{-0.007}$ & $-1$ (fixed)\\
 SNe + BAO + CMB + $H_0$ & $0.272^{+0.015}_{-0.014}$ & $0.002^{+0.007}_{-0.007}$ & $-1.003^{+0.091}_{-0.095}$\\
\hline 
\end{tabular}}
\end{center}
\label{tab:CPFUnion2.1}
\end{table}

%THEORETICAL LANDSCAPE
\section{Theoretical Landscape}
\label{sec:2_3}
The cosmic accelerated expansion has deep consequences for our
understanding of the physical world. From the theoretical side many
plausible explanations have been proposed. The ``simplest'' one is the
traditional cosmological constant,
but as we will see, this solution presents serious theoretical
inconsistencies. To alleviate these problems various solutions have been proposed
which involve either the introduction of an exotic fluid, with negative pressure,
the dynamical consequences of which evolve with time 
(here we call them Dark Energy theories) or a modification of general
relativity.

%The cosmological constant (06/05/10)
\subsection{The cosmological constant}

The Cosmological Constant, $\Lambda$, was introduced by Einstein in
his field equations, in order to obtain a static solution. It is possible since the Einstein tensor, $G^{\mu \nu} = R^{\mu \nu} - 1/2 g^{\mu \nu}R$, 
satisfies the Bianchi identities $\nabla_{\nu}G^{\mu \nu} = 0$ and the energy momentum tensor, $T^{\mu \nu}$, satisfies energy conservation $\nabla_{\nu}T^{\mu \nu} = 0$; furthermore the metric, $g^{\mu \nu}$, is invariant to covariant derivatives $\nabla_{\alpha} g^{\mu \nu} = 0$; then there is freedom to add a constant term to the GR equations: 
\begin{equation}
 R_{\mu \nu} - \frac{1}{2}  g_{\mu \nu} R+ \Lambda g_{\mu \nu} = 8 \pi G T_{\mu \nu}\;,
 \label{GRFE}
\end{equation}
\nomenclature{$G_{\mu \nu}$}{The Einstein tensor. }%
\nomenclature{$R_{\mu \nu}$}{The Ricci tensor. }%
\nomenclature{$R$}{The Ricci scalar. }%
\nomenclature{$g_{\mu \nu}$}{The metric. }%
from which we can obtain equations (\ref{eqF2}) and (\ref{eqF1}).
%\begin{eqnarray*}
% \left(\frac{\dot{a}}{a}\right)^2 &=& \frac{8 \pi G \rho}{3} - \frac{k}{a^2} + \frac{\Lambda}{3}\;, \\
% \frac{\ddot{a}}{a} &=& -\frac{4 \pi G}{3} (\rho + 3 p ) + \frac{\Lambda}{3}\;.
%\end{eqnarray*}
Form eq. (\ref{eqF2}) we can see that:
\begin{equation}
 \rho_{\Lambda} = \frac{\Lambda}{8 \pi G}\;,
\end{equation}
and combining the above with eq.(\ref{eqF1}), we can see that
$p_{\Lambda} = - \rho_{\Lambda}$. As an approximation, in the case
in which the energy density of the cosmological constant dominates the
dynamics of the Universe, and neglecting the matter component, we have that:
\begin{align*}
  \frac{\ddot{a}}{a} &= - \frac{4 \pi G}{3} ( \rho_{\Lambda} + 3 p_{\Lambda})\\
  &= \frac{8 \pi G}{3} \rho_{\Lambda}\;.
 \end{align*} 
 From this rough argument, it becomes evident how the cosmological constant
 explains the phenomenology of the accelerated cosmic expansion, since
 it is clear that we have $\ddot{a} \propto \rho_{\Lambda} a$.%MP2505

 From the previous argument we see that for a cosmological
 constant we have $w = -1$. It is interesting to note that 
 the current high-quality cosmological data strongly suggest
 that the mechanism behind the cosmic acceleration behaves exactly as a cosmological constant. However, we will show that the $\Lambda$-based
 explanation of the accelerating universe presents 
 serious theoretical inconsistencies.% that in any case must be solved. ET

 From the point of view of modern field theories, the cosmological
 constant can be explained as the energy of the vacuum. The possible sources
 for the vacuum energy are basically of two kinds: a bare cosmological
 constant in the general relativity action or the energy density of
 the quantum vacuum.%MP2505 ET

\subsubsection{The cosmological constant problem} 
In this subsection we introduce the cosmological constant (cc) problem or the fine
tuning problem that has a long history \citep{Weinberg1989}, the discussion is 
somewhat standard \citep{Carroll2001} and we roughly follow the work of \citet{Sola2013}.

A bare cosmological constant ($ \Lambda _{0}$) can be
added in the Einstein-Hilbert (EH) action:

          \begin{equation}
            S_{EH} = \frac{-1}{16 \pi G} \int{d^{4}x \sqrt{-g} (R + 2 \Lambda_{0})} =   - \int{d^{4}x \sqrt{-g} \left(\frac{1}{16 \pi G}R + \rho_{\Lambda_{0}} \right)}\;.
          \end{equation}
          \nomenclature{$S$}{The action. }%
          In fact this is the most general covariant action that we
          can construct from the metric and its first and second
          derivatives; we obtain eq.(\ref{GRFE}) varying this action with
          the addition of matter terms. 
          
In the most simple case, the matter sector can be given by a single scalar field ($\phi$).
In order to trigger  Spontaneous Symmetry Breaking (SSB) and then preserve the 
gauge symmetry of the field, we must have a potential of the form:
\begin{equation}
	V(\phi) = \frac{1}{2} m^2\phi^2 + \frac{1}{4!}\lambda \phi^4\ (\lambda > 0)\;,
\end{equation}
where when $m^2 > 0$  we have a single vacuum state and $m$ plays the role of a mass for the free field, whereas when $m^2 < 0 $  we have two degenerate vacuum states, this situation is characteristic of a phase transition. 

The action for the gravitational system including $\phi$ can be given as:
\begin{equation}
	S = S_{EH} +  \int{d^{4}x \sqrt{\abs{g}} \left[\frac{1}{2}g^{\mu \nu}\partial_{\mu} \phi \partial_{\nu} \phi - V(\phi) \right]}\;.
\end{equation}          
If we transfer the bare cc term to the matter sector then the matter action is given by,
\begin{equation}
	S[\phi]  =  \int{d^{4}x \sqrt{\abs{g}} \left[\frac{1}{2}g^{\mu \nu}\partial_{\mu} \phi \partial_{\nu} \phi - V(\phi) - \rho_{\Lambda_{0}} \right]} = \int{d^{4}x \sqrt{\abs{g}} \mathcal{L}_{\phi}}\ \;.
\end{equation}          
Calculating the energy-momentum tensor for the matter Lagrangian, as defined above, we obtain,
\begin{equation}
	\tilde{T}_{\mu \nu}^{\phi} = g_{\mu \nu} \rho_{\Lambda_0} + T_{\mu \nu}^{\phi}\;,
	\label{eqEMTf}
\end{equation}
where $T_{\mu \nu}^{\phi}$ is the scalar field energy-momentum tensor given by
\begin{equation}
	T_{\mu \nu}^{\phi} = \left[\partial_{\mu} \phi \partial_{\nu} \phi - \frac{1}{2} g_{\mu \nu} \partial_{\sigma} \phi \partial^{\sigma} \phi \right] + g_{\mu \nu} V(\phi) \;.
	\label{eqEMTph}
\end{equation}    

The vacuum expectation value for the `total' energy-momentum tensor as given in eq. (\ref{eqEMTf}) is,
\begin{equation}
	\langle \tilde{T}_{\mu \nu}^{\phi} \rangle  = g_{\mu \nu} (\rho_{\Lambda_0} + \langle V(\phi) \rangle ) \;,
	\label{eqTvev}
\end{equation} 
where we note that the kinematical term in eq.(\ref{eqEMTph}) does not play any role. 

As said above, SBB is present when $m^2 < 0$ and then the field vacuum expected value ($vev$) is not trivial and is given by, 
\begin{equation}
	\langle \phi \rangle = \sqrt{\frac{-6m^2}{\lambda}}\;,
\end{equation}
and then the $vev$ for $V(\phi)$ is given by,
\begin{equation}
	\rho_{\Lambda_i} = \langle V(\phi) \rangle = \frac{-3m^4}{2 \lambda} = -\frac{1}{8} M_{H}^2 \langle \phi \rangle^2 = \frac{-1}{8\sqrt{2}}M^2_{H}M^2_{F} \;,
	\label{eqLin}
\end{equation}
where we have introduced $M_{H}$, the physical mass of the Higgs boson, since this is just the process that happens (at the classical level) in the electroweak phase transition generated by the Higgs potential. The value of $M_{H}$ is given by, 
\begin{equation}
	M_{H}^2 = \frac{\partial^2 V(\phi)}{\partial \phi^2} \bigg|_{\phi = \langle \phi \rangle} = -2m^2 > 0 \;.
\end{equation}
Above, we also introduced the Fermi scale, $M_{F} = G^{-1/2}_F \simeq 293\ \mathrm{GeV}$. The value of $G_F$ is given by
\begin{equation}
	\frac{G_F}{\sqrt{2}} = \frac{g^2}{8M^2_{W}} = \frac{1}{2\langle \phi \rangle}\;,
\end{equation}  
where $g$ is the weak gauge coupling and $M_W$ is the mass of the $W^{\pm}$ gauge boson. Then, we have a direct measure of the Higgs $vev$ given as:
\begin{equation}
	\langle \phi \rangle = 2^{-1/4} G^{-1/2}_F \simeq  246\ \mathrm{GeV}.
	\label{eqGFv}
\end{equation}

From eq.(\ref{eqTvev}) it is clear that the $vev$ for $V(\phi)$ plays the role of an \emph{induced} vacuum energy, hence we have identified it in this way in eq.(\ref{eqLin}). At this point the physical value of the cc is given by,
\begin{equation}
	\rho_{\Lambda} = \rho_{\Lambda_0} + \rho_{\Lambda_i}\;.
\end{equation}
At this stage, we already can compare the above calculations with observations, combining the results in eq.(\ref{eqGFv}) and the recently measured value for $M_H \simeq 125\ \mathrm{GeV}$
\citep{Aad2012}, we obtain a value for $\rho_{\Lambda_i} \simeq -1.2 \times 10^8\ \mathrm{GeV^4}$. The observed value of the cc is given by $\rho_{\Lambda}^o  \sim 10^{-47}\ \mathrm{GeV^4}$, thus it is clear that, 
\begin{equation}
	\bigg| \frac{\rho_{\Lambda_i}}{\rho_{\Lambda}^o} \bigg| = \mathcal{O}(10^{55}) \;.
\end{equation}
The last result implies that we must choose the value of $\rho_{\Lambda_0}$ with a precision of 55 orders of magnitude in order to reconcile the above two results, which is clearly a severe fine-tuning problem. 

The importance of the above result lies in the fact that the mass of the Higgs boson has been already measured and then at least in this, the simplest of cases, the reality of the vacuum energy density and hence the cc problem seems  unavoidable.  

In the most general case, and discussing the problem in a simplified way, the energy density of the quantum vacuum arises
          from the fact that for each mode of the quantum field there
          is a zero-point energy $\hbar \omega /2$. Formally the total
          energy would be infinite unless we discard the very high
          momentum modes on the ground that we trust the theory only
          to a certain ultraviolet momentum cutoff $k_{max}$, then we
          have
\begin{equation}
 \rho_{\Lambda} = \frac{1}{2} \sum_{fields}{g_{i} 
\int_{0}^{\infty}{\frac{d^{3}k}{(2 \pi)^{3}}\sqrt{k^{2} + m^{2}} }} 
\simeq \sum_{fields}{\frac{g_{i} k^{4}_{max}}{16 \pi^{3}}},
\end{equation}
where $g_{i}$ accounts for the degrees of freedom of the field (its
sign is $+$ for bosons and $-$ for fermions). From the last equation
we can see that $\rho_{\Lambda} \sim k^{4}_{max}$ , then imposing as a
cutoff the energies where the known symmetry breaks, we have, in addition to the electroweak symmetry breaking discussed above, that:

\begin{itemize}
\item The potential arising from the breaking of chiral symmetry is due
  to the nonzero expectation value of the quark bilinear $q \bar{q}$
  with a potential $M_{QCD} \sim 0.3\ \rm{GeV}$ and then its
  contribution to the vacuum energy is $\rho_{\Lambda}^{QCD} \sim
  (0.3\ \rm{GeV})^4 \sim 1.6 \times 10^{36}\ \rm{erg/cm^3}$.
\item For the Planck scale transition we have a potential $M_{Pl} = (8
  \pi G)^{-1/2} \sim 10^{18}\ \rm{GeV}$ and then its contribution to
  the vacuum energy is $\rho_{\Lambda}^{Pl} \sim (10^{18}\ \rm{GeV})^4
  \sim 2 \times 10^{110}\ \rm{erg/cm^3}$.
\end{itemize}

Then, the observed value of the vacuum energy density is $10^{55} -10^{120}$ times smaller than any theoretical prediction.

%Dark Energy Theories 
\subsection{Dark energy theories}
Due to the extreme fine tuning problem of the cc, several alternatives for the observed accelerated cosmic expansion have been proposed, a class of them postulates one or more 
dynamical fields with  an effective value for the equation of state parameter, $w$, either different from $-1$ or
changing with the redshift, in general they are called dark energy models. Over the years many different  such models have been proposed, for a recent review see \citet{Copeland2006}.

In the dark energy approach the vacuum energy, arising from the ground states of the quantum fields, has a value exactly equal to zero due to e.g. some renormalization procedure. Then the cc problem does not arise at all. 

The simplest dark energy proposal is a scalar field, in general this kind of models have been named \emph{quintessence}. The action for this model is given by
\begin{equation}
 S = \int{d^{4}x \sqrt{-g} \left(\frac{R}{16 \pi G} + 
\mathcal{L}_{SM} + \mathcal{L}_{Q} \right)}\;,
\end{equation}
 \nomenclature{$\mathcal{L}$}{The Lagrangian density or the likelihood estimator. }%
where $R$ is the Ricci scalar, $g$ is the determinant of the metric,
$\mathcal{L}_{SM}$ is the Lagrangian for Standard Model particles and
the quintessence Lagrangian is given by
\begin{equation}
 \mathcal{L}_{Q} = -\frac{1}{2} (\nabla_{\mu} Q) (\nabla^{\mu} Q) - V(Q)\;.
\end{equation}
The field obeys the Klein-Gordon equation:
\begin{equation}
 \Box Q = V_{, Q};
\end{equation}
 \nomenclature{$\Box$}{The D'Alembert operator. }%
and its stress-energy tensor is given by
\begin{equation}
 T_{\mu \nu} = (\nabla_{\mu} Q) (\nabla_{\nu} Q) + g_{\mu \nu} \mathcal{L}_{Q}\;,
\end{equation}
 \nomenclature{$T_{\mu \nu}$}{The stress-energy tensor. }%

%The scalar field obeys the equation of motion:
%\begin{equation}
% \ddot{Q} + 3H\dot{Q} + V_{,Q} = 0
%\end{equation}
with energy density and pressure given by:
\begin{align}
 \rho_{Q} = \frac{1}{2} \dot{Q}^{2} + V(Q), & &    p_{Q} = \frac{1}{2} \dot{Q}^{2} - V(Q)\;.
\end{align}
Then its equation of state parameter, $w=p/\rho$, is given by:
\begin{equation}
 w = \frac{\dot{Q}^{2} / 2 - V(Q)}{\dot{Q}^{2} /2 + V(Q)} = 
\frac{-1 + \dot{Q}^{2} / 2V}{1 + \dot{Q}^{2} / 2V}\;,
\end{equation}
from which it is obvious that if the evolution of the field is slow, we have
$\dot{Q}^{2} / 2V \ll 1$, and the field behaves like a slowly varying
vacuum energy, with $w<0$, $\rho_{Q}(t) \propto V[Q(t)]$ and
$p_{Q}(t) \propto -V[Q(t)]$.

%Modified Gravity Theories
\subsection{Modified gravity theories} 
As it was mentioned earlier, an alternative explanation of the cosmic
acceleration is through a modification to the laws of gravity. This
implies a modification to the geometry side of the GR field equations,
instead of the modification of the stress-energy tensor. Many ideas
have been explored in this direction, some of them based on models
motivated by higher-dimensional theories and string theory
\citep[e.g.][]{Dvali2000, Deffayet2001} and others as phenomenological
modifications to the Einstein-Hilbert action of GR \citep[e.g.][]{Carroll2004,
  Song2007}.
  
%\subsection{General Relativity as an effective field theory}
%An interesting approach to the cc problem is through considering GR as an effective field theory valid to low energies but needing corrections at higher energies or shorter distances. In general we accept that GR must need some corrections at that energy scale in order to be compatible with quantum mechanics, hence this is an appealing proposal. 

%PROBES OF COSMIC ACCELERATION
\section{Probes of Cosmic Acceleration}

The accelerated expansion of the Universe appears to be a well
established fact, while the
dark energy density has been determined apparently to a precision of a few
percent. However, measuring its equation of state parameter and
determining if it is time-varying is a significantly more difficult
task. The primary consequence of dark energy is its effect 
on the expansion rate of the universe and thus
on the redshift-distance
relation and on the growth-rate of cosmic structures. 
Therefore, we have basically two
kinds of probes for dark energy, one geometrical and the other one based
on the rate of growth of density perturbations.%ET

The {\em Growth} probes are related to the rate of growth of
matter density perturbations, a typical example being 
the spatial clustering of extragalactic sources and its evolution \citep[e.g.][]{Pouri2014}.
The {\em Geometrical} probes are related directly to the metric, a 
typical example being the
redshift-distance relation as traced by SNe Ia \citep[e.g.][]{Suzuki2012}.

In general, in order to use the latter probes, based on any kind of tracers,
one has to measure the redshift which is relatively straightforward,
but also the tracer distance, which in general is quite difficult. In
Appendix \ref{ApC} we review the cosmic distance ladder which allows
the determination of distances to remote sources.

\subsection{Type Ia supernovae}
Type Ia Supernovae have been used as geometrical probes,
they are standard candles \citep{Leibundgut2001}, which through
their determination of the Hubble function have provided 
constrains of cosmological parameters through eq.(\ref{eqDM0}). 
Up to date they are the
most effective, and better understood,  probe of the cosmic acceleration
\citep{Frieman2008}.

The standardisation of SNe Ia became possible after the work of
\citet{Phillips1993} where an empirical correlation was established 
between their peak
brightness and the luminosity decline rate, after peak luminosity 
(in the sense that more luminous SNe Ia decline more slowly). 

The main systematics in the distance determination 
derived from SNe Ia, are uncertainties
in host galaxy extinction correction and  in the SNe Ia
intrinsic colours, luminosity evolution and selection bias in the low
redshift samples \citep{Frieman2008}. The extinction correction is
particularly difficult since having the combination of photometric
errors, variation in intrinsic colours and host galaxy dust properties,
causes distance uncertainties even when using multiband
observations. However, a promising solution to this problem is based
on near infrared observations, where the extinction effects are
significantly reduced.

\citet{Frieman2003} estimated that in order to obtain precise
measurements of $w_{0}$ and $w_{1}$, accounting for SNe Ia systematics,
requires $\sim 3000$ light curves out to $z \sim 1.5$, measured
with great precision and careful control of the systematics.%MP ET

\subsection{Galaxy clusters} 

The utility of galaxy clusters as cosmological probes relies in many
aspects, among which is the determination of their mass to light
ratio, where its comparison with the corresponding cosmic ratio 
can provide the value of $\Omega_m$ 
\citep[e.g.][]{Andernach2005}, the cluster masses
can be also used to derive the cluster mass function to be compared with the
analytic (Press-Schechter) or numerical (N-body simulations) model
expectations 
%of different models for later comparison with predictions from
%N-body simulations, based in a cosmological model, and which are
%capable of accurate number density of cluster sized dark matter halos
%as function of redshift and halo mass predictions 
\citep{Basilakos2009, Haiman2001, Warren2006}. 
The determination of the cluster mass can be done by means of the relation
between mass and other observable, such as X-ray luminosity or
temperature, cluster galaxy richness, Sunyaev-Zel'dovich effect (SZE)
flux decrement or weak lensing shear, etc \citep{Frieman2008}. %MP ET

\citet{Frieman2008} give the redshift distribution of clusters
selected according to some observable $O$, with selection function
$f(O, z)$ as
\begin{equation}
 \frac{d^{2} N(z)}{dz d \Omega} = 
\frac{r^{2}(z)}{H(z)} \int_{0}^{\infty} {f(O, z) dO} \int_{0}^{\infty} 
{p(O | M, z) \frac{dn(z)}{dM}dM}\;,
\end{equation}
where $dn(z)/dM$ is the space density of dark halos in comoving
coordinates and $p(O | M,z)$ is the mass-observable relation, the
probability that a halo of mass $M$, at redshift $z$, is observed as a
cluster with observable property $O$. We can see that this last
equation depends on the cosmological parameters through the comoving
volume element (see equation (\ref{eqCVE})) and the term $dn(z)/dM$
which depends on the evolution of density perturbations.
   
\subsection{Baryon acoustic oscillations}

Gravity drives acoustic oscillations of the coupled photon-baryon
fluid in the early universe. The scale of the oscillations is given by
\begin{equation}
 s = \int_{0}^{t_{rec}}{c_{s} (1+z) dt} = \int_{z_{rec}}^{\infty}{\frac{c_{s}}{H(z)} dz},
\end{equation}
 \nomenclature{$s$}{The scale of acoustic oscillations. }%
 \nomenclature{$c_{s}$}{The sound speed. }%
where $c_{s}$ is the sound speed which is determined by the ratio of
the baryon and photon energy densities, whereas $t_{rec}$ and $z_{rec}$ are the time and redshift when recombination occurred. These acoustic oscillations
leave their imprint on the CMB temperature anisotropy angular power
spectrum but also in the baryon mass-density distribution. From the WMAP measurements we
have $s = 147 \pm 2\ \mathrm{Mpc}$. Since the oscillations scale $s$
provides a standard ruler that can be calibrated by the CMB anisotropies,
then measurements of the BAO scale in the galaxy distribution provides
a geometrical probe for cosmic acceleration \citep{Frieman2008}.

The systematics that could affect the BAO measurements are related to
nonlinear gravitational evolution effects, scale-dependent differences
between the clustering of galaxies and of dark matter (the so-called bias) and
redshift-space distortions of the clustering, which can shift the BAO
features \citep{Frieman2008}.

\subsection{Weak gravitational lensing}

The images of distant galaxies are distorted by the gravitational
potential of foreground collapsed structures, intervening in the line of sight of
the distant galaxies. This distortion can be
used to measure the distribution of dark matter of the intervening
structures and its evolution with
time, hence it provides a probe for the effects of the accelerated
expansion on the growth of structure \citep{Frieman2008}.%MP ET

The gravitational lensing produced by the large scale structure (LSS) can be analysed
statistically by locally averaging the shapes of large numbers of
distant galaxies, thus obtaining the so called cosmic shear field at any
point. The angular power spectrum of shear is a statistical measure of
the power spectrum of density perturbations, and is given by \citep{Hu2004}:
\begin{equation}
 P^{\gamma}_{l} (z_{s}) = \int_{0}^{z_{s}}{dz
   \frac{H(z)}{D^{2}_{A}(z)} 
|W (z, z_{s})|^{2} P_{\rho} \left(k = \frac{l}{D_{A}(z)} ; z \right)},
\end{equation}
where $l$ is the angular multipole of the spherical harmonic expansion, $W(z, z_{s})$ is the lensing efficiency
of a population of source galaxies and it is determined by the
distance distributions of the source and lens galaxies, and
$P_{\rho}(k , z)$ is the power spectrum of density perturbations.

Some systematics that could affect weak lensing measurements are, obviously,
incorrect shear estimates, uncertainties in the galaxy photometric
redshift estimates (which are commonly used), intrinsic correlations
of galaxy shapes and
theoretical uncertainties in the mass power spectrum on small scales
\citep{Frieman2008}.%MP ET

\subsection{H \textsc{ii} galaxies}
\hii\ galaxies are dwarf galaxies with a strong burst of star formation
which dominates the luminosity of the host galaxy and allows it to be seen
at very large distances. The $L(\mathrm{H}\beta) - \sigma$ relation of
\hii\ galaxies allows distance modulus determination for these objects
and therefore the construction of the Hubble diagrams. Hence, \hii\
galaxies can be used as geometrical probes of the cosmic acceleration.

Previous analyses \citep{Terlevich1981, Melnick1987}, have shown
that the \hii\ galaxy oxygen abundance affects systematically
its $L(\mathrm{H}\beta) - \sigma$ relation. The distance
indicator proposed by the authors takes into account such effects
\citep{Melnick1988}, and was defined as:
\begin{equation}
 M_{z} = \frac{\sigma^{5}}{\mathrm{O/H}}\;,
\end{equation}
  \nomenclature{$M_{z}$}{The \hii\ galaxies distance indicator. }%
where $\sigma$ is the galaxy velocity dispersion and $\mathrm{O/H}$ is
the oxygen abundance relative to hydrogen. From this distance
indicator, the distance modulus can be calculated as:
\citep{Melnick2000}
\begin{equation}
   \mu = 2.5 \log_{10} \frac{\sigma^{5}}{F(\rm{H}\beta)} - 
2.5 \log_{10} (\rm{O/H}) - A_{\rm{H}\beta} - 26.44,
\end{equation}
  \nomenclature{$\mu$}{The distance modulus. }%
where $F(\rm{H}\beta)$ is the observed $\rm{H}\beta$ flux and
$A_{\rm{H}\beta}$ is the total extinction in $H\beta$.

Some possible systematics that could affect the $L(\mathrm{H}\beta) -
\sigma$ relation, are related to the reddening, the 
age of the stellar burst, as well as the local environment
and morphology.

Through the next chapter we will explore carefully the use of \hii\
galaxies as tracers of the Hubble function and the systematics that
could arise when calibrating the $L(\mathrm{H}\beta) - \sigma$
relation for these objects.

\section{Summary}
The observational evidence for the Universe accelerated expansion is now overwhelming. The best to date data from SNe Ia, BAOs, CMB and many other tracers, all accord that we are living during an epoch in which the evolution of the Universe is dominated  by some sort of dark energy.

Many different models have been proposed to explain the observed dark energy. The cosmological constant is a good candidate in the sense that all current observations are consistent with it, although suffers from severe fine tuning and coincidence problems that have given place to the proposal of dynamical vacuum energy models. 

In this work we will explore an alternative probe to trace the expansion history of the Universe. \hii\ galaxies are a promising new way to explore the nature of dark energy since they can be observed to larger redshifts that many of the currently  best known cosmological probes. 

%% file: ch03/ch03.tex
%%ch03.tex----\hii\ GALAXIES PROPERTIES
\chapter {\hii\ Galaxies}

\epigraph{\emph{\hspace{5cm} Pauca sed matura.}}{--- \textup{C.F. Gauss}, Motto}

\lettrine{I}{n the} search for white dwarfs, \citet{Humason1947}, using the 18 inch Schmidt telescope at Palomar, developed the technique of using multiply exposed large scale plates, each exposure covering a distinct region of the optical spectrum with the intention of identifying the target objects from the relative intensities in the different plates.

\citet{Haro1956}, while searching for emission line galaxies, using a variation of the technique pioneered by Humason and Zwicky (using an objective prism), discovered some compact galaxies with strong emission lines. Some years later \citet{Sargent1970} found in what was to become the \citet{Zwicky1971} catalogue, some compact galaxies whose spectra were very similar to those of giant \hii\ regions in spiral galaxies. They called them \emph{isolated extragalactic} \hii\ \emph{regions}. After analysing their spectra they conclude that the galaxies are ionised by massive clusters of OB stars \citep{Searle1972, Bergeron1977} and are metal poor systems \citep{Searle1972, Lequeux1979, French1980, Kunth1983}.

Since \hii\ galaxies were easily recognised in objective prism plates, due to their strong narrow emission lines, many were discovered by objective prism surveys during the following years \citep{Markarian1967, Smith1976, MacAlpine1977, Markaryan1981}. 

\citet{Terlevich1981} and \citet{Melnick1988} analysed the dynamical properties of \hii\ galaxies and proposed their usefulness as distance indicators; the data used for their analysis was published subsequently as a spectrophotometric catalogue \citep{Terlevich1991} that has been used since in \hii\ galaxies research.

Throughout the first section of the current chapter we will explore the main properties of \hii\ galaxies, then we will discuss their $L(\mathrm{H}\beta) - \sigma $ relation and their possible systematics, ending with an analysis of their use to constraint the dark energy equation of state parameters.

\section {\hii\ Galaxies Properties}

\subsection{Giant extragalactic \hii\ regions and \hii\ galaxies}

One of the defining  characteristics of both \hii\ galaxies and Giant Extragalactic \hii\ Regions (GEHRs), is that the turbulent motions of their gaseous component are supersonic \citep{Melnick1987}.

GEHRs are zones of intense star formation in late type spirals (Sc) and irregular galaxies. Ionising photons are generated by clusters of OB stars at a rate of $10^{51} - 10^{52}\ \mathrm{s^{-1}}$, ionising large amounts ($10^4 - 10^6\ \mathrm{M_{\odot}}$) of low density ($N_e \approx 10 - 100\ \mathrm{cm^{-3}}$), inhomogeneously distributed gas. GEHRs have typical dimensions of the order $10^2 - 10^3\ \mathrm{pc}$ and diverse morphologies \citep{Shields1990, Garcia2009}.  

\hii\ galaxies are dwarf starforming galaxies that have undergone a recent episode of star formation, and their interstellar gas is ionised by one or more massive clusters of OB stars. This type of galaxies have total masses of less than $10^{11}\ \mathrm{M_{\odot}}$ and a radius of less than $2\ \mathrm{kpc}$ with a surface brightness $\mu_{V} \geq 19\ \mathrm{mag\ arcsec^{-2}}$ \citep{Garcia2009}. 

\hii\ galaxies, being active starforming dwarf galaxies, are also a subset of the blue compact dwarf (BCD) galaxies, although in general the term ``\hii\ galaxy'' is used when the objects have been selected for their strong, narrow emission lines \citep{Terlevich1991} while BCD galaxies are selected for their blue colours and compactness. Furthermore, only a fraction of BCDs are dominated by \hii\ regions, being then \hii\ galaxies.

\subsection{Morphology and structure}

\hii\ galaxies are compact objects with high central surface brightness. \citet{Telles1997} have classified \hii\ galaxies in two classes: Type I which have irregular morphology and higher luminosity, and Type II which have symmetric and regular outer structure. This regular outer structure could indicate large ages since the relaxation time is $\sim 10^8\ \mathrm{yr}$ unless the stars have been formed in an already relaxed gaseous cloud \citep{Kunth2000}.  

The determination of the surface brightness profile for \hii\ galaxies has given many apparently contradictory results and both exponential \citep{Telles1997b} and $r^{1/4}$ \citep{Doublier1997} models have been claimed as best fits to the data. 

The central part of \hii\ galaxies is dominated by one or more knots of star formation, giving rise in most cases to excess surface brightness. 

\subsection{Starburst in \hii\ galaxies}

\hii\ galaxies have a high star formation rate \citep{Searle1972}. Recent studies suggest that the recent star formation is concentrated in super star clusters (SSC) with sizes of $\sim 20\ \mathrm{pc}$ \citep{Telles2003}.

One of the open questions about \hii\ galaxies is the star formation triggering mechanism. Studies of environmental properties of \hii\ galaxies have shown that, in general, these are isolated galaxies \citep{Telles1995, Vilchez1995, Telles2000, Campos-Aguilar1993, Brosch2004} hence the star formation could not be triggered by tidal interactions with another galaxies. As an alternative, it has been proposed that interaction with other dwarf galaxies or intergalactic \hi\ clouds could be the cause of the star formation in \hii\ galaxies \citep{Taylor1997}. However, the evidence is not conclusive \citep{Pustilnik2001}.

\subsection{Ages of \hii\ galaxies}
The ages of \hii\ galaxies (and starburst [SB] in general) are estimated from the $\mathrm{H}\beta$ equivalent width, as was suggested initially by \citet{Dottori1981}. In general two models of star formation time evolution are used:
\begin{itemize}
 \item An instantaneous SB model, which assumes that all stars are formed at the same time in a short starburst episode, this model is generally applied to individual, low star-mass clusters.
 \item A continuous SB model, which assumes that the star formation is constant in time, this model is assumed to be an average characteristic of a system. 
\end{itemize}
Both models are simply the limiting cases for the possible star formation evolution. The second model can be thought of as a localized succession of short duration bursts separated by a small interval of time. \citet{Terlevich2003} showed that a continuous SB model fits better the observations of \hii\ galaxies, which indicates that these are not truly young systems and that they have  probably undergone considerable star formation previous to the present burst. This idea is also consistent with the fact that until now no \hii\ galaxy with metal abundance below 1/50th of solar has been found.  

\subsection{Abundances of \hii\ galaxies}
%Figure ---- \hii\ galaxies abundances
\begin{figure}[ht]
\centering
\resizebox{0.5\textwidth}{!}{\includegraphics{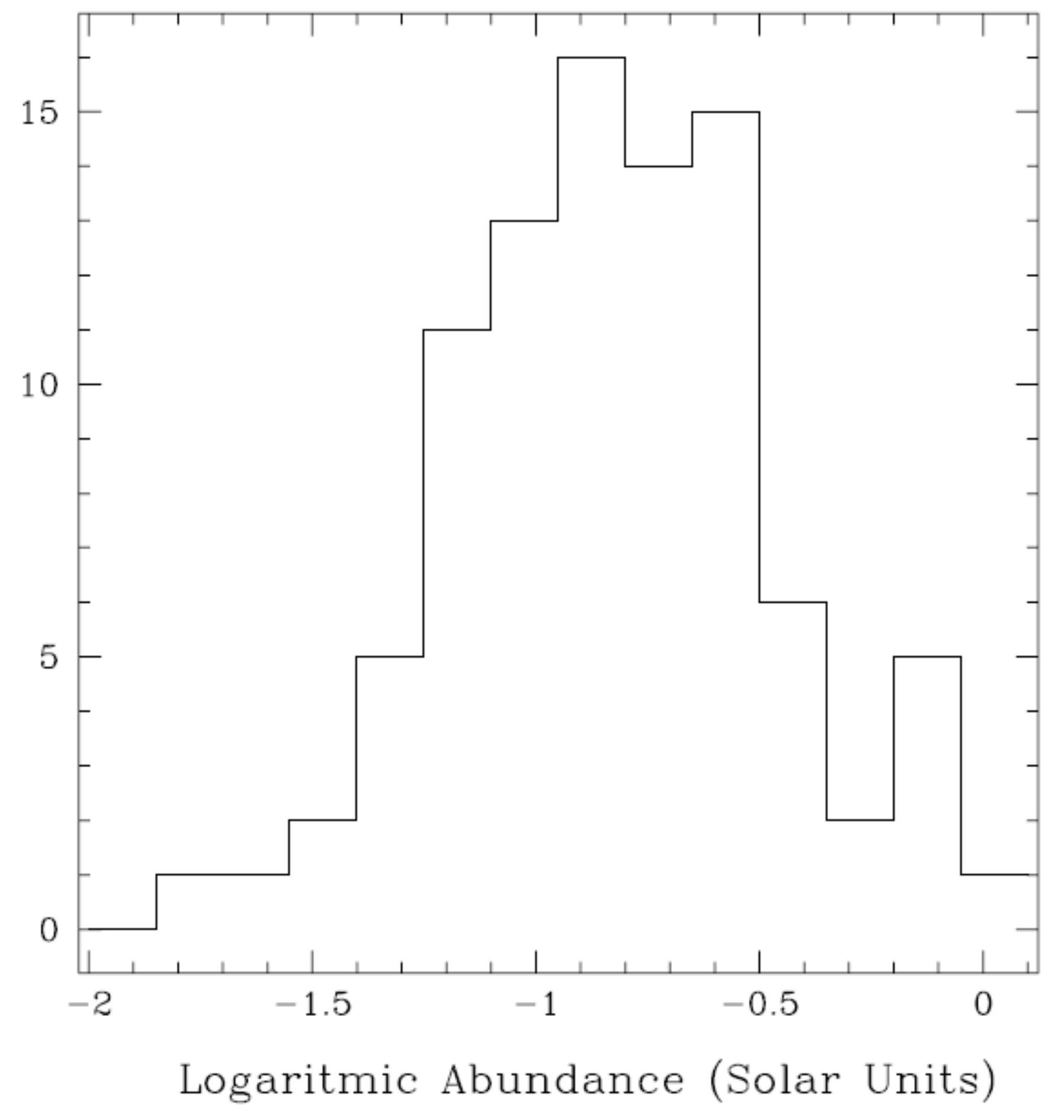}}
\caption[\hii\ galaxies abundances] {\small The metallicity distribution of \hii\ galaxies from \citet{Terlevich1991}, as measured form the oxygen abundances. Taken from \citet{Kunth2000}.}
\label{fig:H2gab}
\end{figure}

The metallicity of \hii\ galaxies was first analysed by \citet{Searle1972}; they showed that oxygen and neon abundances for I Zw18 and II Zw40 were sub-solar, whereas He abundances where about solar supporting He as a primordial element. Subsequently, many works have addressed this issue \citep[e.g.][]{Alloin1978, Lequeux1979, French1980, Kinman1981, Kunth1983, Terlevich1991, Pagel1992}.

\hii\ galaxies are metal poor systems, the abundance of metals in these systems ranges between $1/2\ \mathrm{Z_{\odot}}$ and $1/50\ \mathrm{Z_{\odot}}$. Figure \ref{fig:H2gab} shows the oxygen abundances distribution for a sample of \citet{Terlevich1991} \hii\ galaxies.

The oxygen abundance is normally considered as representative of the metallicity of \hii\ galaxies, as
oxygen is the most abundant of the metals that constitute them. However, the abundances of other elements can be obtained too. Particularly interesting is the fact that since, in general, \hii\ galaxies are chemically unevolved systems, the analysis of helium abundances in these systems is a good method for  determining primordial helium abundances \citep[e.g.][]{Pagel1992}.

\section{The $L(\mathrm{H}\beta) - \sigma $ Relation for \hii\ Galaxies} 
\label{secLSR}
\citet{Melnick1978} found a correlation between the average turbulent velocity of \hii\ regions in late spirals and irregular galaxies and the parent galaxy absolute magnitude, however at that moment the physics behind the correlation was not clear.

\citet{Terlevich1981} analysed the relation between $\mathrm{H}\beta$ luminosity, linewidth, metallicity and size for giant \hii\ regions and \hii\ galaxies finding correlations of the form:
\begin{align*}
 luminosity &\propto (linewidth)^{4}\\
 size &\propto (linewidth)^{4},
\end{align*}
which are of the kind encountered in pressure supported systems, then they conclude that \hii\ galaxies (and giant \hii\ regions) are self-gravitating systems in which the observed emission-line profile widths represent the velocity dispersion of discrete gas clouds in the gravitational potential. Furthermore, they found that the scatter in the $ L - \sigma$ relation was correlated to metallicity.

\citet{Melnick1987} analysed the properties of GEHRs. They found that the turbulent motions of the gaseous component of those systems are supersonic. Furthermore, they obtain correlations of the form:
\begin{align*}
 R_{c} &\sim \sigma^{2.5 \pm 0.5}\\
 L(\mathrm{H}\beta) &\sim \sigma^{5.0 \pm 0.5};
\end{align*}
\nomenclature{$R_{c}$}{The core radius.}%
\nomenclature{$L$}{The luminosity.}%
\nomenclature{$\sigma$}{The emisson-line width.}%
\nomenclature{$W$}{The line equivalent width.}%
\nomenclature{$F$}{The flux.}%
and they confirm the correlation between the scatter in the relations and the metallicity (from oxygen abundance). They concluded that the encountered relations are an indication of the virialized nature of discrete gas fragments  forming the structure of the giant \hii\ regions and  being ionised by a central star cluster. However, they recognise the possibility that stellar winds could have some, then unknown, effect on the velocity dispersion of the nebular gas. 

\citet{Melnick1988} studied the $L(\mathrm{H}\beta) - \sigma $ relation for \hii\ galaxies in a sample of objects that later would be part of the Spectrophotometric Catalogue of \hii\ Galaxies \citep{Terlevich1991}; they found a relation of the form:
\begin{align}
 \logd L(\mathrm{H}\beta) = (4.70 \pm 0.30) \logd \sigma + (33.61 \pm 0.50) & & \delta \logd L(\mathrm{H}\beta) = 0.29.
  \label{eq:LS88}
\end{align}
After a Principal Component Analysis (PCA) for the data, in which the oxygen abundance was used as parameter, they found that the metallicity, $(\mathrm{O / H})$, is effectively  an important component of the scatter in the previous relation. Consequently, they proposed as a distance indicator:
\begin{equation}
 M_{z} = \frac{\sigma^{5}}{(\mathrm{O / H})},
\end{equation}
from which they obtain a new relation:
\begin{align}
 \logd L(\mathrm{H}\beta) = (1.0 \pm 0.04) \logd M_{z} + (41.32 \pm 0.08) & & \delta \logd L(\mathrm{H}\beta) = 0.271
\end{align}
We must note that this last relation uses the distance scale of \citet{Aaronson1986} ($H_{0} \sim 90\ \mathrm{Km\ s^{-1}\ Mpc^{-1}}$).%ET1506

%Figure ---- L - Sigma 2000
\begin{figure}[ht]
\centering
\resizebox{0.8\textwidth}{!}{\includegraphics{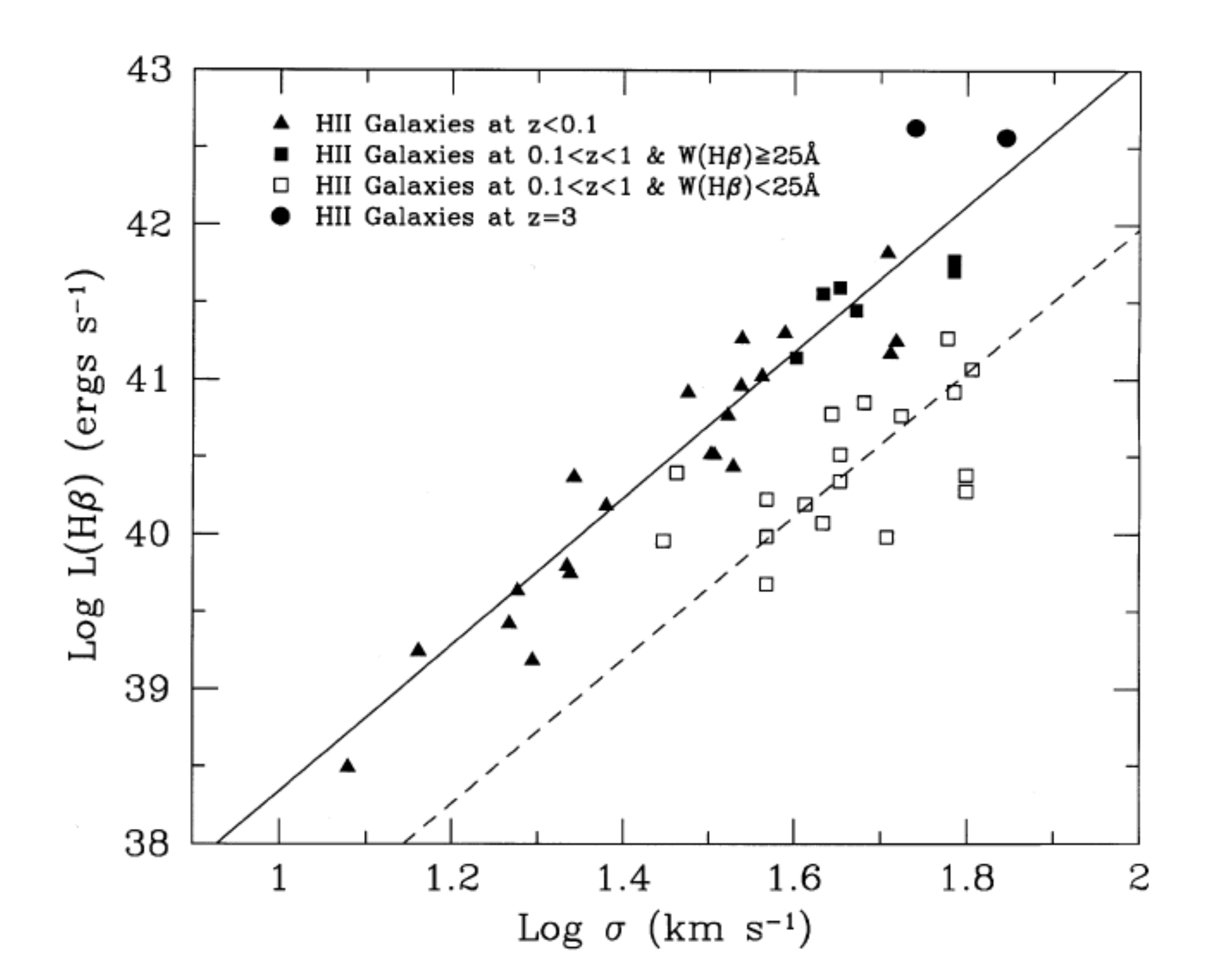}}
\caption[$L - \sigma$ relation from \citet{Melnick2000}] {\small The $L(\mathrm{H}\beta) - \sigma $  relation for \hii\ galaxies at intermediate redshifts. The solid line shows the maximum-likelihood fit to the young \hii\ galaxies in the local Universe. The dashed line shows the predicted $L(\mathrm{H}\beta) - \sigma $ relation for an evolved population of \hii\ galaxies. The cosmology is $H_{0} = 65\  \mathrm{km\ s^{-1}\ Mpc^{-1}}$. Taken from \citet{Melnick2000}. }
\label{fig:LS2000}
\end{figure}

\citet{Melnick2000} selected a sample of intermediate redshift ($z < 1$) \hii\ galaxies from the literature, using as selection criterion the emission lines strength. The objects with strongest emission lines (i.e. largest equivalent widths) were selected in order to avoid the evolved ones \citep{Copetti1986}, which can introduce a systematic error in the $L(\mathrm{H}\beta) - \sigma $ relation due to the effect of the underlying old population over the line widths. Using this sample, they found the  $L(\mathrm{H}\beta) - \sigma $ relation shown in Figure \ref{fig:LS2000}; we can see clearly the effect of the stellar population evolution over the relation. In this work the distance indicator was re-calibrated with the then available distances for the sample. They found
\begin{equation}
 \logd L(\mathrm{H}\beta) =  \logd M_{z} + 29.5,
\end{equation} 
from which they derived the distance modulus as
\begin{equation}
 \mu = 2.5 \logd \frac{\sigma^{5}}{F(\mathrm{H}\beta)} - 2.5 \logd (\mathrm{O/H}) - A_{\mathrm{H}\beta} - 26.44,
 \label{eq:mu2000}
\end{equation}
where $F(\mathrm{H}\beta)$ is the observed $\mathrm{H}\beta$ flux and $A_{\mathrm{H}\beta}$ is the total extinction.

\begin{figure}[ht]
\centering
\resizebox{0.7\textwidth}{!}{\includegraphics{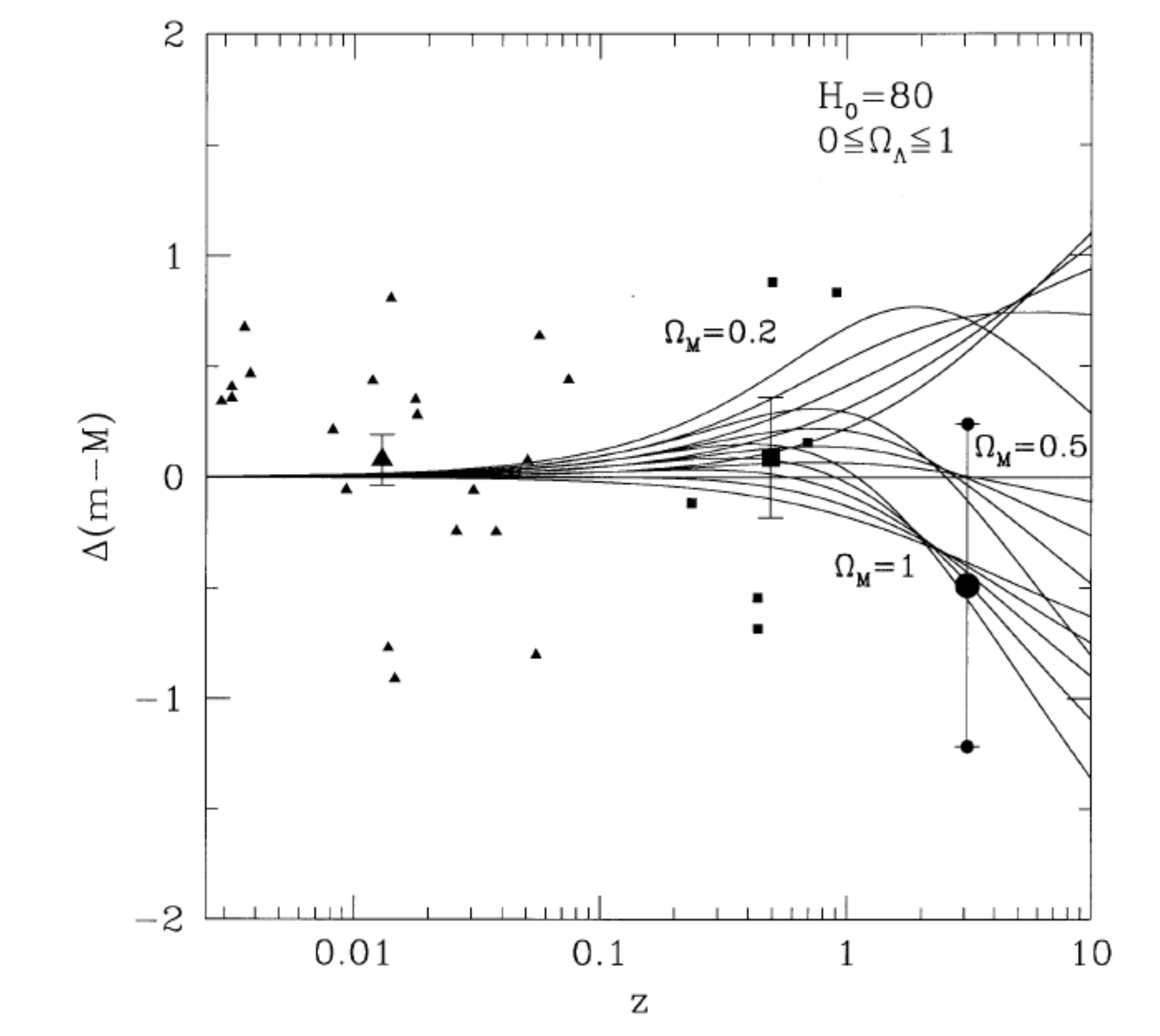}}
\caption[Differential Hubbel diagram for \hii\ galaxies] {\small The differential Hubble diagram for \hii\ galaxies with a wide range of redshifts. Three families of curves for distinct values of $\Omega_m$ are shown, for every one $\Omega_{\Lambda} = (0, 0.25, 0.5, 0.75, 1.0)$. The large symbols represent the average redshift and distance modulus for each subsample. The error bars show the mean error in distance modulus assuming that each point is an independent measurement and ignoring observational errors. $H_0 = 80\ \mathrm{km\ s^{-1}\ Mpc^{-1}}$ was used to normalize the data points. The model lines are independent of $H_0$. From \citet{Melnick2000}.}
\label{fig:MTTHD}
\end{figure}

The differential Hubble diagram for \hii\ galaxies derived by \citet{Melnick2000} is shown in Figure \ref{fig:MTTHD}. From the figure it is clear that the data present large scatter with respect to the models differences. For the local sample ($z < 0.1$), they derived an rms dispersion in distance modulus of $\sigma(\Delta \mu) = 0.52\ \mathrm{mag}$. \citet{Melnick1988} claim that typical errors are about 10\% in flux and 5\% in $\sigma$, adding 10\% in extinction and 20\% in abundances, while \citet{Melnick2000} expect a scatter of about $0.35\ \mathrm{mag}$ in $\mu$ from observational errors. Hence, improvement in measurements is required in order to obtain better constraints.

%Figure ---- Differential Hubbel diagram for \hii\ galaxies 

\citet{Siegel2005} have constrained the value of $\Omega_{m}$ using a sample of 15 high-$z$ \hii\ galaxies ($2.17 <  z  <  3.39$) obtaining a best fit of $\Omega_{m} = 0.21_{-0.12}^{+0.30}$ for a $\Lambda$-dominated universe, which is consistent with other recent determinations. Their sample has been selected using the criterion of emission line strength, as was selected in the \citet{Melnick2000} sample. For the $\mu$ determination they have used (\ref{eq:mu2000}) with a modification in the zero point (they used $26.18$ in place of $26.44$) due to the fact that they have taken $H_{0} = 71\ \mathrm{km\ s^{-1}\ Mpc^{-1} }$. 
\\

\subsection{The physics of the $L(\mathrm{H}\beta) - \sigma $ relation}

%Figure ---- \hii\ galaxies Fundamental Plane
\begin{figure}[ht]
\centering
\resizebox{0.9\textwidth}{!}{\includegraphics{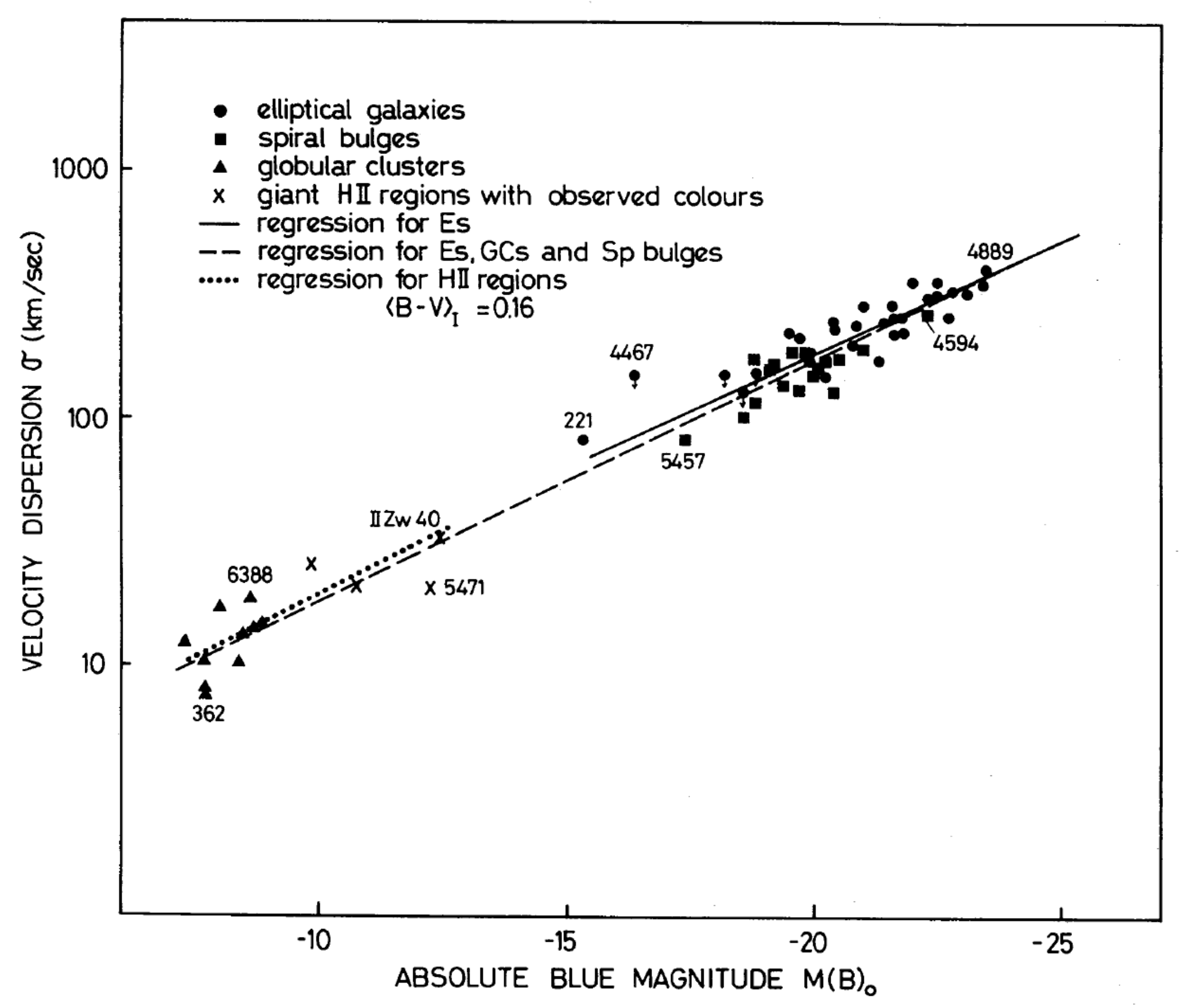}}
\caption[$M(B)_0 - \sigma$ Relation] {\small $M(B)_0 - \sigma$ correlation for elliptical galaxies, bulges of spiral galaxies, globular clusters and GEHR. The dashed line is a linear fit for all the data. The solid line is a fit for elliptical galaxies. The dotted line is a fit to the GEHR. Taken from \citet{Terlevich1981}.}
\label{fig:Msig}
\end{figure}

\begin{figure}[ht]
\centering
\resizebox{0.9\textwidth}{!}{\includegraphics{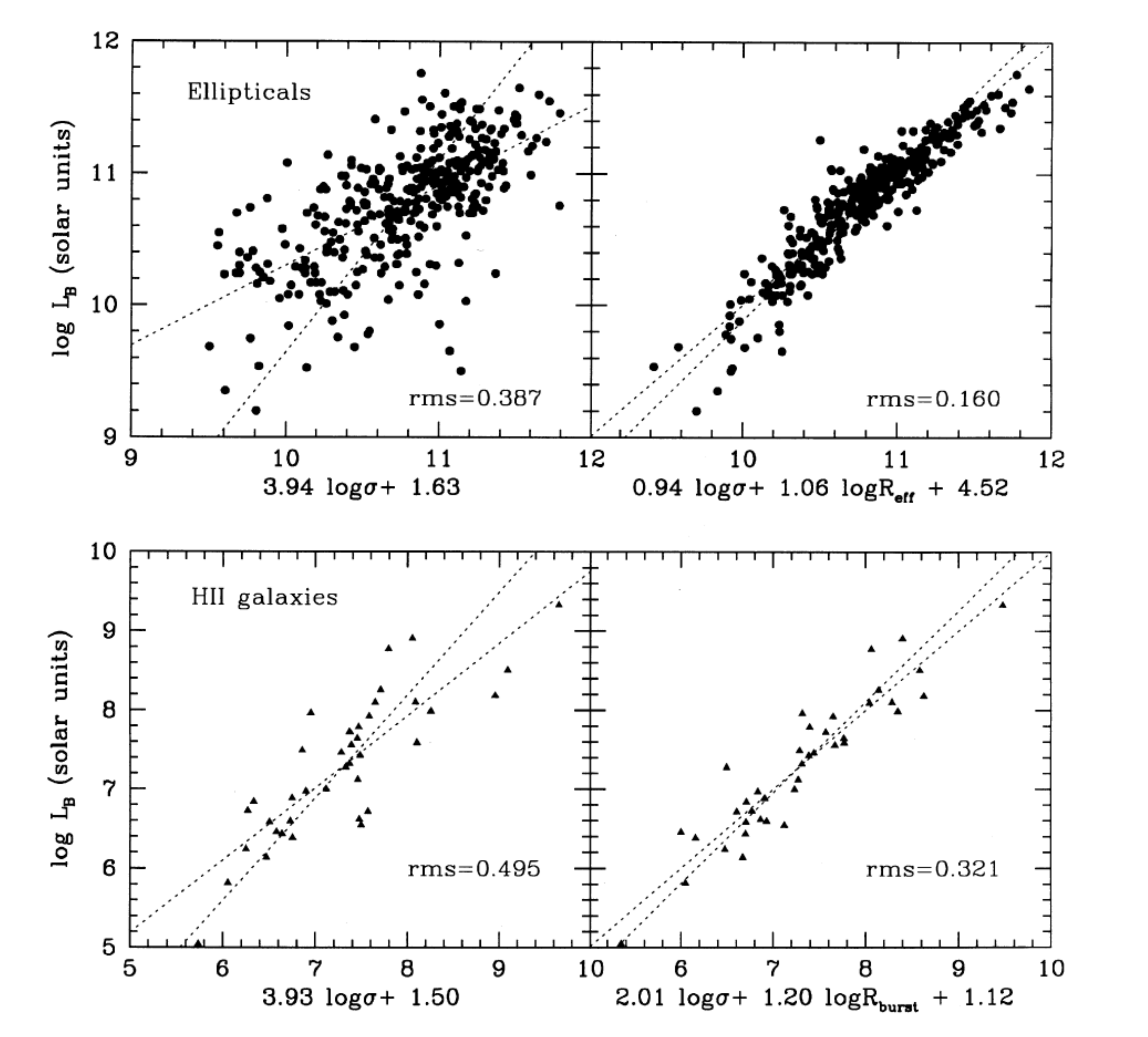}}
\caption[\hii\ galaxies Fundamental Plane] {\small The fundamental plane of \hii\ galaxies and normal elliptical galaxies from \citet{Telles1995b}. The radii and magnitudes of \hii\ galaxies are measured from continuum images. The velocity dispersions are the widths of the emission lines.}
\label{fig:FPH2}
\end{figure}

\citet{Melnick1987} found that \hii\ galaxies present supersonic motions in their gaseous component. In order to explain the motions of the \hii\ galaxies gaseous component, \citet{Terlevich1981} had proposed a model in which its nature is explained as being of gravitational origin. The basis for this argument is that correlations of the kind $L(\mathrm{H}\beta) \propto \sigma^4$ and $R \propto \sigma^2$ were observed in \hii\ galaxies. These correlations are expected for virialized systems and in fact are observed in elliptical galaxies, spiral bulges and globular clusters. 

In order to compare GEHR with globular clusters, bulges of spirals and elliptic galaxies, and thus test the hypothesis of the gravitational origin for the \lsig\ relation, \citet{Terlevich1981} evolved the ionising stellar clusters following the single burst of star formation model by \citet{Larson1978}. The resulting $M(B)_0 - \sigma$ relation is shown in Figure \ref{fig:Msig}. From the figure it is clear that the $M(B)_0 - \sigma$ relation for GEHR is consistent, within uncertainties, with the relations for the other spheroidal systems, strongly suggesting a mainly gravitational origin for the correlation. 

Another factor that contributes to the origin of the supersonic turbulent motions in the gaseous component of \hii\ galaxies is the stellar winds generated by massive  evolved stars. It has been shown that, unlike the case for evolved GEHRs  where this effect dominates \citep{Melnick1999}, for \hii\ galaxies it appears not to be dominant.

A strong support for the gravitational origin idea came from \citet{Telles1995b}, where it was shown that these objects define a fundamental plane that is very similar to that defined by elliptical galaxies (see Figure \ref{fig:FPH2}). However, the scatter observed in the $L(\mathrm{H}\beta) - \sigma $ may be due to the presence of a second parameter, perhaps possible variations in the initial mass function (IMF), rotation or the duration of the burst of star formation that powers the emission lines \citep{Melnick2000}.

It has been shown that the scatter in the $L(\mathrm{H}\beta) - \sigma $ relation can be reduced if objects with $\sigma > 65\ \mathrm{km s^{-1}}$ are rejected from the analysis \citep{Melnick1988, Koo1995}. This can be understood if one assumes that \hii\ galaxies are powered by clusters of stars, and thus the above condition is equivalent to say that the time required for the clusters to form must be smaller than the main sequence lifetime of the most massive stars \citep{Melnick2000}.

\subsection{Age effects}

Around $3\ \mathrm{Myr}$ to $6\  \mathrm{Myr}$ after a starburst, the emission line flux decays fast and continuously whereas the continuum flux is roughly constant. Thus, the equivalent widths ($W$) of emission lines are a good estimator of the starburst age \citep{Copetti1986}. In order to minimize  systematic effects over the $L(\mathrm{H}\beta) - \sigma $ relation it is necessary to consider this effect by restricting the sample to objects with high $W(\mathrm{H}\beta)$ in order to select young starbursts and minimize the effects of a possible old underlying population over the equivalent width of the emission lines. 

\subsection{Extinction effects}

Due to its effect over the flux of the $\mathrm{H}\beta$ line, the extinction or reddening is one important systematic for the $L(\mathrm{H}\beta) - \sigma $ relation. Two possible sources of extinction must be considered: dust in our Galaxy and dust in the \hii\ galaxies themselves.  It has been shown that the extinction correction for \hii\ galaxies can be determined from Balmer decrements \citep{Melnick1987, Melnick1988}.

\subsection{Metallicity effects}
The metallicity has an important effect over  the $L(\mathrm{H}\beta) - \sigma $ relation as was pointed out in the analysis of \citet{Terlevich1981} where it was shown that the residuals of this relation are correlated with metallicity. Furthermore, using PCA,  \citet{Melnick1987} showed that one of the two principal components with the larger weight was mostly determined by the oxygen abundance. 

%\subsection{Environmental effects}

\section{\hii\ Galaxies as Cosmological Probes}
\label{sec:3_3}

%This work main aim is to constrain the space of parameters of the  dark energy equation of state. We will begin by a brief theoretical analysis of the parameters involved. %ET1506

This work's main aim is to constrain the parameter space of the dark energy equation of state and therefore we will review briefly the theoretical analysis of the parameters involved. 

From (\ref{eqHZ}) we know that the Hubble function depends on the cosmological parameters following the relation:
\begin{equation}
 H^2(z) = H_0^2 \left[ \Omega_m(1+z)^3 + \Omega_w \exp \left( 3 \int_{0}^{z}{\frac{1 + w(z')}{1 + z'} dz'} \right) \right],
 \label{eqHZ3}
\end{equation}
where we are neglecting the minuscule contribution of the radiation to the total energy density and we are assuming a flat universe. From (\ref{eqDM}) we also know that:
\begin{equation}
 \mu = 5 \logd D_L + 25,
 \label{eqDM3}
\end{equation}
where $D_L$, the luminosity distance, is given by (\ref{eqLD}) and is expressed in Mpc.%ET1506

Using (\ref{eqHZ3}) we can define a nominal \emph{reference} $\Lambda$-cosmology with $\Omega_m = 0.27, \Omega_{\Lambda} = 0.73 $ and $w = -1$. And then we can compare different models to the reference one. For this purpose we define:
\begin{equation}
 \Delta \mu = \mu_{\Lambda} - \mu_{model}\; ,
\end{equation}
where $\mu_{\Lambda}$ is the distance modulus given by the \emph{reference} $\Lambda$-cosmology and $\mu_{model}$ is the one given by any another model. 

%Figure \ref{fig:DMm} shows the difference between some cosmological models for which their parameters are indicated. It can be seen that the relative magnitude deviations between dark energy models is $\leq 0.1\ \mathrm{mag}$, which indicates the necessary high accuracy in the photometry  of any object used as a tracer. Furthermore, it is clear that the larger relative deviations of the distance moduli are present at $z \geq 1.5$, then High-z tracers are needed to constrain the value of the equation of state parameters, in fact at higher redshift than those at which SNe Ia have been observed.%ET1506

%Figure ---- Distance Modulus Differences
\begin{figure}[ht]
\centering
\resizebox{0.8\textwidth}{!}{\includegraphics{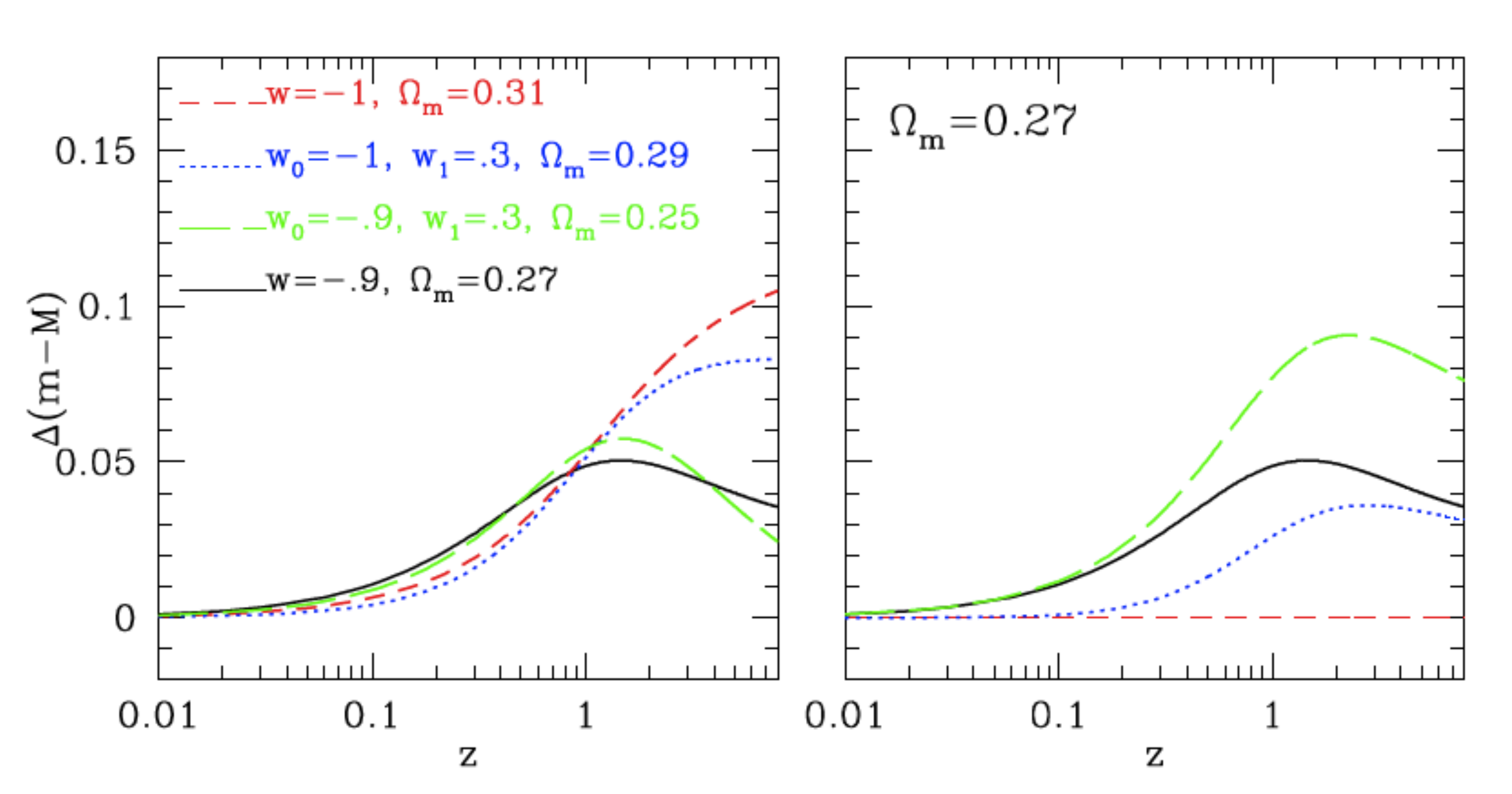}}
\caption[Distance Modulus Differences] {\small \emph{Left Panel:} The expected distance modulus difference between the dark energy models shown and the reference $\Lambda$-model. \emph{Right Panel:} The expected distance modulus differences once that the $\Omega_m - w(z)$ degeneracy is broken (imposing a unique $\Omega_m$ to all models). Taken from \citet{Plionis2009}.}
\label{fig:DMm}
\end{figure}

Figure \ref{fig:DMm} shows the difference between some cosmological models for which their parameters are indicated. It can be seen that the relative magnitude deviations between dark energy models is $\leq 0.1\ \mathrm{mag}$, which indicates the necessary high accuracy in the photometry  of any object used as a tracer. Furthermore, it is clear that larger relative deviations of the distance moduli are present at $z \geq 1.5$, and therefore high-$z$ tracers are needed to effectively constrain the values of the equation of state parameters, in fact at redshifts higher than those currently probed by SNe Ia.
       
%Another important factor, that we can see in Figure \ref{fig:DMm}, is that there are strong degeneracies between different cosmological models at $z \leq 1$ (in some cases even at higher redshifts). This fact shows the necessity of at least two independent cosmological probes in order to break the degeneracies. If we additionally consider that we have abundant evidence for $0.2 \leq \Omega_m \leq 0.3$, we can expect that the degeneracies would be considerably reduced, as in fact is shown in the right hand panel of Figure \ref{fig:DMm}, even though for another value of $\Omega_m$.%ET1506

Another important factor, that we can see in Figure \ref{fig:DMm}, is that there are strong degeneracies between different cosmological models at $z \leq 1$ (in some cases even at higher redshifts), this due to the known $\Omega_m - w(z)$ degeneracy. This fact shows the necessity of at least two independent cosmological probes in order to break the degeneracies. If we additionally consider that we have abundant evidence for $0.26 \leq \Omega_m \leq 0.3$, we can expect that the degeneracies would be considerably reduced, as in fact is shown in the right hand panel of Figure \ref{fig:DMm}, where we have fixed the value of  $\Omega_m = 0.27$.

%Table --- Cosmological Parameters from fit 
\begin{table}[tb]
\caption[Cosmological parameters fits from SNe Ia]{\small Cosmological parameters fits using the SNe Ia data within flat cosmologies. Note that for the case where $\mathbf{p} = (\Omega_m, w)$ (last row), the errors shown are estimated after marginalizing with respect to the other fitted parameters. Taken from \citet{Plionis2009b}.  }
\begin{center}
\resizebox{1.0\textwidth}{!}{
\begin{tabular} {| c c c| c c c|}
\hline
\multicolumn{3}{| c |}{\emph{D07}} & \multicolumn{3} {|c|} {\emph{Constitution}} \\
\hline
$w$  & $\Omega_m$ & $\chi^2_{min} / df$ & $w$ & $\Omega_m$ & $\chi^2_{min} / df$ \\
\hline
$ \mathbf{-1} $ (fixed) & $0.280_{-0.015}^{+0.025}$ & $187.03 / 180$ & $\mathbf{-1}$ (fixed) & $0.286_{-0.018}^{+0.012} $ & $439.78 / 365$ \\
$ -1.025_{-0.045}^{+0.060}$ & $0.292 \pm 0.018$ & $187.02 / 179$ & $-1.025 \pm 0.030$ & $0.298 \pm 0.012$ & $439.79/ 364$\\
\hline 
\end{tabular}}
\end{center}
\label{tab:CPF}
\end{table}

%As  previously said, the unique available direct test for cosmic acceleration is the SNe Ia distance-redshift relation, therefore it is useful to test how the constraints over the cosmological parameters change when the SNe Ia sample is increased. \citet{Plionis2009b} analyse two SNe Ia data sets,  the \citet{Davis2007} [hereafter \emph{D07}] compilation of 192 SNe Ia and the \emph{Constitution} compilation of 397 SNe Ia \citep{Hicken2009}, which are not independent since most of the \emph{D07} is included in the \emph{Constitution} sample.%ET1506 

As  previously mentioned, the single available direct test for cosmic acceleration is based on the SNe Ia distance-redshift relation, and therefore it is useful to test how the constraints of the cosmological parameters change when the SNe Ia sample is increased. \citet{Plionis2009b} analyse two SNe Ia data sets,  the \citet{Davis2007} [hereafter \emph{D07}] compilation of 192 SNe Ia and the \emph{Constitution} compilation of 397 SNe Ia \citep{Hicken2009}, which are not independent since most of the \emph{D07} is included in the \emph{Constitution} sample.%ET1506 MP(16/6/10)

%In order to do the data analysis, the likelihood estimator\footnote{Likelihoods are normalized to their maximum values.} (see Appendix \ref{ApB}) was defined as:
%\begin{equation}
% \mathcal{L}^{SNIa}(\mathbf{p}) \propto \exp[ - \chi^{2}_{SNIa}(\mathbf{p})/2],
%\end{equation}
%%\nomenclature{$\mathcal{L}$}{The likelihood estimator.}%
%where $\mathbf{p}$ is a vector containing the cosmological parameters that we want to fit for, and
%\begin{equation}
% \chi^{2}_{SNIa}(\mathbf{p}) = \sum_{i = 1}^{N} {\left[\frac{\mu^{th}(z_{i}, \mathbf{p}) - \mu^{obs}(z_{i})}{\sigma_{i}} \right]^{2}},
%\end{equation}
%where $\mu^{th}$ is given by (\ref{eqDM3}) and (\ref{eqHZ3}), $z_{i}$ is the observed redshift, $\mu^{obs}$ is the observed distance modulus and $\sigma_{i}$ is the observed distance modulus uncertainty. A flat universe was assumed   for the analysis so $\mathbf{p} \equiv (\Omega_{m}, w_{0}, w_{1})$. Finally, since only SNe Ia with $z > 0.02$ were used in order to avoid redshift uncertainties due to peculiar motions, the final samples were of 181 (\emph{D07}) and 366 (\emph{Constitution}) SNe Ia. %ET1506

%Figure ---- Cosmological Parameters Solution Space
\begin{figure}[h]
\centering
\resizebox{0.9\textwidth}{!}{\includegraphics{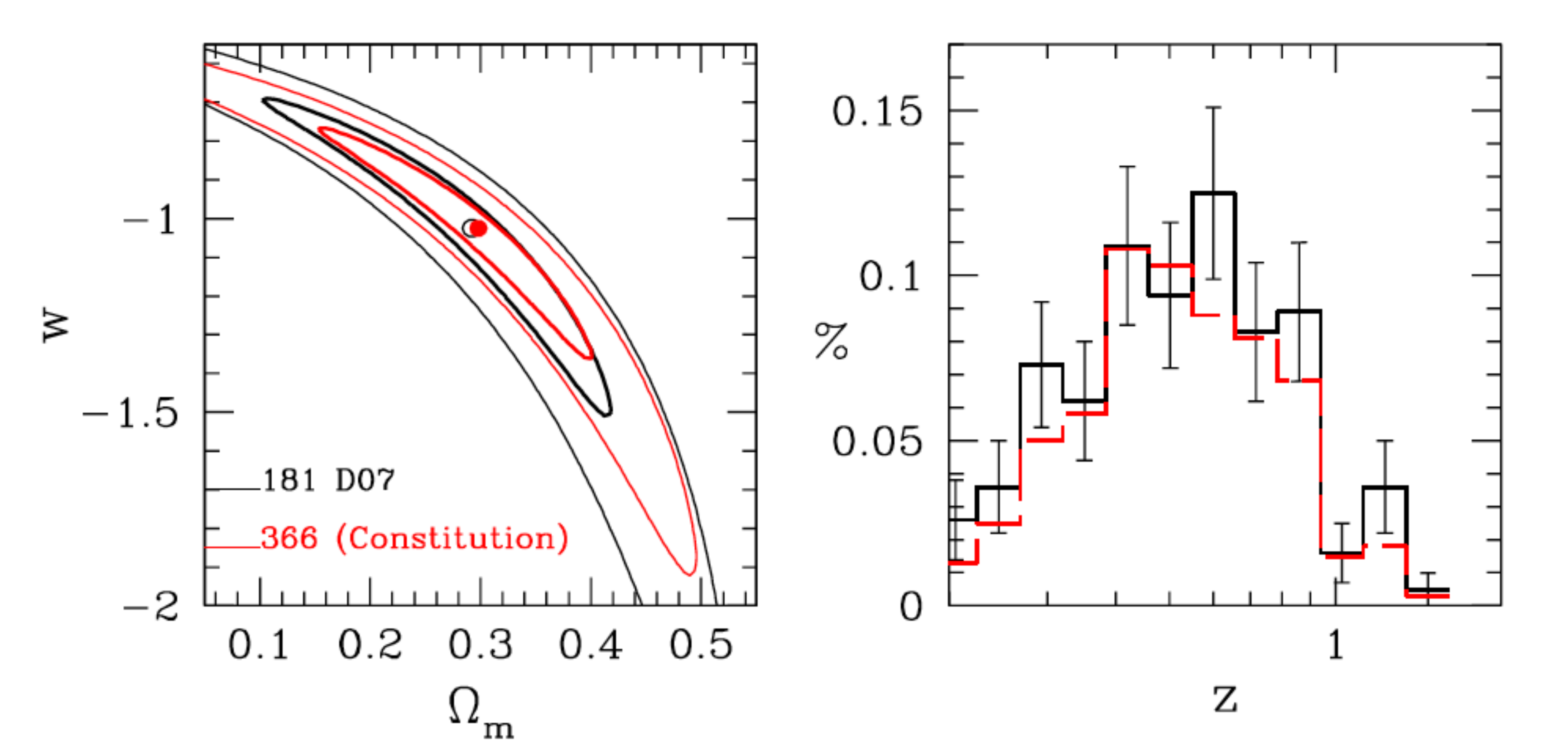}}
\caption[Cosmological Parameters Solution Space] {\small \emph{Left Panel}: Cosmological parameters solution space using either of the two SNe Ia data sets (\emph{Constitution:} red contours and \emph{D07:} black contours). Contours corresponding to the $1$ and $3 \sigma$ confidence levels are shown. \emph{Right Panel:} Normalized redshift distribution of the two SNe Ia data sets. Taken from \citet{Plionis2009b}.}
\label{fig:CPSS}
\end{figure}

In order to perform the data analysis, a likelihood estimator\footnote{Likelihoods are normalized to their maximum values.} (see Appendix \ref{ApB}) was defined as: %MP(16/6/10)
\begin{equation}
 \mathcal{L}^{SNIa}(\mathbf{p}) \propto \exp[ - \chi^{2}_{SNIa}(\mathbf{p})/2],
\end{equation}
%\nomenclature{$\mathcal{L}$}{The likelihood estimator.}%
where $\mathbf{p}$ is a vector containing the cosmological parameters that we want to fit for, and
\begin{equation}
 \chi^{2}_{SNIa}(\mathbf{p}) = \sum_{i = 1}^{N} {\left[\frac{\mu^{th}(z_{i}, \mathbf{p}) - \mu^{obs}(z_{i})}{\sigma_{i}} \right]^{2}},
\end{equation}
where $\mu^{th}$ is given by (\ref{eqDM3}) and (\ref{eqHZ3}), $z_{i}$ is the observed redshift, $\mu^{obs}$ is the observed distance modulus and $\sigma_{i}$ is the observed distance modulus uncertainty. A flat universe was assumed   for the analysis so $\mathbf{p} \equiv (\Omega_{m}, w_{0}, w_{1})$. Finally, since only SNe Ia with $z > 0.02$ were used in order to avoid redshift uncertainties due to peculiar motions, the final samples were of 181 (\emph{D07}) and 366 (\emph{Constitution}) SNe Ia. %ET1506

%Figure ---- Cosmological Parameters Solution Space from Monte-Carlo Analysis
\begin{figure}[h]
\centering
\resizebox{0.9\textwidth}{!}{\includegraphics{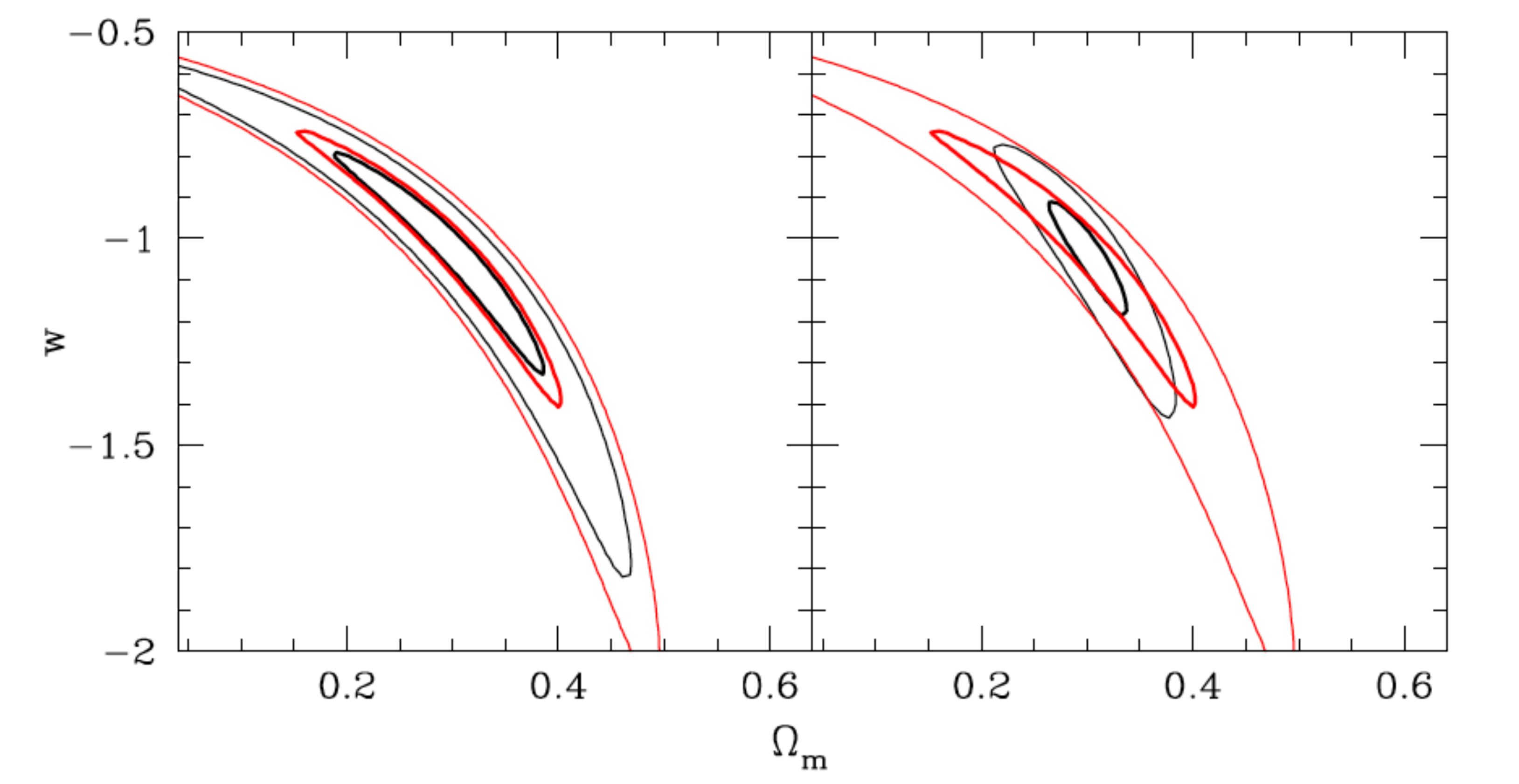}}
\caption[Cosmological Parameters Solution Space from Monte-Carlo Analysis] {\small \emph{Left Panel}: Comparison of the \emph{Constitution} data set derived constraints (red contours) with those derived by reducing to half their uncertainties (black contours). \emph{Right Panel:} Comparison of constrains from \emph{Constitution} (red contours) with those derived  by adding a sample of $82$ high-$z$ tracers ($2.7 \lesssim z \lesssim 3.5$) with distance modulus mean uncertainty of $\sigma_{\mu} \simeq 0.38$ (black contours). Taken from \citet{Plionis2009b}.}
\label{fig:RMC1}
\end{figure}

%Figure ---- Cosmological Parameters Solution Space from Monte-Carlo Analysis. Evolving DE equation of state. 
\begin{figure}[h]
\centering
\resizebox{0.9\textwidth}{!}{\includegraphics{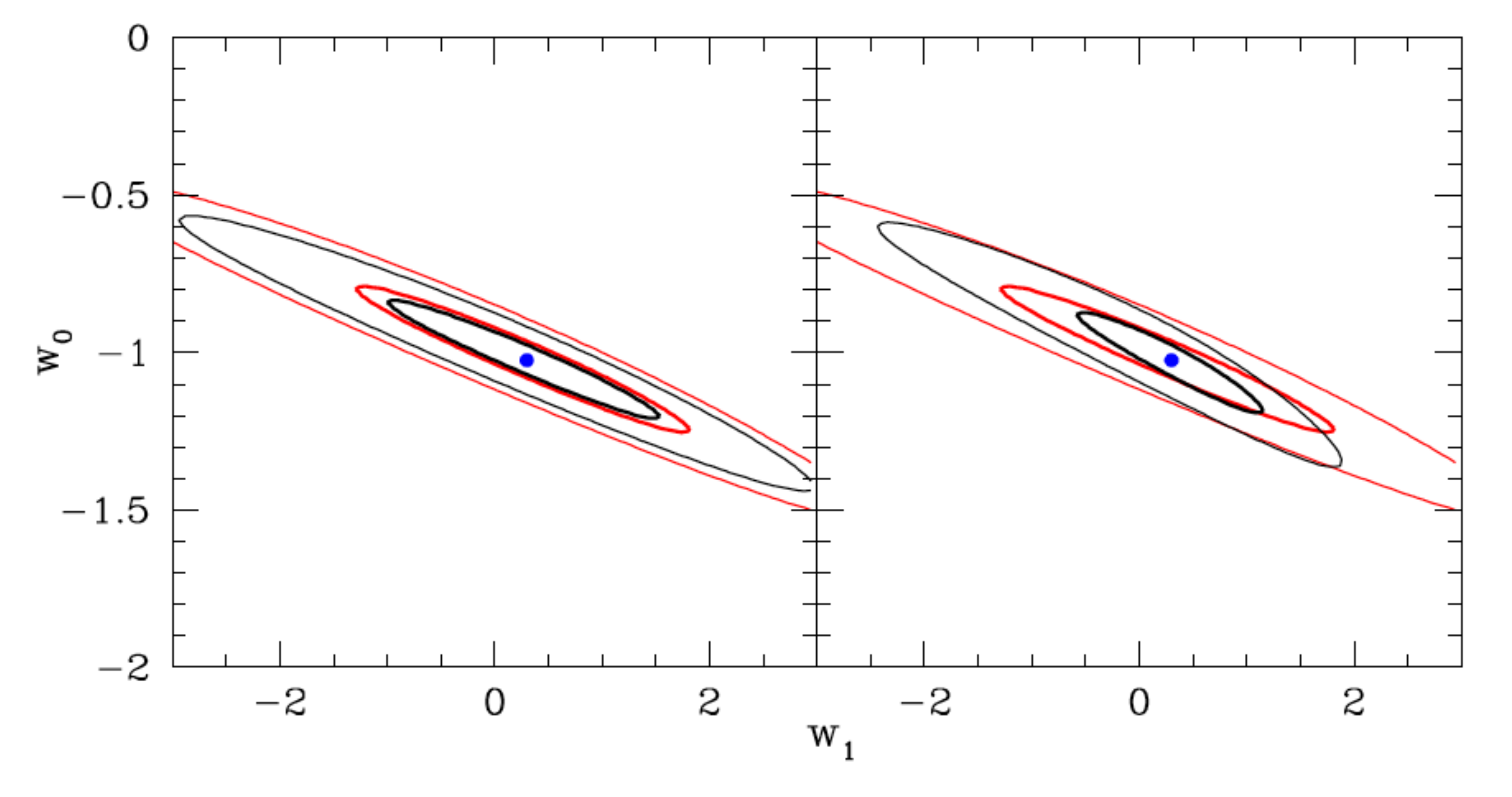}}
\caption[Cosmological Parameters Solution Space from Monte-Carlo Analysis 2] {\small As in Figure \ref{fig:RMC1}, but allowing for an evolving Dark Energy equation of state and after marginalizing with respect to $\Omega_m$. The input cosmological model has $(w_0, w_1) = (-1.025, 0.3)$ and is represented by the red contours. Taken from  \citet{Plionis2009b}.}
\label{fig:RMC2}
\end{figure}

%Table \ref{tab:CPF} presents solutions using the previous mentioned data sets. We can see that the cosmological parameters derived are consistent between both data sets and with the results derived from the \emph{Union2} data set (see Section \ref{sec:SNeres}). %ET1506

Table \ref{tab:CPF} presents solutions using the previous mentioned data sets. We can see that the cosmological parameters derived are consistent between both data sets. %ET1506, MP(16/6/10)

%%MANOLIS: the previous statement is not correct. I have analysed the UNION2 data set and they have the same problem
%% as the UNION set, ie., the unconstrained fit to the (Om,w) space gives w=-1.2 (of course within 1 sigma it coincides with
%% Constitution, but there is a systematic shift to smaller w's which needs to be understood, because basically the SN's are
%% the same as Constitution but reduced differently....

%Figure \ref{fig:CPSS} shows the cosmological parameters solution space for the two above mentioned data sets. We can see that although the \emph{Constitution} data set has twice as many data points as \emph{D07}, the constraints obtained form the former are similar to those obtained from the latter. This fact indicates that, for Hubble function tracers, increasing the number of data points covering the same redshift range and with the current uncertainty level for SNe Ia, does not provide better constraints for cosmological parameters.%ET1506

Figure \ref{fig:CPSS} shows the cosmological parameters solution space for the two above mentioned data sets. We can see that although the \emph{Constitution} data set has twice as many data points as \emph{D07}, the constraints obtained form the former are similar to those obtained from the latter. This fact indicates that, for Hubble function tracers, increasing the number of data points covering the same redshift range and with the current uncertainty level for SNe Ia, does not provide significantly better constraints for cosmological parameters.%ET1506 MP(16/6/10)

%From the previous discussion it becomes clear that we have two possible options to obtain more stringent constraints to cosmological parameters:
%\begin{itemize}
% \item Trace the same redshift range ($ z \lesssim 1.5$), that has been traced until now using SNe Ia, but reducing the distance modulus uncertainties or
% \item Trace at higher redshifts, where the different theoretical models show the largest deviations, maintaining the distance modulus uncertainties now obtained for high-$z$ SNe Ia ($\langle \sigma_{\mu} \rangle \simeq 0.4$).    
%\end{itemize}%ET1506

From the previous discussion it becomes clear that we have two possible options to obtain more stringent constraints of cosmological parameters:
\begin{itemize}
 \item Trace the same redshift range ($ z \lesssim 1.5$), that has been traced until now using SNe Ia, but reducing significantly the distance modulus uncertainties or
 \item Trace at higher redshifts, where the different theoretical models show the largest deviations, maintaining or if possible reducing the distance modulus uncertainties now obtained for high-$z$ SNe Ia ($\langle \sigma_{\mu} \rangle \simeq 0.4$).    
\end{itemize}%ET1506 MP(16/6/10)

%\citet{Plionis2009b} analysed both alternatives by means of a Monte-Carlo procedure, and as is shown in Figure \ref{fig:RMC1}, when the uncertainties are reduced by half, compared with \emph{Constitution} data, the reduction in the range of the solution space is quite small; however, when a high-$z$ mock 82 objects subsample, with distance modulus uncertainties comparable to the occurring in actual high-$z$ SNe Ia data, was added to the \emph{Constitution} data set, a significantly reduced solution space was found. It is important to note that the redshift distribution ($2.68 \lesssim z \lesssim 3.55$) for the added mock subsample is in the range where the largest deviations between different cosmological models are expected (see Figure \ref{fig:DMm}). The same behavior was found when an evolving dark energy equation of state model was implemented (see Figure \ref{fig:RMC2}). %ET1506 

\citet{Plionis2009b} analysed both alternatives by means of a Monte-Carlo procedure, and as it is shown in Figure \ref{fig:RMC1}, when the \emph{Constitution} data uncertainties are reduced by half, the reduction in the range of the solution space is quite small; however, when a high-$z$  82 mock-object subsample, with distance modulus uncertainties comparable to those of actual high-$z$ SNe Ia data, is added to the \emph{Constitution} data set, a significantly reduced solution space is found. It is important to note that the redshift distribution ($2.68 \lesssim z \lesssim 3.55$), for the added mock subsample, is in the range where the largest deviations between different cosmological models are expected (see Figure \ref{fig:DMm}). The same behavior was found when an evolving dark energy equation of state model was implemented (see Figure \ref{fig:RMC2}). %ET1506 MP(16/6/10)

%From the previous discussion it is clear that in order to obtain more stringent constrains to cosmological parameters using the Hubble relation, a better strategy is to use standard candles which trace the redshift range where larger differences between cosmological models are expected ($2 \lesssim z \lesssim 4$).%ET1506

From the previous discussion it is clear that in order to obtain more stringent constraints to the cosmological parameters, using the Hubble relation, a better strategy is to use standard candles which trace a redshift range where larger differences between the cosmological models are expected ($2 \lesssim z \lesssim 4$).%ET1506 MP(16/6/10)

%Near infrared surveys \citep{Pettini2001, Erb2003} have shown that \hii\ galaxies can be observed at much larger redshifts than SNe Ia and since they can be used as standard candles, due to their  $L(\mathrm{H}\beta) - \sigma $, they are excellent candidates for high-$z$ tracers of cosmological parameters.%ET1506

Near infrared surveys \citep{Pettini2001, Erb2003} have shown that \hii\ galaxies can be observed at much larger redshifts than SNe Ia and since they can be used as standard candles, due to their  $L(\mathrm{H}\beta) - \sigma $ relation, they are excellent candidates for high-$z$ tracers to be used to constrain the cosmological parameters.%ET1506 MP(16/6/10)

%In he next chapter we will explore the actions needed to recalibrate the $L(\mathrm{H}\beta) - \sigma$ relation for \hii\ galaxies an to construct their Hubble diagram in order to constrain the cosmological parameters solution space.%ET1506

\section{Summary}
\hii\ galaxies are young massive bursts of star formation whose spectra is dominated by strong emission lines. These objects have been proposed as an useful standard candle, because their velocity dispersion is correlated with the luminosity of their Balmer emission lines.

The origin of the \hii\ galaxies \lsig\ relation has been studied in many works, with the general conclusion that virialization of the cluster gas appears to be the main cause, although other factors, as metallicity and age, also contribute to the relation.

Since \hii\ galaxies can be observed to $z \sim 3$, they constitute a promising new cosmic tracer which may allow to obtain better constraints to cosmological parameters than currently obtained with lower redshift tracers.  

In the next chapter we will explore in depth the $L(\mathrm{H}\beta) - \sigma$ relation for \hii\ galaxies and the systematics affecting it.

%% file: ch04/ch04.tex
%chapter ch04.tex ------- The $L - \sigm$ relation for massive bursts of star formation 
\chapter {The $L - \sigma$ Relation for Massive Bursts of Star Formation}
\label{ch:4}

\epigraph{\emph{It is not knowledge, but the act of learning, not possession but the act of getting there, which grants the greatest enjoyment.}}{--- \textup{C. F. Gauss}, Letter to F. Bolyai (1808)}

\lettrine{T}{o build} a robust model of
the Universe it is necessary not only to set the strongest possible constraints on the cosmological parameters, applying joint analyses of a variety of distinct methodologies, but also  to confirm the results through
extensive consistency checks, using independent measurements and
different methods, in order to identify and remove possible
systematic errors, related to either the methods themselves or the tracers used.

It is accepted  that young massive star clusters, like those responsible for the ionisation in  giant extragalactic \hii\ regions (GEHR)
and \hii\ galaxies display a correlation between the luminosity and the width of their emission lines, the $L(\mathrm{H}\beta)-\sigma$
relation \citep{Terlevich1981}.
The scatter in the  relation is small enough that it can be used to determine cosmic distances independently of redshift 
\citep{Melnick1987, Melnick1988, Siegel2005, Bordalo2011, Plionis2011, Chavez2012, Chavez2014}.

Recently \citet{Bordalo2011} have explored the  $L(\mathrm{H}\alpha) - \sigma$ correlation and its
systematic errors using a  nearby sample selected from the
\citet{Terlevich1991} spectrophotometric catalogue of \hii\ galaxies ($0 \lesssim  z
\lesssim  0.08$). They conclude that considering only the objects with
clearly gaussian profiles in their emission lines, they obtain something close to an $ L(\ha)
\propto \sigma^4$ relation  with an rms scatter of $\delta \logd
L(\ha) \sim 0.30$.  
 It is important to emphasise that the observed properties of \hii\ galaxies, in
particular the derived \lsig\ \footnote{L(H$\beta$)  is related to L(H$\alpha$) by the theoretical Case B recombination ratio = 2.86.} relation, are mostly those of the young burst 
and not those of the parent galaxy.
This is particularly true if one selects  those systems with the
largest equivalent width (EW) in their emission lines, i.e. EW(H$\beta)>50$\AA\,. 
The selection of those \hii\ galaxies having the strongest emission lines
minimises the evolutionary effects in their luminosity \citep*{Copetti1986},  which would introduce a systematic shift in the \lsig\ relation due to the rapid drop of the ionising flux after 5 Myr of evolution. This  selection  minimises also any possible contamination in the observable due to the stellar populations of the parent galaxy.

 A feature of the \hii\ galaxies optical spectrum, their
strong and narrow emission lines,   makes them readily observable with present instrumentation out to $z\sim 3.5$.
Regarding such distant systems,   \citet{Koo1995} and also  \citet{Guzman1996}
have shown that a large fraction 
of the numerous compact star forming galaxies found at intermediate redshifts have kinematical properties similar to 
those of luminous local \hii\ galaxies. They exhibit fairly narrow emission line widths ($\sigma$ from 30 to 150 km/s) rather than the 200 km/s typical for 
galaxies of similar luminosities. In particular galaxies with $\sigma < $ 65 km/s seem to follow the same relations in $\sigma$, M$_B$ and $L(\mathrm{H}\beta)$ 
as the local ones. 

From spectroscopy of Balmer emission lines in a few Lyman break galaxies at $z\sim$ 3  \citet{Pettini1998} suggested that these systems adhere to the 
same relations but  that the conclusions had to be confirmed for a larger sample. 
These results opened the important possibility of applying the distance estimator and mapping 
the Hubble flow up to extremely high redshifts and simultaneously to study the behaviour of starbursts of similar luminosities over a very large redshift range.

Using a  sample of intermediate and high redshift  \hii\ galaxies \citet*{Melnick2000}  investigated the use of  the \lsig\ correlation as a high-$z$ distance indicator. They  found a 
good correlation between the luminosity and velocity dispersion confirming that the  \lsig\  correlation for local \hii\ galaxies is valid up to  $z\sim$3.
 Indeed, our group \citep{Plionis2011} 
showed that the \hii\ galaxies $L(\mathrm{H}\beta)-\sigma$ relation constitutes a viable alternative cosmic probe to SNe~Ia. We also presented a general strategy to use \hii\ galaxies to trace the high-$z$ Hubble expansion in order
to put stringent constraints on the dark energy
equation of state and test its possible evolution with redshift.
A first attempt by \citet{Siegel2005}, using a sample of 15 high-$z$ \hii\ galaxies ($2.1<z<3.4$), selected as in \citet{Melnick2000}, with the original \lsig\ calibration of \citet{Melnick1988},  found a mass content of the universe of
 $\Omega_{m} =0.21_{-0.12}^{+0.30}$ for a flat $\Lambda$-dominated universe. 
Our recent reanalysis of the \citet{Siegel2005} sample 
\citep{Plionis2011}, using a
revised zero-point of the original $L(H\beta)-\sigma$ relation,
provided a similar value of $\Omega_m=0.22^{+0.06}_{-0.04}$  but with substantially smaller errors \citep[see also][]
{Jarosik2011}.

Recapitulating, we reassess the \hii\ galaxies \lsig\ relation using new data obtained with modern instrumentation with the aim of reducing the impact of observational random and systematic errors onto the \hii\ galaxies Hubble diagram. To
achieve this goal, we selected from the SDSS catalogue a sample of 128 local
($z<0.2$), compact \hii\ galaxies with the highest equivalent width of their
Balmer emission lines. We obtained  high S/N high-dispersion echelle 
spectroscopic data with the VLT  and Subaru
telescopes to accurately measure the ionized gas velocity dispersion.
We also obtained integrated H$\beta$ fluxes using low
dispersion wide aperture spectrophotometry from the 2.1m telescopes at
Cananea and San Pedro M\'artir in Mexico, complemented with data from the  SDSS
spectroscopic survey.

This chapter follows closely our paper \citet{Chavez2014}, its layout is as follows: we describe the sample selection
procedure in \S 4.1, observations and data reduction in \S 4.2;
an analysis in depth of the data error budget (observational and
systematic) and  the  method for analysing  the data are discussed in \S 4.3.
The effect that different intrinsic physical parameters of the star-forming regions could 
have on the  \lsig\  relation is studied in \S4.4. The results for the
\lsig\ relation is presented in \S 4.5, together with possible second parameters and  systematic effects. Summary and conclusions are given in \S 4.7.

\section{Sample Selection}
\label{sec:4_1}
\begin{figure}
\centering
\resizebox{8.4cm}{!}{\includegraphics{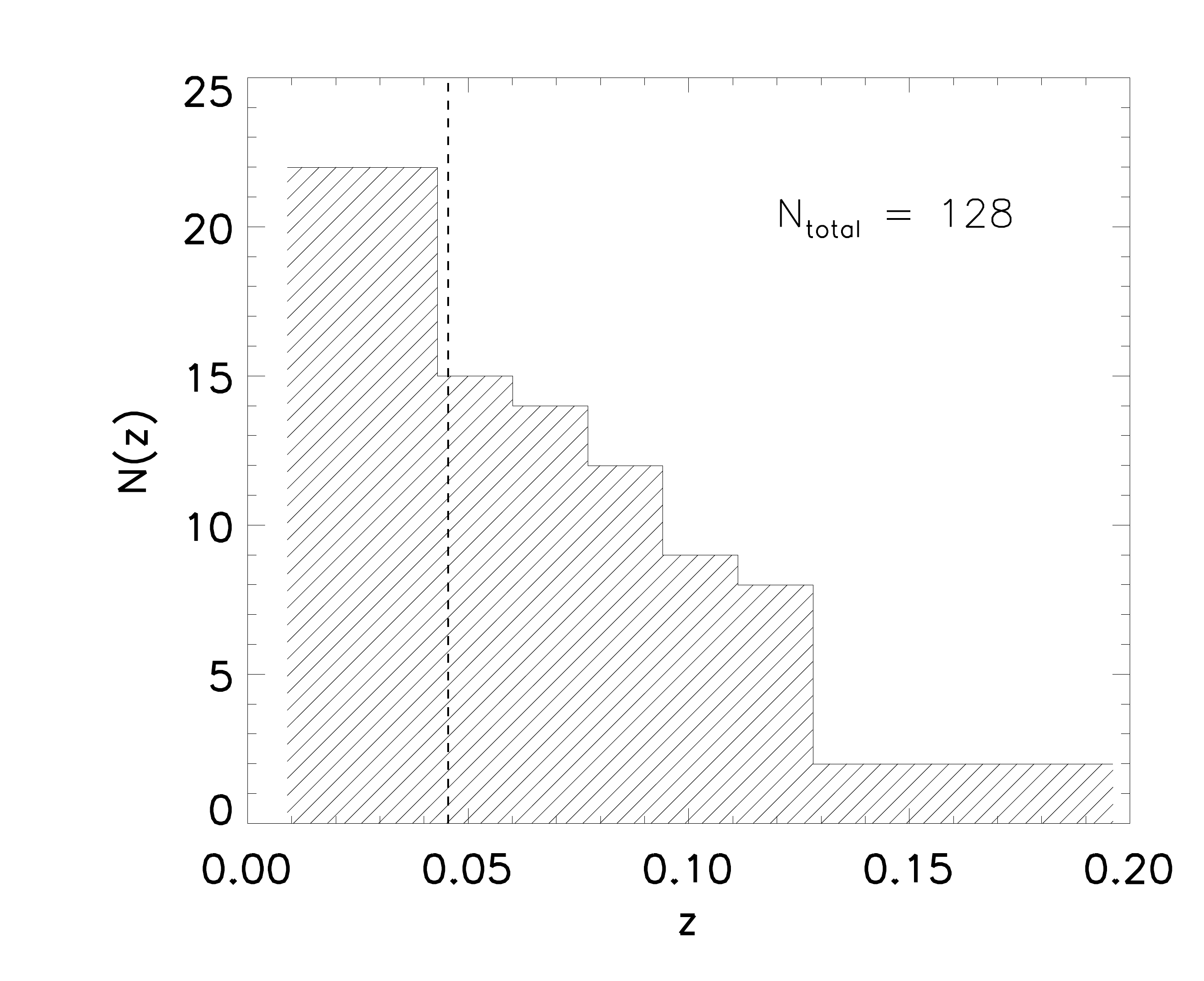}}
\caption[Redshift distribution of the sample] {\small Redshift distribution of the sample. The dashed line marks the median.}
\label{fig:HRshn}
\end{figure}

We observed 128 \hii\ galaxies selected from the
 SDSS DR7 spectroscopic catalogue \citep{Abazajian2009} for having the  strongest  emission
lines relative to the continuum (i.e. largest equivalent widths) and in the redshift range $0.01 < z
< 0.2$. The lower redshift limit  was selected to avoid nearby objects that are more affected by local peculiar
motions relative to the Hubble flow and the upper limit was set to minimize the cosmological
non-linearity effects. Figure \ref{fig:HRshn} shows the redshift
distribution for the sample. The median of the distribution is also
shown as a dashed line at $z \sim 0.045$, the corresponding recession
velocity is $\sim 13500\ \mathrm{km\ s^{-1}}$.

\begin{figure}
\centering
\resizebox{8.4cm}{!}{\includegraphics{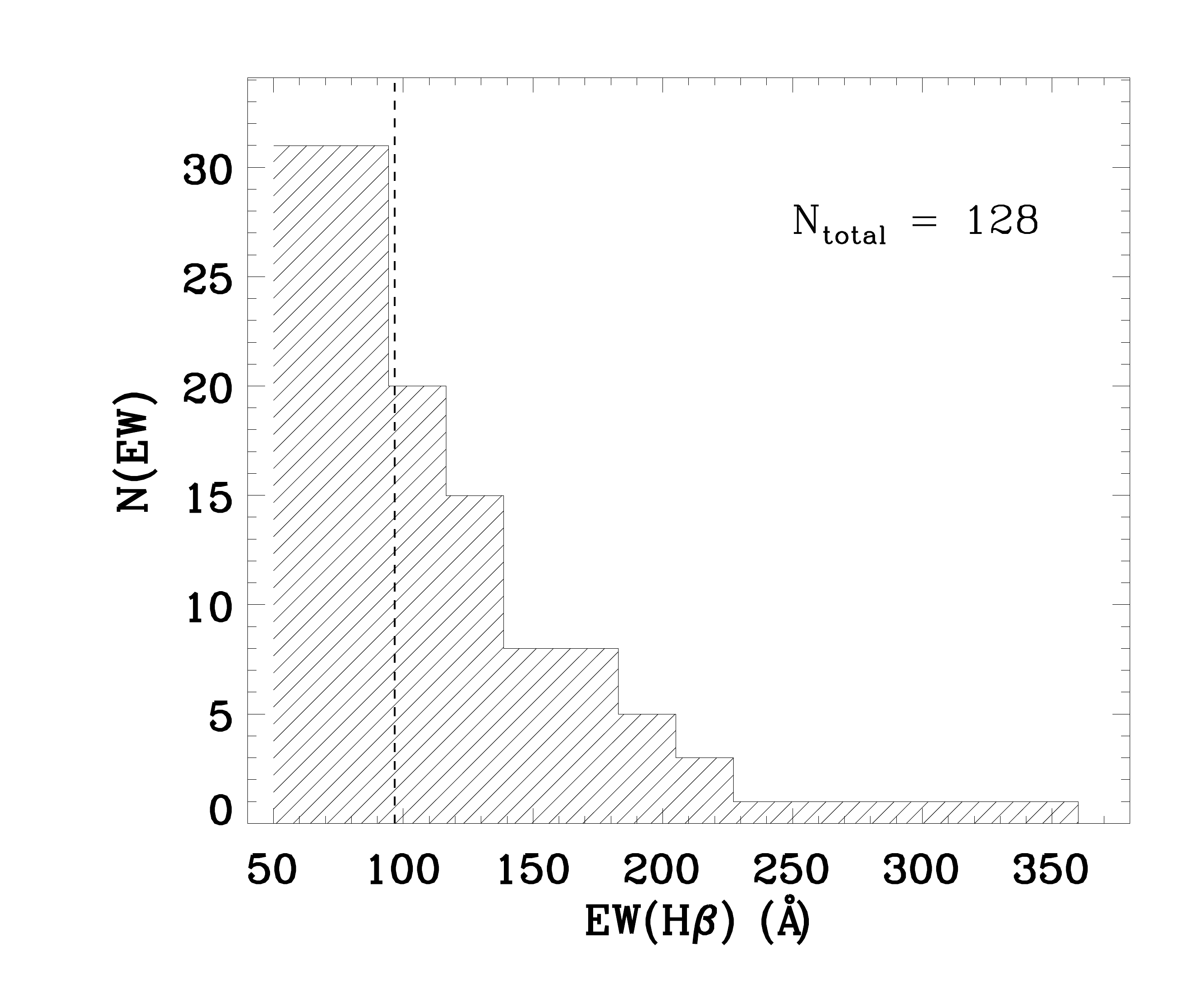}}
\caption [$\hb$ equivalent width distribution for the  sample.]{\small $\hb$ equivalent width distribution for the  sample.
 The dashed line marks the median.}
\label{fig:HW}
\end{figure}

Only those \hii\ galaxies with the largest equivalent width in their
$\hb$ emission lines, $EW(\mathrm{H}\beta) > 50\ \mathrm{\AA}$ were
included in the sample. This relatively high lower limit in the observed equivalent width 
of the recombination hydrogen lines is of fundamental importance 
to guarantee that the sample is composed by systems in which
a single very young starburst dominates the total luminosity. 
This selection criterion also minimizes the posible contamination due to an
underlying older population or older clusters inside the spectrograph
aperture \citep[cf.][]{Melnick2000, Dottori1981, Dottori1981b}. 
Figure \ref{fig:HW} shows the  $EW(\mathrm{H}\beta)$ distribution for the
sample; the dashed line marks the median of the distribution, its
value is $EW(\hb) \sim 87\ \mathrm{\AA}$.  

\begin{figure}
\centering
\resizebox{8.4cm}{!}{\includegraphics{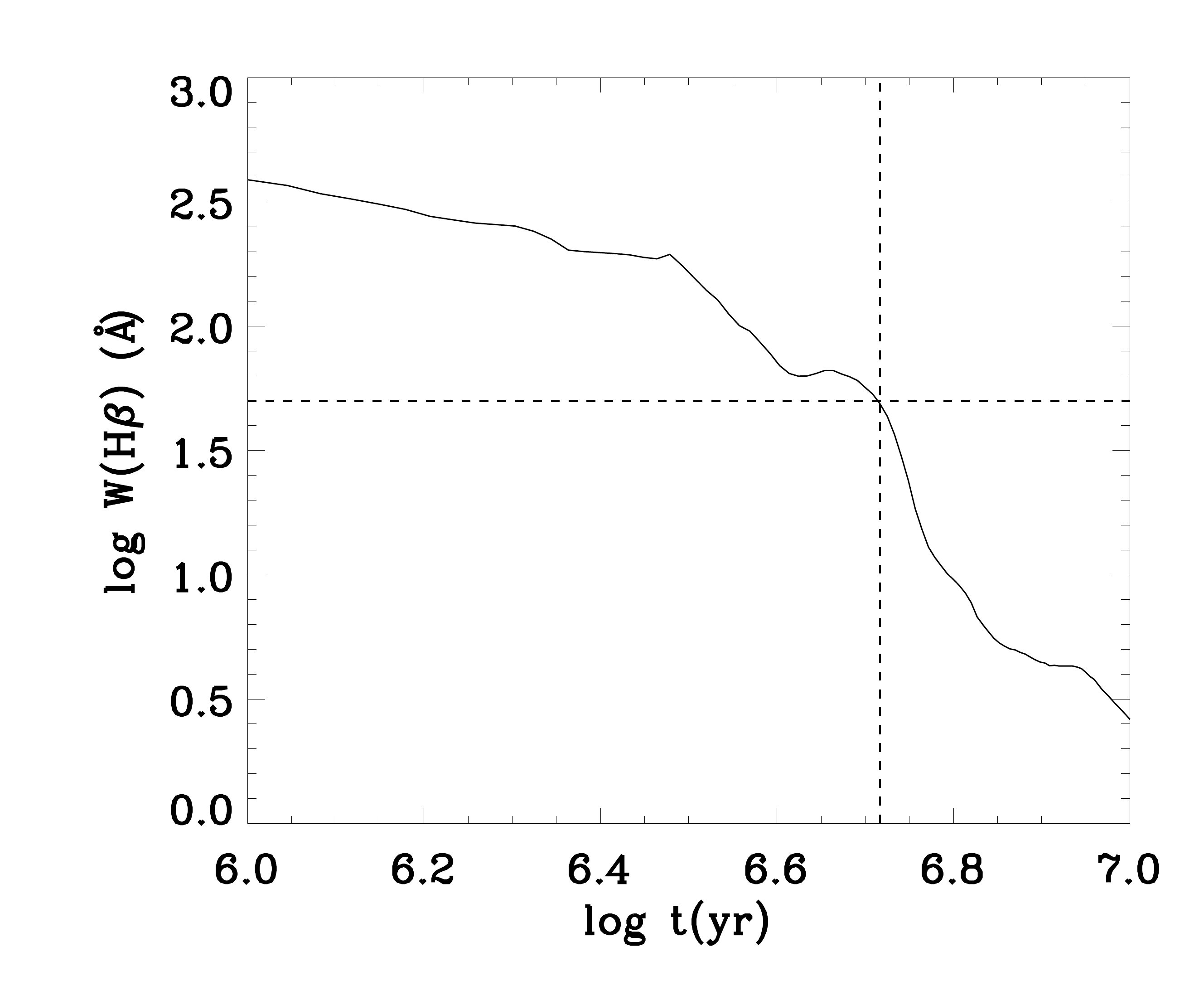}}
\caption [Evolution of the $\hb$ equivalent width for an instantaneous starburst]{\small The evolution of the $\hb$ equivalent width for an instantaneous burst
with metallicity Z$=0.004$ and a Salpeter IMF with upper limit of 100 \Msol\ \citep[]{Leitherer1999}. The horizontal
line marks  the $\hb$ equivalent width of 50 \AA , while the vertical line indicates 
the corresponding age of $\sim$ 5 Myrs.
}
\label{fig:WEv}
\end{figure}

\emph{Starbusrt99} \citep[][SB99]{Leitherer1999} models indicate that 
an instantaneous burst with  $EW(\mathrm{H}\beta) > 50\ \mathrm{\AA}$
and Salpeter IMF has to be younger than about 5 Myr (see Figure \ref{fig:WEv}). This is a strong upper limit because in the case that part of the continuum is produced by an underlying older stellar population, the derived cluster age will be even smaller.

%%%%%%%%%%%% figure 1 %%%%%%%%%%%%%%%%%%%%%%%%
%Figure --- Redshift Distribution
%%%%%%%%%%%%%%%%%%%%%%%%%%%%%%%%%%%%

%%%%%%%%%%%%%  figure 2 %%%%%%%%%%%%%%%%%%%%%%%
%Figure --- EW Distribution
%%%%%%%%%%%%%%%%%%%%%%%%%%%%%%%%%%%%

%%%%%%%%%%%%%%%  figure 3 %%%%%%%%%%%%%%%%%%%%%
%Figure --- EW Evolution
%%%%%%%%%%%%%%%%%%%%%%%%%%%%%%%%%%%%

The sample is also flux limited as it was selected from SDSS for having an $\hb$ line core
$\mathrm{h_{c}}(\hb) > 100\times10^{-17}\ \mathrm{erg\ s^{-1}
  cm^{-2} \AA^{-1}}$. To discriminate against  high velocity
dispersion objects and also to avoid those that are dominated by
rotation, we have selected only those objects with $0.7 <
\sigma(\mathrm{H}\beta) < 2.0\ \mathrm{\AA} $. From the values of the
line core and $\sigma$ of the $\hb$ line we can calculate that the
flux limit in the $\hb$ line is $F_{lim}(\hb) \sim 5 \times 10^{-15}\
\mathrm{erg\ s^{-1}\ cm^{-2}}$ which corresponds to an emission-free continuum magnitude 
of $m_{B, lim} \simeq 19.2$ [cf. \citet*{Terlevich1981} for the conversion].

\begin{figure*}
\centering
\resizebox{14cm}{!}{\includegraphics{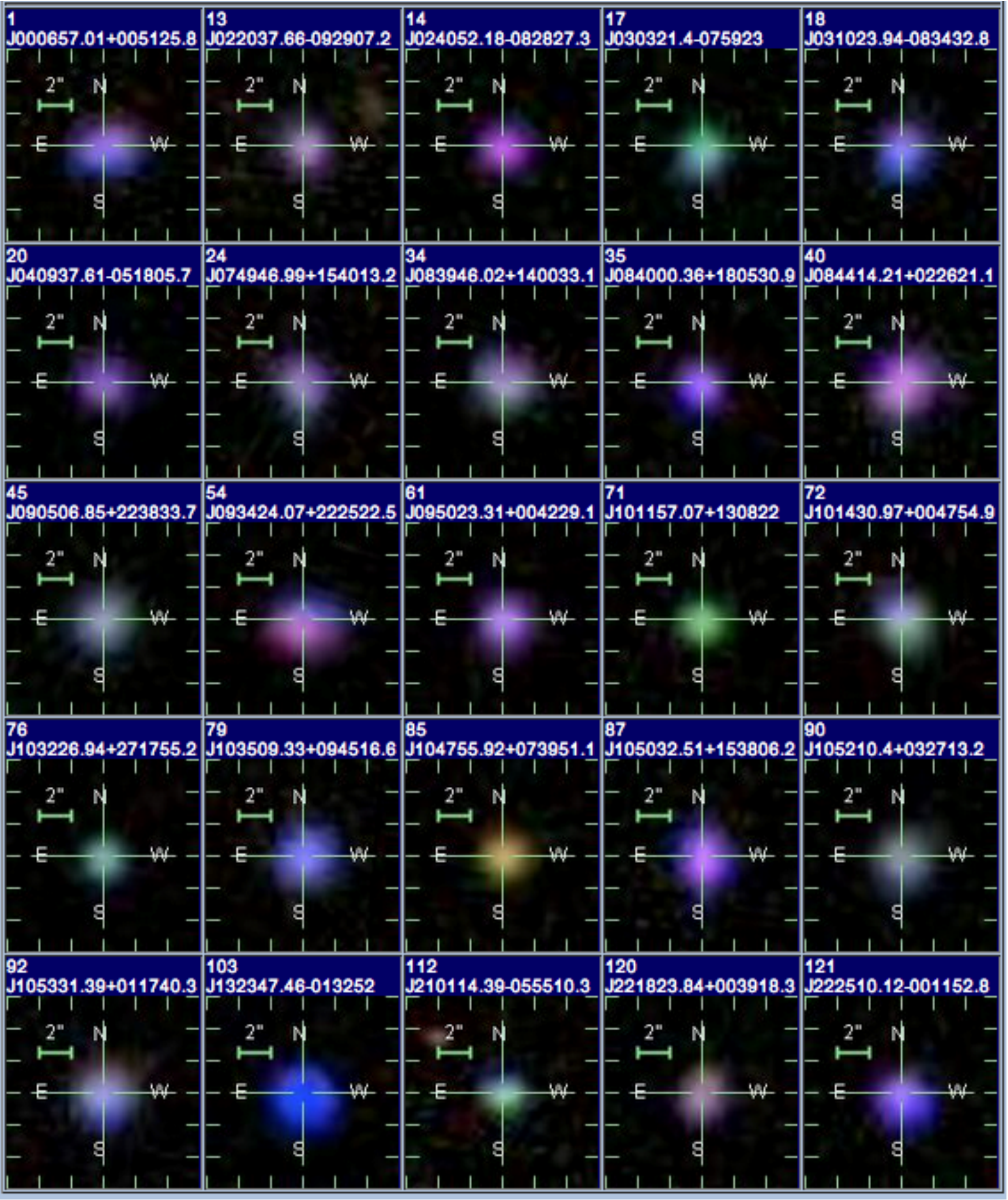}}
\caption[Selection of colour images of \hii\ galaxies] {\small A selection of colour images of \hii\ galaxies from our sample. The SDSS name and our index number are indicated in the
  stamps. The changes in colour are related to the redshift of the object}
\label{fig:ms}
\end{figure*}

To guarantee the best integrated spectrophotometry, only objects
with Petrosian diameter less than 6\arcsec \ were selected. In addition a
visual inspection of the SDSS images was performed to avoid systems composed of   
multiple  knots or extended haloes. Colour images  from SDSS for a subset of objects in the sample are shown
 in Figure \ref{fig:ms}. The range in colour is related to the redshifts span of the 
 objects and is due mainly to the dominant  [OIII]$\lambda \lambda $4959,5007 doublet moving 
 from the g to the r SDSS filters and to the RGB colour definition. The compactness of the
 sources can be appreciated in the figure.

%%%%%%%%%%%%%%%%%%%% Figure 4 %%%%%%%%%%%%%%%%%%%%%%%%%%
%Figure --- Mosaic of SDSS images 
%%%%%%%%%%%%%%%%%%%%%%%%%%%%%%%%%%%%

\section{Observations and Data Reduction}

The data required for determining the  \lsig\ relation are of two kinds:
\begin{enumerate}
 \item Wide slit low resolution spectrophotometry to obtain accurate
   integrated emission line fluxes.

 \item High resolution spectroscopy to measure the velocity
   dispersion from the $\rm{H}\beta$ and [OIII] line profiles. Typical
   values of the FWHM range  from 30 to about 200$\ \rm{km\ s^{-1}}$.
\end{enumerate}
 
 A journal of observations is given in Table \ref{tab:tab01} where column (1) gives
the observing  date, column (2) the telescope, column (3) the
instrument used, column (4) the detector and column (5) the  projected slit
width in arc seconds.

%Table --- Journal of Observations
\input{ch04/tab/tab01}

\subsection{Low resolution spectroscopy}
The low resolution spectroscopy for the line fluxes was performed with two identical
Boller \& Chivens 	Cassegrain spectrographs  (B\&C) in long slit
mode  at similar 2 meter class telescopes, one of them at the
Observatorio Astron\'omico Nacional (OAN) in San Pedro M{\'a}rtir
(Baja California) and the other one at the Observatorio Astrof\'isico
Guillermo Haro (OAGH) in Cananea (Sonora) both in M{\'e}xico. 

The observations at OAN were performed using a  $600\ \mathrm{gr\
  mm^{-1}}$ grating with a blaze angle of $8^{\circ}38'$. The grating was
centred at $\lambda\sim 5850\mathrm{\AA}$ and the slit width was 10\arcsec
. The resolution obtained with this configuration is $R \sim 350$
($\sim$ 2.07 \AA / pix) and the spectral coverage is $\sim 2100\
\mathrm{\AA}$. The data from OAGH was obtained using a  $150\
\mathrm{gr\ mm^{-1}}$ grating with a blaze angle of $3^{\circ}30'$
centred at $\lambda\sim 5000\mathrm{\AA}$.  With this configuration and a
slit width of 8.14\arcsec, the spectral resolution  is  $R
\sim 83$ ($\sim$ 7.88 \AA / pix).

At least four observations of three spectrophotometric standard stars were performed each night. 
Futhermore,
to secure the photometric link between different nights at least one \hii\ galaxy was repeated
every night during each run. All objects were observed at
small zenith distance, but for optimal determination of the atmospheric extinction the first
and the last standard stars  of the night were also observed at high zenith distance.

The wide-slit spectra obtained at OAN and OAGH were reduced using  standard IRAF\footnote{IRAF 
is distributed by the National Optical Astronomy Observatory, which is operated by the
  Association of Universities for Research in Astronomy, Inc., under
  cooperative agreement with the National Science Foundation.} tasks. 
The reduction procedure entailed the following steps: (1)  bias, flat field and  
cosmetic corrections, (2) wavelength calibration, 
(3) background subtraction, (4) flux calibration
and (5) 1d spectrum extraction. The 
spectrophotometric standard stars for each
night were selected among  $\mathrm{G191-B2B}$, Feige 66, Hz
44, $\mathrm{BD+33d2642}$, GD 50, Hiltner 600, HR 3454, Feige 34 and
GD 108.

We complemented our own wide-slit spectrophotometric observations with
the  SDSS DR7 spectroscopic data when available. SLOAN spectra are obtained with 3\arcsec 
diameter fibers, covering a range from
$3200 - 9200\ \mathrm{\AA}$ and a resolution $R$ of $1850  - 2200$. The comparison between 
our own and SDSS spectrophotometry is discussed later on in \S \ref{sec:elf}.

\subsection{High resolution spectroscopy}\label{sec:hrss}
High spectral resolution spectroscopy for the line widths was obtained using echelle 
spectrographs at 8 meter class telescopes.
The telescopes and instruments used are the Ultraviolet and Visual
Echelle Spectrograph (UVES) at the European Southern Observatory (ESO)
Very Large Telescope (VLT) in Paranal, Chile, and the High Dispersion
Spectrograph (HDS) at the National Astronomical Observatory of Japan
(NAOJ) Subaru Telescope in Mauna Kea, Hawaii (see
Table \ref{tab:tab01} for the journal of observations). 

UVES is a two-arm cross-disperser echelle spectrograph  located at the
Nasmyth B focus of ESO-VLT Unit Telescope 2 (UT2; Kueyen)
\citep{Dekker2000}. The spectral range goes from $3000\ \mathrm{\AA}$
to $11000\ \mathrm{\AA}$. The maximum spectral resolution is $80000$
and $110000$ in the blue and red arm respectively. We used the red arm
($31.6\ \mathrm{gr\ mm^{-1}}$ grating, $75.04^{\circ}$ blaze angle)
with cross disperser 3 configuration ($600\ \mathrm{gr\ mm^{-1}}$
grating)  centred at $5800\ \mathrm{\AA}$. The width of the slit was
2\arcsec, giving a spectral resolution of $\sim 22500$ (0.014 \AA /pix). 

HDS is a high resolution cross-disperser echelle spectrograph located
at the optical Nasmyth platform of NAOJ-Subaru Telescope
\citep{Noguchi2002, Sato2002}. The instrument covers from $3000\
\mathrm{\AA}$ to $10000\ \mathrm{\AA}$. The maximum spectral
resolution is $160000$. The echelle grating used has $31.6\
\mathrm{gr\ mm^{-1}}$ with a blaze angle of $70.3^{\circ}$. We used
the  red  cross-disperser  ($250\ \mathrm{gr\ mm^{-1}}$ grating,
$5^{\circ}$ blaze angle) centred at $\sim 5413\ \mathrm{\AA}$ and a
slit width of  4\arcsec, that provided a spectral resolution of $\sim 9000$
(0.054 \AA / pix).

%%%%%%%%%%% figure 5 %%%%%%%%%%%%%%%
%Figure --- Mosaic of spectra
\begin{figure*}
\centering
%\resizebox{15cm}{!}{\includegraphics{ch04/fig/compspec.pdf}}
\subfloat{\includegraphics[width=15cm]{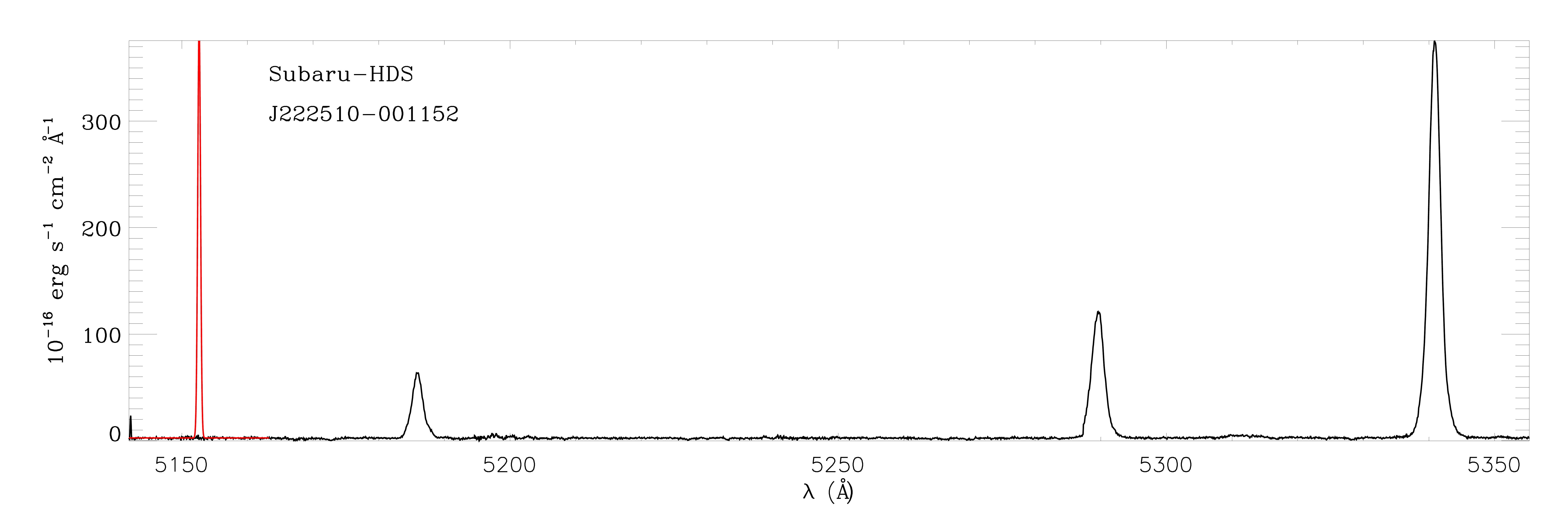}}\\
\subfloat{\includegraphics[width=15cm]{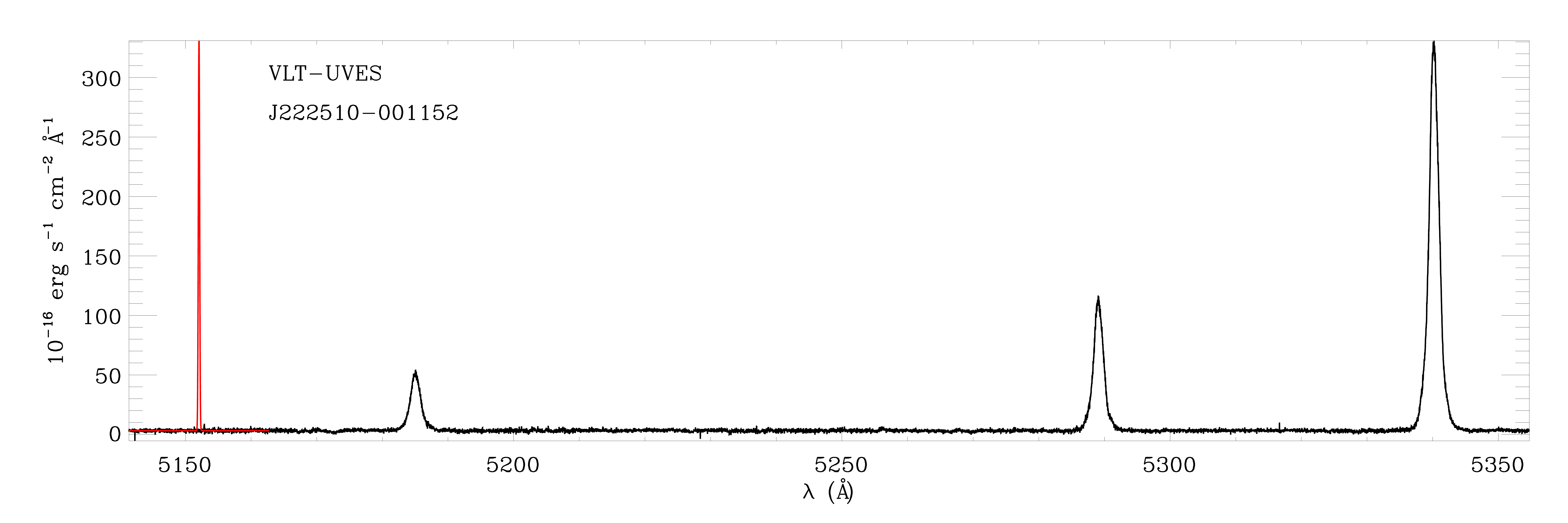}}

\caption[Examples of the high dispersion spectra] {\small Examples of the high dispersion spectra obtained for the same object with Subaru HDS (top) and VLT 
UVES (bottom), 
showing the region covering $\hb\ $ and the [OIII] lines  at $\lambda\lambda$ 4959,5007 \AA . 
The instrumental profile is shown in red at the left of each spectrum. }
\label{fig:CSpec}
\end{figure*}
%%%%%%%%%%%%%%%%%%%%%%%%%%

57 objects were observed with UVES and 76 with HDS. Five of them were 
observed with both instruments. During the UVES observing run 16 objects were observed more than once 
(three times for four objects and four times for another one) in order to estimate better the
observational errors, and to link the different nights of the run. Two objects were observed twice
with the HDS. The five galaxies observed at both telescopes also served as a link between the 
observing runs and to compare the performance of both telescopes/instruments and the quality of the nights.

Similarly, 59  sources were observed at OAGH and 59 at OAN, of which 15
were observed at both telescopes.

The UVES data reduction was carried out using the UVES pipeline V4.7.4
under the GASGANO V2.4.0 environment\footnote{GASGANO is a JAVA based
  Data File Organizer developed and maintained by ESO.}. The reduction
entailed the following steps and tasks: (1) master bias generation 
(\verb|uves_cal_mbias|), (2) spectral orders reference table
generation (\verb|uves_cal_predict| and
\verb|uves_cal_orderpos|), (3) master flat generation 
(\verb|uves_cal_mflat|), (4) wavelength calibration 
(\verb|uves_cal_wavecal|), (5) flux calibration 
(\verb|uves_cal_response|) and (6) science objects reduction 
(\verb|uves_obs_scired|).

The HDS data were reduced using  IRAF 
packages and a script for overscan removal and detector linearity
corrections provided by the NAOJ-Subaru telescope team. The reduction
procedure entailed the following steps: (1) bias subtraction, (2)
generation of spectral order trace template, 
 (3) scattered light removal, (4) flat fielding, (5) 1d spectrum extraction 
and (6) wavelength calibration.

Typical examples of the high dispersion spectra are shown in Figure
\ref{fig:CSpec}. The instrumental profile of each setup is also
shown on the left.

\section{Data Analysis.}
We have already mentioned in \S \ref{sec:4_1} that we observed 128 \hii\ galaxies with EW(H$\beta$) $> 50$ $\mathrm{\AA}$. 
From the observed sample we  have removed 13 objects which
presented problems in the data (low S/N) or showed evidence for a prominent underlying Balmer absorption. We also removed an extra object that
presented highly asymmetric emission lines. After this we were left with 114 objects that  comprise our 
`initial' sample (S2).

It was shown by \citet{Melnick1988} that imposing an 
upper limit  to the velocity dispersion such as $\log \sigma($H$\beta) < 1.8$ $\mathrm{km\ s^{-1}}$, 
minimizes the probability of including  rotationally  supported systems and/or objects with 
multiple young ionising clusters  contributing to the total flux and 
affecting the line profiles.
Therefore from S2 we selected all  objects having $\log \sigma($H$\beta) < 1.8$ $\mathrm{km\ s^{-1}}$ thus creating 
sample S3 -- our `benchmark' sample -- composed of 107 objects. 

A summary of the characteristics of the subsamples used in this chapter can be found in Table  \ref{tab:tab10} and is further discussed in section 6.
Column (1) of Table \ref{tab:tab10} gives the reference name of the sample, column (2) lists its descriptive name, 
column (3) gives the constraints that led to the creation of the subsample and column (4) gives the number of objects left in it.

%%Table  10 --- H II galaxies samples %%%%%%%%%%%%%%%%%%%%%%%%%%%%%%%%%%%%%
\input{ch04/tab/tab10}
%%%%%%%%%%%%%%%%%%%%%%%%%%%%%%%%%%%%%

\subsection{Emission line fluxes.}\label{sec:elf}
Given the importance of accurate measurements for our results, we will describe in  
detail our methods.

Total flux and equivalent width of the strongest emission lines were measured from our  
low dispersion wide-slit spectra. Three methods were used, we
have obtained the  total flux and equivalent width from  single gaussian fits  to the line 
profiles using both the IDL routine \verb|gaussfit| and the IRAF 
task  \verb|splot|, and we also measured the fluxes integrated under the line, in order to have
a measurement independent of the line  shape.

\begin{figure}
\centering
\resizebox{8cm}{!}{\includegraphics{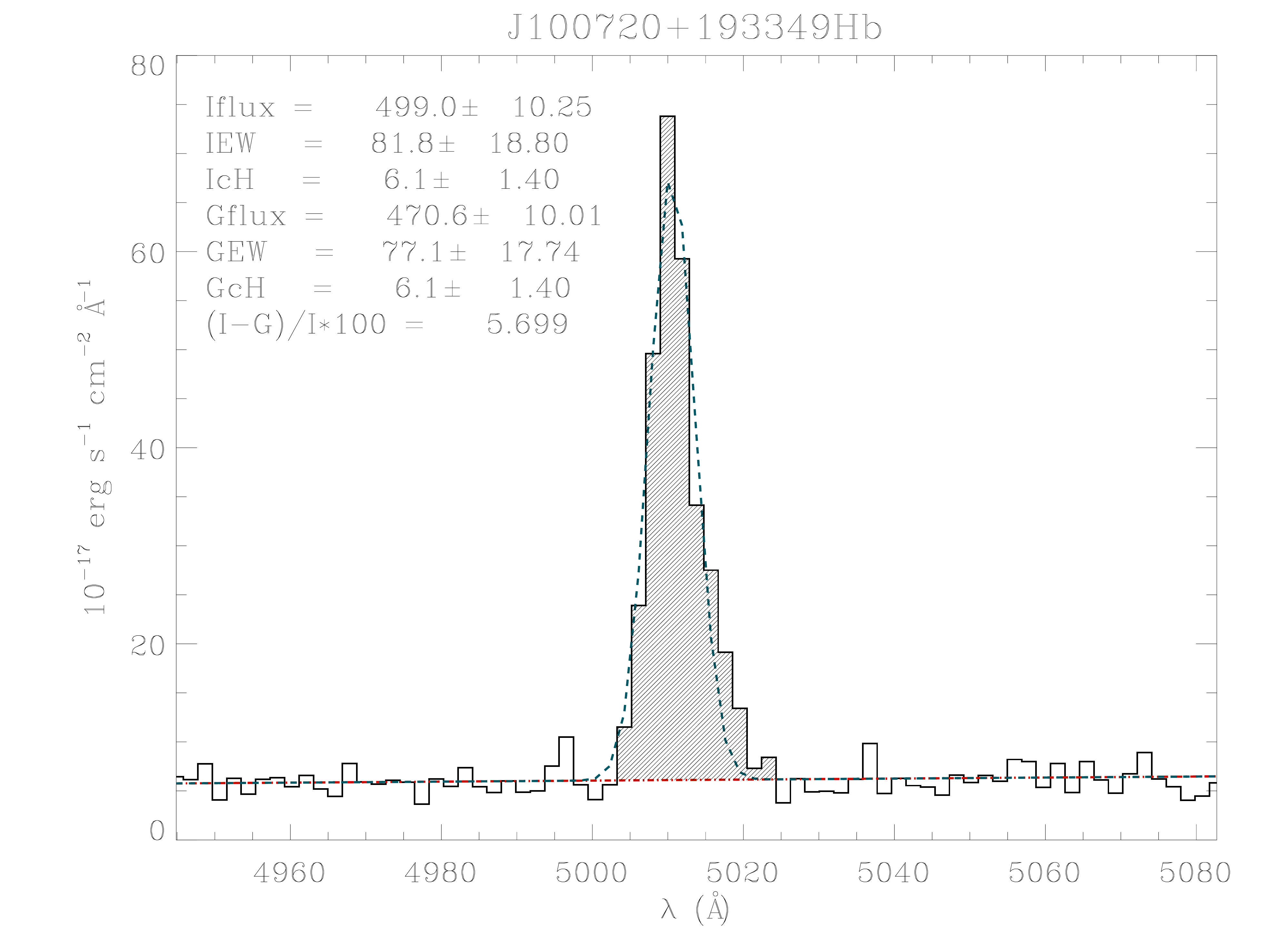}}
\caption[An example  of gaussian fit and integration under the emission line] {\small An example  of gaussian fit (dashed line) and integration under the line
  (shaded area) for an $\hb$ line from the low dispersion data. The
  parameters for both fits are shown in the inset.}
\label{fig:LSfit}
\end{figure}

Figure \ref{fig:LSfit} shows a
gaussian fit and the corresponding integrated flux measurement for an
$\hb$ line from our low dispersion data. It is clear from the figure
that in the cases when the line is asymmetric, the gaussian fit would not provide
a good estimate of the actual flux. In the example shown the
difference between the gaussian fit and the integration is  $\sim
5.7\%$ in flux.

Table \ref{tab:tab03} 
shows the results of our wide-slit  low resolution
spectroscopy measurements. The data listed have not been corrected for internal 
extinction. Column (1) is our  index number, column (2) is the SDSS name,
column (3) is the integrated $\hb$ flux measured by us from  the SDSS published spectra,  
columns (4) and  (5) are the $\hb$ line fluxes as measured from a gaussian fit to the 
emission line and integrating the line respectively, columns (6) and (7) are the 
$[\Oiii]\ \lambda \lambda 4959$  and $5007$ line fluxes measured from a gaussian fit, column (8) gives the EW of the $\hb$ line as measured from the SDSS spectra and 
column (9) is a flag that indicates the origin of the data and is described in the table
caption. 

\begin{figure}
\centering
\resizebox{8cm}{!}{\includegraphics{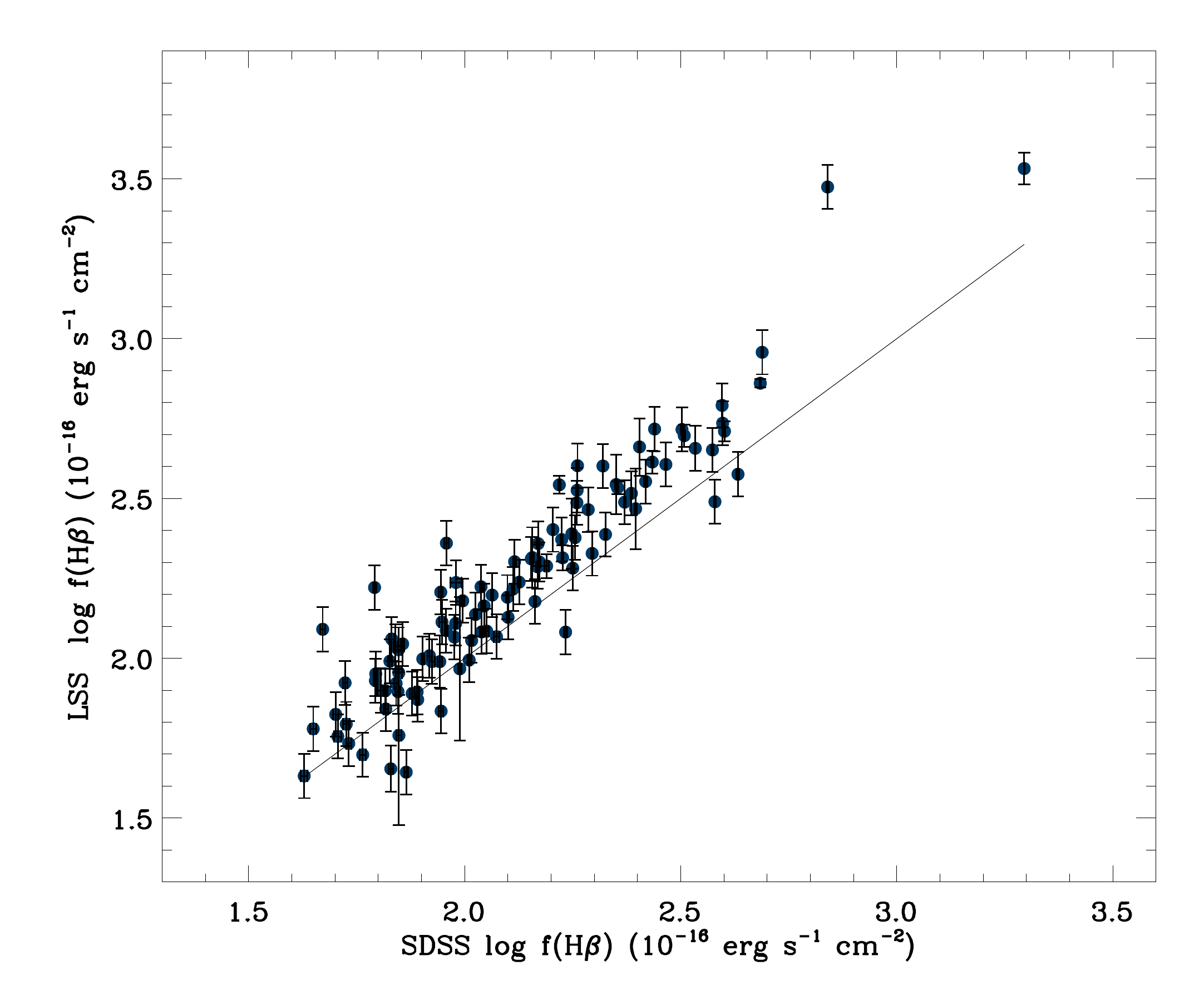}}
\caption[Fluxes measured from SDSS] {\small Fluxes measured from SDSS spectra compared with those
  measured from our low dispersion spectra (LS),                          
 the line shows  the one-to-one correspondence.}
\label{fig:SDSSvLS}
\end{figure}

Figure
\ref{fig:SDSSvLS} shows the comparison between 
SDSS and our low resolution spectra. Clearly  most of the
objects show an excess  flux in our data
which could  easily be explained as an aperture effect, as the 
 $3 \arcsec$ diameter fiber of SDSS in many
cases does not cover all the object whereas our spectra were taken
with apertures of $8\arcsec - 13 \arcsec$ in width, hence covering the entire
compact object in all cases. 

%%%%%%%%%%%%%%%%%%%%%%%%%%%%%%% Figure 6 %%%%%%%%%%%%%%%%%%%%%%%%%%%%%
%Figure --- Hb LS data fit 

%%%%%%%%%%%%%%%% Figure 7 %%%%%%%%%%%%%%%%%%%%
%Figure --- SDSS vs LS spectra 

Fluxes and equivalent
widths of $[\Oii]\ \lambda \lambda 3726,3729$, $[\Oiii]\ \lambda\lambda 4363,4959,5007$, 
$\hg$, $\ha$, $[\Nii]\ \lambda \lambda
6548,6584$ and  $[\Sii]\ \lambda \lambda 6716,6731$ were also measured from the SDSS spectra
when available.
We have fitted single gaussians to the line profiles using both
the IDL  routine  \verb|gaussfit| and the IRAF task  \verb|splot|  and,
when necessary, we have de-blended lines by multiple gaussian
fitting. 

Table \ref{tab:tab04} 
shows the results for the SDSS spectra line
flux measurements as intensity  relative to $\hb = 100$. Columns are:
(1) the index number, (2) the SDSS name, (3) and
(4) the intensities of  $[\Oii]\ \lambda  3726$ and $\lambda 3729$,
(5), (6) and (7) the intensities  of  $[\Oiii]\ \lambda
4363, \lambda 4959$ and $\lambda 5007$, (8)   $\hg$  intensity, (9)
 $\ha$ intensity, (10) and (11) are the intensities of
$[\Nii]\ \lambda 6548$ and $\lambda 6584$ and (12) and (13) the
intensities  of  the $[\Sii]\ \lambda 6716$ and $\lambda 6731$  lines. The
values given are as measured, not corrected for extinction.  The 1$\sigma$ uncertainties for the fluxes are given in percentage.

In all cases, unless otherwise stated in the tables, the  uncertainties and equivalent flux of the lines have
been estimated from the expressions  \citep{Tresse1999}:
\begin{eqnarray}
 \sigma_{F} &=& \sigma_{c} D \sqrt{2 N_{pix} + EW/D}, \\
 \sigma_{EW} &=&\frac{EW}{F} \sigma_{c} D \sqrt{EW/D + 2 N_{pix} + (EW/D)^{2}/N_{pix}},
\end{eqnarray}
where $\sigma_{c}$ is the mean standard deviation per pixel of the
continuum at each side of the line, $D$ is the spectral dispersion in
$\mathrm{\AA\ pix^{-1}}$, $N_{pix}$  is the number of pixels covered
by the line, $EW$ is the line equivalent width in $\mathrm{\AA}$, $F$
is the flux in units of $\mathrm{erg\ s^{-1}\ cm^{-2}}$. When more
than one observation was available, the 1$\sigma$ uncertainty was 
given as the standard deviation of the individual determinations. 

\begin{figure}
\centering
\resizebox{8cm}{!}{\includegraphics{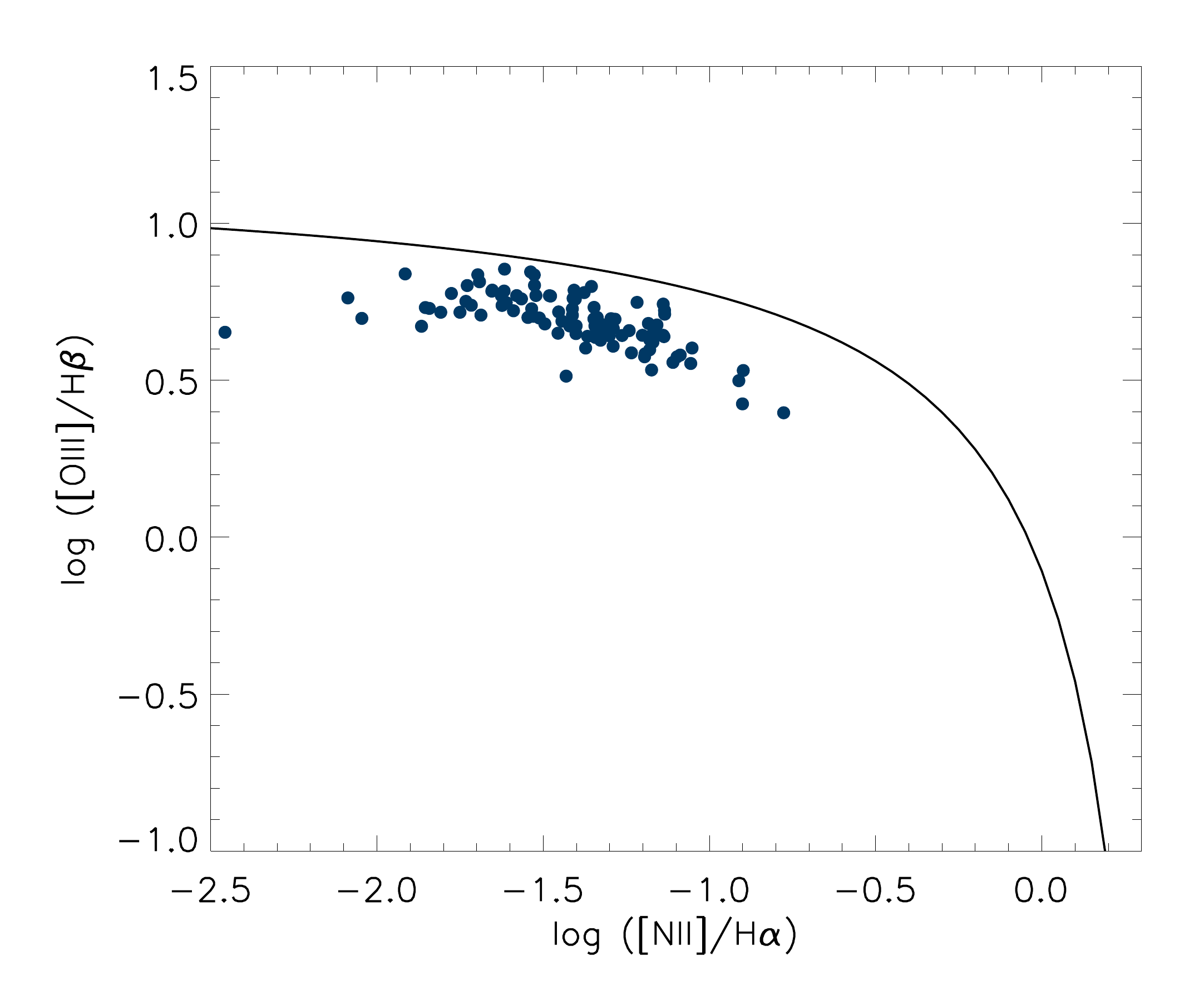}}
\caption[BPT diagram of a sample of \hii\ galaxies] {\small BPT diagram showing the high excitation level of a sample of \hii\ galaxies selected mainly as 
having high equivalent width in their Balmer emission lines. The solid line represents the upper limit for stellar photoionization, from \citet{Kewley2001}.The plot shows 99 points from the S3 sample (see text).}

\label{fig:BPT}
\end{figure}

In order to characterise further the sample, a BPT diagram  was drawn for the 99 objects
of S3 that have a good measurement 
of  $[\Oiii] \lambda$ 5007/ $\hb$\  and  $[\Nii] \lambda$ 6584/$\ha$\  ratios.
The diagram is shown in Figure  \ref{fig:BPT} where it can be seen that clearly,
all objects are located in a narrow strip just below the transition line 
\citep{Kewley2001} indicating high excitation and suggesting low metal content and photoionisation 
by hot main sequence stars, consistent with the expectations for young \hii\ regions.

%%%%%%%%%%%%%%%%%%%%% Figure 8 %%%%%%%%%%%%%%%%%
%Figure --- BPT diagram%%%%%%%%%%%%%%%%%%%%%%%%%%%%%%

\subsection{Line profiles}
From the two dimensional  high dispersion spectra we have obtained the total flux, the position 
and  the full width at
half maximum (FWHM) of $\hb$ and  $[\Oiii]\ \lambda \lambda 4959,5007$\AA \ in each spatial increment 
 i.e.~along the slit.

These measurements were used to map the trends in intensity, position, centroid wavelength 
and FWHM of those emission lines. 
The intensity or brightness distribution across the object  provides information 
about the sizes of the line  and  continuum emitting regions.
The brightness distribution was used to determine the centroid and FWHM of the line emitting region.
On the other hand the trend in the central wavelength of the spectral profile along the
 spatial direction was used to determine the amount of rotation present. 

The trend in FWHM along the slit help us also to verify that there is no FWHM gradient across 
the object; any  important change along the slit could  affect the global measurements. 
In general it was found that the FWHM of the non-rotating systems is almost constant.
Those systems with significant gradient or change, 
were removed from  S3 leaving us with the sample used in \citet{Chavez2012} paper (S5) (see Chapter \ref{CH:5}).
We call this procedure the `kinematic analysis' of the emission line profiles.
 
The observed spatial FWHM of the emitting region was  used to extract the one dimensional 
spectrum of each object.
Three different fits were performed on the 1D spectra profiles (FWHM) of $\hb$ and  the 
$[\Oiii]\ \lambda \lambda 4959,5007$\AA\ lines: 
a single gaussian, two asymmetric gaussians and 3 gaussians (a core plus a blue and a 
red wing). These fits were performed using the IDL routines \verb|gaussfit|, 
\verb|arm_asymgaussfit| and \verb|arm_multgaussfit| respectively. Figure \ref{fig:Mgf} 
shows a typical fit to $\hb$; the best fitting to all the sample objects is presented in 
Appendix \ref{ApD}. 

Multiple fittings with no initial restrictions are not unique, so we computed using 
an automatized IDL code, a grid of fits each with slightly different initial conditions. 
From this set of solutions we chose those that had the minimum $\chi^2$. We 
begin with a blind grid of parameters from which the multiple gaussian fits are constructed, 
hence some of the resulting fits with small $\chi^2$ are not reasonable due to numerical divergence 
in the fitting procedure. We have eliminated unreasonable results by visual inspection.

%%%%%%%%%%%%%%%%%%%%%%%%%% Figure 9 %%%%%%%%%%%%%%%
%Figure --- Hb multiple gaussian fit
\begin{figure*}
\centering
\resizebox{14cm}{!}{\includegraphics{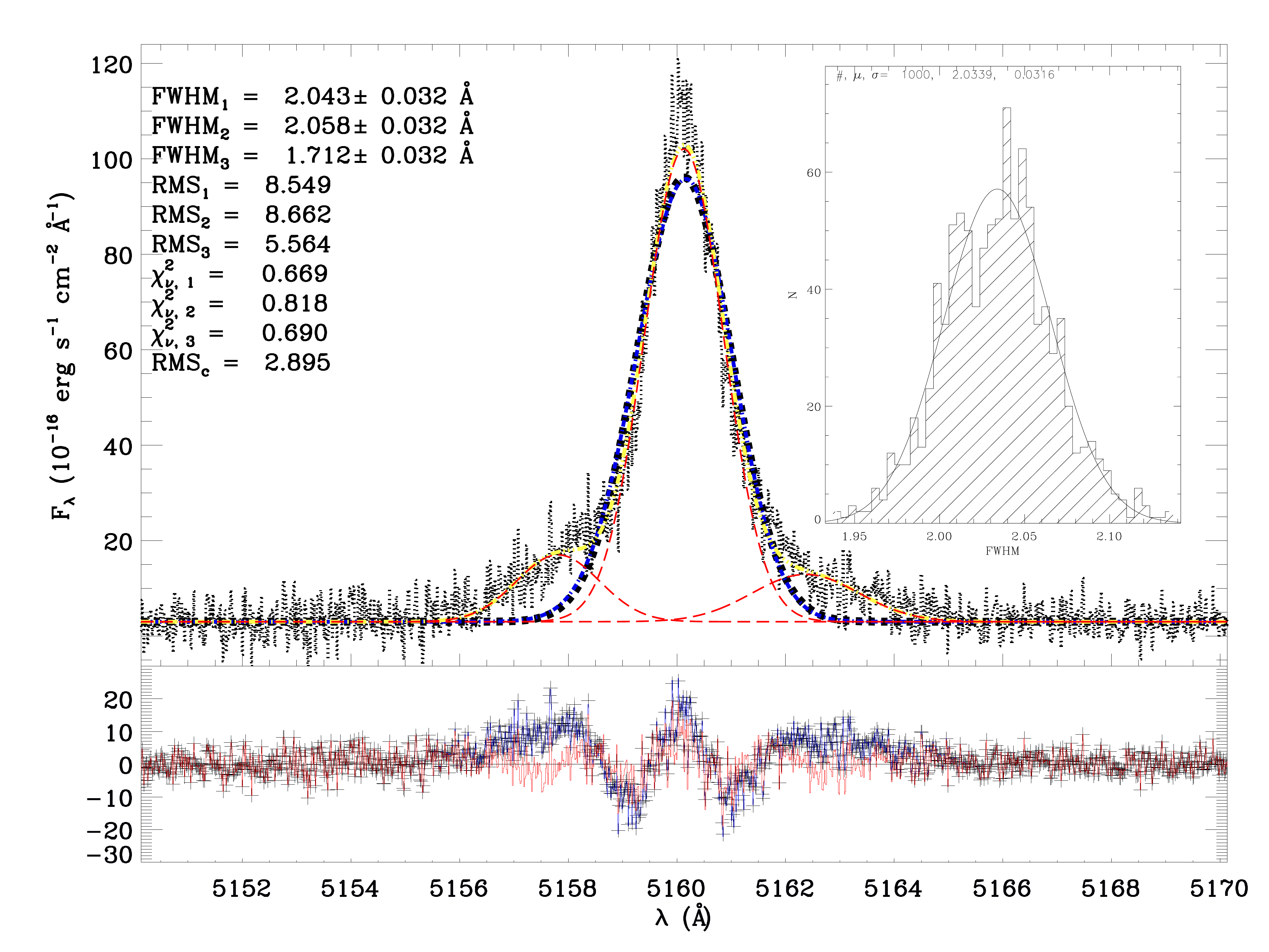}}
\caption[Typical multiple gaussian fit to an $\hb$ line] {\small Typical multiple gaussian fit to an $\hb$ line. \emph{Upper
    panel:} The single gaussian fit is shown with a  dashed line (thick black).
  The asymmetric gaussian fit is indicated by the dash-dotted line (blue).
  In the  three gaussians fit, every gaussian is indicated by long-dashed
  lines (red) and the total fit by a dash-double-dotted line (yellow).
  The parameters of the fits are shown in the top left corner. The inset shows the results 
  from the Montecarlo
  simulation  to estimate the errors in the parameters of the best fit. See further 
  details in the text. \emph{Lower
    panel:} The residuals from the fits follow the same colour code; the
  plusses are the residuals from the single gaussian fit whereas the
  continuous lines are the residuals from the asymmetric and three
  gaussian fits. }

\label{fig:Mgf}
\end{figure*}

The 1$\sigma$ uncertainties of the FWHM were estimated using a Montecarlo 
analysis. A set of 
random realizations of every spectrum was generated 
using the data poissonian 1$\sigma$ 1-pixel uncertainty. Gaussian fitting for every 
synthetic spectrum in the set was performed afterwards, and we obtained a distribution of 
FWHM measurements
from which the 1$\sigma$ uncertainty for the FWHM measured in the 
spectra follows. Average values obtained are 6.3\% in H$\beta$ and 3.6\% in [OIII].

Table \ref{tab:tab02} lists the FWHM measurements for the high
resolution observations prior to any correction such as instrumental
or thermal broadening. Column (1) is the index number, column (2) is the SDSS
name, columns (3) and (4) are the right ascension and
declination in degrees,  column (5) is the heliocentric redshift as
taken from the SDSS DR7 spectroscopic data, columns (6) and (7) are
the measured $\hb$  and $[\Oiii]\ \lambda 5007$ FWHM  in \AA . 

%%%%%%%%%%%%%%TABLES 

\subsection{Emission line widths}
The observed velocity dispersions ($\sigma_o$) -- and their 1$\sigma$ uncertainties -- have  
been derived from  the FWHM measurements of the $\hb$ and $[\Oiii]  \lambda 5007$ 
lines on the high resolution spectra as: 
\begin{equation}
\sigma_o \equiv \frac{FWHM}{2  \sqrt{2\ ln(2)} }
\end{equation}
Corrections for thermal ($\sigma_{th}$), instrumental ($\sigma_{i}$) and fine structure 
($\sigma_{fs}$) broadening have been applied. The corrected value  is given by the expression:
\begin{equation}
 \sigma = \sqrt{\sigma_o^2 - \sigma_{th}^2 - \sigma_i^2 - \sigma_{fs}^2} 
\end{equation}
We have adopted the value of $\sigma_{fs}(\mathrm{H}\beta)  =  2.4\ \mathrm{km\ s^{-1}}$ 
as published in \citet{Garcia2008}. The 1$\sigma$ uncertainties for the velocity dispersion 
have been propagated from the  $\sigma_o$ values. 

\begin{figure}
\centering
\resizebox{8.5cm}{!}{\includegraphics{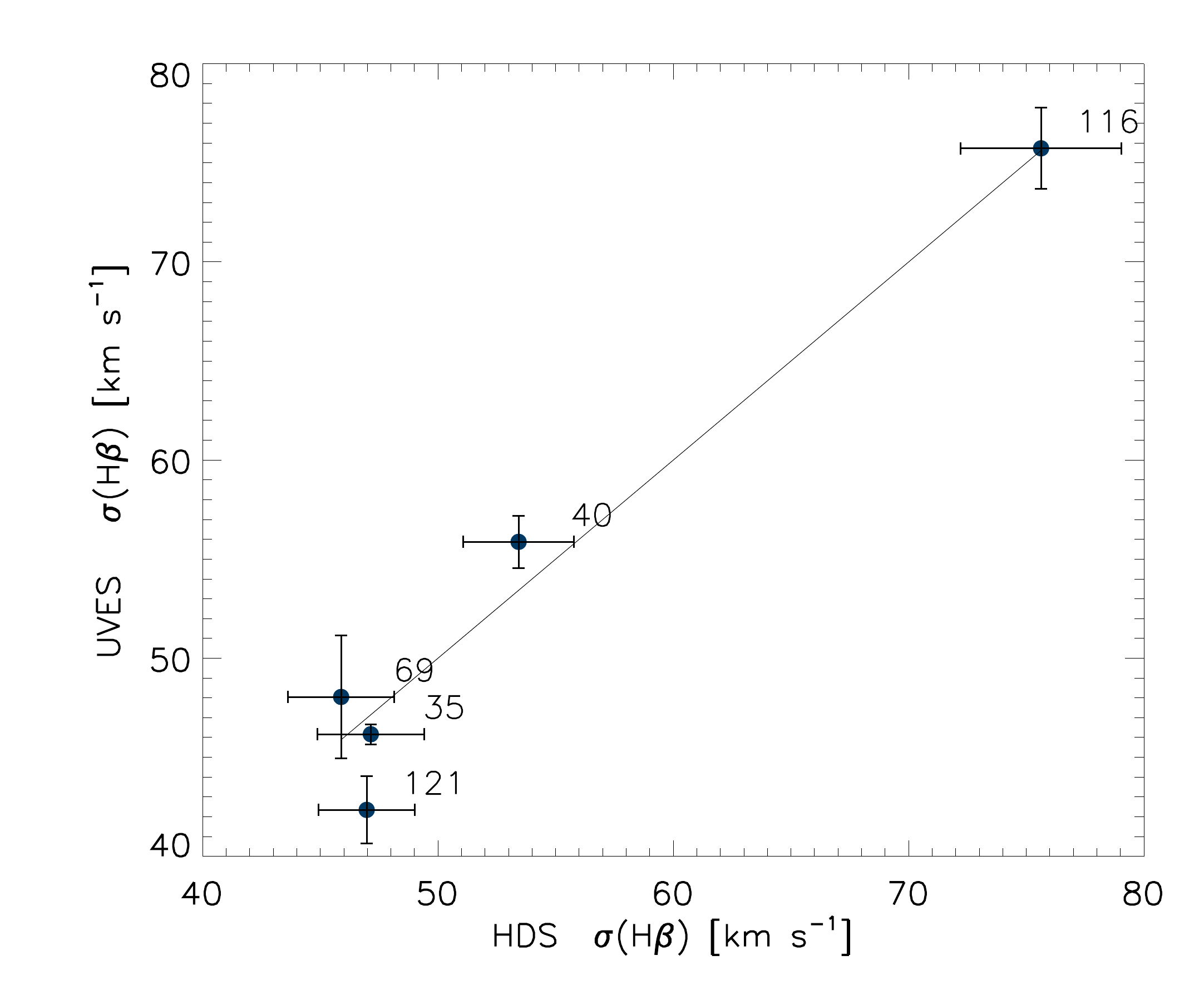}}
\caption[Comparison of $\sigma$ values after applying broadening corrections] {\small Comparison of $\sigma$ values after applying broadening corrections, as described in the text, for the 5 objects observed with both telescopes. The labels are the object indices as in the tables.}
\label{fig:SigHDSvUVES}
\end{figure}

The high resolution spectra were obtained with two different slit widths. The slit size was initially defined as to cover part of the Petrosian diameter of the objects. For UVES data, for which the slit width was 2$''$ and the slit was uniformly illuminated, $\sigma_i$ was directly 
estimated from sky lines, as usual. The Subaru observations have shown that the 4$''$ slit size used, combined with the excellent seeing during 
our observations has the unwanted consequence that the slit was not uniformly illuminated 
for  the most compact \hii\ galaxies that tend to be also the most distant ones. 
Thus we have devised a simple procedure to calculate the instrumental broadening correction 
for  the Subaru data. In this case, $\sigma_i$ was estimated from the target size;
we positioned a rectangular area representing the slit  over the corresponding 
SDSS \emph{r} band image and measured from the image the FWHM of the object along the dispersion direction.
In Figure \ref{fig:SigHDSvUVES} we plot $\sigma$ (after applying the broadening corrections as  described above) for the five objects that have been observed with both instruments. It is clear that the results using both methods are consistent.

%&&&&&&&&&&&&&&&&&&&&&&&&&&&&&&&&&&&&&&&&&&&&&
%Figure --- Log (sigma Hb) distribution %%%%%%%% Figure 10 %%%%%%%%%%%
%%%%%%%%%%%%%%%%%%%%%%%%%%%%%%%%%%%% 

The thermal broadening was calculated assuming a Maxwellian velocity distribution of the 
hydrogen and oxygen ions, from the expression:
\begin{equation}
 \sigma_{th}  \equiv \sqrt{\frac{k T_e}{m}},
\end{equation}
where $k$ is the Boltzmann constant, $m$ is the mass of the ion in question and $T_e$ is 
the electron temperature in degrees Kelvin as discussed in \S 4.4.3. 
For the H lines, an object with the sample median $\sigma_0$=37km/s, thermal broadening represents about 10\%, $\sigma_{fs}$=0.3\% and $\sigma_{inst-UVES}$=2\% while $\sigma_{inst-HDS}$=9\%. For the [OIII] lines, thermal broadening is less than 1\%, typically 0.3\%.

\begin{figure}
\centering
\resizebox{8cm}{!}{\includegraphics{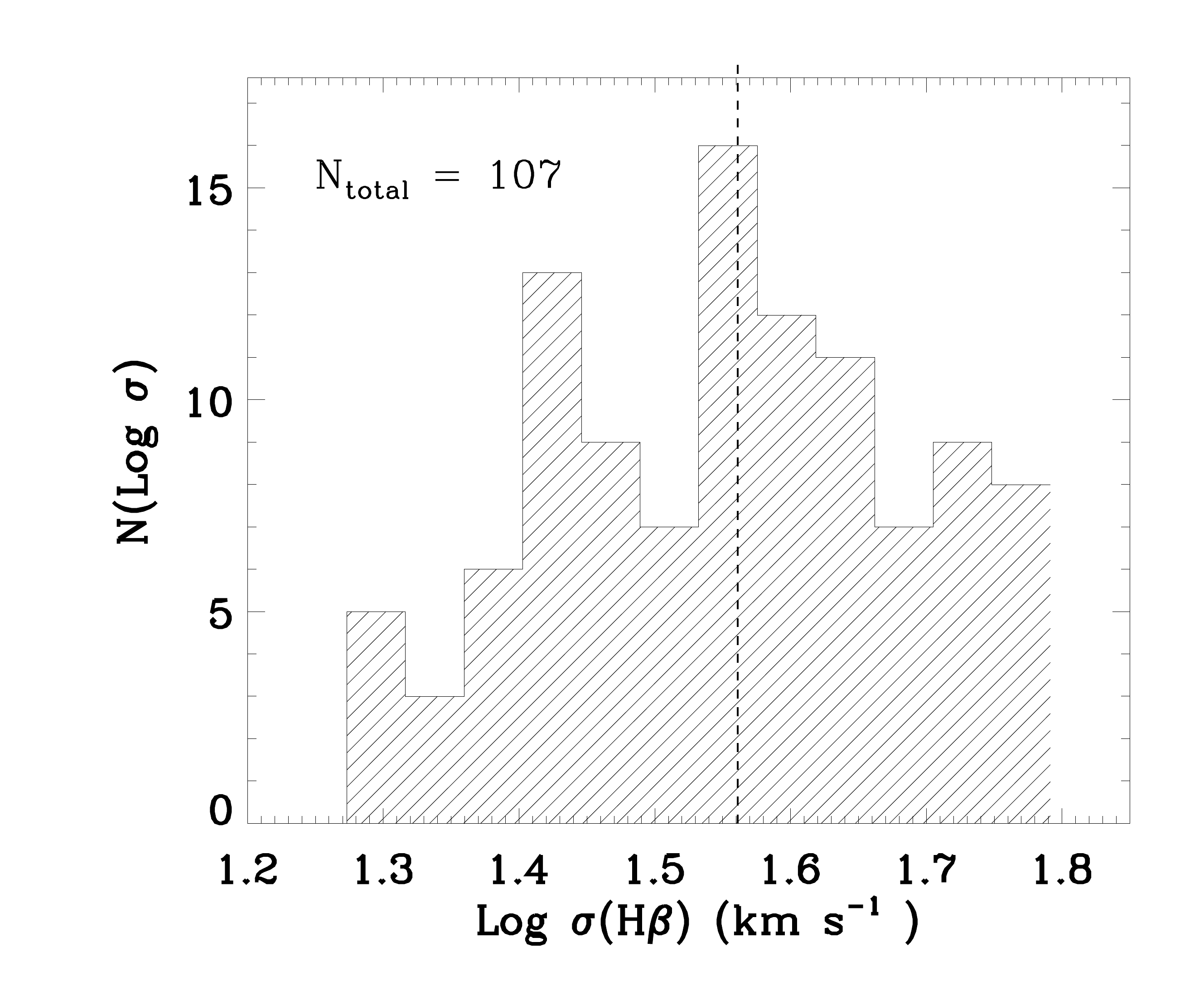}}
\caption[Distribution of the $\hb$  velocity dispersion] {\small Distribution of the $\hb$  velocity dispersion for the sample S3. 
The dashed line shows the median of the distribution.}
\label{fig:HLS}
\end{figure}

The obtained velocity dispersions for the $\hb$ and $[\Oiii]  \lambda 5007$ 
lines are shown in Table \ref{tab:tab05}, in columns (7) and (8) respectively. 
Figure \ref{fig:HLS} shows the distribution of the $\hb$  velocity dispersions for the S3
sample (see Table \ref{tab:tab10}). 

%&&&&&&&&&&&&&&&&&&&&&&&&&&&&&&&&&&&&&&&&&&&&&
%Figure --- Log (sigma Hb) distribution %%%%%%%% Figure 10 %%%%%%%%%%%
%%%%%%%%%%%%%%%%%%%%%%%%%%%%%%%%%%%% 

\subsection{Extinction and underlying absorption}
Reddening correction was performed using the coefficients derived from the 
Balmer decrement, with $\rm{H}\alpha$, $\rm{H}\beta$ and $\rm{H}\gamma$ fluxes obtained
from the SDSS DR7 spectra. However, contamination  by the underlying stellar population 
produces  Balmer  stellar absorption lines under the Balmer nebular emission lines. This 
fact alters the observed emission line ratios in such a way that the Balmer decrement and 
the internal extinction are overestimated    \citep[see e.g.][]{Olofsson1995}.

To correct the extinction determinations for  underlying absorption, we use the technique 
proposed by \citet{RosaGonzalez2002}. The first step is to determine the underlying Balmer 
absorption ($Q$) and  the ``true'' visual extinction ($A_{V}$) from the observed one 
($A_{V}^{*}$).

The ratio between a specific line intensity, $F(\lambda)$, and that of
$\hb$,  $F(\hb)$, is given by
\begin{equation}
	\frac{F(\lambda)}{F(\hb)} = \frac{F_{0}(\lambda)}{F_{0}(\hb)} 10^{-0.4 A_{V}[k(\lambda) - k(\hb)] / R_{V}},
\end{equation}
where $k(\lambda) = A(\lambda) / E(B - V)$ is given by the adopted extinction law, 
$R_{V} = A_{V} / E(B - V)$ is the optical total-to-selective extinction ratio and the 
subscript $0$ indicates unreddened intrinsic values.

We used as reference the theoretical ratios for Case B recombination $F_{0}(\ha) / F_{0}(\hb) = 2.86$ and $F_{0}(\hg) / F_{0}(\hb) = 0.47$\citep{Osterbrock1989}. In the absence of underlying absorption, the observed flux ratios can be expressed as a function of the theoretical ratios and the visual extinction:
\begin{align}
 \logd \frac{F(\ha)}{F(\hb)} &= \logd 2.86 - 0.4[k(\ha) - k(\hb)]A_{V}/R_{V}\; , \label{eqAV1}\\
 \logd \frac{F(\hg)}{F(\hb)} &= \logd 0.47 - 0.4[k(\hg) - k(\hb)]A_{V}/R_{V}\; . \label{eqAV2}
\end{align}

Including the underlying absorption and assuming that the absorption and emission 
lines have the same widths \citep{Gonzalez1999}, the observed ratio between $\mathrm{H}\alpha$ and 
$\mathrm{H}\beta$ is given by
\begin{equation}
 \frac{F(\mathrm{H}\alpha)}{F(\mathrm{H}\beta)} = \frac{2.86\{ 1 - P Q [W_{+}(\mathrm{H}\beta) / W_{+}(\mathrm{H}\alpha)] \}}{1 - Q}, \label{eqQ1}
\end{equation}
where $W_{+}(\mathrm{H}\alpha)$ and $W_{+}(\mathrm{H}\beta)$ are the equivalent widths in 
emission for the lines, $Q = W_{-}(\mathrm{H}\beta)/ W_{+}(\mathrm{H}\beta)$ is the ratio 
between the equivalent widths of $\mathrm{H}\beta$ in absorption and in emission and 
$P = W_{-}(\mathrm{H}\alpha) / W_{-}(\mathrm{H}\beta)$ is the ratio between $\mathrm{H}\alpha$ 
and $\mathrm{H}\beta$ equivalent widths in absorption.

The value $P$ can be obtained theoretically from spectral evolution models. 
\citet{Olofsson1995} has shown that for solar abundance and stellar mass in the range 
$0.1\ \mathrm{M_{\odot}} \leq M \leq 100\ \mathrm{M_{\odot}}$ using a Salpeter IMF, the 
value of $P$ is close to $1$ with a dispersion $\sim 0.3$ for ages between 
$1 - 15\ \mathrm{Myr}$. Since the variation of $P$ produces a change in the  
$ F(\mathrm{H}\alpha) / F(\mathrm{H}\beta) $ ratio of less than 2 \% that, given the low 
extinction in \hii\ galaxies, translates in a flux uncertainty well below 1 \%, we have assumed 
$P = 1$. 

The ratio between $\mathrm{H}\gamma$ and $\mathrm{H}\beta$ is
\begin{equation}
 \frac{F(\mathrm{H}\gamma)}{F(\mathrm{H}\beta)} = \frac{0.47 - G Q}{1 - Q}, \label{eqQ2}
\end{equation} 
where $G = W_{-}(\mathrm{H}\gamma) / W_{-}(\mathrm{H}\beta)$ is the ratio between the 
equivalent widths in absorption of $\mathrm{H}\gamma$ and $\mathrm{H}\beta$.  \citet[][ ; Tables 3a,b ]{Olofsson1995} and \citet[][; Table 1]{Gonzalez1999}  suggest that the value of the parameter $G$ can also be taken as $1$.

When the theoretical values for the ratios $ \logd [F(\ha) / F(\hb)]  = 0.46$  and 
$\logd[ F(\hg) / F(\hb)] = -0.33$, are chosen as the origin, the observed ratios can define  
a vector for the observed visual extinction  ($\mathbf{A_{V}^{*}}$). From equations 
(\ref{eqAV1}) and (\ref{eqAV2}) and a set of values for $A_{V}$, we define a vector for the 
``true'' visual extinction, whereas  from equations (\ref{eqQ1}) and (\ref{eqQ2}) and a set 
of values of $Q$, we define a vector for the underlying absorption $\mathbf{Q}$. Assuming 
that the vector  relation  $\mathbf{Q} + \mathbf{A_{V}} =\mathbf{A_{V}^{*}}$ is satisfied, 
by minimizing the distance between the position of the vector $\mathbf{A_{V}^{*}}$ and the  
sum  $\mathbf{Q} + \mathbf{A_{V}}$ for every pair of parameters $(Q, A_{V})$, we obtain 
simultaneously the values for $Q$ and $A_{V}$ that correspond to the observed visual extinction.

The de-reddened fluxes were obtained from the expression
\begin{equation} 
  F_{o}(\lambda) = F_{obs}(\lambda) 10^{0.4A_{V}k(\lambda)/R_{V}},
\end{equation}
where the  extinction law was  taken from  \citet{Calzetti2000}. The 1$\sigma$ uncertainties 
were propagated by means of a Monte Carlo procedure.

Finally, the de-reddened fluxes were corrected for underlying absorption. For $\hb$ the 
correction is given by:
\begin{equation}
	F(H\beta) = \frac{F_{o}(H\beta)}{1 - Q}
\end{equation}
The 1$\sigma$ uncertainties were propagated straightforwardly. The results  are shown in 
Table \ref{tab:tab05}, columns (4), (5) and (6) where we give the values 
for $A_V$, $Q$ and $C_{\hb}$ respectively.

\subsection{Redshifts and distances}
Redshifts have been transformed from the heliocentric to the local
group frame following \citet{Courteau1999} by the expression:
\begin{equation}
 z_{lg} = z_{hel} - \frac{1}{c} (79 \cos l \cos b - 296 \sin l \cos b + 36 \sin b),
\end{equation}
where $z_{lg}$ is the redshift in the local group reference frame,
$z_{hel}$ is the redshift in the heliocentric reference frame, $c$ is
the speed of light and $l$ and $b$ are the galactic coordinates of
the object. 

We also corrected by bulk flow effects following the method proposed
in \citet{Basilakos1998} and \citet{Basilakos2006}. 
For this correction and since the objects in our sample have low redshifts, the distances  
have been calculated from the expression:
\begin{equation}
 D_{L}  \approx \frac{cz}{H_{0}},
\end{equation}
where $z$ is the redshift and $ D_{L}$ is the luminosity distance. For
the Hubble constant we used a value  of $H_0 = 74.3 \pm 4.3\
\mathrm{km\ s^{-1} Mpc^{-1}} $ \citep{Chavez2012}. The 1$\sigma$
uncertainties for the distances were calculated using error
propagation from the uncertainties in $z$ and $H_0$.  Column (3) in
Table \ref{tab:tab05} (where we show all the parameters derived from
the measurements) gives the corrected redshift.

\subsection{Luminosities}
The H$\beta$ luminosities were calculated from the expression:
\begin{equation}
 L(\mathrm{H}\beta) = 4 \pi  D_L ^2 F(\mathrm{H}\beta),
\end{equation}
where $D_L$ is the previously calculated luminosity distance and $F(\mathrm{H}\beta)$ is 
the reddening and underlying absorption corrected H$\beta$ flux. The 1$\sigma$ uncertainties  
were  obtained by error propagation. 

\begin{figure}
\centering
\resizebox{8.5cm}{!}{\includegraphics{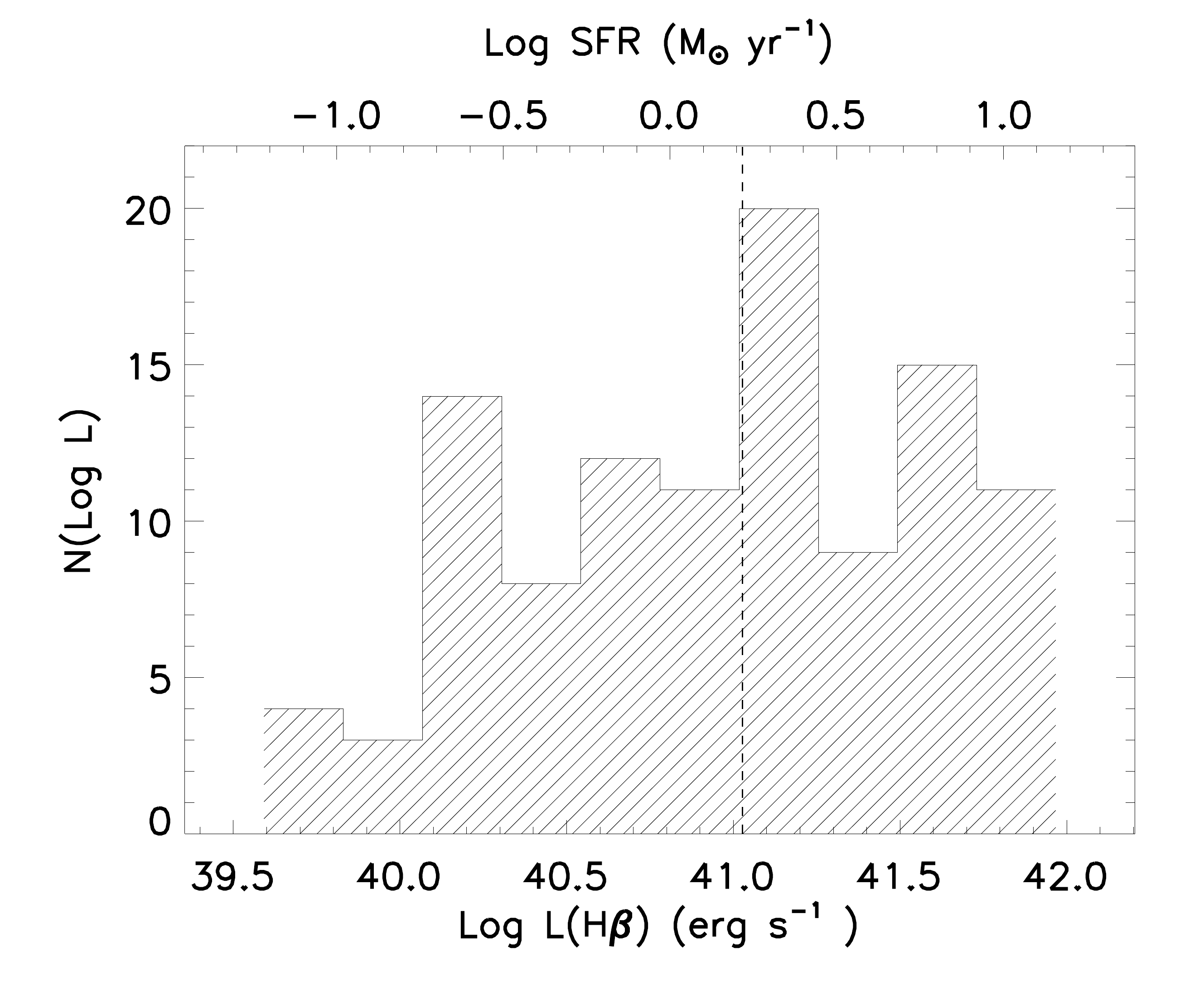}}
\caption[Distribution  of the $\hb$ emission line luminosities] {\small Distribution  of the $\hb$ emission line luminosities (and SFR as labelled on the top of the figure) for the 107 objects in he sample S3. 
The dashed line shows the median of the distribution.}
\label{fig:HLL}
\end{figure}

Table \ref{tab:tab05}, column (9) shows the corrected $\hb$ luminosities 
obtained for the objects in the sample. Figure \ref{fig:HLL} shows the distribution of 
luminosities for the objects in S3. The median of the distribution is log(L($\hb$)) = 41.03 
and the range is from  39.6 to  42.0. 

%%%%%%%%%%%%%%%%%%%%%%%%%%%%%%%%%%%%
%Figure --- Log L Hb distribution %%%%%%%%%%% figure 11 %%%%%%%%%%
%%%%%%%%%%%%%%%%%%%%%%%%%%%%%%%%%%%% 

%&&&&&&&&&&&&&&&&&&&&&&&&&&&&&&&&&&&&&&&&&&&&&
%%Table  3 --- H II galaxies local sample low dispersion measurements. 
\input{ch04/tab/tab03}

%%Table 4 --- H II galaxies local sample SDSS DR7 measurements
\input{ch04/tab/tab04}

%Table  2 --- H II galaxies local sample high dispersion measurements. 
\input{ch04/tab/tab02}

%%Table 5 --- H II galaxies local sample SDSS DR7 Derived data
\input{ch04/tab/tab05}

%&&&&&&&&&&&&&&&&&&&&&&&&&&&&&&&&&&&&&&&&&&&&&

\section{Physical Parameters of the Sample}
In what follows 
 we estimate the different intrinsic parameters that characterise our sample.
 
\subsection{Luminosity function}
The luminosity function (LF) is perhaps the most commonly used statistical tool to compare 
populations. The starforming region or \hii\ regions LF has been usually fitted by a function of the form:
\begin{equation}
 N(dL) = AL^{\alpha} dL,
\end{equation}
where $A$ is a constant and $\alpha$ is the power law index.

In order to test the completeness of our sample we have performed the $V / V_{max}$ test 
\citep[cf.][]{Schmidt1968, Lynden-Bell1971}, obtaining a value of $V/V_{max} = 0.25$ indicating 
that we have a partially incomplete sample, as expected considering the selection criteria 
adopted. 

\begin{figure}
\centering
\resizebox{8.5cm}{!}{\includegraphics{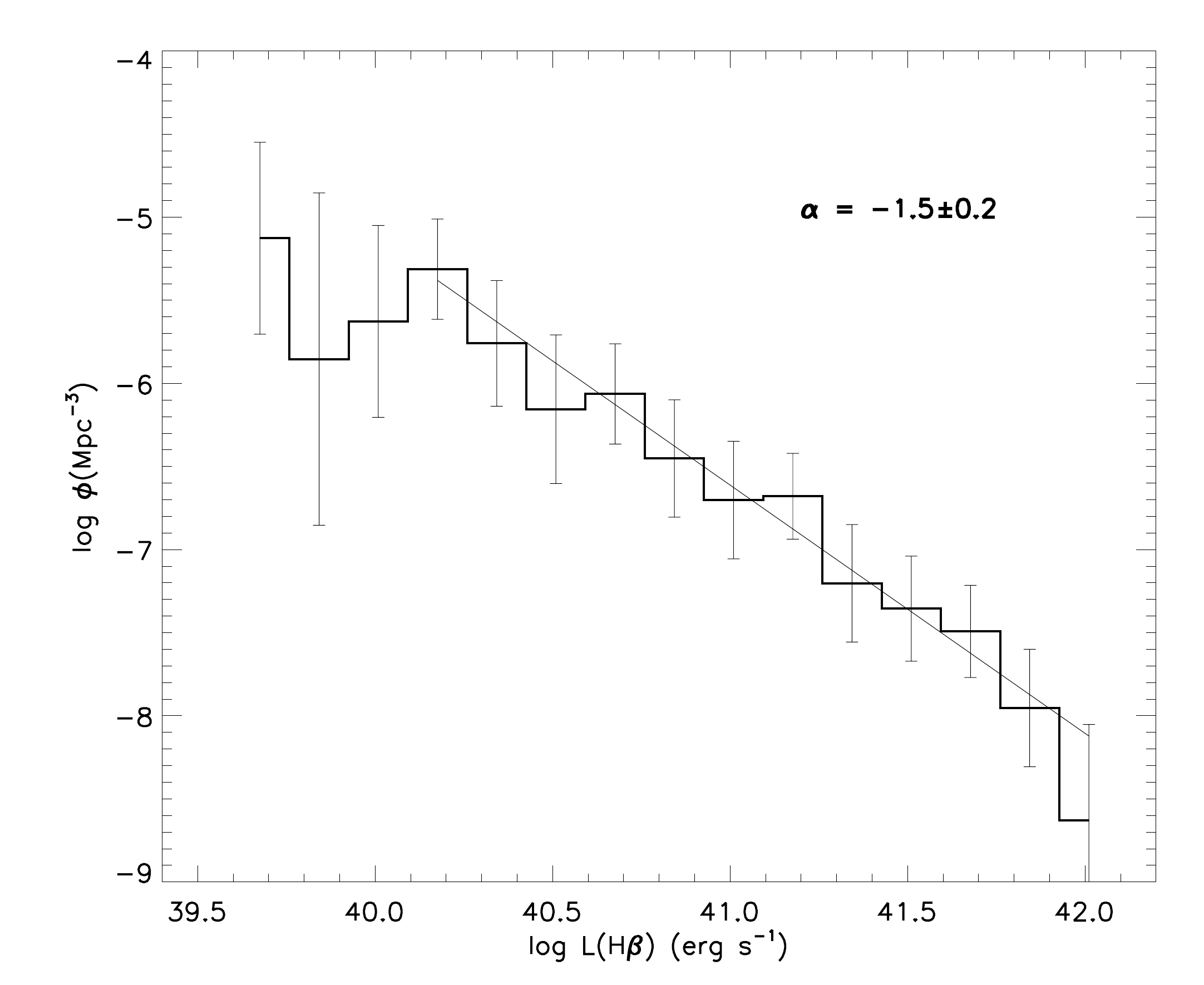}}
\caption[Luminosity function for our sample of \hii\ galaxies] {\small Luminosity function for our sample of \hii\ galaxies. The line is the least squares fit and has a slope of -1.5. The errors are Poissonian.}
\label{fig:LumF}
\end{figure}

The LF for our sample was calculated following the $V_{max}$ method 
\citep{Rowan-Robinson1968, Schmidt1968}. Since we have a flux limited sample with an  
$f_{lim} = 6.9 \times 10^{-15}\ \mathrm{erg\ s^{-1}\ cm^{-2}}$, we have binned the luminosities 
and calculated the maximum volume for each bin as:
\begin{equation}
 V_{max, i} = \frac{4 \pi}{3} \left( \frac{L_i}{4 \pi f_{lim}} \right)^{3/2},
\end{equation}
where $L_i$ is the i$^{th}$ bin maximal luminosity. The density of objects at each luminosity 
is obtained as:
\begin{equation}
 \Phi(L_i) = \frac{N(L_i)}{V_{max, i}},  
\end{equation}
where $N(L_i)$ is the number of objects in the ith bin. The resulting LF is shown in 
Figure \ref{fig:LumF} where it is clear that incompleteness affects only the less luminous 
objects ($\logd L (\hb) \leq 40.2$) which were excluded from the determination of $\alpha$. 

We obtained a value of $\alpha = - 1.5 \pm 0.2$ for the slope of the LF, consistent with the slope found for the luminosity function of \hii\ regions in spiral  and irregular galaxies.

\citet{Kennicutt1989} 
find  $\alpha = -2.0 \pm 0.5$  for the $\ha$ LF of  \hii\ regions 
in 30 nearby  galaxies.  
\citet{Oey1998}  have identified a break in the LF for $\logd L(\ha ) \sim 38.9 $ with the 
slope ($\alpha$) being steeper in the bright part than in the faint end.  \citet{Bradley2006}  
found a value for $\alpha = -1.86 \pm 0.03$ in the bright end of the LF, using a sample of 
$\sim 18,000$ \hii\ regions in 53 galaxies. 
Our result extends the analysis to higher luminosities although the choice of $\logd \sigma < 1.8$ limits the sample to
objects with $\logd L(\ha ) < 42.5$.  We therefore conclude that our sample is representative of the bright-end population of star-forming regions in the nearby universe. 

%Figure --- Luminosity Function %%%%%%%Figure 12 %%%%%%%%%%%%
%%%%%%%%%%%%%%%%%%%%%%%%%%%%%%%%%%%% 

 \subsection{Star formation rates}
The concept of star formation rate  (SFR)  is normally applied to whole galaxies where the 
SFR does not suffer rapid changes. In general the SFR is a parameter that  is difficult to 
define for an instantaneous burst and has limited application. Nevertheless, to allow comparison 
with other starforming galaxies we have estimated the SFR for the objects in our sample. To 
this end we used the expression \citep[cf.][]{Kennicutt2012}:
\begin{equation}
 \logd \dot{M} = \logd L(\hb) - 40.81, 
\end{equation}
where $\dot{M}$ is the star formation rate in $\mathrm{M_{\odot}\ yr^{-1}}$ and $L(\hb)$ 
is the $\hb$ luminosity in $\mathrm{erg\ s^{-1}}$. The 1$\sigma$ uncertainties were propagated 
straightforwardly. The SFR values obtained are given in Table \ref{tab:tab05}, 
column (15) and their distribution is given in Figure \ref{fig:HLL}. The values range from 0.05 to 19.6 $\mathrm{M_{\odot}\ yr^{-1}}$ with a 
mean of 3.7 $\mathrm{M_{\odot}\ yr^{-1}}$. This result is similar to that found in  SFR determinations of 
Blue Compact Dwarf Galaxies \citep{Hopkins2002}. High redshift samples \citep[e.g.][]{Erb2006} where the luminosity of the objects is not limited by design, span a SFR between 2.5 and 100  $\mathrm{M_{\odot}\ yr^{-1}}$ and the maximum value of the distribution is 20 $\mathrm{M_{\odot}\ yr^{-1}}$. Although there is  a wide superposition in the SFR range of our and the high redshift samples, our nearby sample has an upper limit in the luminosities ( corresponding to the upper limit in $\logd \sigma = 1.8$) and therefore in the SFR at around $20\ \mathrm{M_{\odot}\ yr^{-1}}$.

\subsection{Electron densities and temperatures }
We calculated the corresponding electron densities, electron temperatures and oxygen 
abundances for all the objects for which the relevant data was available. We used the
extinction and underlying absorption corrected 
line intensities  as described in Section 4.4.

Electron densities are derived from the ratio $[\Sii]\ \lambda 6716/ \lambda 6731$ 
following \citet{Osterbrock1988} assuming initially an electron temperature $T_e = 10^4\ \rm{K}$. 

We calculate the electron temperature as \citep{Pagel1992}:
\begin{align*}
 t \equiv t(\Oiii) = 1.432[\logd R - 0.85 + 0.03 \logd t \\
 + \logd(1 + 0.0433 x t^{0.06})]^{-1}, \nonumber
\end{align*}
where $t$ is given in units of $10^4\ \rm{K}$, $x = 10^{-4}N_e t_2^{-1/2}$, $N_e$ is the 
electron density in $\rm{cm^{-3}}$ and 
\begin{align*}
 R &\equiv \frac{I(4959) + I(5007)}{I(4363)},\\
 t_2^{-1} &= 0.5 (t^{-1} + 0.8);
\end{align*} 
The temperatures found are between 10,000 and 18,000$^{\circ}\mathrm{K}$

\subsection{Ionic and total abundances }

The  ionic oxygen abundances were calculated following \citet{Pagel1992} from:
\begin{align*}
 12 + \logd (\mathrm{O^{++} / H^{+} }) &= \logd \frac{I(4959) + I(5007)}{\hb}\\
    & + 6.174 + \frac{1.251}{t} - 0.55 \logd t\ , \\
 12 + \logd (\mathrm{O^{+} / H^{+}}) &= \logd \frac{I(3726) + I(3729)}{\hb} + 5.890 \\
    & + \frac{1.676}{t_2} - 0.40 \logd t_2  + \logd (1 + 1.35 x) ;
\end{align*}
 and the  oxygen total abundance  is derived by adding these last two equations.
The errors are propagated by means of a Monte Carlo procedure.

\begin{figure}
\centering
\resizebox{8.5cm}{!}{\includegraphics{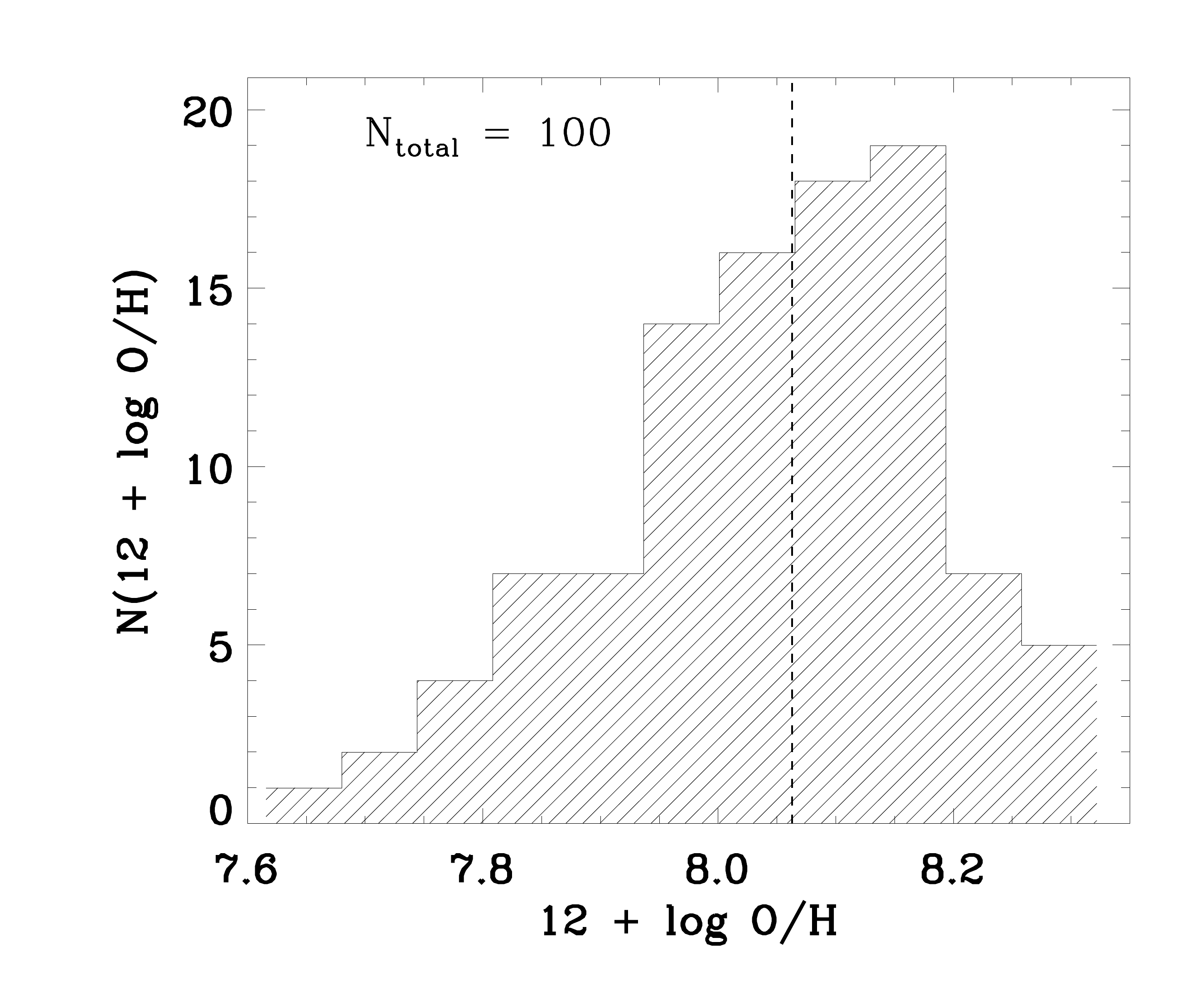}}
\caption[Distribution of oxygen abundances for the sample S3] {\small Distribution of oxygen abundances for the sample S3. The dashed line shows 
the median.}
\label{fig:HAb}
\end{figure}

Table \ref{tab:tab05}, column (10) shows the total oxygen abundance 
as 12+log(O/H). Figure \ref{fig:HAb} shows the distribution of oxygen abundances for the 
S3 sample. The median value is 12+log(O/H) = 8.08. For the very low redshift objects 
where  $[\mathrm{O II}]\ \lambda 3727$ \AA\  falls outside the SDSS observing window we have adopted 
I($[\mathrm{O II}]\ \lambda 3727$) = I($\hb $), reasonable for high excitation \hii\ regions \citep[e.g.][]{Terlevich1981}.

Additionally, as a consistency check and in order to investigate whether we can use a proxy for metallicity 
for future work, we have calculated  the N2 and R23 bright lines metallicity indicators \citep{Storchi-Bergmann1994, Pagel1979}
 given by:
\begin{equation}
	N2 = \frac{I(\mathrm{[NII]}\lambda 6584)}{I(\mathrm{H}\alpha)}
\end{equation}
\begin{equation}
	R_{23} = \frac{I(\mathrm{[OII]}\lambda 3727) + I(\mathrm{[OIII]}\lambda 4959) + I(\mathrm{[OIII]}\lambda 5007)}{I(\mathrm{H}\beta)}.
\end{equation}
In what follows, and to avoid including errors due to different calibrations, we just use the N2 and R23 parameters as defined, without actually  estimating metallicities from them. The metallicities used in the final analysis are only those derived using the direct method.

%Figure --- Abundances Distribution %%%% Figure 13 %%%%%
%%%%%%%%%%%%%%%%%%%%%%%%%%%%%%%%%%%% 
 
\subsection{The ionizing cluster masses}
One of the most fundamental parameters that can be obtained for a stellar system is its total mass.
In the case of the \hii\ galaxies, the  knowledge of the object mass could give us a better 
understanding of the physical nature of the  \lsig\ relation. 

\subsubsection{ The ionizing cluster photometric mass }
 We  estimated the mass  of the ionising star cluster (M$_{cl}$) from the observed emission 
 line luminosity following two different routes:\\
 
1 - Using  
the expression:
\begin{equation}
	M_{cl} = 7.1 \times 10^{-34} L(\hb) ,
\end{equation}
where, $M_{cl}$ is the total photometric mass (in  $M_{\odot}$) of the ionizing star cluster 
and the $\hb$ luminosity [$L(\hb)$] is in $\mathrm{erg\ s^{-1}}$. This expression was 
calibrated  using a \emph{SB99} model of an instantaneous burst of star formation 
with a stellar mass of $3 \times 10^6\ \mathrm{M_{\odot}}$ and a Salpeter initial mass 
function  \citep[][IMF]{Salpeter1955}  integrated 
in the range $(0.2\ \mathrm{M_{\odot}}, 100\ \mathrm{M_{\odot}})$.  The 
equivalent width in the model was taken as $EW(\hb) = 50\  \mathrm{\AA}$, the lower limit 
for our sample selection.
This limit for the equivalent width implies an upper limit for the cluster 
age of about 5.5 Myr, and therefore the derived cluster masses are in general upper limits.\\

2 - We  also estimated  the mass of the ionising star cluster including a correction  for evolution. 
To this end we used \cite{Garcia1995} 
single burst models of solar metallicity. 
These models provide the number of ionising Lyman continuum photons [$Q(H_0)$] per unit mass 
of the ionising cluster [$Q(H_0)/M_{cl}$] computed for a single slope  Salpeter IMF. We 
fixed the values for the lower and upper mass limits at 0.2 and 
100 M$_\odot$.
The decrease of [$Q(H_0)/M_{cl}$]   with increasing age  of the stellar population is 
directly related to the decrease of the equivalent width of the H$\beta$ line 
\citep[e.g.\ ][]{Diaz2000} as,

\[ \logd \left[ Q(H_0)/M_{cl} \right] = 44.0 + 0.86 ~ \logd \left[ EW(\mathrm{H}\beta)\right] \]

The total number of ionising photons for a given region has been derived from the H$\alpha$
luminosity \citep{Leitherer1995}:

\[
Q(H_0)\,=\,2.1\,\times\,10^{12}\,L(\mathrm{H}\beta)
\]
and the mass of  the ionising cluster  M$_{cl}$ is:

\begin{equation}
M_{cl}\,=\, 7.3 \times 10^{-34} \left(\frac{EW(\mathrm{H}\beta)}{50\ \mathrm{\AA }}\right)^{-0.86}
\end{equation}

Given that the EW(H$\beta$)  may be affected by an underlying older 
stellar continuum not belonging to the ionizing cluster, the listed masses for these 
clusters should be considered upper limits. 

The two estimates give similar results for the masses of the ionizing clusters, with the ratio  of the uncorrected to corrected mass  being  about 1.6 on average. 
It is necessary to emphasize that these cluster mass estimates do not include effects such as the escape or absorption by dust of ionizing photons that, if included, would make both estimates lower limits. We assume that the least biased equation is the first one, and that is the one we used to calculate the values given in column (13) of Table \ref{tab:tab05}.

 \subsubsection{ The mass of ionised gas }

The photometric mass of ionised gas ({\it M}$_{ion}$) associated to each star-forming
region complex was derived from their H$\beta$ luminosity 
and electron density ({\it N$_e$})  
using the expression: 
\begin{equation}
	M_{ion} \simeq 5 \times 10^{-34} \frac{L(\hb) m_p}{\alpha^{eff}_{\hb} h \nu_{\hb} N_e} \simeq 6.8 \times 10^{-33} \frac{L(\hb)}{N_e}, 
\end{equation}
where $M_{ion}$ is  given in $M_{\odot}$, $L(\hb)$ 
is the observed $\hb$  luminosity in $\mathrm{erg\ s^{-1}}$, $m_p$ it the proton mass in g, 
$\alpha^{eff}_{\hb}$ is the effective $\hb$ line recombination coefficient in 
$\mathrm{cm^3\ s^{-1}}$  for case B in the low-density limit and  $T = 10^4\ \mathrm{K}$, $h$ 
is the Planck constant in $\mathrm{erg\ s}$, $\nu_{\hb}$ is the frequency corresponding to 
the $\hb$ transition in $\mathrm{s^{-1}}$ and $N_e$ is the electron density in $\mathrm{cm^{-3}}$. 
The values obtained for $M_{ion}$ are given in column (14) of Table \ref{tab:tab05}.

%Figure --- Masses comparison %%%%%%% figure 14 %%%%%%%%%%%%%
\begin{figure}
\centering
\resizebox{8.5cm}{!}{\includegraphics{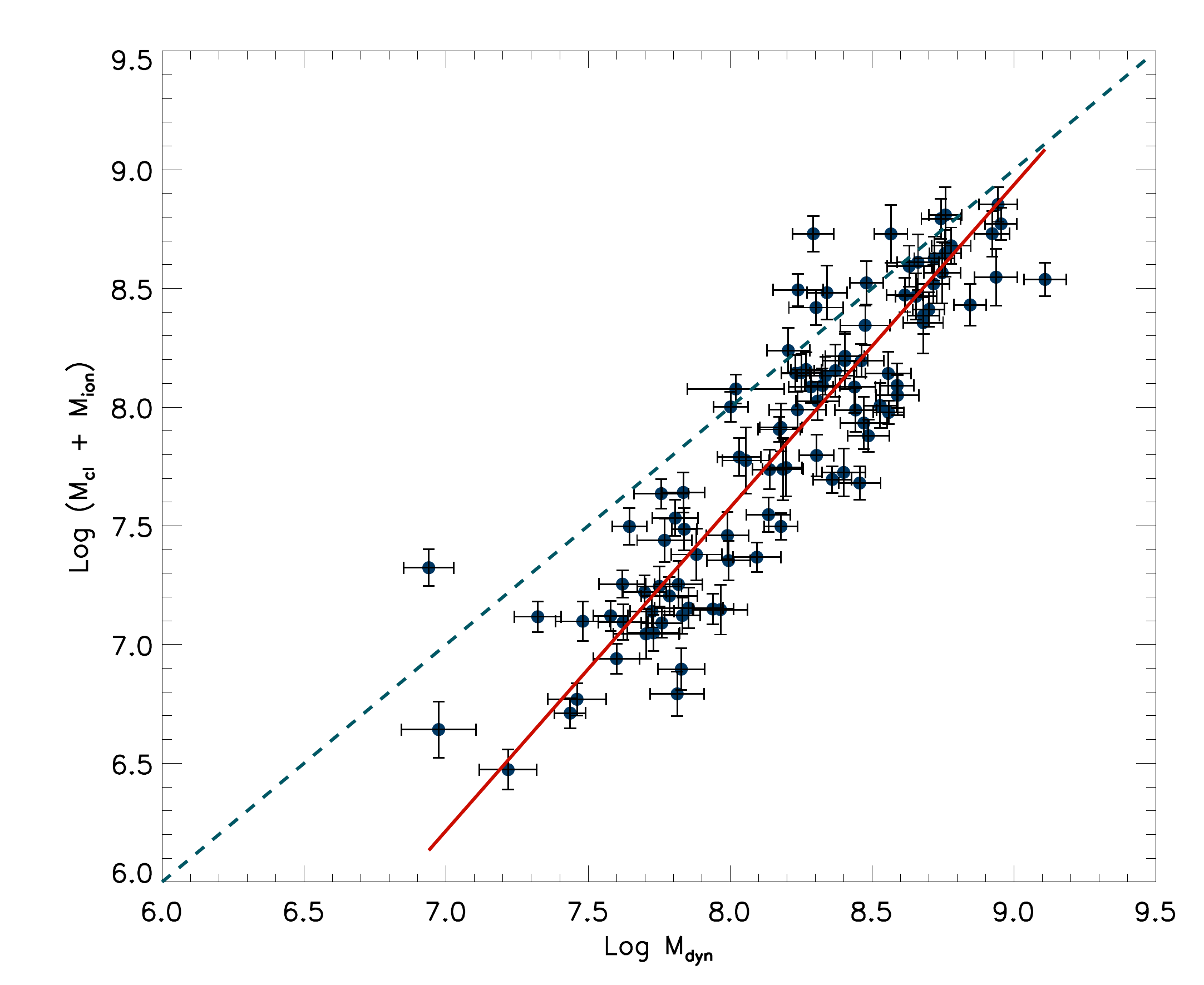}}
\caption[Comparison between $M_{cl} + M_{ion}$  and $M_{dyn}$] {\small Comparison between $M_{cl} + M_{ion}$  and $M_{dyn}$. The continuos thick line represents the best fit to the data. The dashed line shows the one-to-one relation.  }
\label{fig:CrM}
\end{figure}
%%%%%%%%%%%%%%%%%%%%%%%%%%%%%%%%%%%% 

\subsubsection{Dynamical masses}

The dynamical masses were calculated following the expression \citep[cf.][]{Binney1987}:

\begin{equation}
   M_{dyn} = 10^3 R \sigma^2, 
\end{equation}
where $\sigma$ is the velocity dispersion in $\mathrm{km\ s^{-1}}$, $M_{dyn}$ is given 
in $M_{\odot}$ and $R$ is the cluster effective  radius in 
parsecs (i.e. such that $1/2$ of the mass lies inside it).

To obtain an unbiased estimate of the dynamical mass a good measurement of the effective size 
of the ionising massive cluster is necessary. As discussed above regarding the high dispersion 
observations, we have evidence that many of the objects in the sample are perhaps 
unresolved even under very good seeing conditions. 
We have searched the HST database for high resolution images of objects in our sample and found only 2
\hii\ galaxies with HST WFC3 images: J091434+470207 and J093813+542825.

A quick analysis of the  HST images for these two objects shows  
that they are only marginally resolved and have effective radius of just a few parsecs. 
In order to improve the small number statistics we searched the HST high resolution database for star-forming nearby objects  using the same selection criteria as for the objects in this chapter, and found 18 \hii\ galaxies and GEHR that also have SDSS images.
Comparing the HST angular size  with the Petrosian radius obtained 
from the SDSS \emph{u} band photometry (corrected for seeing) 
we have found that the ionising cluster radius measured from the  HST images is on average more than a factor of 5 smaller 
than the SDSS Petrosian radius. 
For estimating the dynamical mass we assumed that this factor applies to all \hii\ galaxies and 
therefore we have used a  HST `corrected'  Petrosian radius as a proxy for the cluster radius. The values of the seeing corrected Petrosian to 50\% of light
radius are listed in column (11) of Table \ref{tab:tab05}.
The calculated $M_{dyn}$ is given in  column (12). 
The masses of the clusters, both photometric, i.e. $M_{cl} + M_{ion}$  and dynamical,  are  large and at the same 
time their size is very compact. The masses range over three decades from about 2 $\times$ 10$^6$ \Msol\ 
to 10$^9$ \Msol\ while the HST corrected Petrosian radius ranges from few tens 
of parsecs  to a few hundred parsecs.

In Figure 
\ref{fig:CrM} we compare  the sum of $M_{cl} + M_{ion}$  with  $M_{dyn}$. 
It is clear from the figure that the value of $M_{dyn}$, computed assuming that the Petrosian 
radius is  on average 5 times larger than the effective radius of the ionising cluster, is slightly larger than  the sum of the 
photometric stellar and ionised gas  components particularly for the lower mass objects. Also the fit to the data has a slope of 1.3 and not 1.0.
Considering the uncertainties in the determination of the three parameters involved, the small level of the disagreement is surprising.

It is not clear  at this stage what is the mass of the cold gas, both atomic and molecular, that remains from the starformation event. To further investigate this important  question, in addition to high resolution optical and NIR images to measure the size of the ionizing clusters, high resolution observations  in HI and CO or other molecular gas indicator are needed.

\subsection{The metallicity -- luminosity relation}
\begin{figure}
\centering
\resizebox{8cm}{!}{\includegraphics{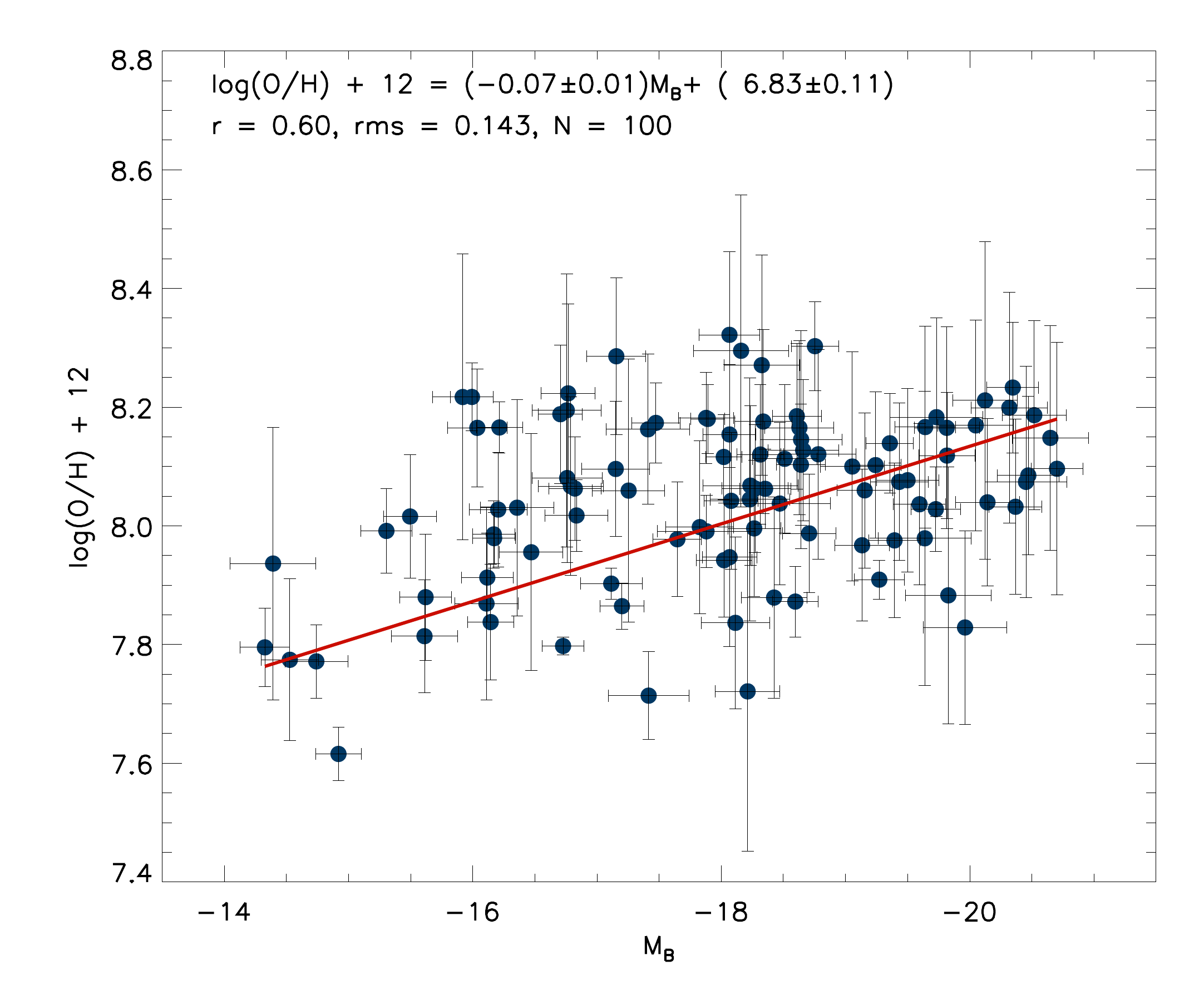}}
\caption[The continuum luminosity-metallicity relation  for S3] {\small The continuum luminosity-metallicity relation  for S3. The red 
line shows the best fit, which is described in the inset text. }
\label{fig:LStM}
\end{figure}

In order to test the possible existence of a  metallicity - luminosity relation for \hii\ galaxies, 
we have performed a least squares fit for the 100 objects with direct metallicity determination in the S3 sample using the continuum luminosity as 
calculated from the relation given by  \citet{Terlevich1981} and the metallicity as calculated before. The results,  shown in Figure \ref{fig:LStM},  clearly indicate 
that a correlation exists albeit weak.

\begin{figure}
\centering
\resizebox{8cm}{!}{\includegraphics{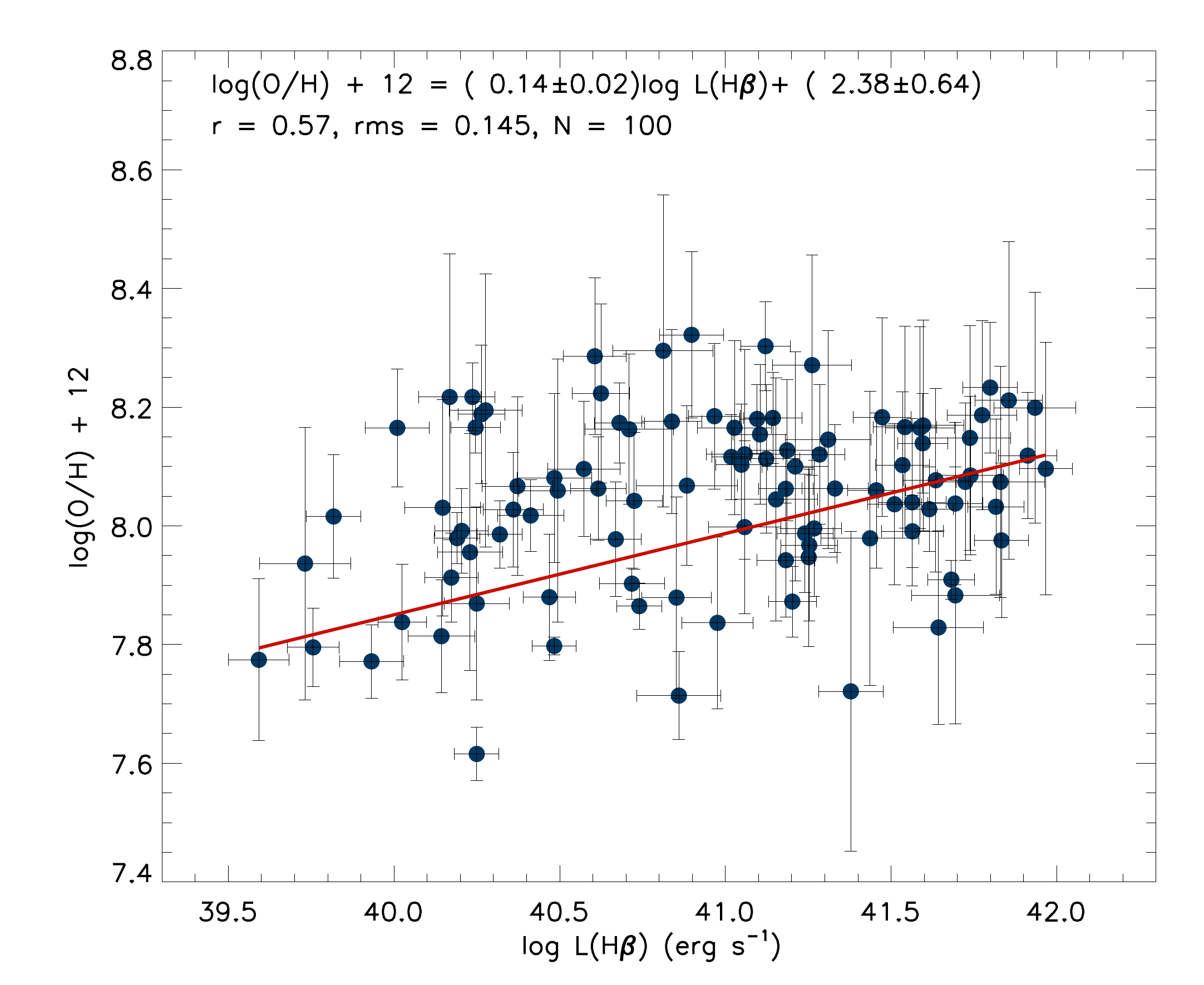}}
\caption[The $\hb$ luminosity-metallicity relation  for S3] {\small The $\hb$ luminosity-metallicity relation  for S3. The red line 
shows the best fit, which is described in the inset text.}
\label{fig:LSt2}
\end{figure}

We have performed also a least squares fit using the H$\beta$ luminosity and the metallicity for the same sample. 
The results are shown in Figure \ref{fig:LSt2} where a similarly weak correlation between both parameters can be seen.

%Figure --- LuminosidadCont.-O/Hrelation %%%%% Figure 15 %%%%%%%%%%%%%%%%%%%%%%

%%%%%%%%%%%%%%%%%%%%%%%%%%%%%%%%%%%%%

%Figure --- LuminosidadHbeta-O/Hrelation %%%% Figure 16 %%%%%%%%%%%%%%%%%%%%%%%%%%

%%%%%%%%%%%%%%%%%%%%%%%%%%%%%%%%%%%%%

\subsection{The metallicity -- equivalent width relation }
\begin{figure}
\centering
\resizebox{8cm}{!}{\includegraphics{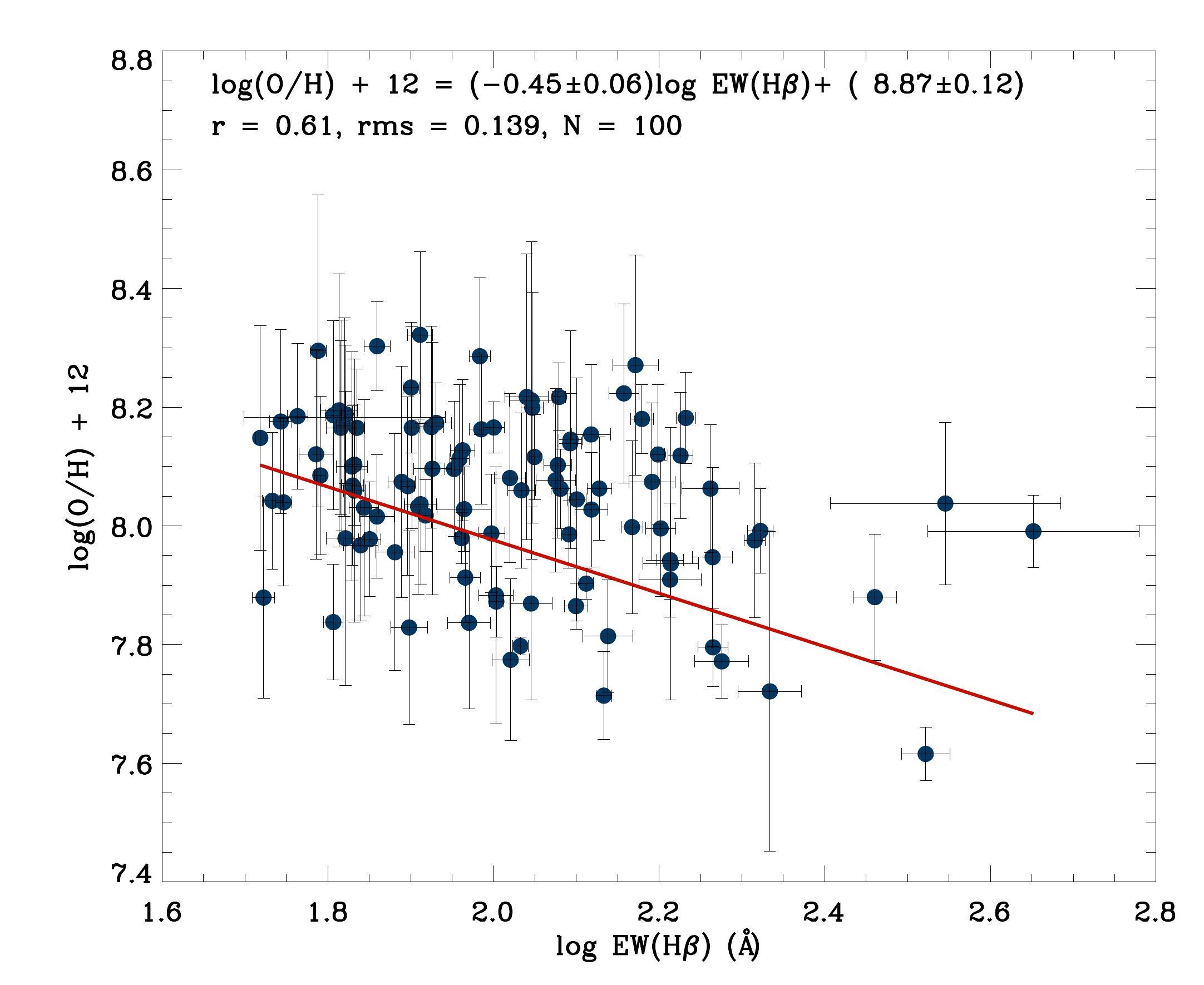}}
\caption[The EW($\hb$) - metallicity relation  for S3] {\small The EW($\hb$) - metallicity relation  for S3. The red line shows 
the best fit, which is described in the inset text. }
\label{fig:LSt3}
\end{figure}

We tested the possibility that a relation exists between the metallicity and 
the equivalent width of the H$\beta$ emission line acting as a proxy for the age of the starburst. 
We have performed a least squares fit to these two parameters for the S3 sample. The results 
are shown in Figure \ref{fig:LSt3} where a trend can  be seen clearly. This correlation between EW$(\hb ) $ and metallicity for a large sample of \hii\ galaxies covering a wider spectrum of ages and metallicities, has already been discussed in 
\citet{Terlevich2004} [see their Figure 5]. They interpreted the results as being consistent with two different timescales for the evolution of \hii\ galaxies on the metallicity -- EW$(\hb )$ plane. The idea is that the observed value of the EW$(\hb ) $ results from the emission produced in the present burst superposed on the continuum generated by the present burst plus all previous episodes of star formation that also contributed to enhance the metallicity.

%Figure --- EW-O/Hrelation %Figure 17 %%%%%%%%%%%%%%%%%%%%%%%%%
%%%%%%%%%%%%%%%%%%%%%%%%%%%%%%%%%%%%%

\section{The L -- $\sigma$ Correlation}
The main objective of this chapter is to assess the validity of the $L - \sigma$ relation 
and its use as a distance estimator.  

As discussed by e.g. \cite {Bordalo2011}, rotation and multiplicity of \hii\ regions in the sample objects  
can cause additional broadening of the emission lines which in turn may introduce  scatter 
in the $L - \sigma$ relation.

 In this context \citet{Chavez2012} performed a selection based on direct visual inspection 
 of the H$\beta$, H$\alpha$ and [OIII]$\lambda\ 4959$ and $\lambda\ 5007$  line 
 profiles  combined with the kinematic analysis mentioned in \S 4.2.1. At 
 the end of this process  only 69 objects (subsample S5) of the observed 128 were left with symmetric 
 gaussian profiles and no evidence of rotation or multiplicity.  
 %These objects are marked  with a `G' in column (2) of Table \ref{tab:tab05}. 
 This turned up as being a very expensive process in terms of observing time.

\subsection{Automatic profile classification}
To evaluate objectively the `quality' of the emission line profiles and to avoid 
possible biases associated with a subjective selection of the objects such as the ones 
performed by \citet{Bordalo2011} or \citet{Chavez2012} we developed a blind 
testing algorithm that can `decide' from the high dispersion data, 
which are the objects that have truly gaussian profiles in their emission lines. 
The  algorithm uses $\delta_{FWHM}(\hb) < 10$ 
 and $\delta_{flux} (\hb) < 10$ as selection criteria. These quantities
are defined as follows: 
\begin{equation}
	\delta_{FWHM} = \frac{\Delta_{FWHM}} {\mu_{FWHM}} \times 100,
\end{equation}
where $\mu_{FWHM}$ is the mean of the FWHM as measured from a single  and 
triple gaussian fitting to  a specific high resolution line profile and $\Delta_{FWHM}$ 
is the absolute value of the difference between these measurements. 
And
\begin{equation}
	\delta_{flux} = \frac{\Delta_{flux}}  {\mu_{flux}} \times 100,
\end{equation}  
where $\mu_{flux}$ is the mean of the fluxes as measured from the integration and gaussian 
fitting  to the same spectral line in low resolution and $\Delta_{flux}$ is the absolute value 
of the difference between those measurements. 

The rationale behind 
this approach is that these two quantities will measure departures from a single gaussian 
fitting of the actual profile. A  large  deviation is an indication of strong 
profile contamination due to second order effects such as large asymmetries and/or bright 
extended wings.  

\begin{figure}
\centering
\resizebox{8.5cm}{!}{\includegraphics{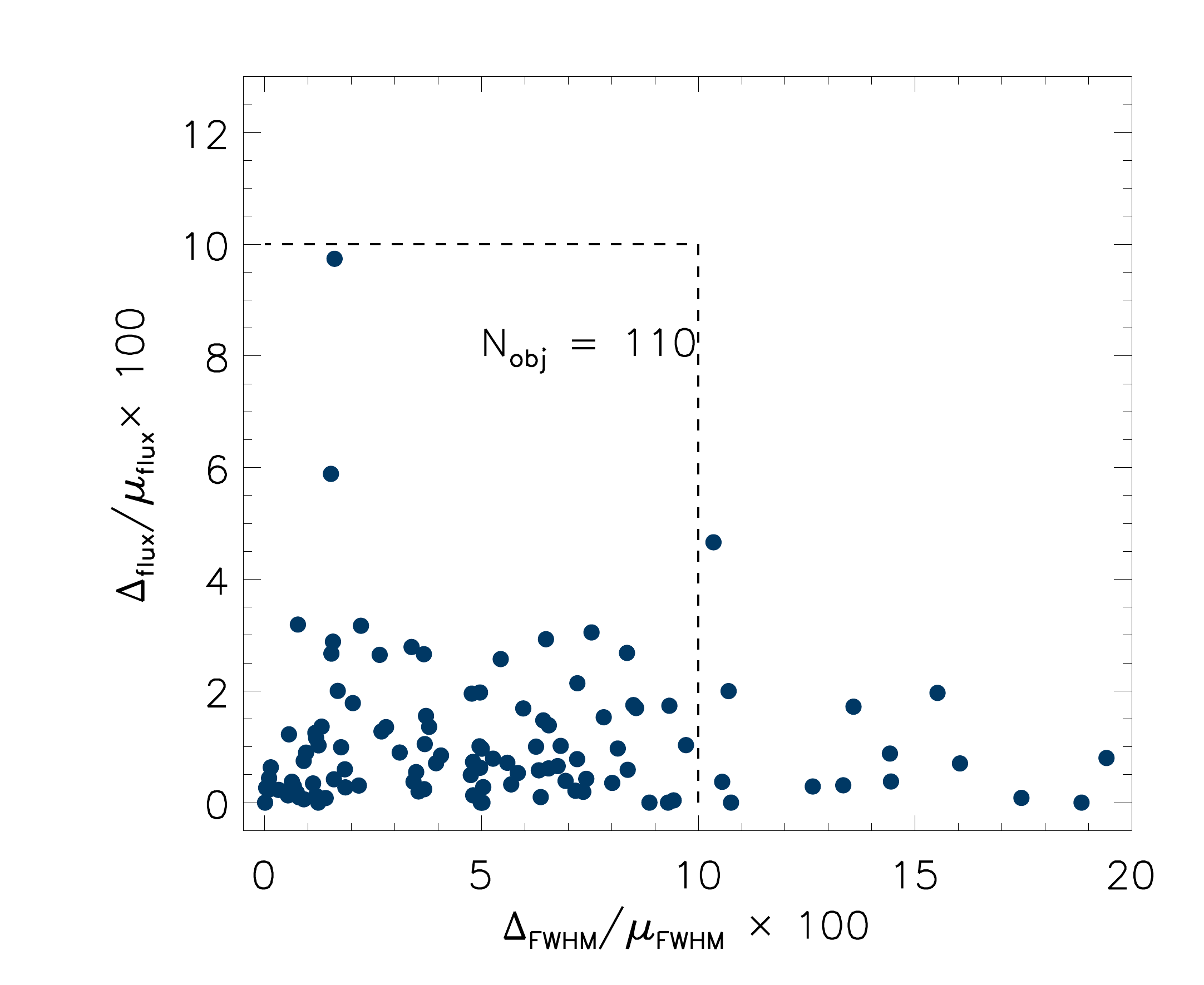}}
\caption[Automatic profile selection] {\small Automatic profile selection. Objects 
inside the box delimited  by a dashed line have $\delta_{flux} (\hb) < 10$ and $\delta_{FWHM}(\hb) < 10$. 
This condition plus  $\log (\sigma) < 1.8$ define the S4 sample of 93 objects.}
\label{fig:SS10}
\end{figure}

Figure \ref{fig:SS10} illustrates the parameters of the automatic selection. Objects 
inside the box delimited  by a dashed line have $\delta_{flux} (\hb) < 10$ and $\delta_{FWHM}(\hb) < 10$. 
This, plus the condition  $\log (\sigma) < 1.8$, define the S4 sample of 93 objects.

%Figure --- Sample Selection  %%%%%%%%% Figure 18 %%%%%%%%
%%%%%%%%%%%%%%%%%%%%%%%%%%%%%%%%%%%%

\begin{figure*}
\centering
\resizebox{14cm}{!}{\includegraphics{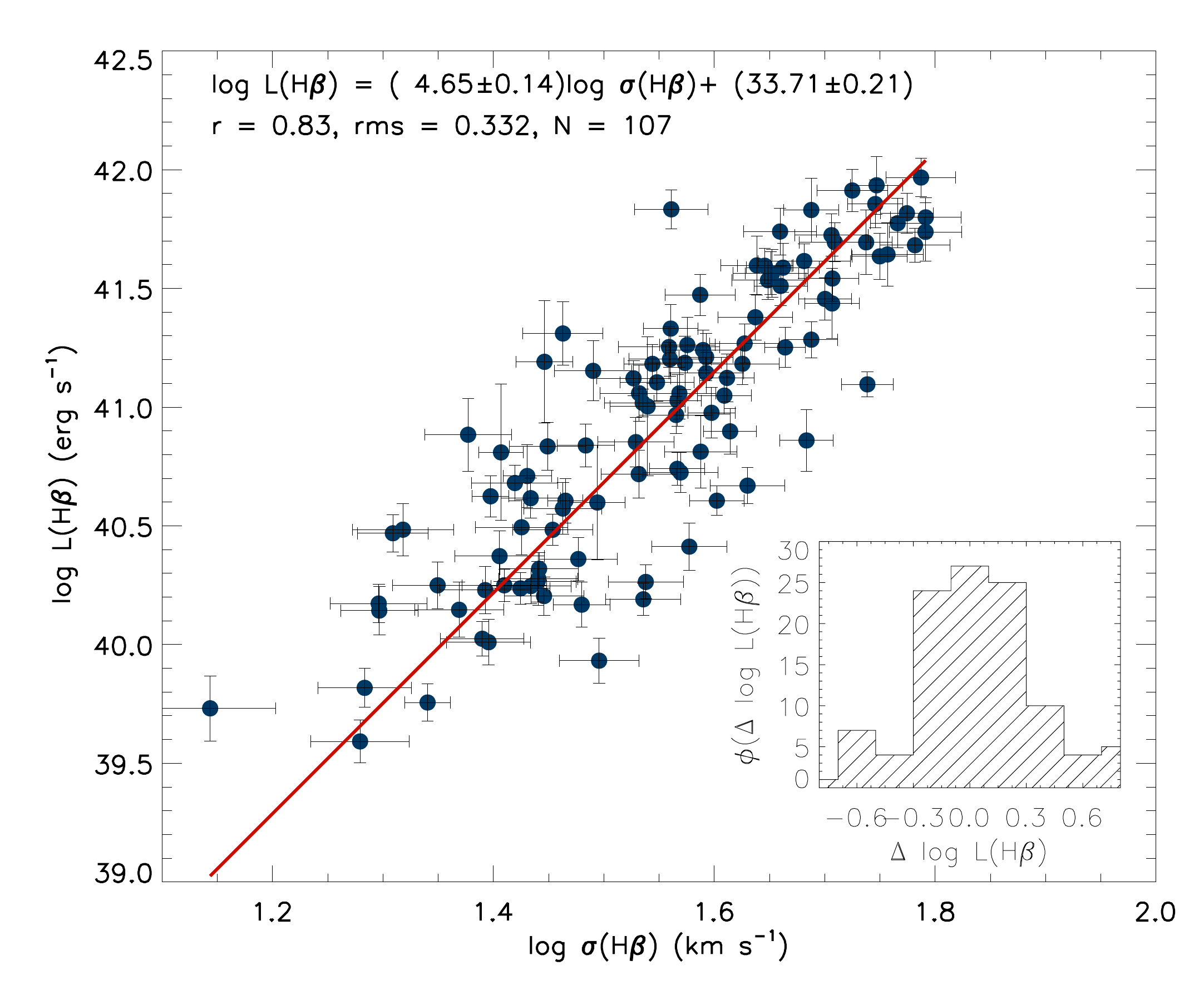}}
\caption[$L - \sigma$ relation for all the \hii\ galaxies] {\small $L - \sigma$ relation for all the \hii\ galaxies  with good determination of  
Luminosity and $\sigma$ (S3).
The inset shows the distribution of the residuals of the fit.}
\label{fig:H2gxAll}
\end{figure*}

The $L - \sigma$ relation for the 107 objects in S3 
for which we have a good estimate of their luminosity and velocity dispersion is shown in Figure \ref{fig:H2gxAll}.
It follows
the expression:

\begin{equation}
 \logd L(\hb) =  ( 4.65 \pm 0.14 ) \logd \sigma + (33.71 \pm 0.21), 
 \label{eq:LS1}
\end{equation}
with an rms scatter of $\delta \logd L(\hb) = 0.332$.

For the 69  objects of the restricted sample (S5) we obtained:
\begin{equation}
 \logd L(\hb) =  ( 4.97 \pm 0.17 ) \logd \sigma + (33.22 \pm 0.27), 
 \label{eq:LS2}
\end{equation}

An important conclusion of the comparison of the results obtained from S3 and S5 is that while 
the $L - \sigma$ relation scatter is reduced from an rms of 0.332 to an rms of 0.25 for 
S5, the errors in both the slope and zero points are slightly larger for the latter as a result 
of reducing the number of objects by about 2/3.

%%%%%%%%%%%%%%%%%%%%%%%%%%%%%%%Figure 19 %%%%%%%%%%%%%%%%%%%%%
%Figure ---  $L - \sigma$ relation for all the objects in the sample  -- H2gxAll
%%%%%%%%%%%%%%%%%%%%%%%%%%%%%%%%%%%%

\subsection{Further restricting the sample by the quality of the line profile fits}
We have also investigated the sensitivity of the \lsig\ relation  to  changes in the emission 
line profiles as determined by the quality of  the gaussian fit. The definition of quality 
is related to the automatic profile classification described in the previous section and illustrated 
in Figure \ref{fig:SS10}. As discussed previously, objects inside the box delimited by the dashed  lines have $\delta_{flux} (\hb) < 10$ 
and $\delta_{FWHM}(\hb) < 10$. By adding the condition that  $\log (\sigma) < 1.8$ we obtain
the S4 sample of 93 objects. We have selected five subsamples with increasing 
restricted definition of departure from a gaussian fit, i.e. with differences smaller  than 10,
8, 5, 3 and 1 percent. The criteria are arbitrary and different cuts could have been justified, but the procedure was just used as a test and as such, any reasonable cut is valuable.
The results of the fits are shown in Table 
\ref{tab:tab09}.

%%Table  --- Restricted Sample %%%%%%%%%%%%%%%%%%%%%%%%%%%%%%%%%%%%%
\input{ch04/tab/tab09}
%%%%%%%%%%%%%%%%%%%%%%%%%%%%%%%%%%%%%

We can see from the table, that more restrictive gaussian selection still gives very similar values 
of the slope and the zero point of the  \lsig\ relation. 
It achieves a small improvement in
the rms but at the cost of a significantlly reduced sample which results in a substantial  
increase of  the uncertainties of the slope and zero point roughly as the inverse of the square 
root of the number of objects. 

It is interesting to compare these results with those using  S3  with 107 
objects some of them with profiles that clearly depart from gaussian. The 
least squares fit for  S3 (see equation \ref{eq:LS1}) gives  
coefficients  $33.71 \pm 0.21$ and $ 4.65 \pm 0.14 $  for the zero point and slope of the 
relation respectively. These values are very similar to those at the 10 percent cut 
 but the rms and errors in the coefficients are smaller, consistent with  a sample containing a larger number of objects. 

The important conclusion from this exercise is that the  \lsig\ relation is robust against profile 
selection. Selecting only those objects with the best gaussian profiles  makes no change in the relation 
coefficients but substantially increases the uncertainties and the rms of the fit due to the reduction in the number of objects.
We therefore strongly suggest the use of the \lsig\ relation without a finer line profile selection.

Furthermore, when applying the \lsig\ distance estimator to  high redshift \hii\ galaxies where the data is bound to have  a  lower S/N, and a selection based 
on details of the emission line profile  will be difficult to perform.
Ideally we would like to reduce the distance estimator scatter without reducing the number 
of objects, i.e. with only a small percentage of rejects from the original observed sample. 

Furthermore, it is clear from an inspection of Figure \ref{fig:H2gxAll} (for S3) that the error 
bars  are somehow smaller than the 
observed scatter in the relation, suggesting the presence of a second parameter in the 
correlation. As we will show below, this is indeed the case and thus it is possible to reduce 
substantially the scatter of the relation by including  additional independent observables
without a drastic reduction of the number of objects in the sample.

\subsection{Search for a second parameter in the \lsig\ relation}
In this section we explore the possibility that the scatter -- at least  part of it -- in 
the \lsig\ relation is due to a second parameter. 

Let us assume that the \lsig\ relation is a reflection of the  virial 
theorem and a constant M/L ratio for the stellar population of these very young stellar 
clusters. Given that the virial theorem is bi-parametric, with the mass of the cluster depending 
on cluster's velocity dispersion and size, one would expect the size of the system to
be a second parameter in the \lsig\ relation. 

The ionising flux  in  these young clusters evolve very rapidly, therefore it is also 
expected that age should play a role in the luminosity scatter. Thus parameters like the 
equivalent width of the Balmer lines or continuum colours, that are good age indicators, may 
also play a role in the scatter. 
\cite {Melnick1987} proposed chemical composition, in fact the oxygen abundance, as a second 
parameter in the \lsig\ relation. 

In what follows we will analyse one by one these potential second parameters.

%%Table 6 --- Second parameter-S3-- %%%%%%%%%%%%%%%%%%%%%%%%%%%%%%%%%%%%%
\input{ch04/tab/tab06A18}
%%%%%%%%%%%%%%%%%%%%%%%%%%%%%%%%%%%%%

%%Table 6 --- Second parameter - S4--%%%%%%%%%%%%%%%%%%%%%%%%%%%%%%%%%%%%%
\input{ch04/tab/tab061018}
%%%%%%%%%%%%%%%%%%%%%%%%%%%%%%%%%%%%%

%%Table 6 --- Second parameter - S3- Bayesian--%%%%%%%%%%%%%%%%%%%%%%%%%%%%%%%%%%%%%
\input{ch04/tab/tab06A18B}
%%%%%%%%%%%%%%%%%%%%%%%%%%%%%%%%%%%%%

\subsubsection{Size}
If the  \lsig\ correlation is a consequence of these young massive clusters being at (or close to)
virial equilibrium, then the strongest  candidate for a second parameter is the size of 
the star forming region \citep{Terlevich1981, Melnick1987}. 
We have explored this possibility using the SDSS measured radii for our sample in all the 
available bands. The general form of the correlation is:
\begin{equation}
	 \logd L(\hb) =  \alpha + \beta \logd \sigma + \gamma \logd R_i\,
\end{equation}
where $\alpha$, $\beta$ and $\gamma$ are the correlation coefficients and i runs over the 
SDSS bands (u, g, r, i, z). In all cases we have used the SDSS measured effective Petrosian 
radii and corrected for seeing also available from SDSS.  Tables \ref{tab:tab06A18} and \ref{tab:tab061018} 
show the correlation coefficients and the scatter obtained by means of a $\chi^2$ reduction 
procedure for the S3 and S4 samples respectively.

Consistent with what we found above regarding the profile selection, the results of the fits 
of the `10\% cut'  sample S4 are not better than those of  S3.
Therefore in what follows we will only consider S3 taking it as the `benchmark' sample.

Using the method proposed 
by \citet{Kelly2007} and his publicly available IDL routines we performed also a bayesian 
multi-linear fit. The reason to use this additional  analysis is to obtain better estimates 
of the uncertainties in every one of the correlation coefficients. The results of the 
analysis are shown in table \ref{tab:tab06A18B} for S3. Comparing these results with those obtained 
previously (Tables \ref{tab:tab06A18} and \ref{tab:tab061018})
it is clear that there are only small differences in the coefficients and their uncertainties 
which are attributable to the better treatment of errors in the bayesian procedure. The bayesian 
zero point tends to be smaller while the  slopes tend to be slightly larger.

Since at high redshifts the O[III]$\lambda$ 5007 line must be easier to observe than the H$\beta$ line, we have repeated the previous analysis using the values of velocity dispersion 
as measured from the O[III]$\lambda$ 5007 line instead of that of the $\hb$ 
line. The results for S3 are shown in Tables \ref{tab:tab06A18O} and \ref{tab:tab06A18OB} 
for the  $\chi^2$ reduction and the bayesian analysis respectively. 
After comparing the results presented in tables 8 and 11 we found that the use of $\sigma$(O[III]) 
introduces only a small extra dispersion in the relation.

At this stage we conclude that the size is indeed a second parameter of the correlation and in particular
the size in the u band shows the best results. The corresponding relation for the S3 sample is given by:
\begin{equation}
\begin{split}
 \logd & L(\hb) =  ( 3.08 \pm 0.22 ) \logd \sigma  + (0.76 \pm 0.13)  \log (R_u) + \\
 &+ (34.04 \pm 0.20), 
 \label{eq:LS1}
 \end{split}
\end{equation}
with an rms scatter of $\delta \logd L(\hb) = 0.261$. \\

Still, we have to be aware that the contribution of the size
to  the reduction of the scatter of the correlation is limited probably due to the fact to be  
discussed in \S 5.7, that the Petrosian radius of \hii\ galaxies is not a good estimator of the 
cluster dimension, but instead a measure of the size of the whole system, but indeed the positive reduction of the scatter of the \lsig\ relation when using even this measure of size, indicates that there is a subtle relation between both sizes. 

\subsubsection{Metallicity}
\citet{Terlevich1981} proposed that oxygen abundance is a good indicator of the long term 
evolution of the system. They proposed a simple `closed box' chemical evolution model with 
many successive cycles of star formation in which, for each cycle, evolution is traced by 
the $EW(\hb)$ 
whereas the long term evolution of the system, spanning two or more cycles, 
could be traced by the oxygen abundance, which then becomes  a plausible second parameter in the \lsig\ 
correlation. When metallicity is used as a second parameter the resulting correlation is given by:
\begin{equation}
	 \logd L(\hb) =  \alpha + \beta \logd \sigma + \gamma [ 12 + \logd \mathrm{(O/H)}]\,
\end{equation}
where $\alpha$, $\beta$ and $\gamma$ are the correlation coefficients  %as well as the corresponding PCA results 
shown in Tables \ref{tab:tab06A18},  \ref{tab:tab061018}, \ref{tab:tab06A18B}, % \ref{tab:tab061018B}, 
\ref{tab:tab06A18O} and \ref{tab:tab06A18OB} following the same procedure as described in the previous section for the radii. It is clear that the 
metallicity plays a role as a 
second parameter albeit relatively small. We must not forget, though, that because of 
the nature of  the sample objects, the dynamical range of  metallicity is very narrow 
(see Figure \ref{fig:HAb}), not  enough to affect significantly the \lsig\ 
correlation. 

%%Table 6 --- Second parameter  S3 with 5007%%%%%%%%%%%%%%%%%%%%%%%%%%%%%%%%%%%%%
\input{ch04/tab/tab06A18O}
%%%%%%%%%%%%%%%%%%%%%%%%%%%%%%%%%%%%%

%%Table 6 --- Second parameter S3 with 5007 -- Bayesian--%%%%%%%%%%%%%%%%%%%%%%%%%%%%%%
\input{ch04/tab/tab06A18OB}
%%%%%%%%%%%%%%%%%%%%%%%%%%%%%%%%%%%%%

We have repeated the  analysis using  the strong line metallicity indicators  N2 and R23.
The results are also given in Tables  \ref{tab:tab06A18},  \ref{tab:tab061018}, \ref{tab:tab06A18B}, %\ref{tab:tab061018B}, 
\ref{tab:tab06A18O} and \ref{tab:tab06A18OB}. They are similar 
to those obtained using T$_e$ based direct metallicity but surprisingly, showing slightly less dispersion 
when using N2.

\subsubsection{Age}
The age of the starburst is also a  second parameter candidate for the \lsig\ correlation.
We used   the $EW (\mathrm{H}\beta)$ as a starburst age indicator \citep{Dottori1981, Dottori1981b}.
The 
resulting correlation is given as:
\begin{equation}
	 \logd L(\hb) =  \alpha + \beta \logd \sigma + \gamma \logd EW(\mathrm{H}\beta)\,
\end{equation}
and the coefficients are shown in Tables \ref{tab:tab06A18},  \ref{tab:tab061018}, \ref{tab:tab06A18B}, 
\ref{tab:tab06A18O} and \ref{tab:tab06A18OB}.

Another possible age indicator is the continuum colour. We consider the (u - i) colour as 
a second parameter, the resulting correlation is given by:
\begin{equation}
	 \logd L(\hb) =  \alpha + \beta \logd \sigma + \gamma (u - i)\,
\end{equation}
The  coefficients are also shown in Tables \ref{tab:tab06A18},  \ref{tab:tab061018}, \ref{tab:tab06A18B}, 
\ref{tab:tab06A18O} and \ref{tab:tab06A18OB}. 

From the above results, it is clear that age should play  a role in the scatter of the 
\lsig\ correlation albeit very small. As with metallicity,  by design the sample covers 
a narrow dynamic range of ages, consequence of the selection of equivalent widths of the emission lines,
chosen such that only bursts younger than about 5 Myr are used in our study.

As already mentioned, we find that limiting the sample to objects with gaussian profiles does 
not improve  the fit
but limiting the sample to objects with log($\sigma$) $<$ 1.8, does.
The second parameter with largest variance is the \emph{u} size. Including it does improve 
radically the fit.

It is interesting to note that in the absence of a size determination, the best second parameter is the oxygen abundance O/H (or its proxy N2 or R23) in line with the early results of 
\citet{Terlevich1981,Melnick1987, Melnick1988}. This result is critical for future work with very distant systems where the Petrosian radius will be difficult to determine. 

We therefore conclude  that the best second parameter is the size in particular $R_u$.
The use of the other observables [O/H, N2, R23, EW(H$\beta$),  and (u - i)] also lead to a 
reduction of the scatter in the relation but to a lesser extent than  what is achieved by 
using the size.  Still they are useable in the absence of a size determination.

\subsection{Multiparametric fits}
The theoretical expectation that the emitted luminosity per unit mass in a young cluster 
should rapidly evolve with age and should also have some dependence on the metallicity of 
the stars, suggests that more parameters (other than the velocity dispersion and size of the 
cluster, e.g. its mass) may be playing a role in the \lsig\ relation.

 We have 
explored the possibility that a third or even a fourth  parameter are present in the correlation;
the general expression for the fit is:
\begin{equation}
	 \logd L(\hb) =  \alpha + \beta \logd \sigma + \gamma  A + \delta B + \epsilon C\,
\end{equation}
where $\alpha$, $\beta$, $\gamma$, $\delta$ and $\epsilon$ are the correlation coefficients and 
A, B and C are different combinations of parameters. Tables \ref{tab:tab07A18} and \ref{tab:tab07A18B}
show the parameter combinations that give the least scatter in the multi-parametric 
correlation for the sample S3 for a $\chi^2$ and a Bayesian methodology respectively.
Tables \ref{tab:tab07A18O} and \ref{tab:tab07A18OB} show the results when using the 
[OIII]$\lambda$ 5007 velocity dispersion.

\begin{figure*}
\centering
\resizebox{14cm}{!}{\includegraphics{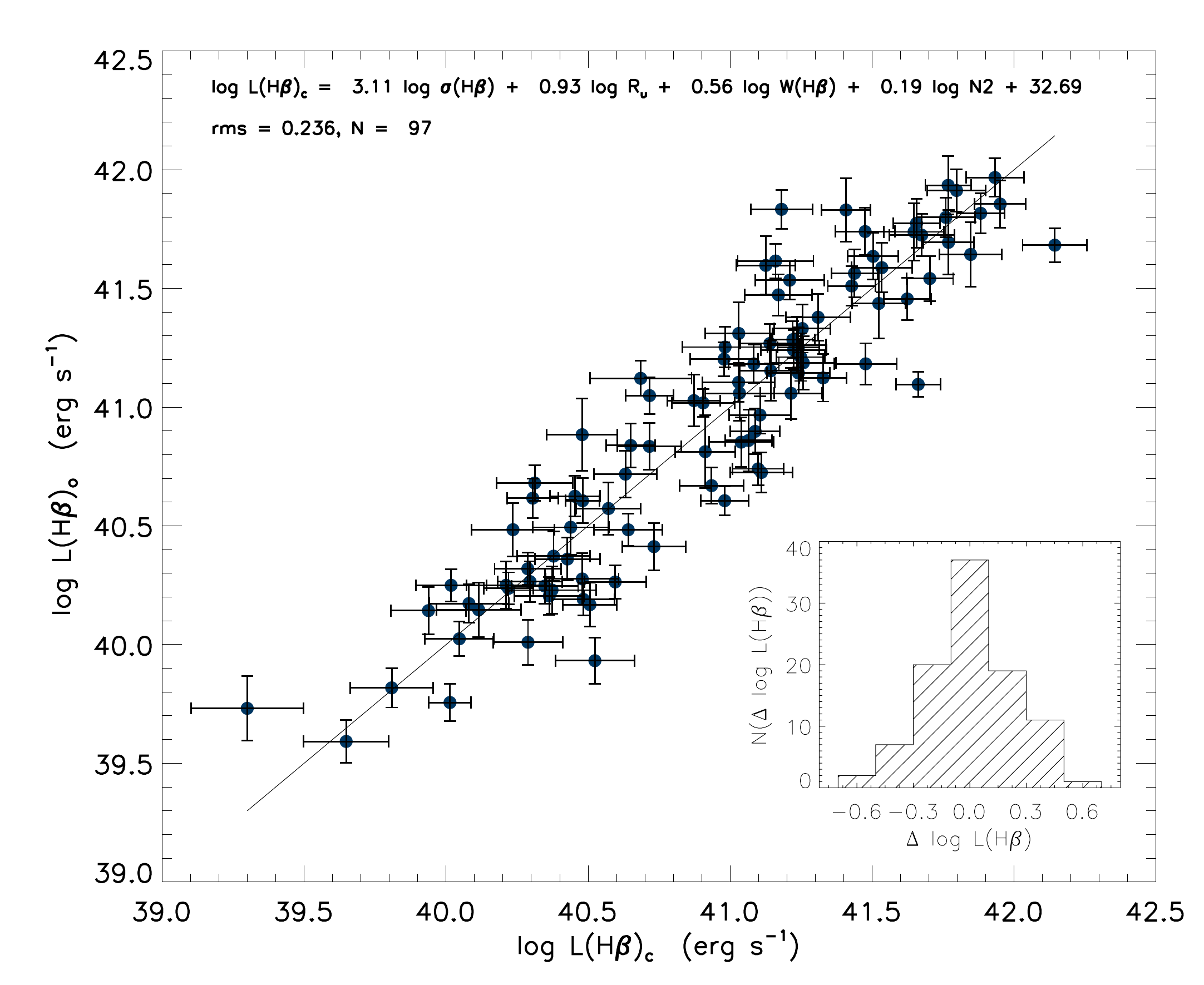}}
\caption[Observed L(H$\beta$) vs. calculated using the best Bayesian multi-parametric fitting] {\small Observed L(H$\beta$) [L(H$\beta)_o$] vs.~L(H$\beta$) calculated using the best Bayesian multi-parametric fitting corresponding to the expression displayed on the top of the figure. The 1:1 line is shown. The inset panel shows the luminosity residuals distribution.
}
\label{fig:LSt}
\end{figure*}

A summary  of the results indicates that  when the \lsig\ relation is combined with the radius in 
the u band, the $(u-i)$ colour and the metallicity, the scatter is significantly reduced. 
The best result is:
\begin{equation}
\begin{split}
\logd &L(\hb) =  ( 2.79 \pm 0.23 ) \logd \sigma  + (0.95 \pm 0.13)  \log R_u  +\\ 
 + (&0.63 \pm 0.19)  \log  EW(H\beta) + (0.28 \pm 0.13)  \log N2  + \\
 + (&33.15 \pm 0.22), 
\end{split}
 \label{eq:LS1}
\end{equation}with an rms scatter of $\delta \logd L(\hb) = 0.233$. This best solution is 
illustrated in Figure \ref{fig:LSt}. 

%%%%%%%%%%%%%%%%%%%%%%%%%%%%%%%%%%%%%
%%Table 7 --- Fourth and fifth parameters
\input{ch04/tab/tab07A18}
%%%%%%%%%%%%%%%%%%%%%%%%%%%%%%%%%%%%%

%%%%%%%%%%%%%%%%%%%%%%%%%%%%%%%%%%%%%
%%Table 7 --- Fourth and fifth parameters
\input{ch04/tab/tab07A18B}
%%%%%%%%%%%%%%%%%%%%%%%%%%%%%%%%%%%%%

%%%%%%%%%%%%%%%%%%%%%%%%%%%%%%%%%%%%%
%%Table 7 --- Fourth and fifth parameters
\input{ch04/tab/tab07A18O}
%%%%%%%%%%%%%%%%%%%%%%%%%%%%%%%%%%%%%

%%%%%%%%%%%%%%%%%%%%%%%%%%%%%%%%%%%%%
%%Table 7 --- Fourth and fifth parameters
\input{ch04/tab/tab07A18OB}
%%%%%%%%%%%%%%%%%%%%%%%%%%%%%%%%%%%%%

%%%%%%%%%%%%%%%%% Figure 20 %%%%%%%%%%%%%%%%%%%
%%%%%%%%%%%%%%%%%%%%%%%%%%%%%%%%%%%%%

It seems reasonable to infer that the resulting coefficients support the scenario of a virial origin of the \lsig\ 
relation, in that the $\logd \sigma$ coefficient is smaller than 3, the size coefficient is close to 1 and that 
other effects like the age and metallicity of the burst alter the virial nature of the relation.

%%%%%%%%%%%%%%%%%%%%%%%%%%%%%%%%%%%%%
%%Table 7 --- Fourth and fifth parameters
\input{ch04/tab/tab07HDS}
%%%%%%%%%%%%%%%%%%%%%%%%%%%%%%%%%%%%%

%%%%%%%%%%%%%%%%%%%%%%%%%%%%%%%%%%%%%
%%Table 7 --- Fourth and fifth parameters
\input{ch04/tab/tab07UVES}
%%%%%%%%%%%%%%%%%%%%%%%%%%%%%%%%%%%%%

\subsubsection{Comparing the scatter between UVES and HDS data}
 We discussed 
in \S \ref{sec:hrss} the different setups used for the HDS and UVES observations. 
We show in Tables \ref{tab:tab07HDS} and \ref{tab:tab07UVES}  the 
regression coefficients calculated separately for both sets of observations 
and the  combination of parameters that renders the least scatter. 

It can be seen that the scatter of the HDS data is larger than that of the UVES data. We 
interpret this as an effect of the wider slit used in the HDS 
observations combined with  the compact size of the sources and the excellent seeing prevailing 
during the observations. All these effects put together plus unavoidable fluctuations  in 
the auto guiding procedure may have contributed in increasing the uncertainties in the 
observed  emission line profiles.

Although a similar but smaller effect cannot at this stage be ruled out from the UVES data, 
given that the slit used was also larger than the seeing disk, we can conclude that the `true' 
scatter of the relation is probably closer -- if not even smaller -- to that observed in 
the UVES data, 
i.e.~r.m.s.~$\lesssim $0.2.

\section{Summary}
 We have carefully constructed a sample of 128 compact local \hii\ galaxies, with high equivalent widths of their Balmer emission lines, with the 
objective of assessing the validity of the $L({H\beta}) - \sigma$ relation and its use as an accurate distance estimator. 
To this end we obtained high S/N high-dispersion 
ESO VLT and Subaru echelle spectroscopy, in order to accurately measure the ionized gas velocity dispersion.  Additionally, we obtained integrated H$\beta$ fluxes from low
dispersion wide aperture spectrophotometry, using the 2.1m telescopes at
Cananea and San Pedro M\'artir in Mexico, complemented with data from the  SDSS
spectroscopic survey.

 After  further restricting the sample to include only those systems with $\logd \sigma<1.8$ and removing objects with low quality data, 
the remaining sample  consists of 107
`bonafide' \hii\ galaxies. These systems have indeed luminosities and metallicities typical of \hii\ galaxies  and their position in the diagnostic diagram is typical of high excitation, low metallicity and extremely young \hii\ regions.

Using this sample we have found that:
 
\begin{enumerate}
\item  The \lsig\ relation is strong and stable against changes in the sample definition,  based on the characteristics  of the emission line profiles.
In particular we have investigated the role that the `gaussianity' of the  line profile plays on the relation. 
This was tested to destruction with both objective and subjective 
methods of profile classification and assessment to define several subsets.
 
In agreement with previous work we find that
the \lsig\ relation for \hii\ galaxies with gaussian emission line profiles has 
a smaller scatter than that of the complete sample.
On the other hand this is achieved at the cost of substantially reducing the sample. 
The rejected fraction  in  \citet{Bordalo2011} or \citet{Chavez2012} is close to or larger than 50\%, which is not compensated by the gain in rms. 
The use of the complete sample, i.e. without a profile classification, is a far more practical 
proposal given that, in order to perform a proper selection of gaussian profiles, we need  data 
that have S/N and resolution much higher than  that required to measure just the FWHM. 
Therefore it is far more costly in terms of observing time and instrumentation requirements to determine departures from gaussianity than to just accurately measure the FWHM of an emission line. 
It is  shown in section 4.5.2 that while the r.m.s. errors are indeed reduced on the fits to the subset of \hii\ galaxies with Gaussian profiles, the value of the coefficients hardly change at all, although their errors are substantially larger than those of the complete sample.

In conclusion, selecting the best gaussian profiles improves the rms but at a very heavy cost in terms of rejects and hence of telescope time, which is neither practical nor justified for a distance estimator, as we showed in section 4.5.

Therefore, the use of the full sample, limited only by the $\logd \sigma<1.8$ selection, is strongly recommended.
Our best \lsig\ relation is:
$${ \logd L(\hb) =  4.65  \logd \sigma + 33.71 \;, }$$
with an rms scatter of ${\delta \logd L(\hb) = 0.332}$. 

\item  We searched for the presence of a second parameter in the \lsig\ relation. We found that using as second parameter either  size, oxygen abundance O/H  or its proxy N2 or R23, EW or continuum colour, the scatter is considerably reduced. Including the size as a second parameter produces the best fits, and among them the size in the u-band shows the smallest scatter, 

$$ \logd L(\hb) =  3.08 \logd \sigma  + 0.76 \log R_u + 34.04 \;, $$
with an rms scatter of $\delta \logd L(\hb) = 0.261$. 

This result points clearly to the existence of  a Fundamental Plane in \hii\ galaxies suggesting that the main mechanism of line broadening is linked to the gravitational potential of the young massive cluster. 
It is important  to underline that in the absence of a size measurement, the best second parameter is the abundance O/H  or its proxy N2 or R23, a result that is crucial for the application to very distant systems where the size will be difficult to determine. 

\item We also investigated which parameters in addition to the size can further reduce the scatter. We found, using multi parametric fits ,
that including as a third parameter the $(u-i)$ colour or the equivalent width, and as a fourth parameter the 
metallicity does significantly reduce further the scatter.

Our best multiparametric estimator is:\\
\begin{eqnarray}
\logd L(\hb) =  2.79 \logd \sigma  + 0.95 \log R_u + 0.63 \log EW(H\beta) + \nonumber \\
 0.28 \log N2  + 33.15 \nonumber 
\end{eqnarray}
with an rms scatter of $\delta \logd L(\hb) = 0.233$. \\

The argument could be sustained that the value of the  coefficients of the fit provides further support for the virial origin of the \lsig\ 
relation since the $\logd \sigma$ coefficient is smaller than 3. It is quite possible  that such 
virial nature is altered by other effects like the age (EW) and metallicity (N2)  of the burst.
Thus the coefficients in the best 
estimator (see equation (\ref{eq:LS1})) are very close to what is expected from a young 
virialized ionising cluster and, perhaps even more relevant, the sum of the stellar and ionised gas masses of the cluster are similar to the dynamical mass estimated with the HST `corrected' Petrosian radius.

We conclude that the evidence strongly points to gravity as the main mechanism for the broadening of the emission lines in these very young and massive clusters. 

\item The masses of the clusters, both photometric and dynamical,  are very large  while  their size is very compact. 
Their range covers three decades from about 2 $\times$ 10$^6$ \Msol\  to  10$^9$ \Msol.
Their HST corrected Petrosian radius range from a few tens of parsecs to a few hundred parsecs.
To further investigate this important property of the \hii\ galaxies and its impact on the distance estimator it is crucial to secure high resolution optical and NIR images of this sample of objects.

\item Bayesian and $\chi^2$ fits to the $L(\hb)-\sigma$ correlation give similar results.

\item  The application of  the \lsig\ distance estimator to \hii\ galaxies at cosmological distances, where the size would be difficult to determine, will require the use of a metallicity indicator and the EW of the Balmer lines as a second and third parameter. 
According to our findings, this will result in a predictor with 
$\delta \logd L(\hb) \sim 0.3$ using either $\sigma (H\beta)$  or the easier to determine, at such redshifts, $\sigma [OIII]$.
 
\item Given that the \lsig\ relation is basically a  correlation between the ionising flux, produced by the massive stars, and the velocity field produced by the star and gas potential well, the existence of a narrow  \lsig\ relation puts strong limits on the possible changes in the IMF. 
Any systematic variation in the IMF will affect directly the M/L ratio and therefore the slope and/or  zero point of the relation.  
 A change of 0.1  in the slope of the IMF would be reflected in a change in luminosity scale of the \lsig\ relation of about $\logd L(\hb) \sim\ 0.2 $. This seems to be too large for the derived correlation.

\item An important aspect to remark is that the design of our complete selection criteria guarantees homogeneous samples at all redshifts in the sense that the  imposed EW limit guarantees a sample younger than a certain age and relatively free of contamination by older populations, the upper limit in $\sigma$ guarantees a sample limited in luminosity and the diagnostic diagram selection guarantees that they are starbursts. The limitation in $\sigma$ is particularly important given that this criterion should remove biases associated with samples in which the mean luminosity changes with distance (Malmquist bias). Any  dependence of the luminosity in parameters like age and  metallicity are  included in the multiparametric fits.

Finally, we envisage observations of  \hii\ galaxies having a limiting $\sigma$ of 63 km/s or equivalently an $\ha\  $ luminosity less than $3 \times 10^{43}$ erg/s  at z $\sim$ 2 to 3 with enough S/N with present instrumentation. They will require exposure times of  about 1.5 to 3 hours in an instrument like X-SHOOTER at the VLT in ESO  to obtain line profiles with enough S/N to determine FWHM with less than 10\% rms error. This in turn will allow us to measure the local expansion rate of the Universe, $H_0$, to a percent precision which is a prerequisite  for independent constraints on the mass-energy content and age of the Universe 
as well as to map its behaviour by using
several independent yet accurate tracers of the cosmic expansion over the widest possible range of redshift.
\end{enumerate}

%% file: ch04/tab/tab01.tex
%Table --- Journal of Observations
\begin{table*}
 \centering
 \begin{minipage}{140mm}
  \caption[Local \hii\ galaxies journal of Observations] {\small Journal of observations. }
  \label{tab:tab01}
 \resizebox{1.0 \textwidth}{!}{
\begin{tabular} { l l c c c }
\hline
\hline
(1) & (2) & (3) & (4) & (5) \\ 
Dates & Telescope & Instrument & Detector & Slit-width \\
\hline
5 \& 16 Nov 2008 & NOAJ-Subaru & HDS    &EEV  (2 $\times$ 2K $\times$ 4K)\footnote{$2 \times 4 $ binning.} & 4$''$ \\
16 \& 17 Apr 2009 & ESO-VLT & UVES-Red & EEV (2 $\times$ 2K $\times$ 4K) & 2$''$ \\
15 - 17 Mar 2010 & OAN - 2.12m & B\&C %+ 600 l/mm (8d63m)
                                                                   & SITe3 (1K $\times$ 1K) & 10$''$ \\
10 - 13 Apr 2010 & OAGH - 2.12m & B\&C %+ 150 l/mm (3d30m) 
                                                                   & VersArray (1300 $\times$ 660)& 8.14$''$ \\
8 -10 Oct 2010 & OAN - 2.12m & B\&C & Thompson 2K  & 13.03$''$ \\
7 - 11 Dic 2010 & OAGH - 2.12m & B\&C & VersArray (1300 $\times$ 660)& 8.14$''$ \\
4 - 6 Mar 2011 & OAN - 2.12m & B\&C & Thompson 2K  & 13.03$''$ \\
1 - 4 Apr  2011 & OAGH - 2.12m & B\&C & VersArray (1300 $\times$ 660)& 8.14$''$ \\
\hline 
\end{tabular}
}
\end{minipage}
\end{table*}

%% file: ch04/tab/tab10.tex
%Table --- Linear regressions 3 
\begin{table*}
 \centering
 \begin{minipage}{120mm}
  \caption[Samples Description] {\small Samples Description.}
  \label{tab:tab10}
 \resizebox{1.1 \textwidth}{!}{
\begin{tabular} { c l l r }
\hline
\hline
(1) & (2) & (3) & (4) \\ 
Sample  & Description & Constraints &  N \\
\hline
S1	&	Observed  		&	None						&      128	\\
S2	&	Initial 			&	S1 - all dubious data eliminated		&	114	\\
S3	&	Benchmark  		&	S2 - $\log \sigma$ (H$\beta) > 1.8$	&	107	\\
S4	&	10\% cut 			&	S3 - $\delta_{flux}(\mathrm{H}\beta) > 10$, $\delta_{FWHM}(\mathrm{H}\beta) > 10$	&	93	\\
S5	&	Restricted			&	S3 - kinematical analysis		        &	69	\\

\hline 
\end{tabular}
}
\end{minipage}
\end{table*}

%% file: ch04/tab/tab03.tex
%Table  3 --- H II galaxies local sample low dispersion measurements. 

\onecolumn
\tiny
\begin{center}
\begin{longtable}{@{} l l c c c c c c l @{}}
 
\caption[Local \hii\ galaxies Fluxes and EW]{\small Low resolution and SDSS DR7 $\hb$ and $[\Oiii]\ \lambda \lambda 4959, 5007$ fluxes and EW($\hb$). } 
\label{tab:tab03} \\

%This is the header for the first page of the table...
\hline \hline \\[-2ex]
\multicolumn{1}{c}{(1)} &
\multicolumn{1}{c}{(2)} &
\multicolumn{1}{c}{(3)} &
\multicolumn{1}{c}{(4)} &
\multicolumn{1}{c}{(5)} &
\multicolumn{1}{c}{(6)} &
\multicolumn{1}{c}{(7)} &
\multicolumn{1}{c}{(8)} &
\multicolumn{1}{c}{(9)} \\ 

\multicolumn{1}{c}{\bf{Index}} &
\multicolumn{1}{c}{\bf{Name}} &
\multicolumn{1}{c}{ \textbf{F$^{*}$($\mathbf{\hb}$) }} &
\multicolumn{1}{c}{ \textbf{F($\mathbf{\hb}$) }} &
\multicolumn{1}{c}{ \textbf{F($\mathbf{\hb}$) }} &
\multicolumn{1}{c}{ \textbf{F($\mathbf{[\Oiii]\ \lambda 4959}$) } } &
\multicolumn{1}{c}{ \textbf{F($\mathbf{[\Oiii]\ \lambda 5007}$) } } &
\multicolumn{1}{c}{ \textbf{EW($\mathbf{\hb}$) }} &
\multicolumn{1}{c}{ \textbf{Inst.$^{\dagger}$} } \\

\multicolumn{1}{c}{} &
\multicolumn{1}{c}{} &
\multicolumn{1}{c}{ \textbf{SDSS DR7}} &
\multicolumn{1}{c}{ \textbf{LS Gaussian Fit}} &
\multicolumn{1}{c}{ \textbf{LS Integral}} &
\multicolumn{1}{c}{ \textbf{LS Gaussian Fit} } &
\multicolumn{1}{c}{ \textbf{LS Gaussian Fit} } &
\multicolumn{1}{c}{ \textbf{\AA}} &
\multicolumn{1}{c}{} \\[0.5ex] \hline \\[-1.8ex]

%\multicolumn{1}{c}{} &
%\multicolumn{1}{c}{} &
%\multicolumn{1}{c}{ ($10^{-16}\ \mathrm{erg\ s^{-1} cm^{-2}}$)} &
%\multicolumn{1}{c}{ ($10^{-16}\ \mathrm{erg\ s^{-1} cm^{-2}}$)} &
%\multicolumn{1}{c}{ ($10^{-16}\ \mathrm{erg\ s^{-1} cm^{-2}}$)} &
%\multicolumn{1}{c}{ ($10^{-16}\ \mathrm{erg\ s^{-1} cm^{-2}}$)} &
%\multicolumn{1}{c}{ ($10^{-16}\ \mathrm{erg\ s^{-1} cm^{-2}}$)} &
%\multicolumn{1}{c}{} \\[0.5ex] \hline \\[-1.8ex]

\endfirsthead

%This is the header for the remaining page(s) of the table...
\multicolumn{1}{c}{(1)} &
\multicolumn{1}{c}{(2)} &
\multicolumn{1}{c}{(3)} &
\multicolumn{1}{c}{(4)} &
\multicolumn{1}{c}{(5)} &
\multicolumn{1}{c}{(6)} &
\multicolumn{1}{c}{(7)} &
\multicolumn{1}{c}{(8)} &
\multicolumn{1}{c}{(9)} \\ 

\multicolumn{1}{c}{\bf{Index}} &
\multicolumn{1}{c}{\bf{Name}} &
\multicolumn{1}{c}{ \textbf{F$^{*}$($\mathbf{\hb}$) }} &
\multicolumn{1}{c}{ \textbf{F($\mathbf{\hb}$) }} &
\multicolumn{1}{c}{ \textbf{F($\mathbf{\hb}$) }} &
\multicolumn{1}{c}{ \textbf{F($\mathbf{[\Oiii]\ \lambda 4959}$) } } &
\multicolumn{1}{c}{ \textbf{F($\mathbf{[\Oiii]\ \lambda 5007}$) } } &
\multicolumn{1}{c}{ \textbf{EW($\mathbf{\hb}$) }} &
\multicolumn{1}{c}{ \textbf{Inst.$^{\dagger}$} } \\

\multicolumn{1}{c}{} &
\multicolumn{1}{c}{} &
\multicolumn{1}{c}{ \textbf{SDSS DR7}} &
\multicolumn{1}{c}{ \textbf{LS Gaussian Fit}} &
\multicolumn{1}{c}{ \textbf{LS Integral}} &
\multicolumn{1}{c}{ \textbf{LS Gaussian Fit} } &
\multicolumn{1}{c}{ \textbf{LS Gaussian Fit} } &
\multicolumn{1}{c}{ \textbf{\AA}} &
\multicolumn{1}{c}{} \\[0.5ex] \hline \\[-1.8ex]

\endhead

\multicolumn{2}{l}{{Continued on Next Page\ldots}} \\
\endfoot

\\[-1.8ex] \hline 
\multicolumn{5}{l}{$^*$ All the fluxes are given in units of $10^{-16}\ \mathrm{erg\ s^{-1} cm^{-2}}$.}\\ 
\multicolumn{8}{l}{$^{\dagger}$ The instrument flag indicates the origin of the data. $1:$ Directly measured using long slit as described in the text. $2$: from aperture corrected}\\ 
\multicolumn{8}{l}{SDSS DR7 measurements. $3:$ from \citet{Terlevich1991}, in this case errors in fluxes and EW are taken directly form the cited source. }\\

\endlastfoot
001	&	J000657+005125	&	88.1	$\pm$	1.1	&	112.7	$\pm$	11.6	&	113.0	$\pm$	11.3	&	126.9	$\pm$	8.1	&	381.7	$\pm$	22.6	&	102.2	$\pm$	5.3	&	1	\\
002	&	J001647-104742	&	167.7	$\pm$	1.1	&	231.5	$\pm$	28.2	&	236.1	$\pm$	28.9	&	298.1	$\pm$	19.1	&	882.5	$\pm$	52.2	&	67.6	$\pm$	1.5	&	1	\\
003	&	J002339-094848	&	125.6	$\pm$	0.9	&	153.6	$\pm$	18.7	&	155.1	$\pm$	19.0	&	315.3	$\pm$	20.2	&	955.5	$\pm$	56.5	&	123.9	$\pm$	4.1	&	1	\\
004	&	J002425+140410	&	272.0	$\pm$	1.7	&	407.3	$\pm$	43.2	&	408.1	$\pm$	41.3	&	603.2	$\pm$	48.0	&	1804.8	$\pm$	147.0	&	66.3	$\pm$	1.3	&	1	\\
005	&	J003218+150014	&	254.3	$\pm$	1.4	&	457.0	$\pm$	91.9	&	456.0	$\pm$	96.2	&	671.9	$\pm$	43.1	&	2060.0	$\pm$	121.8	&	82.8	$\pm$	1.7	&	1	\\
006	&	J005147+000940	&	94.8	$\pm$	0.5	&	117.3	$\pm$	14.4	&	116.6	$\pm$	14.3	&	192.1	$\pm$	12.3	&	581.1	$\pm$	34.4	&	107.8	$\pm$	2.7	&	1	\\
007	&	J005602-101009	&	65.7	$\pm$	0.8	&	66.7	$\pm$	8.2	&	66.6	$\pm$	8.2	&	88.8	$\pm$	5.7	&	252.0	$\pm$	14.9	&	52.8	$\pm$	1.7	&	1	\\
008	&	J013258-085337	&	77.9	$\pm$	0.7	&	71.5	$\pm$	8.8	&	73.5	$\pm$	9.1	&	113.5	$\pm$	7.3	&	307.1	$\pm$	18.2	&	72.4	$\pm$	2.3	&	1	\\
\end{longtable}
%\end{ThreePartTable}
\end{center}
\normalsize 
%\twocolumn

%% file: ch04/tab/tab04.tex
%Table 4 --- H II galaxies local sample SDSS DR7 measurements
\onecolumn
\begin{landscape}
\tiny
\begin{center}
\setlength{\tabcolsep}{4pt}
\begin{longtable}{@{} l l c c c c c c c c c c c @{}}
 
\caption[Local \hii\ galaxies line intensity ratios]{\small Line intensity ratios with respect to $\hb = 100$, 1$\sigma$ uncertainties in \% are shown in parenthesis. Our measurements from SDSS DR7.} 
\label{tab:tab04} \\

%This is the header for the first page of the table...
\hline \hline \\[-2ex]
\multicolumn{1}{c}{(1)} &
\multicolumn{1}{c}{(2)} &
\multicolumn{1}{c}{(3)} &
\multicolumn{1}{c}{(4)} &
\multicolumn{1}{c}{(5)} &
\multicolumn{1}{c}{(6)} & 
\multicolumn{1}{c}{(7)} &
\multicolumn{1}{c}{(8)} &
\multicolumn{1}{c}{(9)} &
\multicolumn{1}{c}{(10)} &
\multicolumn{1}{c}{(11)} &
\multicolumn{1}{c}{(12)} & 
\multicolumn{1}{c}{(13)} \\

\multicolumn{1}{c}{\bf{Index}} &
\multicolumn{1}{c}{\bf{Name}} &
\multicolumn{1}{c}{$\mathbf{[\Oii]}$ } &
\multicolumn{1}{c}{$\mathbf{[\Oii]}$ } &
\multicolumn{1}{c}{$\mathbf{[\Oiii]}$ } &
\multicolumn{1}{c}{$\mathbf{[\Oiii]}$ } &
\multicolumn{1}{c}{$\mathbf{[\Oiii]}$ } &
\multicolumn{1}{c}{$\mathbf{\hg}$ } &
%\multicolumn{1}{c}{$\mathbf{\hb}$ } &
\multicolumn{1}{c}{$\mathbf{\ha}$ } &
\multicolumn{1}{c}{$\mathbf{[\Nii]}$ } &
\multicolumn{1}{c}{$\mathbf{[\Nii]}$ } &
\multicolumn{1}{c}{$\mathbf{[\Sii]}$ } &
\multicolumn{1}{c}{$\mathbf{[\Sii]}$ } \\

\multicolumn{1}{c}{} &
\multicolumn{1}{c}{} &
\multicolumn{1}{c}{ $ \lambda  3726$ } &
\multicolumn{1}{c}{ $ \lambda  3729$ } &
\multicolumn{1}{c}{ $ \lambda  4363$ } &
\multicolumn{1}{c}{ $ \lambda  4959$ } &
\multicolumn{1}{c}{ $ \lambda  5007$ } &
\multicolumn{1}{c}{ } &
%\multicolumn{1}{c}{ } &
\multicolumn{1}{c}{ } &
\multicolumn{1}{c}{ $ \lambda  6548$ } &
\multicolumn{1}{c}{ $ \lambda  6584$ } &
\multicolumn{1}{c}{ $ \lambda  6716$ } &
\multicolumn{1}{c}{ $ \lambda  6731$ } \\[0.5ex] \hline \\[-1.8ex]

\endfirsthead

%This is the header for the remaining page(s) of the table...
%\multicolumn{7}{c}{{\tablename} \thetable{} -- Continued} \\[0.5ex]
%\hline \hline \\[-2ex]

\multicolumn{1}{c}{(1)} &
\multicolumn{1}{c}{(2)} &
\multicolumn{1}{c}{(3)} &
\multicolumn{1}{c}{(4)} &
\multicolumn{1}{c}{(5)} &
\multicolumn{1}{c}{(6)} & 
\multicolumn{1}{c}{(7)} &
\multicolumn{1}{c}{(8)} &
\multicolumn{1}{c}{(9)} &
\multicolumn{1}{c}{(10)} &
\multicolumn{1}{c}{(11)} &
\multicolumn{1}{c}{(12)} & 
\multicolumn{1}{c}{(13)} \\

\multicolumn{1}{c}{\bf{Index}} &
\multicolumn{1}{c}{\bf{Name}} &
\multicolumn{1}{c}{$\mathbf{[\Oii]}$ } &
\multicolumn{1}{c}{$\mathbf{[\Oii]}$ } &
\multicolumn{1}{c}{$\mathbf{[\Oiii]}$ } &
\multicolumn{1}{c}{$\mathbf{[\Oiii]}$ } &
\multicolumn{1}{c}{$\mathbf{[\Oiii]}$ } &
\multicolumn{1}{c}{$\mathbf{\hg}$ } &
%\multicolumn{1}{c}{$\mathbf{\hb}$ } &
\multicolumn{1}{c}{$\mathbf{\ha}$ } &
\multicolumn{1}{c}{$\mathbf{[\Nii]}$ } &
\multicolumn{1}{c}{$\mathbf{[\Nii]}$ } &
\multicolumn{1}{c}{$\mathbf{[\Sii]}$ } &
\multicolumn{1}{c}{$\mathbf{[\Sii]}$ } \\

\multicolumn{1}{c}{} &
\multicolumn{1}{c}{} &
\multicolumn{1}{c}{ $ \lambda  3726$ } &
\multicolumn{1}{c}{ $ \lambda  3729$ } &
\multicolumn{1}{c}{ $ \lambda  4363$ } &
\multicolumn{1}{c}{ $ \lambda  4959$ } &
\multicolumn{1}{c}{ $ \lambda  5007$ } &
\multicolumn{1}{c}{ } &
%\multicolumn{1}{c}{ } &
\multicolumn{1}{c}{ } &
\multicolumn{1}{c}{ $ \lambda  6548$ } &
\multicolumn{1}{c}{ $ \lambda  6584$ } &
\multicolumn{1}{c}{ $ \lambda  6716$ } &
\multicolumn{1}{c}{ $ \lambda  6731$ } \\[0.5ex] \hline \\[-1.8ex]

\endhead

\multicolumn{2}{l}{{Continued on Next Page\ldots}} \\
\endfoot

\\[-1.8ex] \hline 
%\insertTableNotes
\endlastfoot
001	&	J000657+005125	&	81.4	(37.0)	&	101.8	(32.3)	&	8.51	(8.9)	&	194	(1.9)	&	581	(1.8)	&	43.9	(3.0)	&	344	(1.8)	&	6.07	(14.1)	&	15.8	(6.6)	&	20.7	(3.6)	&	15.9	(4.8)	\\
002	&	J001647-104742	&	84.1	(3.1)	&	119.9	(2.5)	&	4.70	(10.9)	&	136	(1.2)	&	409	(1.1)	&	40.1	(2.0)	&	358	(1.1)	&	5.88	(3.0)	&	18.0	(1.8)	&	33.3	(1.4)	&	24.9	(1.4)	\\
003	&	J002339-094848	&	64.9	(3.2)	&	99.3	(2.4)	&	7.30	(6.2)	&	207	(1.0)	&	616	(0.9)	&	42.1	(1.8)	&	377	(0.9)	&	3.29	(11.6)	&	9.2	(4.8)	&	19.8	(1.5)	&	15.0	(1.8)	\\
004	&	J002425+140410	&	---	&	---	&	3.08	(12.8)	&	139	(0.9)	&	423	(0.7)	&	42.6	(1.5)	&	335	(0.8)	&	5.26	(5.9)	&	15.6	(2.5)	&	17.6	(1.4)	&	13.3	(1.8)	\\
005	&	J003218+150014	&	---	&	---	&	6.21	(5.9)	&	163	(0.7)	&	481	(0.6)	&	44.7	(1.3)	&	304	(0.6)	&	3.49	(5.9)	&	9.7	(2.6)	&	17.4	(1.0)	&	12.5	(1.3)	\\
006	&	J005147+000940	&	45.1	(6.6)	&	62.6	(5.3)	&	12.37	(3.5)	&	176	(0.9)	&	526	(0.7)	&	47.0	(1.4)	&	283	(0.8)	&	1.38	(16.8)	&	4.1	(7.0)	&	8.7	(2.8)	&	6.6	(3.2)	\\
007	&	J005602-101009	&	107.3	(4.0)	&	168.5	(3.2)	&	5.72	(36.5)	&	102	(1.8)	&	317	(1.6)	&	42.4	(4.3)	&	335	(1.5)	&	13.40	(5.5)	&	41.5	(2.9)	&	34.3	(2.3)	&	27.8	(2.5)	\\
008	&	J013258-085337	&	137.0	(2.6)	&	178.0	(2.3)	&	2.44	(23.4)	&	116	(1.4)	&	352	(1.4)	&	44.2	(2.4)	&	287	(1.4)	&	6.14	(6.7)	&	22.2	(3.4)	&	36.1	(2.2)	&	25.6	(3.1)	\\
\end{longtable}
%\end{ThreePartTable}
\end{center}
\normalsize 
\end{landscape}
%\twocolumn

%% file: ch04/tab/tab02.tex
%Table  2 --- H II galaxies local sample high dispersion measurements. 

\onecolumn
%\begin{landscape}
\tiny
\begin{center}
\begin{longtable}{@{} c l r r l c c @{}}
 
\caption[Local \hii\ galaxies FWHMs]{\small FWHM of $\hb$ and $[\Oiii]\ \lambda 5007$  from the high resolution spectra. } 
\label{tab:tab02} \\

%This is the header for the first page of the table...
\hline \hline \\[-2ex]
\multicolumn{1}{c}{(1)} &
\multicolumn{1}{c}{(2)} &
\multicolumn{1}{c}{(3)} &
\multicolumn{1}{c}{(4)} &
\multicolumn{1}{c}{(5)} &
\multicolumn{1}{c}{(6)} & 
\multicolumn{1}{c}{(7)} \\ 

\multicolumn{1}{c}{\bf{Index}} &
\multicolumn{1}{c}{\bf{Name}} &
\multicolumn{1}{c}{$\mathbf{\alpha}$(J2000) } &
\multicolumn{1}{c}{$\mathbf{\delta }$ (J2000) } &
\multicolumn{1}{c}{$\mathbf{z_{hel}^{*}}$} &
\multicolumn{1}{c}{ \textbf{FWHM ($\mathbf{\hb}$)}} &
\multicolumn{1}{c}{ \textbf{FWHM($\mathbf{[\Oiii]\ \lambda  5007}$)}} \\

\multicolumn{1}{c}{} &
\multicolumn{1}{c}{} &
\multicolumn{1}{c}{(deg) } &
\multicolumn{1}{c}{ (deg)} &
\multicolumn{1}{c}{ } &
\multicolumn{1}{c}{(\AA)} & 
\multicolumn{1}{c}{(\AA)}  \\[0.5ex] \hline \\[-1.8ex]

\endfirsthead

%This is the header for the remaining page(s) of the table...
\multicolumn{1}{c}{(1)} &
\multicolumn{1}{c}{(2)} &
\multicolumn{1}{c}{(3)} &
\multicolumn{1}{c}{(4)} &
\multicolumn{1}{c}{(5)} &
\multicolumn{1}{c}{(6)} & 
\multicolumn{1}{c}{(7)} \\ 

\multicolumn{1}{c}{\bf{Index}} &
\multicolumn{1}{c}{\bf{Name}} &
\multicolumn{1}{c}{$\mathbf{\alpha}$(J2000) } &
\multicolumn{1}{c}{$\mathbf{\delta }$ (J2000) } &
\multicolumn{1}{c}{$\mathbf{z_{hel}^{*}}$} &
\multicolumn{1}{c}{ \textbf{FWHM ($\mathbf{\hb}$)}} &
\multicolumn{1}{c}{ \textbf{FWHM($\mathbf{[\Oiii]\ \lambda  5007}$)}} \\

\multicolumn{1}{c}{} &
\multicolumn{1}{c}{} &
\multicolumn{1}{c}{(deg) } &
\multicolumn{1}{c}{ (deg)} &
\multicolumn{1}{c}{ } &
\multicolumn{1}{c}{(\AA)} & 
\multicolumn{1}{c}{(\AA)}  \\[0.5ex] \hline \\[-1.8ex]

\endhead

\multicolumn{2}{l}{{Continued on Next Page\ldots}} \\
\endfoot

\\[-1.8ex] \hline 
\multicolumn{5}{l}{$^*$ The errors in redshift are given in units of $10^{-5}$.}\\ 

\endlastfoot
001	&	J000657+005125	&	1.73758	&	0.85719	&	0.07370	(0.78)	&	---	$\pm$	---	&	1.69	$\pm$	0.09	\\
002	&	J001647-104742	&	4.19896	&	-10.79506	&	0.02325	(0.78)	&	1.06	$\pm$	0.08	&	0.91	$\pm$	0.05	\\
003	&	J002339-094848	&	5.91508	&	-9.81350	&	0.05305	(0.56)	&	1.32	$\pm$	0.10	&	1.43	$\pm$	0.08	\\
004	&	J002425+140410	&	6.10808	&	14.06961	&	0.01424	(1.06)	&	1.42	$\pm$	0.10	&	1.30	$\pm$	0.07	\\
005	&	J003218+150014	&	8.07746	&	15.00392	&	0.01796	(0.96)	&	1.55	$\pm$	0.11	&	1.69	$\pm$	0.09	\\
006	&	J005147+000940	&	12.94708	&	0.16111	&	0.03758	(1.18)	&	1.26	$\pm$	0.09	&	0.91	$\pm$	0.05	\\
007	&	J005602-101009	&	14.00942	&	-10.16928	&	0.05817	(1.46)	&	1.52	$\pm$	0.11	&	1.30	$\pm$	0.07	\\
008	&	J013258-085337	&	23.24392	&	-8.89378	&	0.09521	(1.80)	&	1.60	$\pm$	0.11	&	1.43	$\pm$	0.08	\\
\end{longtable}
%\end{ThreePartTable}
\end{center}
\normalsize 
%\end{landscape}
%\twocolumn

%% file: ch04/tab/tab05.tex
%Table  5 --- Derived data for the H II galaxies sample
\onecolumn
\begin{landscape}
\tiny
\begin{center}
\setlength{\tabcolsep}{4pt}
\begin{longtable}{@{} l c l c c c c c c c c c c c c@{}}
 
\caption[Local \hii\ galaxies derived values]{\small Derived values for the \hii\ galaxies sample. } 
\label{tab:tab05} \\

%This is the header for the first page of the table...
\hline \hline \\[-2ex]
\multicolumn{1}{c}{(1)} &
\multicolumn{1}{c}{(2)} &
\multicolumn{1}{c}{(3)} &
\multicolumn{1}{c}{(4)} &
\multicolumn{1}{c}{(5)} &
\multicolumn{1}{c}{(6)} & 
\multicolumn{1}{c}{(7)} &
\multicolumn{1}{c}{(8)} &
\multicolumn{1}{c}{(9)} &
\multicolumn{1}{c}{(10)} &
\multicolumn{1}{c}{(11)} &
\multicolumn{1}{c}{(12)} & 
\multicolumn{1}{c}{(13)} &
\multicolumn{1}{c}{(14)} &
\multicolumn{1}{c}{(15)} \\

\multicolumn{1}{c}{\bf{Index}} &
%\multicolumn{1}{c}{\bf{Name}} &
\multicolumn{1}{c}{\bf{Flag$^{\dagger}$}} &
\multicolumn{1}{c}{$\mathbf{z_c^{*}}$} &
\multicolumn{1}{c}{$\mathbf{A_v}$} &
\multicolumn{1}{c}{$\mathbf{100 \times Q}$} &
\multicolumn{1}{c}{$\mathbf{C_{\hb}}$} &
\multicolumn{1}{c}{$\mathbf{\logd \sigma (\hb)}$} &
\multicolumn{1}{c}{$\mathbf{\logd \sigma ([\Oiii])}$} &
\multicolumn{1}{c}{$\mathbf{\logd L (\hb)}$} &
\multicolumn{1}{c}{$\mathbf{12 + \logd \mathrm{O/H}}$} &
\multicolumn{1}{c}{$\mathbf{\logd R_{u}}$} &
\multicolumn{1}{c}{$\mathbf{\logd M_{dyn}}$} &
\multicolumn{1}{c}{$\mathbf{\logd M_{cl}}$} &
\multicolumn{1}{c}{$\mathbf{\logd M_{ion}}$} &
\multicolumn{1}{c}{$\mathbf{\logd SFR}$} \\

\multicolumn{1}{c}{} &
%\multicolumn{1}{c}{} &
\multicolumn{1}{c}{} &
\multicolumn{1}{c}{} &
\multicolumn{1}{c}{ (mag)} &
\multicolumn{1}{c}{} &
\multicolumn{1}{c}{} & 
\multicolumn{1}{c}{($\mathrm{km\ s^{-1}}$)} & 
\multicolumn{1}{c}{($\mathrm{km\ s^{-1}}$)} & 
\multicolumn{1}{c}{($\mathrm{erg\ s^{-1}}$)} & 
\multicolumn{1}{c}{}& 
\multicolumn{1}{c}{($\mathrm{Pc}$)} & 
\multicolumn{1}{c}{($\mathrm{M_{\odot}}$)} & 
\multicolumn{1}{c}{($\mathrm{M_{\odot}}$)} & 
\multicolumn{1}{c}{($\mathrm{M_{\odot}}$)} &
\multicolumn{1}{c}{($\mathrm{M_{\odot}\ yr^{-1}}$)} \\[0.5ex] \hline \\[-1.8ex]

\endfirsthead

%This is the header for the remaining page(s) of the table...
%\multicolumn{7}{c}{{\tablename} \thetable{} -- Continued} \\[0.5ex]
%\hline \hline \\[-2ex]

\multicolumn{1}{c}{(1)} &
\multicolumn{1}{c}{(2)} &
\multicolumn{1}{c}{(3)} &
\multicolumn{1}{c}{(4)} &
\multicolumn{1}{c}{(5)} &
\multicolumn{1}{c}{(6)} & 
\multicolumn{1}{c}{(7)} &
\multicolumn{1}{c}{(8)} &
\multicolumn{1}{c}{(9)} &
\multicolumn{1}{c}{(10)} &
\multicolumn{1}{c}{(11)} &
\multicolumn{1}{c}{(12)} & 
\multicolumn{1}{c}{(13)} &
\multicolumn{1}{c}{(14)} &
\multicolumn{1}{c}{(15)} \\

\multicolumn{1}{c}{\bf{Index}} &
%\multicolumn{1}{c}{\bf{Name}} &
\multicolumn{1}{c}{\bf{Flag$^{\dagger}$}} &
\multicolumn{1}{c}{$\mathbf{z_c^{*}}$} &
\multicolumn{1}{c}{$\mathbf{A_v}$} &
\multicolumn{1}{c}{$\mathbf{100 \times Q}$} &
\multicolumn{1}{c}{$\mathbf{C_{\hb}}$} &
\multicolumn{1}{c}{$\mathbf{\logd \sigma (\hb)}$} &
\multicolumn{1}{c}{$\mathbf{\logd \sigma ([\Oiii])}$} &
\multicolumn{1}{c}{$\mathbf{\logd L (\hb)}$} &
\multicolumn{1}{c}{$\mathbf{12 + \logd \mathrm{O/H}}$} &
\multicolumn{1}{c}{$\mathbf{\logd R_{u}}$} &
\multicolumn{1}{c}{$\mathbf{\logd M_{dyn}}$} &
\multicolumn{1}{c}{$\mathbf{\logd M_{cl}}$} &
\multicolumn{1}{c}{$\mathbf{\logd M_{ion}}$} &
\multicolumn{1}{c}{$\mathbf{\logd SFR}$} \\

\multicolumn{1}{c}{} &
%\multicolumn{1}{c}{} &
\multicolumn{1}{c}{} &
\multicolumn{1}{c}{} &
\multicolumn{1}{c}{ (mag)} &
\multicolumn{1}{c}{} &
\multicolumn{1}{c}{} & 
\multicolumn{1}{c}{($\mathrm{km\ s^{-1}}$)} & 
\multicolumn{1}{c}{($\mathrm{km\ s^{-1}}$)} & 
\multicolumn{1}{c}{($\mathrm{erg\ s^{-1}}$)} & 
\multicolumn{1}{c}{}& 
\multicolumn{1}{c}{($\mathrm{Pc}$)} & 
\multicolumn{1}{c}{($\mathrm{M_{\odot}}$)} & 
\multicolumn{1}{c}{($\mathrm{M_{\odot}}$)} & 
\multicolumn{1}{c}{($\mathrm{M_{\odot}}$)} &
\multicolumn{1}{c}{($\mathrm{M_{\odot}\ yr^{-1}}$)} \\[0.5ex] \hline \\[-1.8ex]

\endhead

\multicolumn{8}{l}{{Continued on Next Page\ldots}} \\
\endfoot

\\[-1.8ex] \hline 
\multicolumn{15}{l}{$^{\dagger}$ The number indicates that the object belongs to a given sample and to the samples up in the hierarchy, e.g. 4 indicates membership to S4 but also to S3, S2 and S1, whereas}\\ 
\multicolumn{15}{l}{3 indicates membership to S3, S2 and S1 etc..}\\ 

\multicolumn{8}{l}{$^*$ The errors in redshift are given in units of $10^{-5}$.}\\ 
\endlastfoot
001	&	1	&	0.07252	(0.78)	&	0.66	$\pm$	0.20	&	0.00	&	0.07	$\pm$	0.02	&	---	&	1.551	$\pm$	0.027	&	41.31	$\pm$	0.10	&	8.06	$\pm$	0.16	&	3.04	$\pm$	0.03	&	---	&	8.16	$\pm$	0.10	&	7.00	$\pm$	0.10	&	0.50	$\pm$	0.10	\\
002	&	4	&	0.02203	(0.78)	&	1.41	$\pm$	0.28	&	0.88	&	0.14	$\pm$	0.03	&	1.377	$\pm$	0.039	&	1.324	$\pm$	0.026	&	40.88	$\pm$	0.15	&	8.06	$\pm$	0.19	&	3.00	$\pm$	0.03	&	8.06	$\pm$	0.08	&	7.74	$\pm$	0.15	&	6.72	$\pm$	0.15	&	0.07	$\pm$	0.15	\\
003	&	4	&	0.05191	(0.56)	&	1.06	$\pm$	0.24	&	0.00	&	0.10	$\pm$	0.02	&	1.463	$\pm$	0.036	&	1.507	$\pm$	0.024	&	41.31	$\pm$	0.13	&	8.15	$\pm$	0.18	&	3.18	$\pm$	0.03	&	8.40	$\pm$	0.08	&	8.16	$\pm$	0.13	&	7.08	$\pm$	0.13	&	0.50	$\pm$	0.13	\\
004	&	4	&	0.01257	(1.06)	&	0.36	$\pm$	0.10	&	6.49	&	0.04	$\pm$	0.01	&	1.538	$\pm$	0.034	&	1.497	$\pm$	0.026	&	40.26	$\pm$	0.07	&	8.20	$\pm$	0.08	&	2.56	$\pm$	0.03	&	7.94	$\pm$	0.07	&	7.11	$\pm$	0.07	&	6.04	$\pm$	0.07	&	-0.55	$\pm$	0.07	\\
005	&	4	&	0.01636	(0.96)	&	0.04	$\pm$	0.06	&	4.73	&	0.00	$\pm$	0.01	&	1.577	$\pm$	0.034	&	1.615	$\pm$	0.023	&	40.41	$\pm$	0.10	&	8.02	$\pm$	0.05	&	2.54	$\pm$	0.04	&	7.99	$\pm$	0.08	&	7.26	$\pm$	0.10	&	6.63	$\pm$	0.10	&	-0.40	$\pm$	0.10	\\
006	&	4	&	0.03637	(1.18)	&	0.01	$\pm$	0.06	&	0.44	&	0.00	$\pm$	0.01	&	1.454	$\pm$	0.036	&	1.325	$\pm$	0.024	&	40.48	$\pm$	0.07	&	7.80	$\pm$	0.05	&	2.89	$\pm$	0.04	&	8.09	$\pm$	0.08	&	7.33	$\pm$	0.07	&	6.24	$\pm$	0.07	&	-0.33	$\pm$	0.07	\\
007	&	4	&	0.05712	(1.46)	&	0.59	$\pm$	0.18	&	0.88	&	0.06	$\pm$	0.02	&	1.529	$\pm$	0.034	&	1.467	$\pm$	0.024	&	40.85	$\pm$	0.11	&	7.89	$\pm$	0.14	&	3.04	$\pm$	0.03	&	8.40	$\pm$	0.08	&	7.70	$\pm$	0.11	&	6.41	$\pm$	0.11	&	0.04	$\pm$	0.11	\\
\end{longtable}
%\end{ThreePartTable}
\end{center}
\normalsize 
\end{landscape}
%\twocolumn

%% file: ch04/tab/tab09.tex
%Table --- L-Sigma relation for different cuts of the sample 4 
\begin{table}
 \centering
 \begin{minipage}{90mm}
  \caption[Local \hii\ galaxies \lsig\ correlation coefficients] {\small Correlation coefficients for the $L-\sigma$ relation for a range of discrimination levels in the automatic selection algorithm. }
  \label{tab:tab09}
 \resizebox{1.0 \textwidth}{!}{
\begin{tabular} { c c c c c }
\hline
\hline
(1) & (2) & (3) & (4) & (5) \\ 
Cut Level & $\alpha$ & $\beta$ & rms & N \\
\hline
10	&	33.69	$\pm$	0.22	&	4.67	$\pm$	0.14	&	0.337	&	93	\\
8	&	33.70	$\pm$	0.23	&	4.66	$\pm$	0.15	&	0.343	&	82	\\
5	&	33.94	$\pm$	0.26	&	4.51	$\pm$	0.17	&	0.317	&	55	\\
3	&	33.55	$\pm$	0.33	&	4.74	$\pm$	0.22	&	0.314	&	34	\\
1	&	33.60	$\pm$	0.49	&	4.63	$\pm$	0.32	&	0.289	&	16	\\
\hline 
\end{tabular}
}
\end{minipage}
\end{table}

%% file: ch04/tab/tab06A18.tex
%Table --- Linear regressions 3 
\begin{table}
 \centering
 \begin{minipage}{140mm}
  \caption[Regression coefficients for S3] {\small Regression coefficients for S3}%. All Profiles - Log$\sigma < $1.80}
  \label{tab:tab06A18}
 \resizebox{0.9 \textwidth}{!}{
\begin{tabular} { c c c c c c }
\hline
\hline
(1) & (2) & (3) & (4) & (5) & (6) \\ 
Parameter & $\alpha$ & $\beta$ & $\gamma$ & rms & N \\
\hline
$R_u$	&	34.04	$\pm$	0.20	&	3.08	$\pm$	0.22	&	0.76	$\pm$	0.13	&	0.261	&	99	\\
$R_g$	&	34.29	$\pm$	0.20	&	3.22	$\pm$	0.22	&	0.59	$\pm$	0.12	&	0.270	&	103	\\
$R_r$	&	34.08	$\pm$	0.21	&	3.29	$\pm$	0.22	&	0.61	$\pm$	0.13	&	0.274	&	101	\\
$R_i$	&	34.08	$\pm$	0.23	&	3.50	$\pm$	0.22	&	0.50	$\pm$	0.14	&	0.286	&	102	\\
$R_z$	&	34.09	$\pm$	0.23	&	3.36	$\pm$	0.23	&	0.56	$\pm$	0.14	&	0.282	&	101	\\
$O/H$	&	32.16	$\pm$	0.32	&	3.71	$\pm$	0.22	&	0.38	$\pm$	0.21	&	0.295	&	100	\\
$N2$	&	35.60	$\pm$	0.19	&	3.63	$\pm$	0.24	&	0.21	$\pm$	0.12	&	0.294	&	103	\\
$R23$	&	34.47	$\pm$	0.24	&	3.85	$\pm$	0.23	&	0.59	$\pm$	0.46	&	0.300	&	102	\\
$W(H\beta)$	&	34.74	$\pm$	0.23	&	3.73	$\pm$	0.22	&	0.22	$\pm$	0.15	&	0.303	&	107	\\
$(u-i)$	&	35.08	$\pm$	0.21	&	3.76	$\pm$	0.22	&	0.05	$\pm$	0.07	&	0.302	&	103	\\

\hline 
\end{tabular}
}
\end{minipage}
\end{table}

%% file: ch04/tab/tab061018.tex
%Table --- Linear regressions 3 
\begin{table}
 \centering
 \begin{minipage}{140mm}
  \caption[Regression coefficients for S4] {\small Regression coefficients for S4.}% Profiles $< 10\%$ - Log$\sigma < $1.80}
  \label{tab:tab061018}
 \resizebox{0.9 \textwidth}{!}{
\begin{tabular} { c c c c c c }
\hline
\hline
(1) & (2) & (3) & (4) & (5) & (6) \\ 
Parameter & $\alpha$ & $\beta$ & $\gamma$ & rms & N \\
\hline
$R_u$	&	34.08	$\pm$	0.20	&	2.96	$\pm$	0.25	&	0.81	$\pm$	0.14	&	0.260	&	88	\\
$R_g$	&	34.33	$\pm$	0.20	&	3.07	$\pm$	0.25	&	0.65	$\pm$	0.13	&	0.268	&	90	\\
$R_r$	&	34.09	$\pm$	0.22	&	3.18	$\pm$	0.25	&	0.67	$\pm$	0.14	&	0.274	&	89	\\
$R_i$	&	33.98	$\pm$	0.23	&	3.33	$\pm$	0.25	&	0.62	$\pm$	0.17	&	0.285	&	89	\\
$R_z$	&	34.13	$\pm$	0.24	&	3.31	$\pm$	0.25	&	0.57	$\pm$	0.16	&	0.285	&	89	\\
$O/H$	&	32.30	$\pm$	0.33	&	3.71	$\pm$	0.24	&	0.36	$\pm$	0.24	&	0.298	&	87	\\
$N2$	&	35.59	$\pm$	0.20	&	3.63	$\pm$	0.27	&	0.19	$\pm$	0.14	&	0.299	&	89	\\
$R23$	&	34.56	$\pm$	0.25	&	3.85	$\pm$	0.25	&	0.49	$\pm$	0.50	&	0.304	&	89	\\
$W(H\beta)$	&	34.77	$\pm$	0.24	&	3.75	$\pm$	0.24	&	0.19	$\pm$	0.18	&	0.308	&	93	\\
$(u-i)$	&	35.09	$\pm$	0.22	&	3.77	$\pm$	0.23	&	0.03	$\pm$	0.08	&	0.305	&	90	\\

\hline 
\end{tabular}
}
\end{minipage}
\end{table}

%% file: ch04/tab/tab06A18B.tex
%Table --- Bayesian Linear regression
\begin{table}
 \centering
 \begin{minipage}{140mm}
  \caption[Bayesian Regression coefficients for S3] {\small Bayesian Regression coefficients for S3}%. All Profiles - Log$\sigma < $1.80}
  \label{tab:tab06A18B}
 \resizebox{0.9 \textwidth}{!}{
\begin{tabular} { c c c c c c }
\hline
\hline
(1) & (2) & (3) & (4) & (5) & (6) \\ 
Parameter & $\alpha$ & $\beta$ & $\gamma$ & rms & N \\
\hline
$R_u$	&	33.75	$\pm$	0.37	&	3.36	$\pm$	0.26	&	0.71	$\pm$	0.14	&	0.263	&	99	\\
$R_g$	&	33.84	$\pm$	0.37	&	3.47	$\pm$	0.25	&	0.61	$\pm$	0.13	&	0.273	&	103	\\
$R_r$	&	33.71	$\pm$	0.41	&	3.57	$\pm$	0.26	&	0.59	$\pm$	0.14	&	0.277	&	101	\\
$R_i$	&	33.61	$\pm$	0.47	&	3.76	$\pm$	0.25	&	0.52	$\pm$	0.16	&	0.289	&	102	\\
$R_z$	&	33.15	$\pm$	0.46	&	3.47	$\pm$	0.25	&	0.82	$\pm$	0.17	&	0.290	&	101	\\
$O/H$	&	30.67	$\pm$	3.07	&	3.96	$\pm$	0.26	&	0.51	$\pm$	0.40	&	0.298	&	100	\\
$N2$	&	34.94	$\pm$	0.55	&	3.99	$\pm$	0.28	&	0.14	$\pm$	0.13	&	0.297	&	103	\\
$R23$	&	33.83	$\pm$	0.67	&	4.16	$\pm$	0.26	&	0.75	$\pm$	0.49	&	0.303	&	102	\\
$W(H\beta)$	&	34.23	$\pm$	0.51	&	4.02	$\pm$	0.24	&	0.23	$\pm$	0.17	&	0.305	&	107	\\
$(u-i)$	&	34.62	$\pm$	0.38	&	4.05	$\pm$	0.24	&	0.04	$\pm$	0.08	&	0.304	&	103	\\

\hline 
\end{tabular}
}
\end{minipage}
\end{table}

%% file: ch04/tab/tab06A18O.tex
%Table --- Linear regressions Oxygen  
\begin{table}
 \centering
 \begin{minipage}{140mm}
  \caption[Regression coefficients for S3 using $\sigma(\lbrack \mathrm{OIII} \rbrack)$] {\small Regression coefficients for S3 using $\sigma([\mathrm{OIII}])$.}
  \label{tab:tab06A18O}
 \resizebox{0.9 \textwidth}{!}{
\begin{tabular} { c c c c c c }
\hline
\hline
(1) & (2) & (3) & (4) & (5) & (6) \\ 
Parameter & $\alpha$ & $\beta$ & $\gamma$ & rms & N \\
\hline
$R_u$	&	34.44	$\pm$	0.17	&	2.78	$\pm$	0.24	&	0.82	$\pm$	0.14	&	0.290	&	99	\\
$R_g$	&	34.77	$\pm$	0.16	&	2.93	$\pm$	0.24	&	0.61	$\pm$	0.13	&	0.303	&	103	\\
$R_r$	&	34.55	$\pm$	0.17	&	3.00	$\pm$	0.24	&	0.64	$\pm$	0.14	&	0.306	&	101	\\
$R_i$	&	34.67	$\pm$	0.18	&	3.23	$\pm$	0.24	&	0.48	$\pm$	0.16	&	0.321	&	102	\\
$R_z$	&	34.53	$\pm$	0.20	&	3.07	$\pm$	0.24	&	0.60	$\pm$	0.15	&	0.312	&	101	\\
$O/H$	&	33.45	$\pm$	0.24	&	3.45	$\pm$	0.23	&	0.28	$\pm$	0.23	&	0.328	&	100	\\
$N2$	&	36.33	$\pm$	0.15	&	3.28	$\pm$	0.25	&	0.26	$\pm$	0.14	&	0.327	&	103	\\
$R23$	&	35.35	$\pm$	0.18	&	3.52	$\pm$	0.24	&	0.29	$\pm$	0.50	&	0.332	&	102	\\
$W(H\beta)$	&	35.02	$\pm$	0.20	&	3.46	$\pm$	0.23	&	0.35	$\pm$	0.17	&	0.329	&	107	\\
$(u-i)$	&	35.64	$\pm$	0.17	&	3.51	$\pm$	0.23	&	-0.02	$\pm$	0.08	&	0.334	&	103	\\

\hline 
\end{tabular}
}
\end{minipage}
\end{table}

%% file: ch04/tab/tab06A18OB.tex
%Table --- Linear regressions Oxygen Bayesian  
\begin{table}
 \centering
 \begin{minipage}{140mm}
  \caption[Bayesian regression coefficients  for S3 using $\sigma(\lbrack \mathrm{OIII} \rbrack)$] {\small Bayesian regression coefficients  for S3 using $\sigma([\mathrm{OIII}])$.}
  \label{tab:tab06A18OB}
 \resizebox{0.9 \textwidth}{!}{
\begin{tabular} { c c c c c c }
\hline
\hline
(1) & (2) & (3) & (4) & (5) & (6) \\ 
Parameter & $\alpha$ & $\beta$ & $\gamma$ & rms & N \\
\hline
$R_u$	&	34.29	$\pm$	0.40	&	2.95	$\pm$	0.27	&	0.78	$\pm$	0.16	&	0.291	&	99	\\
$R_g$	&	34.46	$\pm$	0.39	&	3.08	$\pm$	0.27	&	0.64	$\pm$	0.15	&	0.304	&	103	\\
$R_r$	&	34.34	$\pm$	0.44	&	3.16	$\pm$	0.27	&	0.62	$\pm$	0.16	&	0.307	&	101	\\
$R_i$	&	34.37	$\pm$	0.50	&	3.40	$\pm$	0.26	&	0.49	$\pm$	0.18	&	0.322	&	102	\\
$R_z$	&	33.75	$\pm$	0.50	&	3.08	$\pm$	0.26	&	0.85	$\pm$	0.19	&	0.317	&	101	\\
$O/H$	&	31.87	$\pm$	3.41	&	3.57	$\pm$	0.27	&	0.45	$\pm$	0.44	&	0.330	&	100	\\
$N2$	&	35.94	$\pm$	0.56	&	3.49	$\pm$	0.28	&	0.22	$\pm$	0.15	&	0.328	&	103	\\
$R23$	&	34.92	$\pm$	0.71	&	3.72	$\pm$	0.26	&	0.42	$\pm$	0.54	&	0.333	&	102	\\
$W(H\beta)$	&	34.72	$\pm$	0.55	&	3.65	$\pm$	0.24	&	0.35	$\pm$	0.19	&	0.331	&	107	\\
$(u-i)$	&	35.35	$\pm$	0.38	&	3.70	$\pm$	0.25	&	-0.03	$\pm$	0.09	&	0.335	&	103	\\

\hline 
\end{tabular}
}
\end{minipage}
\end{table}

%% file: ch04/tab/tab07A18.tex
%Table --- Linear regressions Multi
\begin{table*}
 \centering
% \begin{minipage}{140mm}
  \caption[Regression coefficients for S3]{\small Regression coefficients for S3.}
  \label{tab:tab07A18}
 \resizebox{0.9 \textwidth}{!}{
\begin{tabular} { c c c c c c c c }
\hline
\hline
(1) & (2) & (3) & (4) & (5) & (6) & (7) & (8)\\ 
Parameters & $\alpha$ & $\beta$ & $\gamma$ &  $\delta$ & $\epsilon$ &  rms & N \\
\hline
$R_u$, $(u - i)$	&	33.93	$\pm$	0.20	&	2.97	$\pm$	0.22	&	0.91	$\pm$	0.14	&	-0.16	$\pm$	0.08	&	---			&	0.255	&	99	\\
$R_u$, $O/H$	&	32.76	$\pm$	0.24	&	3.10	$\pm$	0.22	&	0.71	$\pm$	0.13	&	0.17	$\pm$	0.19	&	---			&	0.260	&	96	\\
$R_u$, $N2$	&	33.73	$\pm$	0.21	&	3.08	$\pm$	0.22	&	0.88	$\pm$	0.13	&	0.01	$\pm$	0.11	&	---			&	0.247	&	97	\\
$R_u$, $R23$	&	32.96	$\pm$	0.24	&	3.20	$\pm$	0.23	&	0.80	$\pm$	0.13	&	0.91	$\pm$	0.42	&	---			&	0.256	&	98	\\
$R_u$, $W(H\beta)$	&	32.87	$\pm$	0.23	&	3.00	$\pm$	0.22	&	0.90	$\pm$	0.13	&	0.47	$\pm$	0.16	&	---			&	0.250	&	99	\\
$W(H\beta)$, $O/H$	&	30.63	$\pm$	0.38	&	3.69	$\pm$	0.22	&	0.26	$\pm$	0.17	&	0.51	$\pm$	0.22	&	---			&	0.291	&	100	\\
$W(H\beta)$, $N2$	&	35.38	$\pm$	0.19	&	3.42	$\pm$	0.26	&	0.43	$\pm$	0.20	&	0.42	$\pm$	0.16	&	---			&	0.288	&	103	\\
$W(H\beta)$, $R23$	&	34.46	$\pm$	0.24	&	3.83	$\pm$	0.23	&	0.05	$\pm$	0.19	&	0.51	$\pm$	0.54	&	---			&	0.300	&	102	\\
$R_u$, $(u - i)$, $O/H$	&	30.43	$\pm$	0.31	&	2.90	$\pm$	0.23	&	0.90	$\pm$	0.14	&	-0.25	$\pm$	0.09	&	0.46	$\pm$	0.21	&	0.249	&	96	\\
$R_u$, $(u - i)$, $N2$	&	34.17	$\pm$	0.19	&	2.85	$\pm$	0.25	&	0.99	$\pm$	0.14	&	-0.19	$\pm$	0.09	&	0.17	$\pm$	0.13	&	0.241	&	97	\\
$R_u$, $(u - i)$, $R23$	&	33.07	$\pm$	0.23	&	3.09	$\pm$	0.23	&	0.91	$\pm$	0.14	&	-0.13	$\pm$	0.08	&	0.75	$\pm$	0.43	&	0.252	&	98	\\
$R_u$, $W(H\beta)$, $O/H$	&	29.30	$\pm$	0.33	&	2.95	$\pm$	0.22	&	0.85	$\pm$	0.13	&	0.57	$\pm$	0.17	&	0.44	$\pm$	0.19	&	0.244	&	96	\\
$R_u$, $W(H\beta)$, $N2$	&	33.15	$\pm$	0.22	&	2.79	$\pm$	0.23	&	0.95	$\pm$	0.13	&	0.63	$\pm$	0.19	&	0.28	$\pm$	0.13	&	0.233	&	97	\\
$R_u$, $W(H\beta)$, $R23$	&	32.53	$\pm$	0.25	&	3.07	$\pm$	0.23	&	0.89	$\pm$	0.13	&	0.39	$\pm$	0.17	&	0.45	$\pm$	0.46	&	0.249	&	98	\\

\hline 
\end{tabular}
}
%\end{minipage}
\end{table*}

%% file: ch04/tab/tab07A18B.tex
%Table --- Linear regressions Multi Bayesian 
\begin{table*}
 \centering
% \begin{minipage}{140mm}
  \caption[Bayesian Regression coefficients for S3]{\small Bayesian Regression coefficients for S3.}
  \label{tab:tab07A18B}
 \resizebox{0.9 \textwidth}{!}{
\begin{tabular} { c c c c c c c c }
\hline
\hline
(1) & (2) & (3) & (4) & (5) & (6) & (7) & (8)\\ 
Parameters & $\alpha$ & $\beta$ & $\gamma$ &  $\delta$ & $\epsilon$ &  rms & N \\
\hline
$R_u$, $(u - i)$	&	33.60	$\pm$	0.37	&	3.21	$\pm$	0.26	&	0.90	$\pm$	0.16	&	-0.18	$\pm$	0.08	&	---			&	0.257	&	99	\\
$R_u$, $O/H$	&	32.27	$\pm$	2.77	&	3.38	$\pm$	0.27	&	0.65	$\pm$	0.15	&	0.20	$\pm$	0.37	&	---			&	0.262	&	96	\\
$R_u$, $N2$	&	33.16	$\pm$	0.57	&	3.41	$\pm$	0.26	&	0.85	$\pm$	0.14	&	-0.07	$\pm$	0.12	&	---			&	0.250	&	97	\\
$R_u$, $R23$	&	32.40	$\pm$	0.67	&	3.50	$\pm$	0.26	&	0.77	$\pm$	0.14	&	1.11	$\pm$	0.45	&	---			&	0.258	&	98	\\
$R_u$, $W(H\beta)$	&	32.46	$\pm$	0.56	&	3.24	$\pm$	0.25	&	0.87	$\pm$	0.15	&	0.52	$\pm$	0.17	&	---			&	0.251	&	99	\\
$W(H\beta)$, $O/H$	&	27.89	$\pm$	3.96	&	3.89	$\pm$	0.27	&	0.38	$\pm$	0.24	&	0.78	$\pm$	0.47	&	---			&	0.295	&	100	\\
$W(H\beta)$, $N2$	&	34.77	$\pm$	0.54	&	3.81	$\pm$	0.31	&	0.36	$\pm$	0.25	&	0.31	$\pm$	0.18	&	---			&	0.291	&	103	\\
$W(H\beta)$, $R23$	&	33.84	$\pm$	0.67	&	4.14	$\pm$	0.26	&	0.02	$\pm$	0.21	&	0.71	$\pm$	0.57	&	---			&	0.303	&	102	\\
$R_u$, $(u - i)$, $O/H$	&	26.05	$\pm$	4.61	&	2.96	$\pm$	0.35	&	0.95	$\pm$	0.19	&	-0.42	$\pm$	0.18	&	0.99	$\pm$	0.60	&	0.260	&	96	\\
$R_u$, $(u - i)$, $N2$	&	33.53	$\pm$	0.62	&	3.20	$\pm$	0.29	&	0.95	$\pm$	0.15	&	-0.15	$\pm$	0.10	&	0.06	$\pm$	0.15	&	0.244	&	97	\\
$R_u$, $(u - i)$, $R23$	&	32.48	$\pm$	0.67	&	3.38	$\pm$	0.27	&	0.89	$\pm$	0.16	&	-0.14	$\pm$	0.09	&	0.97	$\pm$	0.45	&	0.255	&	98	\\
$R_u$, $W(H\beta)$, $O/H$	&	27.19	$\pm$	3.39	&	3.14	$\pm$	0.27	&	0.82	$\pm$	0.15	&	0.69	$\pm$	0.23	&	0.65	$\pm$	0.40	&	0.248	&	96	\\
$R_u$, $W(H\beta)$, $N2$	&	32.69	$\pm$	0.57	&	3.11	$\pm$	0.28	&	0.93	$\pm$	0.14	&	0.56	$\pm$	0.22	&	0.19	$\pm$	0.16	&	0.236	&	97	\\
$R_u$, $W(H\beta)$, $R23$	&	31.97	$\pm$	0.68	&	3.34	$\pm$	0.27	&	0.87	$\pm$	0.15	&	0.40	$\pm$	0.20	&	0.66	$\pm$	0.50	&	0.252	&	98	\\

\hline 
\end{tabular}
}
%\end{minipage}
\end{table*}

%% file: ch04/tab/tab07A18O.tex
%Table --- Linear regressions Multi
\begin{table*}
 \centering
% \begin{minipage}{140mm}
  \caption[Regression coefficients for S3 using $\sigma(\lbrack \mathrm{OIII} \rbrack)$]{\small Regression coefficients for S3 using $\sigma([\mathrm{OIII}])$.}
  \label{tab:tab07A18O}
 \resizebox{0.9 \textwidth}{!}{
\begin{tabular} { c c c c c c c c }
\hline
\hline
(1) & (2) & (3) & (4) & (5) & (6) & (7) & (8)\\ 
Parameters & $\alpha$ & $\beta$ & $\gamma$ &  $\delta$ & $\epsilon$ &  rms & N \\
\hline
$R_u$, $(u - i)$	&	34.21	$\pm$	0.18	&	2.67	$\pm$	0.23	&	1.03	$\pm$	0.15	&	-0.26	$\pm$	0.08	&	---			&	0.276	&	99	\\
$R_u$, $O/H$	&	33.84	$\pm$	0.19	&	2.81	$\pm$	0.24	&	0.77	$\pm$	0.15	&	0.09	$\pm$	0.21	&	---			&	0.290	&	96	\\
$R_u$, $N2$	&	34.40	$\pm$	0.17	&	2.72	$\pm$	0.24	&	0.90	$\pm$	0.15	&	0.08	$\pm$	0.13	&	---			&	0.280	&	97	\\
$R_u$, $R23$	&	33.81	$\pm$	0.19	&	2.85	$\pm$	0.24	&	0.83	$\pm$	0.14	&	0.56	$\pm$	0.47	&	---			&	0.288	&	98	\\
$R_u$, $W(H\beta)$	&	32.91	$\pm$	0.22	&	2.73	$\pm$	0.22	&	0.98	$\pm$	0.14	&	0.60	$\pm$	0.17	&	---			&	0.273	&	99	\\
$W(H\beta)$, $O/H$	&	31.18	$\pm$	0.33	&	3.43	$\pm$	0.23	&	0.39	$\pm$	0.18	&	0.47	$\pm$	0.25	&	---			&	0.320	&	100	\\
$W(H\beta)$, $N2$	&	35.79	$\pm$	0.17	&	3.02	$\pm$	0.25	&	0.71	$\pm$	0.21	&	0.59	$\pm$	0.16	&	---			&	0.309	&	103	\\
$W(H\beta)$, $R23$	&	35.20	$\pm$	0.19	&	3.49	$\pm$	0.24	&	0.30	$\pm$	0.20	&	-0.15	$\pm$	0.58	&	---			&	0.328	&	102	\\
$R_u$, $(u - i)$, $O/H$	&	30.45	$\pm$	0.29	&	2.60	$\pm$	0.23	&	1.02	$\pm$	0.15	&	-0.35	$\pm$	0.09	&	0.50	$\pm$	0.23	&	0.269	&	96	\\
$R_u$, $(u - i)$, $N2$	&	35.03	$\pm$	0.16	&	2.41	$\pm$	0.24	&	1.08	$\pm$	0.15	&	-0.35	$\pm$	0.09	&	0.35	$\pm$	0.14	&	0.260	&	97	\\
$R_u$, $(u - i)$, $R23$	&	33.90	$\pm$	0.19	&	2.71	$\pm$	0.24	&	1.03	$\pm$	0.15	&	-0.25	$\pm$	0.08	&	0.30	$\pm$	0.46	&	0.275	&	98	\\
$R_u$, $W(H\beta)$, $O/H$	&	29.98	$\pm$	0.30	&	2.69	$\pm$	0.23	&	0.93	$\pm$	0.14	&	0.68	$\pm$	0.18	&	0.37	$\pm$	0.21	&	0.269	&	96	\\
$R_u$, $W(H\beta)$, $N2$	&	33.40	$\pm$	0.20	&	2.42	$\pm$	0.23	&	0.98	$\pm$	0.14	&	0.90	$\pm$	0.20	&	0.45	$\pm$	0.14	&	0.253	&	97	\\
$R_u$, $W(H\beta)$, $R23$	&	33.04	$\pm$	0.21	&	2.72	$\pm$	0.23	&	0.96	$\pm$	0.14	&	0.60	$\pm$	0.19	&	-0.10	$\pm$	0.49	&	0.273	&	98	\\

\hline 
\end{tabular}
}
%\end{minipage}
\end{table*}

%% file: ch04/tab/tab07A18OB.tex
%Table --- Linear regressions Multi
\begin{table*}
 \centering
% \begin{minipage}{140mm}
  \caption[Bayesian regression coefficients for S3 and using $\sigma(\lbrack \mathrm{OIII} \rbrack)$]{\small Bayesian regression coefficients for S3 and using $\sigma([\mathrm{OIII}])$.}
  \label{tab:tab07A18OB}
 \resizebox{0.9 \textwidth}{!}{
\begin{tabular} { c c c c c c c c }
\hline
\hline
(1) & (2) & (3) & (4) & (5) & (6) & (7) & (8)\\ 
Parameters & $\alpha$ & $\beta$ & $\gamma$ &  $\delta$ & $\epsilon$ &  rms & N \\
\hline
$R_u$, $(u - i)$	&	34.03	$\pm$	0.39	&	2.83	$\pm$	0.26	&	1.02	$\pm$	0.17	&	-0.29	$\pm$	0.09	&	---			&	0.277	&	99	\\
$R_u$, $O/H$	&	33.42	$\pm$	3.08	&	2.96	$\pm$	0.28	&	0.73	$\pm$	0.17	&	0.12	$\pm$	0.41	&	---			&	0.290	&	96	\\
$R_u$, $N2$	&	34.04	$\pm$	0.63	&	2.92	$\pm$	0.27	&	0.89	$\pm$	0.17	&	0.02	$\pm$	0.14	&	---			&	0.281	&	97	\\
$R_u$, $R23$	&	33.50	$\pm$	0.71	&	3.03	$\pm$	0.27	&	0.80	$\pm$	0.16	&	0.70	$\pm$	0.49	&	---			&	0.289	&	98	\\
$R_u$, $W(H\beta)$	&	32.68	$\pm$	0.63	&	2.85	$\pm$	0.25	&	0.97	$\pm$	0.16	&	0.63	$\pm$	0.19	&	---			&	0.273	&	99	\\
$W(H\beta)$, $O/H$	&	28.57	$\pm$	4.40	&	3.52	$\pm$	0.28	&	0.50	$\pm$	0.26	&	0.75	$\pm$	0.52	&	---			&	0.323	&	100	\\
$W(H\beta)$, $N2$	&	35.39	$\pm$	0.57	&	3.23	$\pm$	0.30	&	0.72	$\pm$	0.26	&	0.55	$\pm$	0.19	&	---			&	0.310	&	103	\\
$W(H\beta)$, $R23$	&	34.77	$\pm$	0.71	&	3.69	$\pm$	0.26	&	0.29	$\pm$	0.23	&	0.02	$\pm$	0.62	&	---			&	0.329	&	102	\\
$R_u$, $(u - i)$, $O/H$	&	24.62	$\pm$	5.03	&	2.53	$\pm$	0.35	&	1.10	$\pm$	0.20	&	-0.57	$\pm$	0.19	&	1.23	$\pm$	0.65	&	0.286	&	96	\\
$R_u$, $(u - i)$, $N2$	&	34.68	$\pm$	0.62	&	2.60	$\pm$	0.28	&	1.08	$\pm$	0.17	&	-0.36	$\pm$	0.11	&	0.30	$\pm$	0.16	&	0.261	&	97	\\
$R_u$, $(u - i)$, $R23$	&	33.52	$\pm$	0.69	&	2.88	$\pm$	0.27	&	1.03	$\pm$	0.17	&	-0.27	$\pm$	0.09	&	0.46	$\pm$	0.49	&	0.276	&	98	\\
$R_u$, $W(H\beta)$, $O/H$	&	27.64	$\pm$	3.87	&	2.73	$\pm$	0.29	&	0.92	$\pm$	0.17	&	0.80	$\pm$	0.25	&	0.63	$\pm$	0.46	&	0.272	&	96	\\
$R_u$, $W(H\beta)$, $N2$	&	33.10	$\pm$	0.62	&	2.56	$\pm$	0.27	&	0.99	$\pm$	0.16	&	0.89	$\pm$	0.24	&	0.40	$\pm$	0.17	&	0.254	&	97	\\
$R_u$, $W(H\beta)$, $R23$	&	32.72	$\pm$	0.73	&	2.85	$\pm$	0.27	&	0.95	$\pm$	0.16	&	0.62	$\pm$	0.21	&	0.02	$\pm$	0.53	&	0.274	&	98	\\

\hline 
\end{tabular}
}
%\end{minipage}
\end{table*}

%% file: ch04/tab/tab07HDS.tex
%Table --- Linear regressions Multi
\begin{table}
 \centering
 \begin{minipage}{140mm}
  \caption[Regression coefficients-HDS] {\small Regression coefficients-HDS.}
  \label{tab:tab07HDS}
 \resizebox{0.9 \textwidth}{!}{
\begin{tabular} { c c c c c c c c }
\hline
\hline
(1) & (2) & (3) & (4) & (5) & (6) & (7) & (8)\\ 
Parameters & $\alpha$ & $\beta$ & $\gamma$ &  $\delta$ & $\epsilon$ &  rms & N \\
\hline
$R_u$, $(u - i)$, $O/H$	&	28.44	&	2.72	&	1.12	&	-0.23	&	0.66	&	0.256	&	55	\\
$R_u$, $(u - i)$, $N2$	&	34.21	&	2.62	&	1.13	&	-0.19	&	0.25	&	0.258	&	57	\\
$R_u$, $W(H\beta)$, $O/H$	&	27.37	&	2.81	&	1.03	&	0.72	&	0.61	&	0.240	&	57	\\
$R_u$, $W(H\beta)$, $N2$	&	32.95	&	2.43	&	1.04	&	1.01	&	0.49	&	0.232	&	59	\\

\hline 
\end{tabular}
}
\end{minipage}
\end{table}

%% file: ch04/tab/tab07UVES.tex
%Table --- Linear regressions Multi
\begin{table}
 \centering
 \begin{minipage}{140mm}
  \caption[Regression coefficients-UVES] {\small Regression coefficients-UVES.}
  \label{tab:tab07UVES}
 \resizebox{0.9 \textwidth}{!}{
\begin{tabular} { c c c c c c c c }
\hline
\hline
(1) & (2) & (3) & (4) & (5) & (6) & (7) & (8)\\ 
Parameters & $\alpha$ & $\beta$ & $\gamma$ &  $\delta$ & $\epsilon$ &  rms & N \\
\hline
$R_u$, $(u - i)$, $O/H$	&	30.84	&	3.09	&	0.85	&	-0.18	&	0.38	&	0.199	&	38	\\
$R_u$, $(u - i)$, $N2$	&	34.80	&	2.85	&	0.84	&	-0.22	&	0.30	&	0.209	&	38	\\
$R_u$, $W(H\beta)$, $O/H$	&	33.23	&	3.02	&	0.53	&	0.13	&	0.16	&	0.232	&	39	\\
$R_u$, $W(H\beta)$, $N2$	&	34.29	&	3.04	&	0.72	&	0.07	&	0.13	&	0.216	&	38	\\

\hline 
\end{tabular}
}
\end{minipage}
\end{table}

%% file: ch05/ch05.tex
%chapter ch05.tex 
\chapter{The Hubble Constant}\label{ch:5}

\epigraph{\emph{We have found a strange footprint on the shores of the unknown.}}{--- \textup{Sir A. S. Eddington}, Space, Time and Gravitation (1920)}

\lettrine{T}{he} accurate determination of the Hubble constant, $H_0$, is considered one of the most fundamental tasks in the interface between Astronomy and Cosmology. The importance of measuring the expansion rate of the Universe to high precision stems from the fact that $H_0$, besides providing cosmic distances, is also a prerequisite for independent constraints on the mass-energy content of the Universe \citep[e.g.][]{Suyu2012}.

The direct determination of the Hubble constant can only be obtained by measuring cosmic distances and mapping the local expansion of the Universe, since the Hubble relation, $cz=H_0 d$, is valid and independent of the mass-energy content of the Universe only locally ($z\mincir 0.15$). A variety of methods have been used to estimate $H_0$, based on Cepheids, surface brightness fluctuations, masers, the tip of the red giant branch (TRGB), or type Ia supernovae \citep[for general reviews see][]{Jackson2007, Tammann2008, Freedman2010}.
In particular, the use of SNae~Ia to measure the Hubble constant has a long history in astronomy \citep[e.g.][]{Sandage1982, Sandage1990}. The subsequent discovery of the correlation between the magnitude at peak brightness and the rate at which it declines thereafter \citep[e.g.][]{Phillips1993} allowed the reduction of the distance determination intrinsic scatter.
However, one has to remember that SNae~Ia are secondary indicators and their use relies on the determination of well-established local calibrators, like the Large Magellanic Cloud (LMC), Galactic Cepheids, the ``maser'' galaxy NGC 4258, etc. \citep[cf.][]{Riess2011}.

Indirect methods to measure $H_0$ have also been developed, based either on the physics of the hot X-ray emitting inter-cluster gas (Sunyaev-Zeldovich effect), providing values of $H_{0} = 73.7 \pm 4 \pm 9$ km s$^{-1}$ Mpc$^{-1}$ \citep[e.g.][]{Birkinshaw1999, Carlstrom2002, Bonamente2006}, or on gravitational lens time delays \citep[e.g.][]{Schechter2005, Oguri2007, Suyu2009} providing a wide range of $H_0$ values, with the latter study giving $H_0 = 71\pm 3$ km s$^{-1}$ Mpc$^{-1}$. In addition, strategies based on the physics of the Universe just prior to recombination have been proposed \citep[CMB anisotropy and BAO;][and references therein]{Spergel2007, Komatsu2011, Beutler2011}. However, the latter methods use as priors other cosmological parameters, and thus the resulting $H_0$ measurements are strongly model dependent. 

Returning to the direct method to estimate $H_0$, an important breakthrough occurred a decade or so ago by the {\em HST} Calibration program \citep{Saha2001, Sandage2006} who found Cepheids in local galaxies that host SNae~Ia and provided a Cepheid based zero-point calibration, and by the
{\em HST} Key project \citep{Freedman2001} who furnished a value of $H_{0}=72 \pm 2 {\rm (random)} \pm 7 {\rm (systematic)}$ km s$^{-1}$ Mpc$^{-1}$, based on Cepheid distances of external galaxies and the LMC as the first rung of the distance ladder.
This value was recently revised by the same authors, using a new Cepheid zero-point \citep{Benedict2007} and the new SNae~Ia of \citet{Hicken2009}, to a
similar but less uncertain value of $H_{0}=73 \pm 2 {\rm (random)} \pm 4 {\rm (systematic)}$
km s$^{-1}$ Mpc$^{-1}$ \citep[see][]{Freedman2010}. \citet{Tammann2008} used a 
variety of local calibrators to recalibrate the SNae~Ia and found a significantly lower value of $H_{0}=62.3\pm 4$ km s$^{-1}$Mpc$^{-1}$. The difference has since been explained as
being due to a variety of external causes among which the use of heavily reddened Galactic Cepheids and of less accurate photographic data \citep{Riess2009a, Riess2009b}.

The most recent analysis of \citet{Riess2011} uses new {\em HST} optical and infrared 
observations of 600 Cepheid variables to determine the distance to eight galaxies hosting recent SNae~Ia. The resulting best estimate for the Hubble constant is: $H_{0}=73.8\pm 2.4$ km s$^{-1}$ Mpc$^{-1}$ (including random and systematic errors).

From the above discussion it becomes clear that SNae~Ia are the only tracers of the Hubble expansion utilized to-date, over a relatively large redshift range ($0\mincir z \mincir 0.1$). Therefore, due to the great importance of direct determinations of  the Hubble constant for cosmological studies \citep[e.g.][]{Suyu2012} it is highly desirable to independently confirm the SNae~Ia based $H_0$ value by using an alternative tracer.

\hii\ galaxies have been proposed as such an alternative, and a first attempt to estimate $H_0$, using giant \hii\ regions as local calibrators, was presented in \citet{Melnick1988}. Recently, we presented a thorough investigation of the viability of using \hii\ galaxies to constrain the dark energy equation of state, accounting also for the effects of gravitational lensing, which are expected to be non-negligible for very high redshift `standard candles' and we showed that indeed \hii\ galaxies can represent an important cosmological probe \citep{Plionis2011}.

The aim of the current chapter is to use \hii\ galaxies and a local calibration of the \lsig\ relation based on giant \hii\ regions of nearby galaxies, as an alternative direct approach for estimating the Hubble constant over a redshift range of $0.01<z< 0.16$. This chapter follows closely the work presented in \citet{Chavez2012}.

\section{Sample Selection and Observations}
As described in the previous chapter, a sample of 128 \hii\ galaxies was selected from the SDSS DR7 spectroscopic data release \citep{Abazajian2009} within a redshift range $0.01<z<0.16$, chosen for being compact ($D<5\ \mathrm{arcsec}$) having large Balmer  emission line fluxes and  equivalent widths. %An equivalent width lower limit of 50 \AA\ for  $\hb$  
A lower limit for the equivalent width ($W$) of $\hb$  was chosen  to avoid evolved
starbursts, that would present underlying absorption components in their spectra due to an older
stellar population component, thus affecting the emission line fluxes
\citep[cf.][]{Melnick2000}.
The redshift lower limit was chosen to minimise the effects of local
peculiar motions relative to the Hubble flow and the upper limit to 
minimise any possible Malmquist bias and avoid gross 
cosmological effects. %The $z$ distribution of the sample is shown in figure 1.

In order to improve the parameters of the \lsig\  relation obtained 
from previous work, high-resolution echelle spectroscopy for the \hii\ galaxy sample was
performed with 8 meter class telescopes. To obtain accurate total $\hb$ fluxes for the \hii\ galaxy
sample, we performed long slit spectrophotometry  at 2-meter class
telescopes  under photometric conditions and using a slit width  (8
arcsec) larger  than the upper limit of the HII galaxies size in our
sample. 

$\hb$ and  $[\Oiii]\ \lambda \lambda 4959,\ 5007$ line widths were
measured by means of fitting single gaussians to the line profiles. 
\hii\ galaxies with peculiar line profiles due to 
double lines or rotation-broadened ones,
were removed from the sample in order to
avoid systematic errors in their $\hb$ luminosity determination. Our
final sample of \hii\ galaxies was thus reduced from 128 to 69 galaxies (see also the discussion in chapter 4).

\subsection{The anchor sample}
The first step in the determination of $H_0$ is the determination of an accurate value of the zero point for the \lsig\ correlation. In order to do so, we need a sample for which independent distance determinations are available. 

We have used a sample of 22 Giant Extragalactic \hii\ Regions (GEHR) located in 9 nearby galaxies for which distances can be estimated from alternative primary methods, fundamentally Cepheids, RR Lyrae and Eclipsing Binaries. We have reassessed the distance modulus for every object from the measurements reported in the literature since 1995. We have used weighted mean distance values. Table \ref{tab:tab8} shows the values of the distance modulus for every object and the references from which it has been derived.

The values for integrated $\hb$ fluxes, velocity dispersion and absorption coefficients have been obtained from the results published by \citet{Melnick1987}. 

%%Table 8 --- Distance for the anchor sample 
\input{ch05/tab/tab01.tex}
%%%%%%%%%%%%%

\section{Determination of $H_{0}$}
The procedure we use to estimate the Hubble constant comprises the following
steps:
\begin{enumerate}
\item Determine the slope of the $L(\mathrm{H}\beta)-\sigma$ distance indicator,
using the \hii\ galaxy sample,
\item Determine the intercept of the relation (the zero-point) using the
local calibration `anchor'  GEHR sample,
\item Use a $\chi^2$ minimization procedure to find which value of $H_0$
minimizes the difference between the \hii\ galaxy luminosities
predicted from the derived $L(\mathrm{H}\beta)-\sigma$ relation, and
those estimated from the $\hb$ flux and the distance based on the
value of $H_0$.
\end{enumerate}

%%%%%%%%%%%%%%%%%%%%%%%%%%%%%%%%%%%%%
%Figure --- L - Sigma Relation for the GEH2R sample
\begin{figure}
\centering
\resizebox{12cm}{!}{\includegraphics{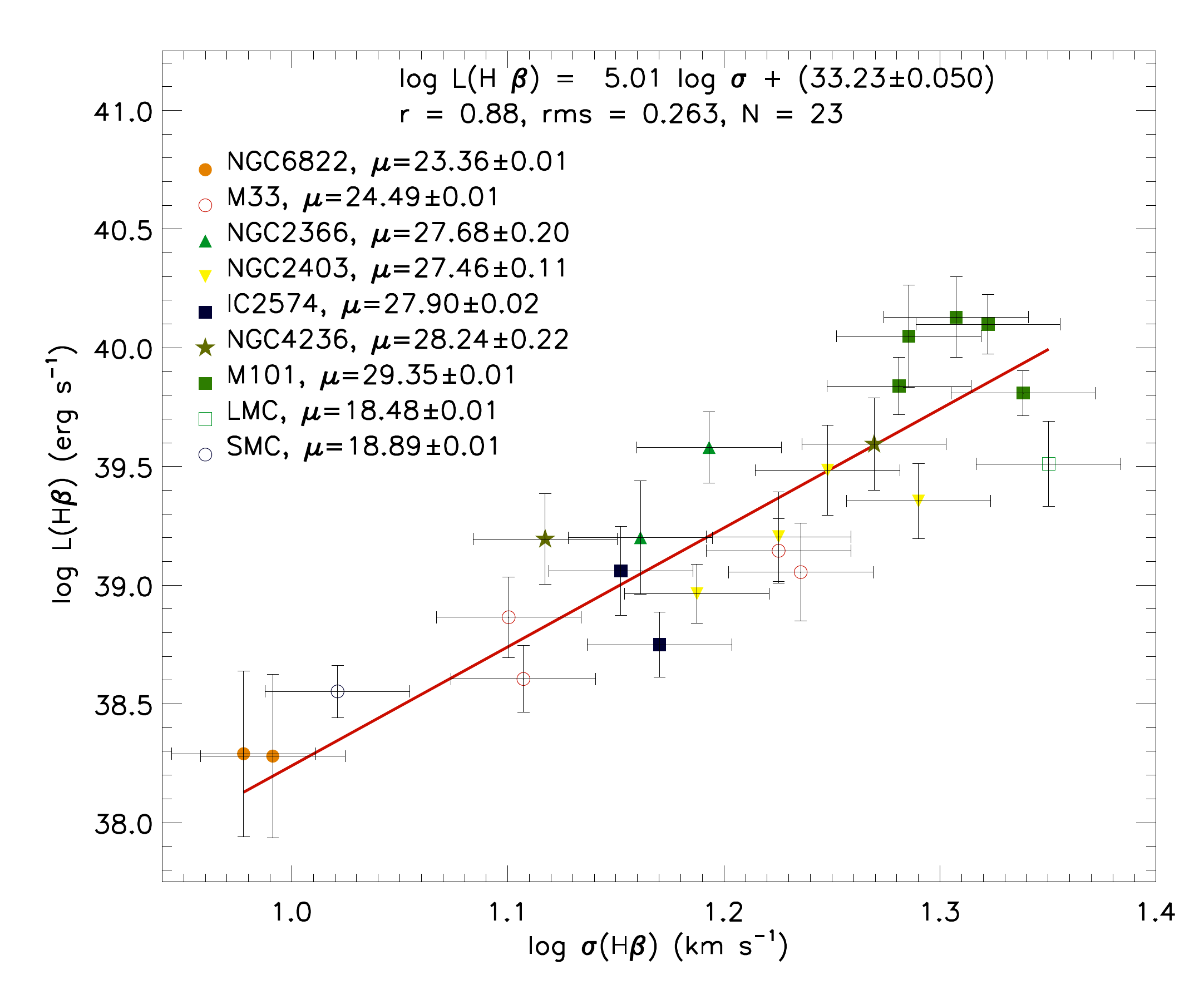}}
\caption[\lsig\ relation for the GEHR sample] {\small \lsig\ relation for the GEHR sample. The correlation parameters and the adopted individual distance moduli are given in the inset. 
The red line is the best fit 
for a fixed value of the slope. }
\label{fig:LSgeh2r}
\end{figure}
%%%%%%%%%%%%%%%%%%%%%%%%%%%%%%%%%%%%%
%%%%%%%%%%%%%%%%%%%%%%%%%%%%%%%%%%%%%
%Figure --- L - Sigma Relation for the GEH2R+HII galaxies sample
\begin{figure}
\centering
\resizebox{12cm}{!}{\includegraphics{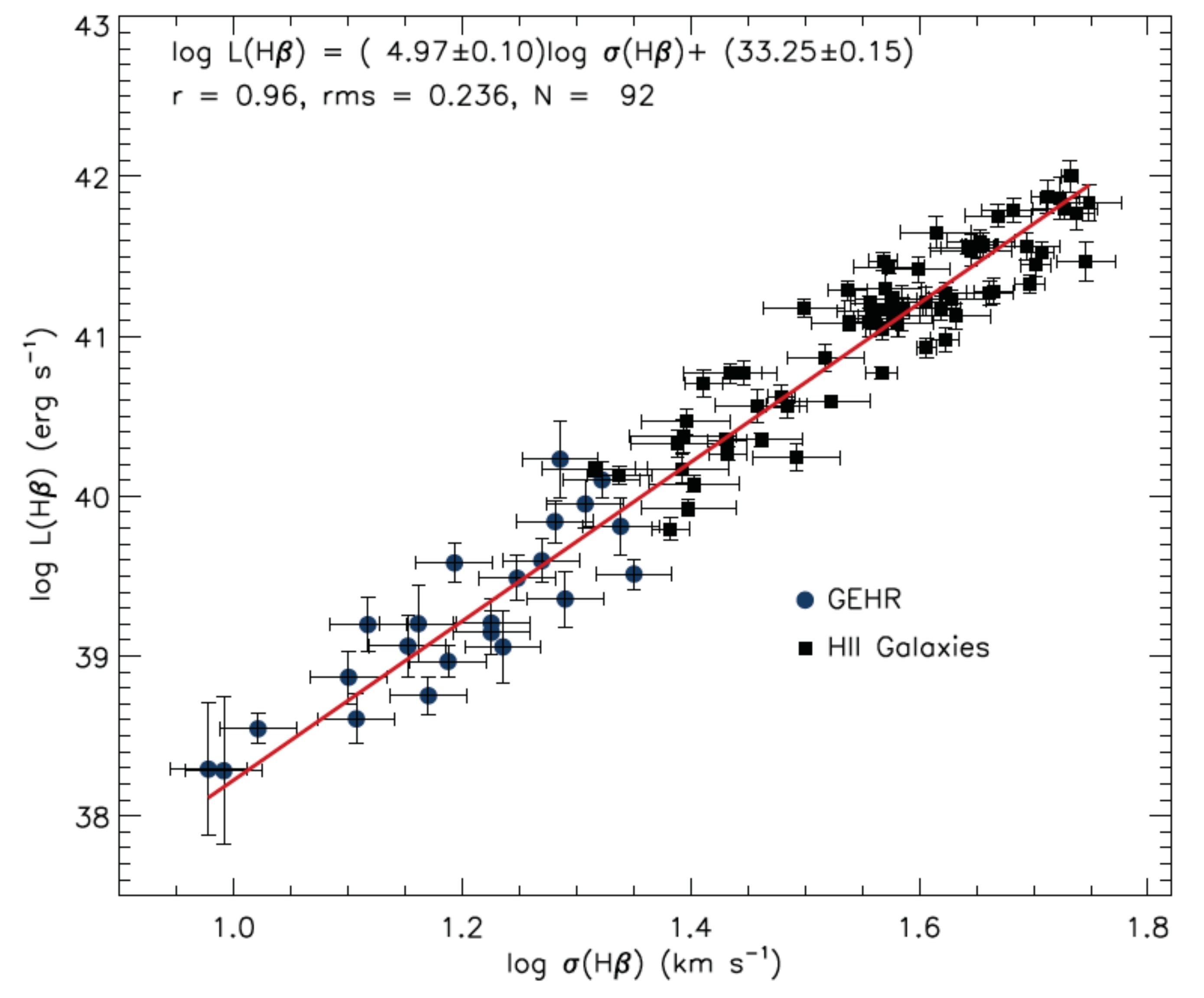}}
\caption[\lsig\ relation for the joint \hii\ galaxies and GEHR sample]{\small \lsig\ relation for the joint \hii\ galaxies and GEHR sample.}
\label{fig:LSigJoin}
\end{figure}
%%%%%%%% %%% Important
%%%%%%%%%Now working in a new version of this plot
%%%%%%%%%%%%%%%%%%%%%%%%%%%%%%%%%%%%%

In detail, we first estimate the slope of the
$L(\mathrm{H}\beta)-\sigma$ relation for \hii\ galaxies, using an arbitary value of $H_0$,
since the slope is independent of $H_0$, to determine luminosities
from the observed $\hb$ flux and the Hubble distance\footnote{We have
  verified that the initial choice for the value of $H_0$ does not alter the
  determined slope value.}. We then
%Once we have determined the value of the slope, the second step
%consists in determining the
determine the intercept of the relation from a fit to the `anchor'
GEHR sample, but fixing the slope to that determined in step one, i.e., that based on \hii\ galaxies.

Figure \ref{fig:LSgeh2r} 
shows the \lsig\ relation for the GEHR
sample. The slope of the correlation has been fixed to the value obtained 
from the \hii\ galaxies sample fitting as explained above.
%The thick blue points show the weighted -- based on the
%measurement accuracy -- mean values for each
%galaxy, whereas the thin points are the values for individual GEHR.
The resulting GEHR sample \lsig\ correlation is:
\begin{equation}
 \logd L(\mathrm{H}\beta) = 5.01 \logd \sigma + (33.23 \pm 0.050) 
\end{equation}
with r.m.s.~$ \logd L(\mathrm{H}\beta)= 0.263$.

The \lsig\ correlation resulting from fitting jointly the \hii\ galaxies sample and the
GEHR sample is shown in Figure \ref{fig:LSigJoin} and is given by:
%The resulting \lsig\ correlation, calculated as above described and
%shown together with the GEHR and \hii\ galaxy samples in 
%Figure 3, is given by:
\begin{equation}
 \logd L(\mathrm{H}\beta) = (4.97 \pm 0.10) \logd \sigma + (33.25 \pm 0.15) 
 \label{eq:ls}
\end{equation}
with r.m.s. $\logd L(\mathrm{H}\beta)= 0.236$.
%\ref{fig:LSjoint}. 
%The parameters of the fit are given as
%labels in both figures, and the procedure is  explained in the
%following section.

The final step of our procedure is to determine the value of $H_0$
by minimizing, over a grid of $H_0$ values, the function:
\begin{equation}
\chi^2(H_0) = \sum\limits_{i=1}^n \frac{[L_i(\sigma_i) -
  \tilde{L}_i(H_0, f_i, z_i)]^2}{\sigma_{L, i}^2 + \sigma_{\tilde{L}, i}^2}, 
\end{equation}
where the summation is over the \hii\ galaxies, 
$\sigma_i$ are the measured velocity dispersions,
$L_i(\sigma)$ are the logarithmic luminosities estimated from the `distance
indicator' as defined in eq.(\ref{eq:ls}), $\sigma_{L, i}$ are their errors
propagated from the uncertainties in $\sigma$, the slope and intercept
of the relation, $\tilde{L}_i(H_o, f_i, z_i)$ are the logarithm of the luminosities obtained from
the measured fluxes and redshifts  by using a particular value of
$H_0$ in the Hubble law to estimate distances, and
$\sigma_{\tilde{L}, i}$ are the errors in this last estimation of
luminosities, propagated from the uncertainties in the fluxes and
redshifts.   

%%%%%%%%%%%%%%%%%%%%%%%%%%%%%%%%%%%%%
%Figure --- Chisq plot
\begin{figure}
\centering
\resizebox{12cm}{!}{\includegraphics{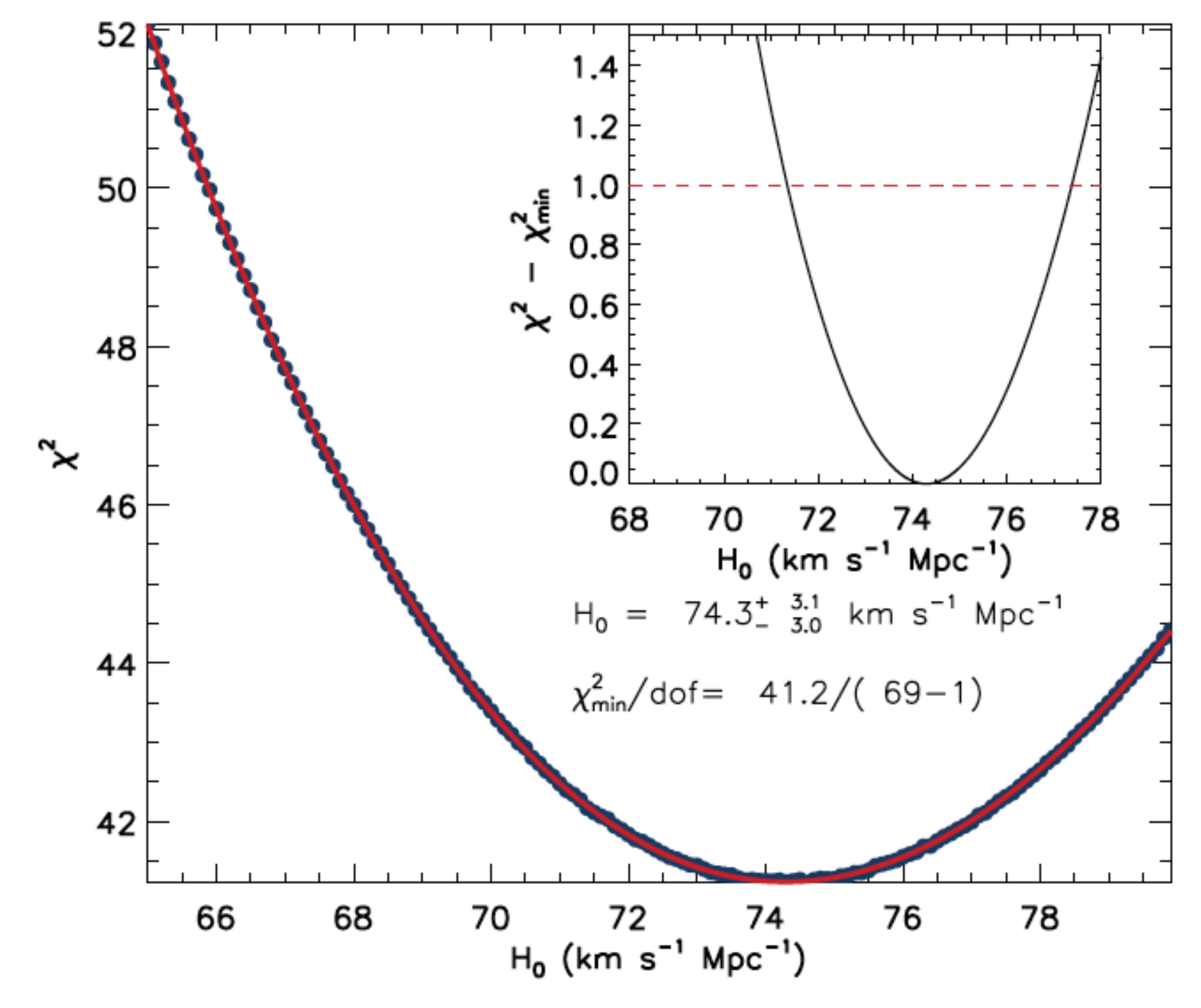}}
\caption[Values of  $\chi^2$  for the grid of $H_0$] {\small Values of  $\chi^2$  for the grid of $H_0$. The solid line is a cubic fit to the points. 
The inset panel shows the value of $\chi^2 - \chi^2_{min}$.}
\label{fig:chsqplot}
\end{figure}
%%%%%%%%%%%%%%%%%%%%%%%%%%%%%%%%%%%%%

The value obtained for $H_0$ using the above described procedure is:
\begin{equation}
  H_0 = 74.3^{+3.1}_{-3.0}\ \mathrm{Km\ s^{-1}\ Mpc^{-1}}\;.
\end{equation}
Figure \ref{fig:chsqplot} 
shows the resulting $\chi^2$ for the grid of
$H_0$ values used, with the solid line being a cubic fit to the points. The inset panel
shows the value of $\chi^2 - \chi^2_{min}$, 
from here we have obtained the $1\sigma$ confidence limits as the values of 
$H_0$ for which  $\chi^2 - \chi^2_{min} = 1$ since the fit has only 1 degree of freedom (dof) . 

\begin{figure}
\centering
%\resizebox{19cm}{!}{\includegraphics{HubbleDiagram.pdf}}
\resizebox{12cm}{!}{\includegraphics{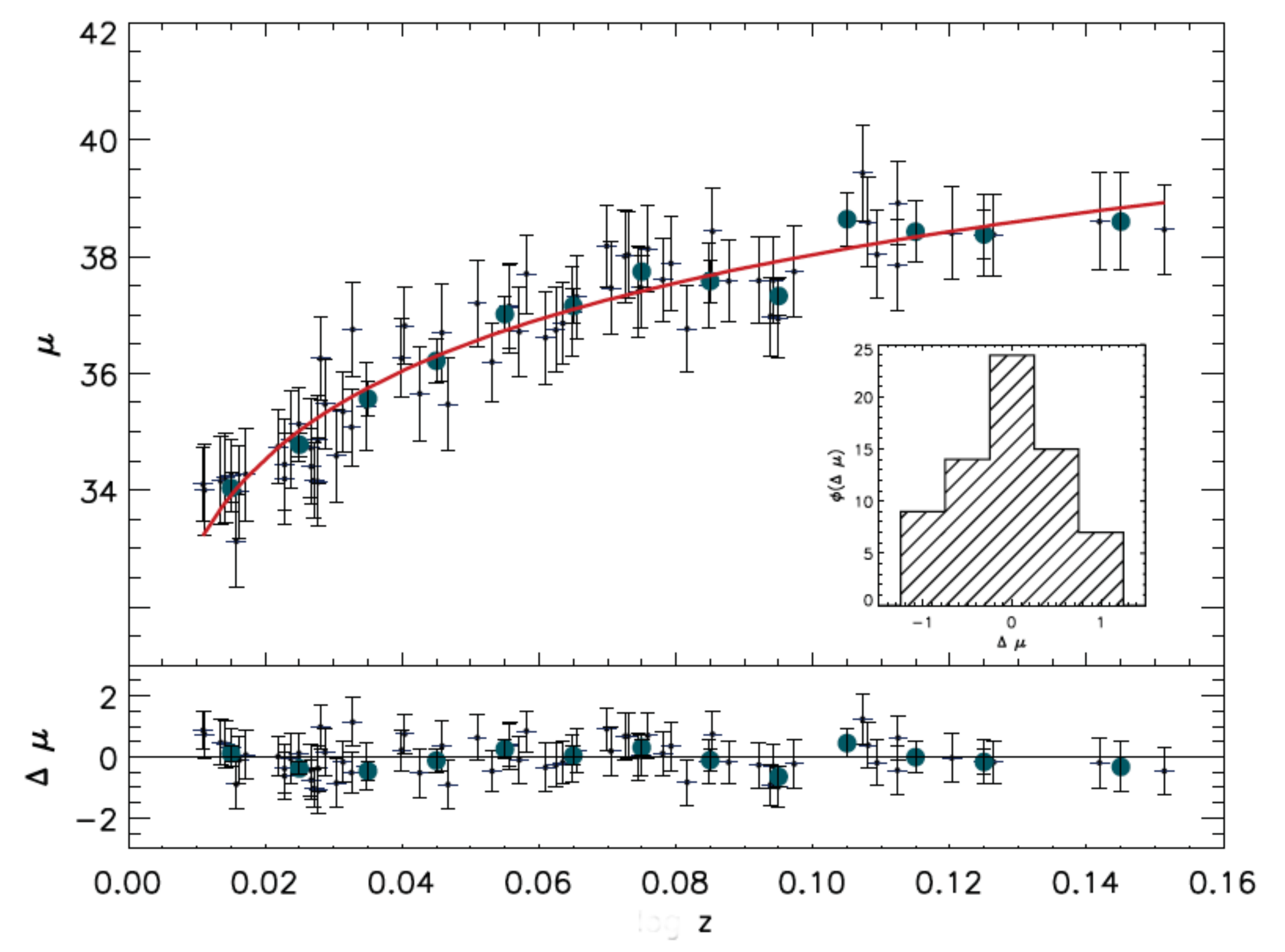}}
\caption[Hubble diagram for our sample of 69 \hii\ galaxies] {\small Hubble diagram for our sample of 69 \hii\ galaxies. The thick points
are the mean values for bins of $0.01$ in redshift. The solid line shows the run with redshift
of the distance modulus for $H_0 = 74.3$. Residuals are plotted in the bottom panel and their distribution
is shown in the inset. The rms value is 0.57 mag.  }
\label{fig:HRsh}
\end{figure}

Figure \ref{fig:HRsh} shows the Hubble diagram
for the sample of \hii\ galaxies used for the $H_0$ value
determination. The continuous line
 shows the redshift run of the distance modulus, obtained from the linear
 Hubble law and the fitted $H_0$ value, whereas the points correspond
 to the individual \hii\ galaxy distance moduli %from the luminosities 
obtained  through the \lsig\ correlation.

\subsection{Systematics}
%%%%%%%%%%%%%%%%%%%%%%%%%%%%%%%%%%%%%
 %Table --- Systematic Errors Budget 
\begin{table*}
\centering
 \begin{minipage}{100mm}
  \caption[Systematic Error Budget on the $H_0$ determination] {\small Systematic Error Budget on the $H_0$ determination.}
  \label{tab:SEB}
  \resizebox{0.8 \textwidth}{!}{
\begin{tabular} { l l l }
\hline
\hline
Symbol & Source & Error (km s$^{-1}$ Mpc$^{-1}$)  \\
\hline
$\sigma_{a,b}$ & Rotation,Multiplicity & 0.7 \\
$\sigma_c$ & Stellar Winds & 1.1\\
$\sigma_d$ & Internal Extinction & 0.7 \\
$\sigma_f $ & Object's Age & 1.4 \\
$\sigma_g$ & Malmquist Bias & 2.1 \\
$\sigma_h$ & IMF & ---- \\
\hline
Total &  &2.9\\
\hline
\end{tabular}
}
\end{minipage}
\end{table*}
%%%%%%%%%%%%%%%%%%%%%%%%%%%%%%%%%%%%%
The most difficult problem when pursuing high precision cosmology is probably recognising and treating  the systematic errors affecting the procedure. Here we will discuss the most important ones in the derivation of $H_0$ by means  of the \lsig\ relation for giant bursts of star formation and HII galaxies. Table \ref{tab:SEB} shows the systematic error budget on the $H_0$
determination.

\subsubsection{Malmquist bias}

Malmquist bias is a selection effect affecting flux limited samples, consisting of the preferential detection of the most luminous objects as a function of distance and limiting flux. A flux limited sample is biased towards the bright end of the luminosity function because as the distance towards an object increases the limiting flux imposes a higher luminosity detection threshold. 

The Malmquist bias for our flux limited sample was calculated following the procedure given by \citet{Giraud1987}. Basically the procedure consists in calculating the mean value for the absolute magnitudes distribution at every kinematic distance ($\langle M(\logd D_v) \rangle$) and compare it to the true mean value of the absolute magnitudes distribution ($M_0$); the resulting value is called $\Delta M(\logd D_v)$ and the bias is given by the expression:
\begin{equation}
	b(\logd D_v) = \frac{\sigma_0^2}{\sigma_{M_0}^2 + \sigma_0^2} \Delta M (\logd D_v),
\end{equation}
where, $D_v$ is given in $\mathrm{km\ s^{-1}}$, $\sigma_0$ is the true dispersion of residuals of the \lsig\ relation and $\sigma_{M_0}$ is the true dispersion of the distribution of absolute magnitudes of the sample. Finally the correction to $H_0$ at every kinematic distance is given by:
\begin{equation}
	\Delta \logd H = 0.2 b(\logd D_v).
\end{equation}
The main difference between our calculation and the one by \citet{Giraud1987} is that whereas he used a gaussian distribution for the absolute magnitudes, we use a power law one.% as  discussed in section 4.4. 

Another consideration is that the true values for the absolute magnitude distribution parameters and the true dispersion of residuals for the \lsig\ relation are obviously not directly known, but since we have corrected the absolute magnitude distribution for selection bias using the $V_{max}$ method and using only the brighter end of the distribution for the parametric fit,  we used directly the parameters obtained in section 4.4. On the other hand the true dispersion of residuals in the \lsig\ correlation can not be obtained, so we used instead the observed dispersion of residuals, hence our results are an upper limit to the true Malmquist bias.   

Our main result is that the value of the Malmquist bias is quite small:  $2.1\  \mathrm{km\ s^{-1}}$ at $z \sim 0.15$, and smaller at lower $z$. 

\subsubsection{Age}
As discussed in  \S 6.1.3, the value of $W(\hb)$ is a good indicator of the age of the starburst.  \citet{Melnick2000} have demonstrated that \hii\ galaxies with $W(\hb) < 25\ \mathrm{\AA}$ do follow an \lsig\ relation with a similar slope but  different intercept than the ones with larger $W(\hb)$. In principle we can conclude that age could be an important systematic effect present in the distances estimated from the \lsig\ relation. 

For the study presented here, the starburst age is a controlled parameter in the sense that we have selected our sample to be composed of very young objects by putting a high lower limit to the value of $\mathrm{W}(\mathrm{H}\beta) > 50\ \mathrm{\AA}$. Only the youngest bursts therefore were considered.

From the previous discussion it is clear that we expect to have only a small systematic effect from the evolution of the starburst on the determination of $H_0$; we have quantified it to be $\sim 1.4\ \mathrm{km\ s^{-1}\ Mpc^{-1}}$.

\subsubsection{Chemical abundance}
%As was described in \S 6.1.2, 
We have investigated the possible influence of chemical abundance over the \lsig\ relation but the results obtained are quite inconclusive. It appears that the abundance does not contribute decisively to the scatter in the relation and then its influence at least for our sample with small dynamic range  is very small.

\subsubsection{Radius}
As we already discussed in chapter 4, the size of the objects definitely contribute as a parameter to explain the scatter in the \lsig\ relation but %as was discussed in \S 5.2 
it appears that sizes as measured from SDSS DR7 data are severely overestimated, although probably systematically. Further investigation is therefore required to determinate the extent of its contribution as source of error. However, the uncertainties derived from not considering the size as a parameter are already considered, since the errors in the \lsig\ relation are propagated to the random error in the determination of $H_0$.

\subsubsection{Extinction}
We have already considered the effect of reddening and underlying absorption on the measurements. %The correction procedure was described in detail in \S4.4. 
Yet, concerned about the Balmer decrement not allowing us to correct completely for internal extinction, we have estimated the corresponding systematic error in the $H_0$ determination.  

\subsubsection{Environment}
In order to explore the possible effect of  the environment in the \lsig\ correlation, we have studied the close environment ($\sim 200h^{-1}_{75}\ \mathrm{kpc}$) for the objects in our sample plus  SDSS DR7 data. We used the available spectroscopic and photometric redshifts to explore the presence of close and possibly interacting companions to our target sample. 

The complete description of the study and the results are presented in  \citet{Koulouridis2013} where we conclude that the \lsig\ correlation is not affected by environmental or host galaxy differences of isolated and paired \hii\ galaxies.

\section{Summary}
It is indisputable that in the epoch of intense studies aimed at measuring the
dark energy equation of state, it is of paramount importance to
minimize the amount of priors needed to successfully complete such a
task. One such prior is the Hubble constant $H_0$  and  its
measurement at the $\sim 1\%$ accuracy level  has been identified as a
necessary prerequisite for putting effective constraints on the dark
energy, on neutrino physics and even on tests, at cosmological scales,
of general relativity \citep[see][]{Suyu2012}. Furthermore, it is highly desirable to have independent
determinations of $H_0$, since this will help understand and control
systematic effects that may affect individual methods and tracers of
the Hubble expansion.

It is within this latter strategy that our current work falls. We have carried out
%It is within this latter strategy that our current work falls, for
%which we have performed 
VLT and Subaru observations of a sample of nearby
\hii\ galaxies, identified in the
SDSS DR7 catalogue and 2m class telescopes spectrophotometry, in order to define their $L(\mathrm{H}\beta)-\sigma$ 
correlation, which we use to estimate the value of the Hubble
constant. This is achieved by estimating the zero-point of the distance
indicator using giant \hii\ regions in
nearby galaxies, for which accurate independent distance measurements
exist (based on Cepheids, RR Lyrae, TRGB  and eclipsing binaries). 

Using our final and clean of systematic effects sample of 69 \hii\ galaxies with 
$z\mincir 0.16$, we obtain: \\ $H_{0}=74.3\pm 3.1$(random) $\pm 2.9$ (systematic)  km s$^{-1}$
Mpc$^{-1}$,\\ independently confirming  the recent SNae~Ia-based
results of \citet{Riess2011}.

%% file: ch05/tab/tab01.tex
%Table  8 --- Zero Point Sample Distance Moduli

\onecolumn
%\begin{landscape}
\scriptsize
\begin{center}
\begin{longtable}{ l l l l  l  }
 
\caption[Distance Moduli for the GEHR sample]{\small Distance Moduli for the host galaxies of the GEHR sample. } 
\label{tab:tab8} \\

%This is the header for the first page of the table...
\hline \hline \\[-2ex]
\multicolumn{1}{c}{(1)} &
\multicolumn{1}{c}{(2)} &
\multicolumn{1}{c}{(3)} &
\multicolumn{1}{c}{(4)} &
\multicolumn{1}{c}{(5)} \\ 

\multicolumn{1}{c}{\bf{Galaxy}} &
\multicolumn{1}{c}{$\mu$} &
\multicolumn{1}{c}{$D$} &
\multicolumn{1}{c}{\bf{Method}} &
\multicolumn{1}{c}{\bf{Reference}} \\

\multicolumn{1}{c}{} &
\multicolumn{1}{c}{(mag)} &
\multicolumn{1}{c}{(kpc)} &
\multicolumn{1}{c}{ } &
\multicolumn{1}{c}{ }  \\[0.5ex] \hline \\[-1.8ex]

\endfirsthead

%This is the header for the remaining page(s) of the table...
\multicolumn{1}{c}{(1)} &
\multicolumn{1}{c}{(2)} &
\multicolumn{1}{c}{(3)} &
\multicolumn{1}{c}{(4)} &
\multicolumn{1}{c}{(5)} \\ 

\multicolumn{1}{c}{\bf{Galaxy}} &
\multicolumn{1}{c}{$\mu$} &
\multicolumn{1}{c}{$D$} &
\multicolumn{1}{c}{\bf{Method}} &
\multicolumn{1}{c}{\bf{Reference}} \\

\multicolumn{1}{c}{} &
\multicolumn{1}{c}{(mag)} &
\multicolumn{1}{c}{(kpc)} &
\multicolumn{1}{c}{ } &
\multicolumn{1}{c}{ }  \\[0.5ex] \hline \\[-1.8ex]

\endhead

\multicolumn{4}{l}{{Continued on Next Page\ldots}} \\
\endfoot

\\[-1.8ex] \hline 
%Notes 

\endlastfoot
LMC	&	18.64	$\pm$	0.02	&	53.46	$\pm$	2.46	&	Cepheids 	&	\citet{1995ApJ...452..195D}	\\
	&	18.42	$\pm$	0.11	&	48.31	$\pm$	12.24	&	Cepheids 	&	\citet{1997AJ....113...13B}	\\
	&	18.57	$\pm$	0.19	&	51.76	$\pm$	22.64	&	RR Lyrae 	&	\citet{1997ApJ...482...89A}	\\
	&	18.58	$\pm$	0.02	&	52.00	$\pm$	2.39	&	Cepheids 	&	\citet{1997ApJ...486...60D}	\\
	&	18.53	$\pm$	0.04	&	50.82	$\pm$	4.68	&	RR Lyrae 	&	\citet{1997MNRAS.284..761F}	\\
	&	18.26	$\pm$	0.15	&	44.87	$\pm$	15.50	&	RR Lyrae 	&	\citet{1998AyA...330..515F}	\\
	&	18.29	$\pm$	0.17	&	45.50	$\pm$	17.81	&	Cepheids 	&	\citet{1998AyA...335L..81L}	\\
	&	18.37	$\pm$	0.23	&	47.21	$\pm$	25.00	&	RR Lyrae 	&	\citet{1998AyA...335L..81L}	\\
	&	18.44	$\pm$	0.35	&	48.75	$\pm$	39.29	&	Cepheids 	&	\citet{1998ApJ...492..110M}	\\
	&	18.50	$\pm$	0.13	&	50.12	$\pm$	15.00	&	Cepheids 	&	\citet{1998ApJ...492..110M}	\\
	&	18.53	$\pm$	0.14	&	50.82	$\pm$	16.38	&	Cepheids 	&	\citet{1998ApJ...492..110M}	\\
	&	18.57	$\pm$	0.11	&	51.76	$\pm$	13.11	&	Cepheids 	&	\citet{1998ApJ...492..110M}	\\
	&	18.59	$\pm$	0.15	&	52.24	$\pm$	18.04	&	Cepheids 	&	\citet{1998ApJ...492..110M}	\\
	&	18.60	$\pm$	0.15	&	52.48	$\pm$	18.13	&	Cepheids 	&	\citet{1998ApJ...492..110M}	\\
	&	18.62	$\pm$	0.18	&	52.97	$\pm$	21.95	&	Cepheids 	&	\citet{1998ApJ...492..110M}	\\
	&	18.74	$\pm$	0.24	&	55.98	$\pm$	30.93	&	Cepheids 	&	\citet{1998ApJ...492..110M}	\\
	&	18.77	$\pm$	0.24	&	56.75	$\pm$	31.36	&	Cepheids 	&	\citet{1998ApJ...492..110M}	\\
	&	18.86	$\pm$	0.36	&	59.16	$\pm$	49.04	&	Cepheids 	&	\citet{1998ApJ...492..110M}	\\
	&	18.46	$\pm$	0.02	&	49.20	$\pm$	2.27	&	Cepheids 	&	\citet{1998ApJ...496...17G}	\\
	&	18.30	$\pm$	0.07	&	45.71	$\pm$	7.37	&	Eclipsing Binary 	&	\citet{1998ApJ...509L..21G}	\\
	&	18.35	$\pm$	0.07	&	46.77	$\pm$	7.54	&	Eclipsing Binary 	&	\citet{1998ApJ...509L..21G}	\\
	&	18.22	$\pm$	0.13	&	44.06	$\pm$	13.19	&	Eclipsing Binary 	&	\citet{1998ApJ...509L..25U}	\\
	&	18.61	$\pm$	0.28	&	52.72	$\pm$	33.99	&	RR Lyrae 	&	\citet{1999AyA...348L..33G}	\\
	&	18.42	$\pm$	0.30	&	48.31	$\pm$	33.37	&	Cepheids 	&	\citet{1999ApJ...512..711B}	\\
	&	18.62	$\pm$	0.12	&	52.97	$\pm$	14.64	&	Cepheids 	&	\citet{1999ApJ...512..711B}	\\
	&	18.62	$\pm$	0.17	&	52.97	$\pm$	20.73	&	Cepheids 	&	\citet{1999ApJ...512..711B}	\\
	&	18.74	$\pm$	0.13	&	55.98	$\pm$	16.76	&	Cepheids 	&	\citet{1999ApJ...512..711B}	\\
	&	18.57	$\pm$	0.05	&	51.76	$\pm$	5.96	&	Cepheids 	&	\citet{1999ApJ...522..250S}	\\
	&	18.72	$\pm$	0.09	&	55.46	$\pm$	11.49	&	Cepheids 	&	\citet{1999PASP..111..775F}	\\
	&	18.52	$\pm$	0.18	&	50.58	$\pm$	20.96	&	Cepheids 	&	\citet{2000AyA...356..849G}	\\
	&	18.54	$\pm$	0.19	&	51.05	$\pm$	22.33	&	Cepheids 	&	\citet{2000AyA...356..849G}	\\
	&	18.60	$\pm$	0.11	&	52.48	$\pm$	13.29	&	Cepheids 	&	\citet{2000AyA...356..849G}	\\
	&	18.66	$\pm$	0.15	&	53.95	$\pm$	18.63	&	Cepheids 	&	\citet{2000AyA...356..849G}	\\
	&	18.52	$\pm$	0.02	&	50.58	$\pm$	2.33	&	RR Lyrae 	&	\citet{2000AyA...363L...1K}	\\
	&	18.40	$\pm$	0.07	&	47.86	$\pm$	7.71	&	Eclipsing Binary 	&	\citet{2000AJ....119.1205N}	\\
	&	18.42	$\pm$	0.10	&	48.31	$\pm$	11.12	&	Cepheids 	&	\citet{2000ApJ...533L.107G}	\\
	&	18.50	$\pm$	0.13	&	50.12	$\pm$	15.00	&	Cepheids 	&	\citet{2000ApJS..128..431F}	\\
	&	18.42	$\pm$	0.08	&	48.31	$\pm$	8.90	&	Eclipsing Binary 	&	\citet{2001AyA...366..752G}	\\
	&	18.46	$\pm$	0.06	&	49.20	$\pm$	6.80	&	Eclipsing Binary 	&	\citet{2001AyA...366..752G}	\\
	&	18.42	$\pm$	0.24	&	48.31	$\pm$	26.69	&	Eclipsing Binary 	&	\citet{2002AyA...381..440S}	\\
	&	18.43	$\pm$	0.24	&	48.53	$\pm$	26.82	&	Eclipsing Binary 	&	\citet{2002AyA...381..440S}	\\
	&	18.35	$\pm$	0.03	&	46.77	$\pm$	3.23	&	Cepheids 	&	\citet{2002AyA...389...19P}	\\
	&	18.37	$\pm$	0.14	&	47.21	$\pm$	15.22	&	Cepheids 	&	\citet{2002AyA...389...19P}	\\
	&	18.38	$\pm$	0.10	&	47.42	$\pm$	10.92	&	RR Lyrae 	&	\citet{2002AJ....123..473B}	\\
	&	18.53	$\pm$	0.10	&	50.82	$\pm$	11.70	&	RR Lyrae 	&	\citet{2002AJ....123..473B}	\\
	&	18.50	$\pm$	0.13	&	50.12	$\pm$	15.00	&	Cepheids 	&	\citet{2002AJ....124.1695B}	\\
	&	18.58	$\pm$	0.15	&	52.00	$\pm$	17.96	&	Cepheids 	&	\citet{2002AJ....124.1695B}	\\
	&	18.31	$\pm$	0.10	&	45.92	$\pm$	10.57	&	Eclipsing Binary 	&	\citet{2002ApJ...564..260F}	\\
	&	18.36	$\pm$	0.10	&	46.99	$\pm$	10.82	&	Eclipsing Binary 	&	\citet{2002ApJ...564..260F}	\\
	&	18.50	$\pm$	0.06	&	50.12	$\pm$	6.92	&	Eclipsing Binary 	&	\citet{2002ApJ...564..260F}	\\
	&	18.52	$\pm$	0.06	&	50.58	$\pm$	6.99	&	Eclipsing Binary 	&	\citet{2002ApJ...564..260F}	\\
	&	18.53	$\pm$	0.05	&	50.82	$\pm$	5.85	&	Cepheids 	&	\citet{2002ApJ...565L..83B}	\\
	&	18.37	$\pm$	0.08	&	47.21	$\pm$	8.70	&	Eclipsing Binary 	&	\citet{2002ApJ...574..771R}	\\
	&	18.38	$\pm$	0.08	&	47.42	$\pm$	8.74	&	Eclipsing Binary 	&	\citet{2002ApJ...574..771R}	\\
	&	18.48	$\pm$	0.13	&	49.66	$\pm$	14.86	&	Cepheids 	&	\citet{2002ApJ...574L..33B}	\\
	&	18.53	$\pm$	0.08	&	50.82	$\pm$	9.36	&	Cepheids 	&	\citet{2002ApJ...574L..33B}	\\
	&	18.55	$\pm$	0.02	&	51.29	$\pm$	2.36	&	Cepheids 	&	\citet{2002ApJ...578..144K}	\\
	&	18.63	$\pm$	0.08	&	53.21	$\pm$	9.80	&	Eclipsing Binary 	&	\citet{2003AyA...402..509C}	\\
	&	18.65	$\pm$	0.10	&	53.70	$\pm$	12.37	&	Cepheids 	&	\citet{2003AyA...410..887G}	\\
	&	18.30	$\pm$	0.04	&	45.71	$\pm$	4.21	&	Cepheids 	&	\citet{2003AJ....125.1261D}	\\
	&	18.32	$\pm$	0.01	&	46.13	$\pm$	1.06	&	Cepheids 	&	\citet{2003AJ....125.1261D}	\\
	&	18.47	$\pm$	0.07	&	49.43	$\pm$	7.97	&	RR Lyrae 	&	\citet{2003AJ....125.1261D}	\\
	&	18.30	$\pm$	0.14	&	45.71	$\pm$	14.73	&	RR Lyrae 	&	\citet{2003AJ....125.1309C}	\\
	&	18.38	$\pm$	0.16	&	47.42	$\pm$	17.47	&	RR Lyrae 	&	\citet{2003AJ....125.1309C}	\\
	&	18.45	$\pm$	0.09	&	48.98	$\pm$	10.15	&	RR Lyrae 	&	\citet{2003AJ....125.1309C}	\\
	&	18.52	$\pm$	0.09	&	50.58	$\pm$	10.48	&	RR Lyrae 	&	\citet{2003AJ....125.1309C}	\\
	&	18.18	$\pm$	0.09	&	43.25	$\pm$	8.96	&	Eclipsing Binary 	&	\citet{2003ApJ...587..685F}	\\
	&	18.23	$\pm$	0.09	&	44.26	$\pm$	9.17	&	Eclipsing Binary 	&	\citet{2003ApJ...587..685F}	\\
	&	18.56	$\pm$	0.03	&	51.52	$\pm$	3.56	&	Cepheids 	&	\citet{2003MNRAS.342L..58K}	\\
	&	18.55	$\pm$	0.03	&	51.29	$\pm$	3.54	&	RR Lyrae 	&	\citet{2003MNRAS.342L..58K}	\\
	&	18.48	$\pm$	0.10	&	49.66	$\pm$	11.43	&	Cepheids 	&	\citet{2003MNRAS.345..269H}	\\
	&	18.51	$\pm$	0.10	&	50.35	$\pm$	11.59	&	Cepheids 	&	\citet{2003MNRAS.345..269H}	\\
	&	18.54	$\pm$	0.10	&	51.05	$\pm$	11.75	&	Cepheids 	&	\citet{2003MNRAS.345..269H}	\\
	&	18.45	$\pm$	0.07	&	48.98	$\pm$	7.89	&	Cepheids 	&	\citet{2004AyA...415..531S}	\\
	&	18.46	$\pm$	0.07	&	49.20	$\pm$	7.93	&	Cepheids 	&	\citet{2004AyA...415..531S}	\\
	&	18.47	$\pm$	0.07	&	49.43	$\pm$	7.97	&	Cepheids 	&	\citet{2004AyA...415..531S}	\\
	&	18.48	$\pm$	0.07	&	49.66	$\pm$	8.00	&	Cepheids 	&	\citet{2004AyA...415..531S}	\\
	&	18.50	$\pm$	0.07	&	50.12	$\pm$	8.08	&	Cepheids 	&	\citet{2004AyA...415..531S}	\\
	&	18.48	$\pm$	0.08	&	49.66	$\pm$	9.15	&	RR Lyrae 	&	\citet{2004AyA...423...97B}	\\
	&	18.43	$\pm$	0.06	&	48.53	$\pm$	6.70	&	RR Lyrae 	&	\citet{2004AJ....127..334A}	\\
	&	18.50	$\pm$	0.05	&	50.12	$\pm$	5.77	&	Cepheids 	&	\citet{2004AJ....128.2239P}	\\
	&	18.50	$\pm$	0.10	&	50.12	$\pm$	11.54	&	Cepheids 	&	\citet{2004ApJ...608...42S}	\\
	&	18.52	$\pm$	0.01	&	50.58	$\pm$	1.16	&	RR Lyrae 	&	\citet{2004ApJ...610..269D}	\\
	&	18.39	$\pm$	0.05	&	47.64	$\pm$	5.49	&	Cepheids 	&	\citet{2005AyA...434.1077M}	\\
	&	18.51	$\pm$	0.02	&	50.35	$\pm$	2.32	&	Cepheids 	&	\citet{2005AyA...434.1077M}	\\
	&	18.54	$\pm$	0.02	&	51.05	$\pm$	2.35	&	RR Lyrae 	&	\citet{2005AJ....129.2257M}	\\
	&	18.56	$\pm$	0.04	&	51.52	$\pm$	4.75	&	Cepheids 	&	\citet{2005ApJ...627..224G}	\\
	&	18.54	$\pm$	0.02	&	51.05	$\pm$	2.35	&	Cepheids 	&	\citet{2006ApJ...642..834K}	\\
	&	18.41	$\pm$	0.10	&	48.08	$\pm$	11.07	&	Cepheids 	&	\citet{2006ApJ...652.1133M}	\\
	&	18.54	$\pm$	0.15	&	51.05	$\pm$	17.63	&	RR Lyrae 	&	\citet{2006MNRAS.372.1675S}	\\
	&	18.42	$\pm$	0.10	&	48.31	$\pm$	11.12	&	Cepheids 	&	\citet{2007AyA...462..599T}	\\
	&	18.45	$\pm$	0.11	&	48.98	$\pm$	12.41	&	Cepheids 	&	\citet{2007AyA...462..599T}	\\
	&	18.42	$\pm$	0.06	&	48.31	$\pm$	6.67	&	Eclipsing Binary 	&	\citet{2007AyA...473..847V}	\\
	&	18.40	$\pm$	0.10	&	47.86	$\pm$	11.02	&	Cepheids 	&	\citet{2007AyA...476...73F}	\\
	&	18.40	$\pm$	0.05	&	47.86	$\pm$	5.51	&	Cepheids 	&	\citet{2007AJ....133.1810B}	\\
	&	18.50	$\pm$	0.03	&	50.12	$\pm$	3.46	&	Cepheids 	&	\citet{2007AJ....133.1810B}	\\
	&	18.34	$\pm$	0.06	&	46.56	$\pm$	6.43	&	Cepheids 	&	\citet{2007ApJ...671.1640A}	\\
	&	18.39	$\pm$	0.05	&	47.64	$\pm$	5.49	&	Cepheids 	&	\citet{2007MNRAS.379..723V}	\\
	&	18.45	$\pm$	0.04	&	48.98	$\pm$	4.51	&	Cepheids 	&	\citet{2007MNRAS.379..723V}	\\
	&	18.47	$\pm$	0.03	&	49.43	$\pm$	3.41	&	Cepheids 	&	\citet{2007MNRAS.379..723V}	\\
	&	18.52	$\pm$	0.03	&	50.58	$\pm$	3.49	&	Cepheids 	&	\citet{2007MNRAS.379..723V}	\\
	&	18.49	$\pm$	0.11	&	49.89	$\pm$	12.64	&	RR Lyrae 	&	\citet{2008AJ....135...83C}	\\
	&	18.58	$\pm$	0.03	&	52.00	$\pm$	3.59	&	RR Lyrae 	&	\citet{2008AJ....136..272S}	\\
	&	18.44	$\pm$	0.11	&	48.75	$\pm$	12.35	&	RR Lyrae 	&	\citet{2008ApJ...676L.135C}	\\
	&	18.45	$\pm$	0.09	&	48.98	$\pm$	10.15	&	Cepheids 	&	\citet{2008ApJ...684..102B}	\\
	&	18.56	$\pm$	0.09	&	51.52	$\pm$	10.68	&	Cepheids 	&	\citet{2008ApJ...684..102B}	\\
	&	18.53	$\pm$	0.13	&	50.82	$\pm$	15.21	&	RR Lyrae 	&	\citet{2009AyA...502..505B}	\\
	&	18.50	$\pm$	0.06	&	50.12	$\pm$	6.92	&	Eclipsing Binary 	&	\citet{2009ApJ...697..862P}	\\
	&	18.48	$\pm$	0.03	&	49.59	$\pm$	3.77	&	Cepheids 	&	\citet{Freedman2012}	\\
	&	18.49	$\pm$	0.05	&	49.97	$\pm$	1.13	&	Cepheids 	&	\citet{2013Natur.495...76P}	\\
\hline													
LMC Final	&	18.482	$\pm$	0.004	&	49.697	$\pm$	0.458	&	---	&	---	\\
\hline													
SMC	&	19.09	$\pm$	0.02	&	65.77	$\pm$	3.03	&	Cepheids 	&	\citet{1995ApJ...452..195D}	\\
	&	18.84	$\pm$	0.10	&	58.61	$\pm$	13.50	&	Cepheids 	&	\citet{1997AJ....113...13B}	\\
	&	19.00	$\pm$	0.03	&	63.10	$\pm$	4.36	&	Cepheids 	&	\citet{1997ApJ...486...60D}	\\
	&	18.83	$\pm$	0.05	&	58.34	$\pm$	6.72	&	Cepheids 	&	\citet{1997ApJ...491...13K}	\\
	&	18.98	$\pm$	0.02	&	62.52	$\pm$	2.88	&	Cepheids 	&	\citet{1997ApJ...491...13K}	\\
	&	18.98	$\pm$	0.28	&	62.52	$\pm$	40.31	&	Cepheids 	&	\citet{1999ApJ...512..711B}	\\
	&	19.14	$\pm$	0.15	&	67.30	$\pm$	23.24	&	Cepheids 	&	\citet{1999ApJ...512..711B}	\\
	&	19.19	$\pm$	0.17	&	68.87	$\pm$	26.96	&	Cepheids 	&	\citet{1999ApJ...512..711B}	\\
	&	19.28	$\pm$	0.17	&	71.78	$\pm$	28.10	&	Cepheids 	&	\citet{1999ApJ...512..711B}	\\
	&	19.05	$\pm$	0.13	&	64.57	$\pm$	19.33	&	Cepheids 	&	\citet{2000AyA...360L...1K}	\\
	&	19.04	$\pm$	0.17	&	64.27	$\pm$	25.16	&	Cepheids 	&	\citet{2000AyA...363..901G}	\\
	&	19.11	$\pm$	0.11	&	66.37	$\pm$	16.81	&	Cepheids 	&	\citet{2000AyA...363..901G}	\\
	&	18.99	$\pm$	0.05	&	62.81	$\pm$	7.23	&	Cepheids 	&	\citet{2000ApJS..128..431F}	\\
	&	19.01	$\pm$	0.13	&	63.39	$\pm$	18.97	&	Cepheids 	&	\citet{2002ApJ...574L..33B}	\\
	&	19.04	$\pm$	0.11	&	64.27	$\pm$	16.28	&	Cepheids 	&	\citet{2002ApJ...574L..33B}	\\
	&	18.88	$\pm$	0.01	&	59.70	$\pm$	1.37	&	Cepheids 	&	\citet{2003AJ....125.1261D}	\\
	&	18.91	$\pm$	0.04	&	60.53	$\pm$	5.58	&	Cepheids 	&	\citet{2003AJ....125.1261D}	\\
	&	18.86	$\pm$	0.07	&	59.16	$\pm$	9.53	&	RR Lyrae 	&	\citet{2003AJ....125.1261D}	\\
	&	18.89	$\pm$	0.04	&	59.98	$\pm$	5.52	&	Eclipsing Binary 	&	\citet{2003MNRAS.339..157H}	\\
	&	18.88	$\pm$	0.12	&	59.70	$\pm$	16.50	&	Cepheids 	&	\citet{2004AyA...415..531S}	\\
	&	18.93	$\pm$	0.24	&	61.09	$\pm$	33.76	&	RR Lyrae 	&	\citet{2004AJ....128..736W}	\\
	&	18.99	$\pm$	0.05	&	62.81	$\pm$	7.23	&	Cepheids 	&	\citet{2004ApJ...608...42S}	\\
	&	18.91	$\pm$	0.03	&	60.53	$\pm$	4.18	&	Eclipsing Binary 	&	\citet{2005MNRAS.357..304H}	\\
	&	18.93	$\pm$	0.02	&	61.09	$\pm$	2.81	&	Cepheids 	&	\citet{2006ApJ...642..834K}	\\
	&	18.97	$\pm$	0.03	&	62.23	$\pm$	4.30	&	RR Lyrae 	&	\citet{2009AJ....138.1661S}	\\
	&	18.83	$\pm$	0.01	&	58.34	$\pm$	1.34	&	RR Lyrae 	&	\citet{2010MNRAS.402..691D}	\\
	&	18.84	$\pm$	0.01	&	58.61	$\pm$	1.35	&	RR Lyrae 	&	\citet{2010MNRAS.402..691D}	\\
	&	18.86	$\pm$	0.01	&	59.16	$\pm$	1.36	&	RR Lyrae 	&	\citet{2010MNRAS.402..691D}	\\
	&	18.87	$\pm$	0.03	&	59.43	$\pm$	4.11	&	RR Lyrae 	&	\citet{2010MNRAS.402..691D}	\\
	&	18.87	$\pm$	0.02	&	59.43	$\pm$	2.74	&	RR Lyrae 	&	\citet{2010MNRAS.402..691D}	\\
	&	18.89	$\pm$	0.01	&	59.98	$\pm$	1.38	&	RR Lyrae 	&	\citet{2010MNRAS.402..691D}	\\
	&	18.89	$\pm$	0.04	&	59.98	$\pm$	5.52	&	RR Lyrae 	&	\citet{2010MNRAS.402..691D}	\\
	&	18.92	$\pm$	0.04	&	60.81	$\pm$	5.60	&	RR Lyrae 	&	\citet{2010MNRAS.402..691D}	\\
\hline													
SMC Final	&	18.885	$\pm$	0.004	&	59.839	$\pm$	0.487	&	---	&	---	\\
\hline													
IC 2574	&	27.89	$\pm$	0.03	&	3784.43	$\pm$	261.42	&	TRGB 	&	\citet{2009ApJS..183...67D}	\\
	&	27.93	$\pm$	0.03	&	3854.78	$\pm$	266.28	&	TRGB 	&	\citet{2009ApJS..183...67D}	\\
	&	28.02	$\pm$	0.22	&	4017.91	$\pm$	2035.35	&	TRGB 	&	\citet{2003AyA...398..479K}	\\
\hline													
IC 2574 Final	&	27.911	$\pm$	0.021	&	3821.23	$\pm$	185.79	&	---	&	---	\\
\hline													
M 101	&	29.34	$\pm$	0.10	&	7379.04	$\pm$	1699.09	&	Cepheids 	&	\citet{1996ApJ...463...26K}	\\
	&	29.05	$\pm$	0.14	&	6456.54	$\pm$	2081.34	&	Cepheids 	&	\citet{1998ApJ...508..491S}	\\
	&	29.21	$\pm$	0.17	&	6950.24	$\pm$	2720.60	&	Cepheids 	&	\citet{1998ApJ...508..491S}	\\
	&	29.20	$\pm$	0.08	&	6918.31	$\pm$	1274.40	&	Cepheids 	&	\citet{2001ApJ...548..564W}	\\
	&	29.04	$\pm$	0.08	&	6426.88	$\pm$	1183.87	&	Cepheids 	&	\citet{2001ApJ...549..721M}	\\
	&	29.37	$\pm$	0.01	&	7481.70	$\pm$	172.27	&	Cepheids 	&	\citet{2001ApJ...549..721M}	\\
	&	29.45	$\pm$	0.08	&	7762.47	$\pm$	1429.90	&	Cepheids 	&	\citet{2001ApJ...549..721M}	\\
	&	29.58	$\pm$	0.09	&	8241.38	$\pm$	1707.88	&	Cepheids 	&	\citet{2001ApJ...549..721M}	\\
	&	29.71	$\pm$	0.18	&	8749.84	$\pm$	3626.50	&	Cepheids 	&	\citet{2001ApJ...549..721M}	\\
	&	29.77	$\pm$	0.09	&	8994.98	$\pm$	1864.05	&	Cepheids 	&	\citet{2001ApJ...549..721M}	\\
	&	29.13	$\pm$	0.11	&	6698.85	$\pm$	1696.71	&	Cepheids 	&	\citet{Freedman2001}	\\
	&	29.13	$\pm$	0.11	&	6698.85	$\pm$	1696.71	&	Cepheids 	&	\citet{Freedman2001}	\\
	&	29.06	$\pm$	0.11	&	6486.34	$\pm$	1642.89	&	Cepheids 	&	\citet{2001ApJ...553..562N}	\\
	&	29.16	$\pm$	0.09	&	6792.04	$\pm$	1407.53	&	Cepheids 	&	\citet{2001ApJ...553..562N}	\\
	&	29.23	$\pm$	0.07	&	7014.55	$\pm$	1130.61	&	Cepheids 	&	\citet{2002AyA...389...19P}	\\
	&	29.26	$\pm$	0.15	&	7112.14	$\pm$	2456.44	&	Cepheids 	&	\citet{2002AyA...389...19P}	\\
	&	29.30	$\pm$	0.07	&	7244.36	$\pm$	1167.65	&	Cepheids 	&	\citet{2002AyA...389...19P}	\\
	&	29.14	$\pm$	0.09	&	6729.77	$\pm$	1394.63	&	Cepheids 	&	\citet{2004ApJ...608...42S}	\\
	&	29.24	$\pm$	0.08	&	7046.93	$\pm$	1298.09	&	Cepheids 	&	\citet{2004ApJ...608...42S}	\\
	&	29.18	$\pm$	0.08	&	6854.88	$\pm$	1262.72	&	Cepheids 	&	\citet{2006ApJS..165..108S}	\\
	&	29.30	$\pm$	0.12	&	7244.36	$\pm$	2008.63	&	TRGB	&	\citet{2012ApJ...760L..14L}	\\
\hline													
M 101 Final	&	29.353	$\pm$	0.009	&	7423.351	$\pm$	154.889	&	---	&	---	\\
\hline													
NGC 6822	&	23.49	$\pm$	0.08	&	498.88	$\pm$	91.90	&	Cepheids 	&	\citet{1996AJ....112.1928G}	\\
	&	23.27	$\pm$	0.18	&	450.82	$\pm$	186.85	&	Cepheids 	&	\citet{2001ApJ...548..564W}	\\
	&	23.22	$\pm$	0.52	&	440.55	$\pm$	527.50	&	Cepheids 	&	\citet{2002AyA...389...19P}	\\
	&	23.30	$\pm$	0.40	&	457.09	$\pm$	420.99	&	Cepheids 	&	\citet{2002AyA...389...19P}	\\
	&	23.38	$\pm$	0.52	&	474.24	$\pm$	567.83	&	Cepheids 	&	\citet{2002AyA...389...19P}	\\
	&	23.36	$\pm$	0.17	&	469.89	$\pm$	183.94	&	RR Lyrae 	&	\citet{2003ApJ...588L..85C}	\\
	&	23.34	$\pm$	0.04	&	465.59	$\pm$	42.88	&	Cepheids 	&	\citet{2004AJ....128.2815P}	\\
	&	23.39	$\pm$	0.08	&	476.43	$\pm$	87.76	&	Cepheids 	&	\citet{2004ApJ...608...42S}	\\
	&	23.31	$\pm$	0.02	&	459.20	$\pm$	21.15	&	Cepheids 	&	\citet{2006ApJ...647.1056G}	\\
	&	23.31	$\pm$	0.03	&	459.20	$\pm$	31.72	&	Cepheids 	&	\citet{2006ApJS..165..108S}	\\
	&	23.49	$\pm$	0.03	&	498.88	$\pm$	34.46	&	Cepheids 	&	\citet{2009ApJ...693..936M}	\\
	&	23.40	$\pm$	0.05	&	478.63	$\pm$	55.10	&	Cepheids 	&	\citet{2012MNRAS.421.2998F}	\\
\hline													
NGC 6822 Final	&	23.358	$\pm$	0.013	&	469.499	$\pm$	13.798	&	---	&	---	\\
\hline													
M33	&	24.85	$\pm$	0.09	&	933.25	$\pm$	193.40	&	Cepheids 	&	\citet{1997ApJ...491...13K}	\\
	&	24.84	$\pm$	0.16	&	928.97	$\pm$	342.24	&	RR Lyrae 	&	\citet{2000AJ....120.2437S}	\\
	&	24.85	$\pm$	0.13	&	933.25	$\pm$	279.36	&	Miras 	&	\citet{2000MNRAS.313..271P}	\\
	&	24.47	$\pm$	0.13	&	783.43	$\pm$	234.51	&	Cepheids 	&	\citet{2001ApJ...548..564W}	\\
	&	24.56	$\pm$	0.08	&	816.58	$\pm$	150.42	&	Cepheids 	&	\citet{Freedman2001}	\\
	&	24.62	$\pm$	0.08	&	839.46	$\pm$	154.63	&	Cepheids 	&	\citet{Freedman2001}	\\
	&	24.70	$\pm$	0.13	&	870.96	$\pm$	260.71	&	Cepheids 	&	\citet{2002AyA...389...19P}	\\
	&	24.77	$\pm$	0.39	&	899.50	$\pm$	807.76	&	Cepheids 	&	\citet{2002AyA...389...19P}	\\
	&	24.83	$\pm$	0.12	&	924.70	$\pm$	255.50	&	Cepheids 	&	\citet{2002AyA...389...19P}	\\
	&	24.52	$\pm$	0.14	&	801.68	$\pm$	258.43	&	Cepheids 	&	\citet{2002ApJ...565..959L}	\\
	&	24.52	$\pm$	0.15	&	801.68	$\pm$	276.89	&	Cepheids 	&	\citet{2002ApJ...565..959L}	\\
	&	24.47	$\pm$	0.11	&	783.43	$\pm$	198.43	&	Cepheids 	&	\citet{2004ApJ...608...42S}	\\
	&	24.52	$\pm$	0.14	&	801.68	$\pm$	258.43	&	TRGB	&	\citet{2004MNRAS.350..243M}	\\
	&	24.67	$\pm$	0.08	&	859.01	$\pm$	158.24	&	RR Lyrae 	&	\citet{2006AJ....132.1361S}	\\
	&	24.92	$\pm$	0.12	&	963.83	$\pm$	266.32	&	Eclipsing Binary 	&	\citet{2006ApJ...652..313B}	\\
	&	24.64	$\pm$	0.06	&	847.23	$\pm$	117.05	&	Cepheids 	&	\citet{2006ApJS..165..108S}	\\
	&	24.55	$\pm$	0.28	&	812.83	$\pm$	524.05	&	Cepheids 	&	\citet{2007ApJ...671.1640A}	\\
	&	24.37	$\pm$	0.02	&	748.17	$\pm$	34.45	&	Cepheids 	&	\citet{2009MNRAS.396.1287S}	\\
	&	24.53	$\pm$	0.11	&	805.38	$\pm$	203.99	&	Cepheids 	&	\citet{2009MNRAS.396.1287S}	\\
	&	24.54	$\pm$	0.03	&	809.10	$\pm$	55.89	&	Cepheids 	&	\citet{2009MNRAS.396.1287S}	\\
\hline													
M33 Final	&	24.492	$\pm$	0.013	&	791.554	$\pm$	24.426	&	---	&	---	\\
\hline													
NGC 2366 Final	&	27.680	$\pm$	0.200	&	3435.579	$\pm$	1582.14	&	Cepheids 	&	\citet{1995AJ....110.1640T}	\\
\hline													
NGC 2403	&	27.48	$\pm$	0.24	&	3133.29	$\pm$	1731.52	&	Cepheids 	&	\citet{Freedman2001}	\\
	&	27.54	$\pm$	0.24	&	3221.07	$\pm$	1780.03	&	Cepheids 	&	\citet{Freedman2001}	\\
	&	27.43	$\pm$	0.15	&	3061.96	$\pm$	1057.56	&	Cepheids 	&	\citet{2006ApJS..165..108S}	\\
\hline													
NGC 2403 Final	&	27.465	$\pm$	0.112	&	3111.842	$\pm$	805.308	&	---	&	---	\\
\hline													
NGC 4236 Final	&	28.240	$\pm$	0.220	&	4446.313	$\pm$	2252.363	&	TRGB 	&	\citet{2002AyA...383..125K}	\\
\hline

\end{longtable}
%\end{ThreePartTable}
\end{center}
\normalsize 
%\end{landscape}
%\twocolumn

%% file: ch06/ch06.tex
%chapter 6
\chapter{The Dark Energy Equation of State}

\epigraph{\emph{The true method of knowledge is experiment.}}{--- \textup{W. Blake}, All Religions are One (1788)}

\lettrine{T}{he} nature of dark energy is one of the outstanding questions in  modern cosmology. Its multiple connections with almost every aspect of physics makes the problem of understanding its inherent features a specially important one. As already said (see \S \ref{sec:2_3}), many theoretical explanations have been advanced to explain the problem, but to date there is no convincing answer. 

Dark energy can be explored in the first instance by means of constraining the value of its equation of state parameter, $w$, in general given as:
\begin{equation}
	p_{Q} =  w(z) \rho_{Q} \;,
\end{equation}
where $p_{Q}$ is the pressure and $\rho_{Q}$ is the density of the postulated dark energy fluid; in the special case of $w  = -1$  we are dealing with a cosmological constant. 

Many of the proposed models can be parametrised by:
\begin{equation}
	w(z) = w_0 + w_1f(z) \;,
\end{equation}
where $w_0 = w(z = 0)$ and $f(z)$ is a function of redshift. An example is the CPL model where $f(z) = z/(1 + z)$ \citep{Chevallier2001, Linder2003, Peebles2003, Dicus2004, Wang2006}.

We can constrain the value of $w$, as well as other cosmological parameters, using high-$z$ cosmological tracers as described in \S \ref{sec:3_3}. In this Chapter we explore the use of \hii\ galaxies as cosmological tracers to high-$z$.

\section{Observations and Data Reduction}
A sample of 9 star forming galaxies with $EW(\mathrm{H}\beta) > 50\ \mathrm{\AA}$ was selected from  \citet{Hoyos2005, Erb2006b, Erb2006} and  \citet{Matsuda2011}. The redshift range covered was $0.64 \leq z \leq 2.33$. The rationale for the selection criteria was the same as described in section \S \ref{sec:4_1}.

The high spectral resolution spectroscopic observations were obtained using the X-Shooter echelle spectrograph \citep{Vernet2011} at the Cassegrain focus of the ESO-VLT in Paranal, Chile. The observations were conducted during the nights of September 29th and 30th, 2013. 

Xshooter is a three arm intermediate resolution echelle spectrograph. We obtained spectra for the three arms though in the end we only used the UVB and NIR arms data, the typical spectral resolution on the UVB arm was $\sim 10000$, whereas in the NIR arm was $\sim 8000$ with an $0.8 \arcsec$ slit.

The data reduction was carried out using the XShooter pipeline V2.3.0 over the GASGANO V2.4.3 environment\footnote{GASGANO is a JAVA based Data File Organizer developed and maintained by ESO.} using the `physical model mode' reduction.

\section{Data Analysis}
\subsection{Emission line widths}
\begin{figure*}
\centering
\resizebox{14cm}{!}{\includegraphics{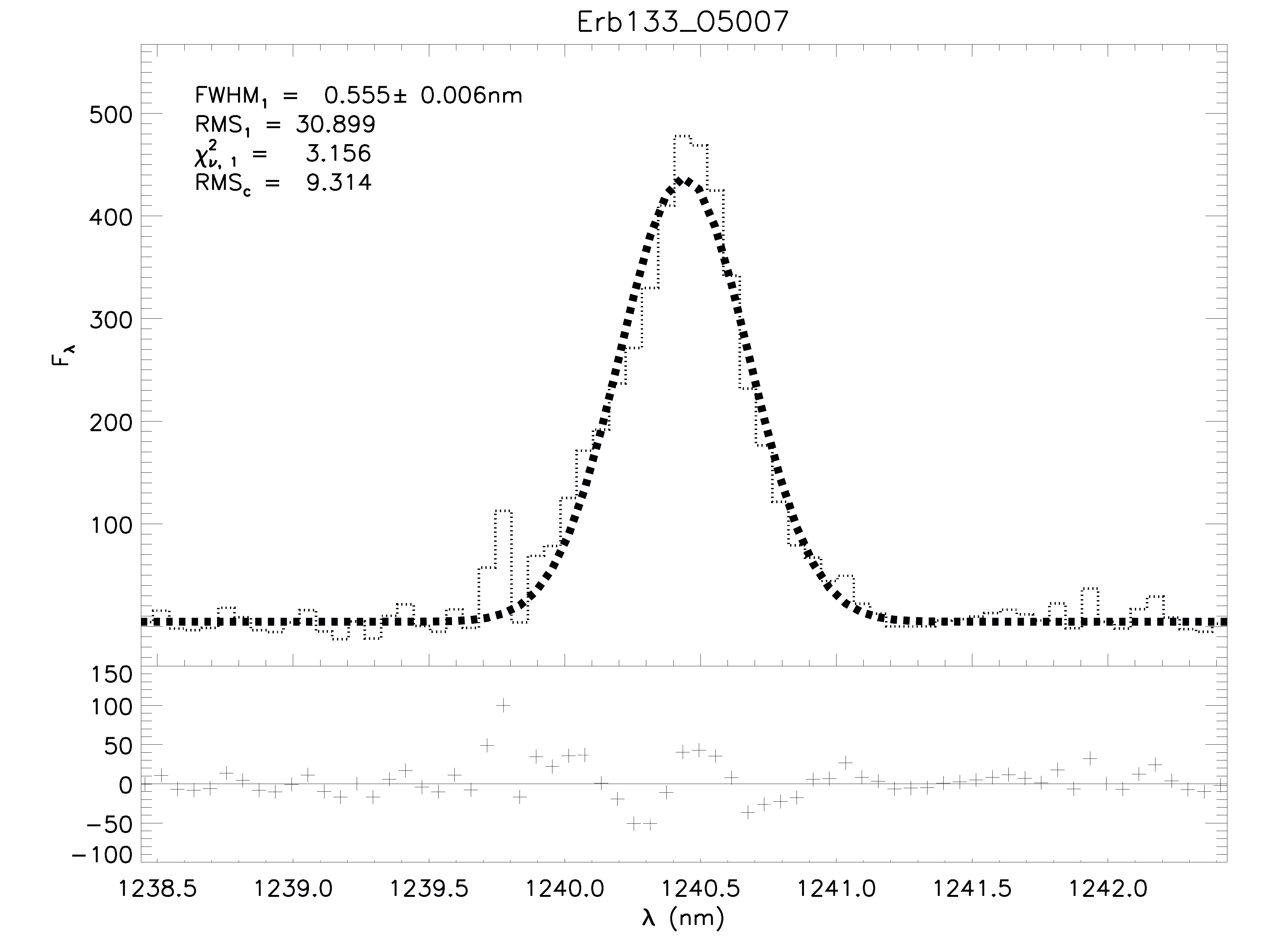}}
\caption[Typical gaussian fit to a resolved emission line]{\small Typical gaussian fit to an $[\Oiii]\ \lambda 5007$\AA\ line. \emph{Upper
    panel:} The single gaussian fit is shown with a  dashed line (thick black).
  The parameters of the fits are shown in the top left corner. \emph{Lower
    panel:} The residuals from the fit. }

\label{fig:O3gf}
\end{figure*}

A gaussian fit was performed on the 1D spectra profiles of the $[\Oiii]\ \lambda 5007$\AA\  and $\ha$ lines when available. These fits were performed using the IDL routine \verb|gaussfit|. Figure \ref{fig:O3gf} shows a typical fit to the  $[\Oiii]\ \lambda 5007$\AA\ line.

The 1$\sigma$ uncertainties of the FWHM were estimated using a Montecarlo 
analysis. A set of 
random realizations of every spectrum was generated 
using the data poissonian 1$\sigma$ 1-pixel uncertainty. Gaussian fitting for every 
synthetic spectrum in the set was performed afterwards, and we obtained a distribution of 
FWHM measurements
from which the 1$\sigma$ uncertainty for the FWHM measured in the 
spectra follows.

%%%%%%%%%%%%%%%%%%%%%%%%%% Figure 9 %%%%%%%%%%%%%%%
%Figure --- Hb multiple gaussian fit
%%%%%%%%%%%%%%%%%%%%%%%%%%%%%%%%%%%%%%%%%%%%%%

%%Table  1--- H II galaxies high-z sample high dispersion measurements. 
\input{ch06/tab/tab01}
%%%%%%%%%%%%%%%%%%%%%%%%%%%%%%%%%%%%

Table \ref{tab:tab601} lists the FWHM measurements for the high
resolution observations prior to any correction such as instrumental
or thermal broadening. Column (1) is an index number, column (2) is the 
name, column (3) is the heliocentric redshift, columns (4) and (5) are 
the measured $[\Oiii]\ \lambda 5007$ central wavelength in $\mathrm{nm}$ and FWHM in $\mathrm{km\ s^{-1}}$ whereas 
columns (6) and (7) are the $\ha$ central wavelength in $\mathrm{nm}$ and FWHM in $\mathrm{km\ s^{-1}}$.

The observed velocity dispersions ($\sigma_o$) -- and their 1$\sigma$ uncertainties -- have  
been derived from  the FWHM measurements of the $[\Oiii]  \lambda 5007$ and $\ha$
lines  as: 
\begin{equation}
\sigma_o \equiv \frac{FWHM}{2  \sqrt{2\ ln(2)} }
\end{equation}
Corrections for thermal ($\sigma_{th}$) and instrumental ($\sigma_{i}$) broadening have been applied. The corrected value  is given by the expression:
\begin{equation}
 \sigma = \sqrt{\sigma_o^2 - \sigma_{th}^2 - \sigma_i^2 } 
\end{equation}
The 1$\sigma$ uncertainties for the velocity dispersion 
have been propagated from the  $\sigma_o$ values. 

The thermal broadening was calculated assuming a Maxwellian velocity distribution of the 
hydrogen and oxygen ions, from the expression:
\begin{equation}
 \sigma_{th}  \equiv \sqrt{\frac{k T_e}{m}},
\end{equation}
where $k$ is the Boltzmann constant, $m$ is the mass of the ion in question and $T_e$ is 
the electron temperature in degrees Kelvin. $T_e$ has been assumed to have a value of $10000$ K. 

In the cases where we do not have $\ha$ velocity dispersions we have corrected the  $[\Oiii]  \lambda 5007$ velocity dispersion by the mean difference calculated from the cases where we have both values. 

As discused previously (see chapter 4) imposing an upper limit  to the velocity dispersion such as $\log \sigma($H$\beta) \lesssim 1.8$ $\mathrm{km\ s^{-1}}$, 
minimizes the probability of including  rotationally  supported systems and/or objects with 
multiple young ionising clusters  contributing to the total flux and 
affecting the line profiles. Therefore from the sample we selected all  objects having $\log \sigma(\ha) < 1.86$ $\mathrm{km\ s^{-1}}$ thus restricting the sample to 6 objects. 

The corrected velocity dispersions for the $\ha$ lines and their 1-$\sigma$ uncertainties are shown in Table \ref{tab:tab602}, in column (4).

\subsection{Fluxes}
In the case of objects taken from \citet{Matsuda2011} and \citet{Erb2006b, Erb2006}, H$\alpha$ fluxes were obtained form the literature directly, then we can easily deduce from the reddened corrected $f(\mathrm{H}\alpha)$, the $f(\mathrm{H}\beta)$ values from the theoretical ratio between both. In the case of objects taken from \citet{Hoyos2005},  luminosities have been derived using \citep[see][]{Terlevich1981}:
\begin{equation}
	B_C = -2.5 \log [L_o(\mathrm{H}\beta) / EW_{\lambda}] + 79.4\ \;,
\end{equation}
where $B_C$ is the absolute blue continuum magnitude, $L_o(\mathrm{H}\beta)$ is the $\hb$ luminosity and $EW_{\lambda}$ is the $\hb$ equivalent width. %From the $\hb$ luminosity, measured redshift and a fiducial cosmological model, $f(\mathrm{H}\alpha)$ can be deduced. 
 Table \ref{tab:tab602}, column (6) shows the adopted $\hb$ fluxes and their 1-$\sigma$ uncertainties.
 
 %%Table  2--- H II galaxies high-z sample adopted values.  
\input{ch06/tab/tab02}
%%%%%%%%%%%%%%%%%%%%%%%%%%%%%%%%%%%%

\subsection{Data from literature}
Since after applying the $\log \sigma(\ha) < 1.86$ $\mathrm{km\ s^{-1}}$ criterium we ended with a sample of only 6 objects, we decided to use also data from the literature. Following the same criterium, we selected a sample of 15 objects from \citet{Erb2006b, Erb2006} for which $\sigma (\ha)$ and $f(\ha)$ are given. Table \ref{tab:tab602}, columns (4) and (6) shows the adopted data from the \citet{Erb2006b, Erb2006} sample. In the same table, the objects with index larger that $200$ are our observed sample  and we show their  final adopted data.

\section{Results and Discussion}
\begin{figure*}
\centering
\resizebox{14cm}{!}{\includegraphics{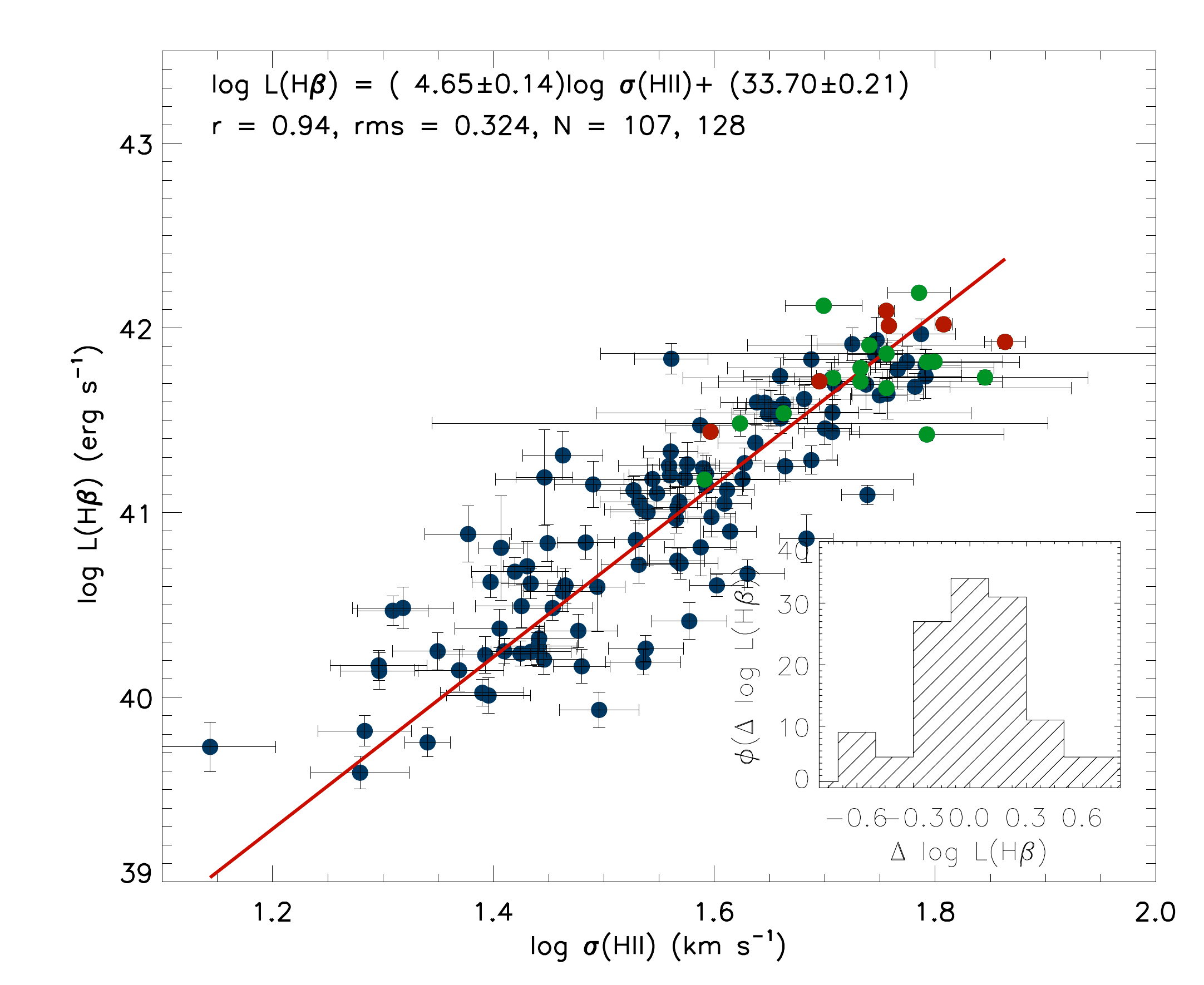}}
\caption[$L-\sigma$ relation for the combined local and high-z samples of \hii\ galaxies] {\small $L-\sigma$ relation for the combined local and high-z samples of \hii\ galaxies, the blue points are the local sample, the red points indicate our high-$z$ observations, and the green ones, the data from \citet{Erb2006b, Erb2006}. The inset shows the distribution of the residuals of the fit.}

\label{fig:LSTOT}
\end{figure*}

Using the concordance $\Lambda$CDM cosmology as fiducial model we can calculate the luminosities for all the objects in our new sample and then construct the corresponding $L-\sigma$ correlation. The values of the luminosities calculated following this procedure are shown in Table \ref{tab:tab602}, column (5). Figure \ref{fig:LSTOT} shows the $L - \sigma$ relation for the high-z sample of \hii\ galaxies combined with the local sample (S3) from Figure \ref{fig:H2gxAll}. It is clear that there is a remarkable similarity between both.

%%%%%%%%%%%%%%%%%%%%%%%%%% Figure 9 %%%%%%%%%%%%%%%
%Figure --- Hb multiple gaussian fit
%%%%%%%%%%%%%%%%%%%%%%%%%%%%%%%%%%%%%%%%%%%%%%

Using the $L-\sigma$ relation for the `best sample' that we used to estimate the value of $H_0$ given in eq. (\ref{eq:ls}), we can calculate a value for $L(\hb)$ that can be compared with the value obtained from $f(\hb)$, $z$ and a cosmological model. In general we can minimise the function:
\begin{equation}
\chi^2(\mathbf{p}) = \sum\limits_{i=1}^n \frac{[L_i(\sigma_i) -
  \tilde{L}_i(\mathbf{p}, f_i, z_i)]^2}{\sigma_{L, i}^2 + \sigma_{\tilde{L}, i}^2}, 
\end{equation}  
where the summation is over the combined sample of local and high-$z$ \hii\ galaxies, 
$\sigma_i$ are the measured velocity dispersions,
$L_i(\sigma)$ are the logarithmic luminosities estimated from the `distance
indicator' as defined in equation (\ref{eq:ls}), $\sigma_{L, i}$ are their errors
propagated from the uncertainties in $\sigma$, the slope and intercept
of the relation, $\tilde{L}_i(\mathbf{p}, f_i, z_i)$ is the logarithm of the luminosities obtained from
the measured fluxes and redshifts  by using a particular set of cosmological parameters $\mathbf{p}$, and
$\sigma_{\tilde{L}, i}$ are the errors in this last estimation of
luminosities, propagated from the uncertainties in the fluxes and
redshifts.   

The most general set of cosmological parameters, assuming a flat universe and an insignificant value of $\Omega_r$ and using the CPL model for parametrising the value of $w(z)$, is given by $\mathbf{p} = \{H_0, \Omega_m, w_0, w_1\}$, in this case the value of the luminosity distance used to estimate the value of $\tilde{L}_i(\mathbf{p}, f_i, z_i)$, is given by,
\begin{equation}
	 D_L = \frac{c(1+z)}{H_0} \int_{0}^{z}{\frac{dz'}{\sqrt{ \Omega_m(1+z')^3  + (1 - \Omega_m) (1+z')^{3(1 + w_0 + w_1)} \exp[-3 w_1 z'/(z'+1)]}}}\;.
\end{equation}

\begin{figure*}
\centering
\resizebox{14cm}{!}{\includegraphics{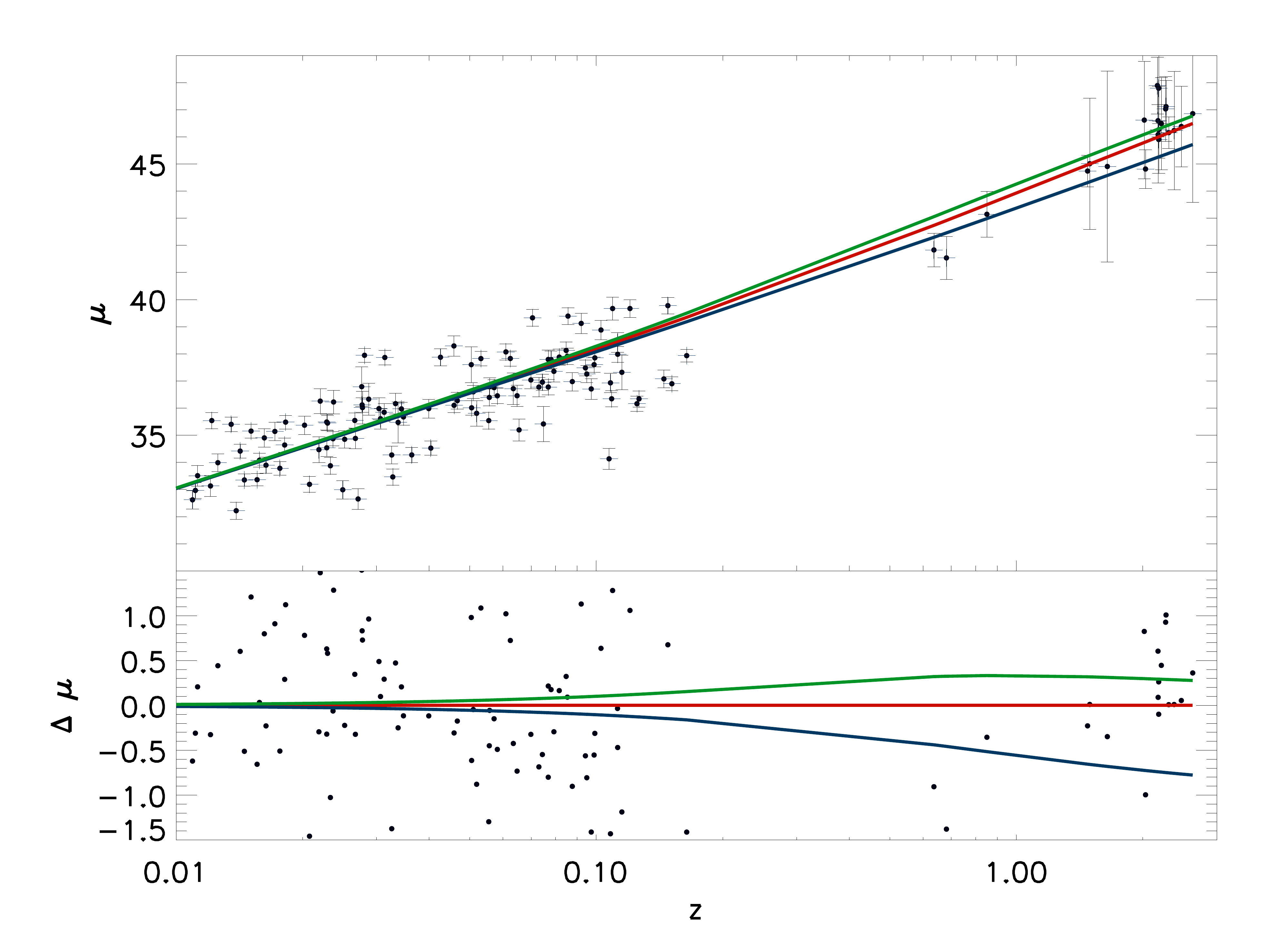}}
\caption[Typical gaussian fit to a resolved emission line]{\small \emph{Upper
    panel:} Hubble diagram for the joint local and high-z samples, the red line is the concordance $\Lambda$-CDM model, the green line is a model with $w_0 = -2$ and $w_1 = 0.0$, the blue line is a model without cosmological constant. \emph{Lower
    panel:} The Differential Hubble diagram showing the same data points and models. }

\label{fig:HD}
\end{figure*}

In Figure \ref{fig:HD} we show a Hubble diagram for the joint local and high-$z$ samples of \hii\ galaxies, together with the $\Lambda$-CDM model, a model without cosmological constant and a phantom energy model, all of them assuming the value of $H_0$ obtained in the previous chapter. From the figure it is clear that our data favors the concordance $\Lambda$-CDM model.

\begin{figure}
  \centering
  \subfloat[]{\label{test:1}\includegraphics[width=80mm]{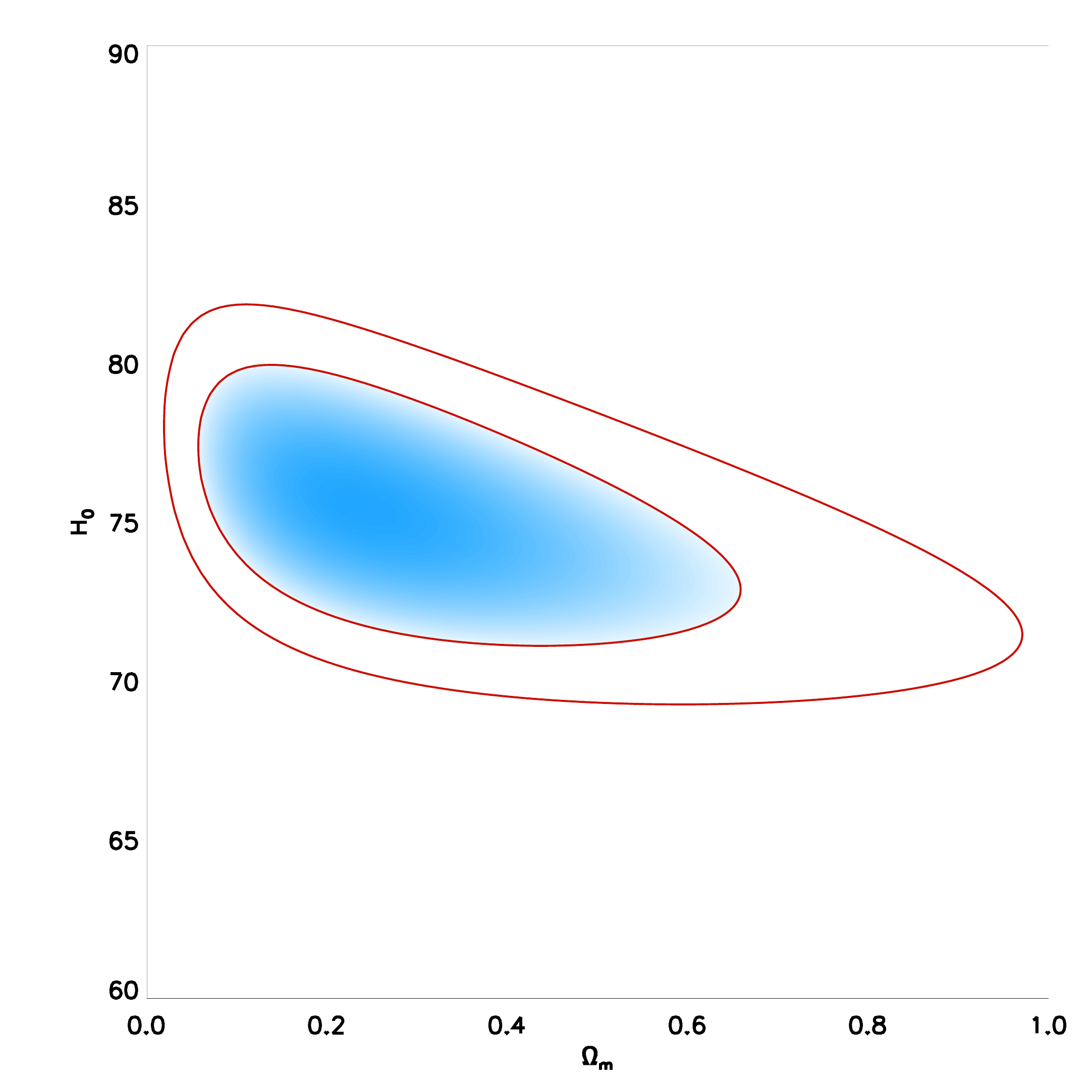}}
  %\subfloat[]{\label{test:2}\includegraphics[width=80mm]{ch06/fig/OmH0XSH+LSR_A.pdf}}
  %\\
  \subfloat[]{\label{test:3}\includegraphics[width=80mm]{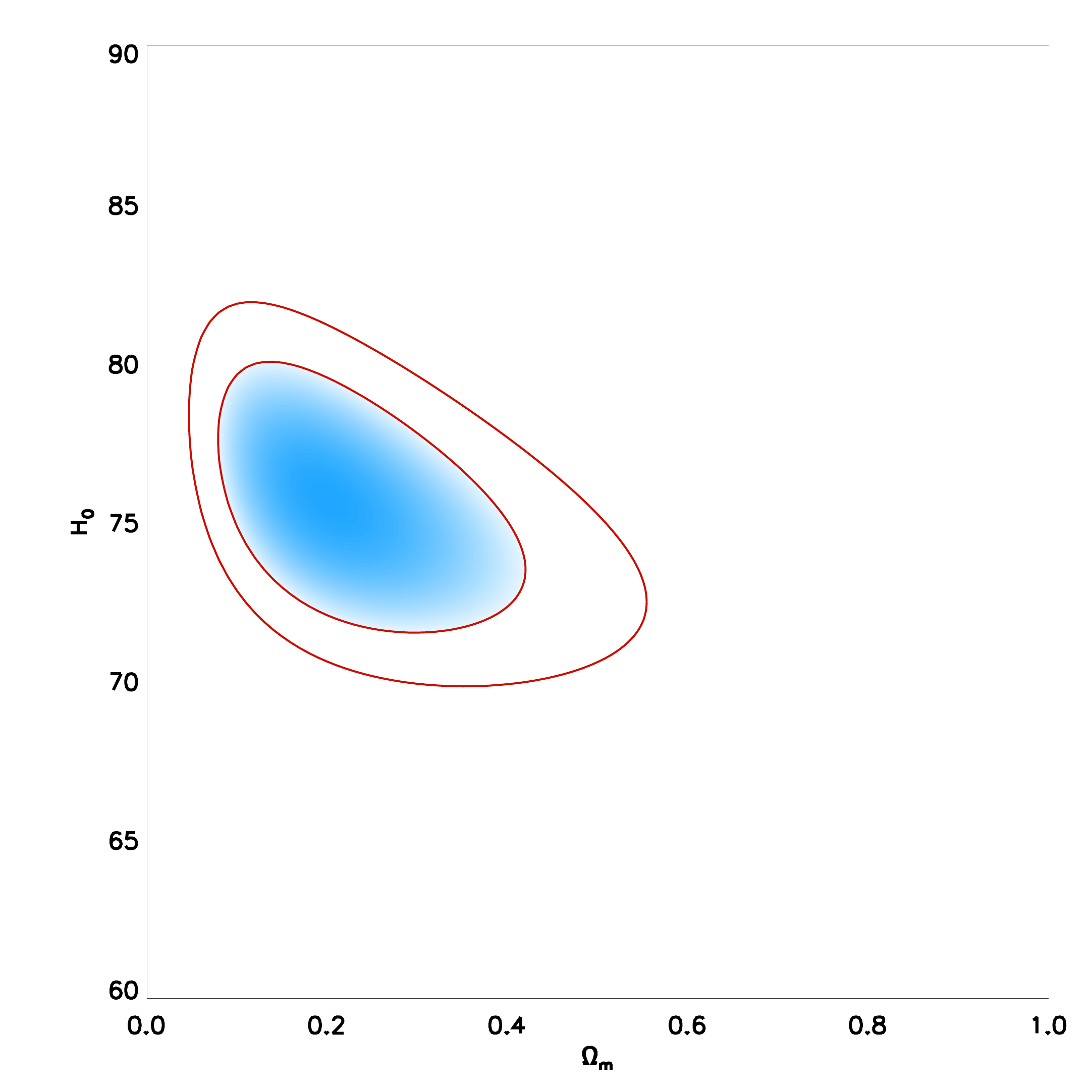}}
  %\subfloat[]{\label{test:4}\includegraphics[width=80mm]{ch06/fig/OmH0XSH+LS_A.pdf}}
  \caption[Solution spaces for $\mathbf{p} = \{H_0, \Omega_m\}$]{\small Solution spaces for $\mathbf{p} = \{H_0, \Omega_m\}$. Panel (a) shows the solution space when we are using our observed sample of 6 high-$z$ \hii\ galaxies combined with the 107 local  \hii\ galaxies from (S3). We are assuming a cosmological constant and using asymmetrical errors in the logarithm of luminosity. Panel (b) shows the same but using the combined 21 objects high-$z$ sample and the local (S3) sample, also using asymmetrical errors. In both panels we show the 1 and 2-$\sigma$ contours.}
  \label{fig:HOmLS}
\end{figure}

If we assume a cosmological constant and minimise for $\mathbf{p} = \{H_0, \Omega_m\}$, using different combinations of our high-$z$ sample  and our local `benchmark' sample (S3) of 107  \hii\ galaxies, we obtain the results shown in Figure \ref{fig:HOmLS}, we have used asymmetrical errors in the logarithm of luminosity following the procedure detailed in \citet{Barlow2004} to weight the $\chi^2$ function:
\begin{equation}
	\sigma_{\tilde{L}, i} = \sigma_1 + \sigma_2[\tilde{L}_i(\mathbf{p}, f_i, z_i) - L_i(\sigma_i)]\ ,
\end{equation}
where $\sigma_1 = 2 \sigma_p \sigma_n / (\sigma_p + \sigma_n)$, $\sigma_2 = (\sigma_p - \sigma_n) / (\sigma_p + \sigma_n)$ and $\sigma_p$ is the upper error whereas $\sigma_n$ is the lower one. 

It is clear form Figure \ref{fig:HOmLS} that the solutions are not very restrictive for $\Omega_{m}$ but are in good accord with other determinations for $H_0$. It is also clear that when we use the full 21 high-$z$ sample instead of only the best 6 objects, there are obtained better restrictions in the $\{H_0, \Omega_m\}$ plane, even when many of these objects have large errors in their measured values of $\sigma$.

%%%%%%%%%%%%%%%%%%%%%%%%%% Figure 9 %%%%%%%%%%%%%%%
%Figure --- Hb multiple gaussian fit
%%%%%%%%%%%%%%%%%%%%%%%%%%%%%%%%%%%%%%%%%%%%%%

%%%%%%%%%%%%%%%%%%%%%%%%%% Figure %%%%%%%%%%%%%%%
%\begin{figure}
%  \centering
%  \subfloat[]{\label{test:1}\includegraphics[width=80mm]{ch06/fig/OmH0XSH.pdf}}
%  \subfloat[]{\label{test:2}\includegraphics[width=80mm]{ch06/fig/OmH0XSH_A.pdf}}
%  \\
%  \subfloat[]{\label{test:3}\includegraphics[width=80mm]{ch06/fig/OmH0XSH+E06.pdf}}
%  \subfloat[]{\label{test:4}\includegraphics[width=80mm]{ch06/fig/OmH0XSH+E06_A.pdf}}
%  \caption{Panel (a) shows the solution space for $\mathbf{p} = \{H_0, \Omega_m\}$, where we are using our observed sample of  6 high-$z$ \hii\ galaxies and are assuming a cosmological constant, we are using symmetrical errors in the logarithm of luminosity.  Panel (b) shows the same but using asymmetrical errors as detailed in the text. Panel (c) shows the same solution space but using the 21 objects combined high-$z$ sample and symmetrical errors. Panel (d) is as (c) but using asymmetrical errors. In all panels we show the 1 and 2-$\sigma$ contours.}
%  \label{fig:HOm}
%\end{figure}

%%%%%%%%%%%%%%%%%%%%%%%%%%%%%%%%%%%%%%%%%%%%%%

\begin{figure}
  \centering
  \subfloat[]{\label{test:1}\includegraphics[width=80mm]{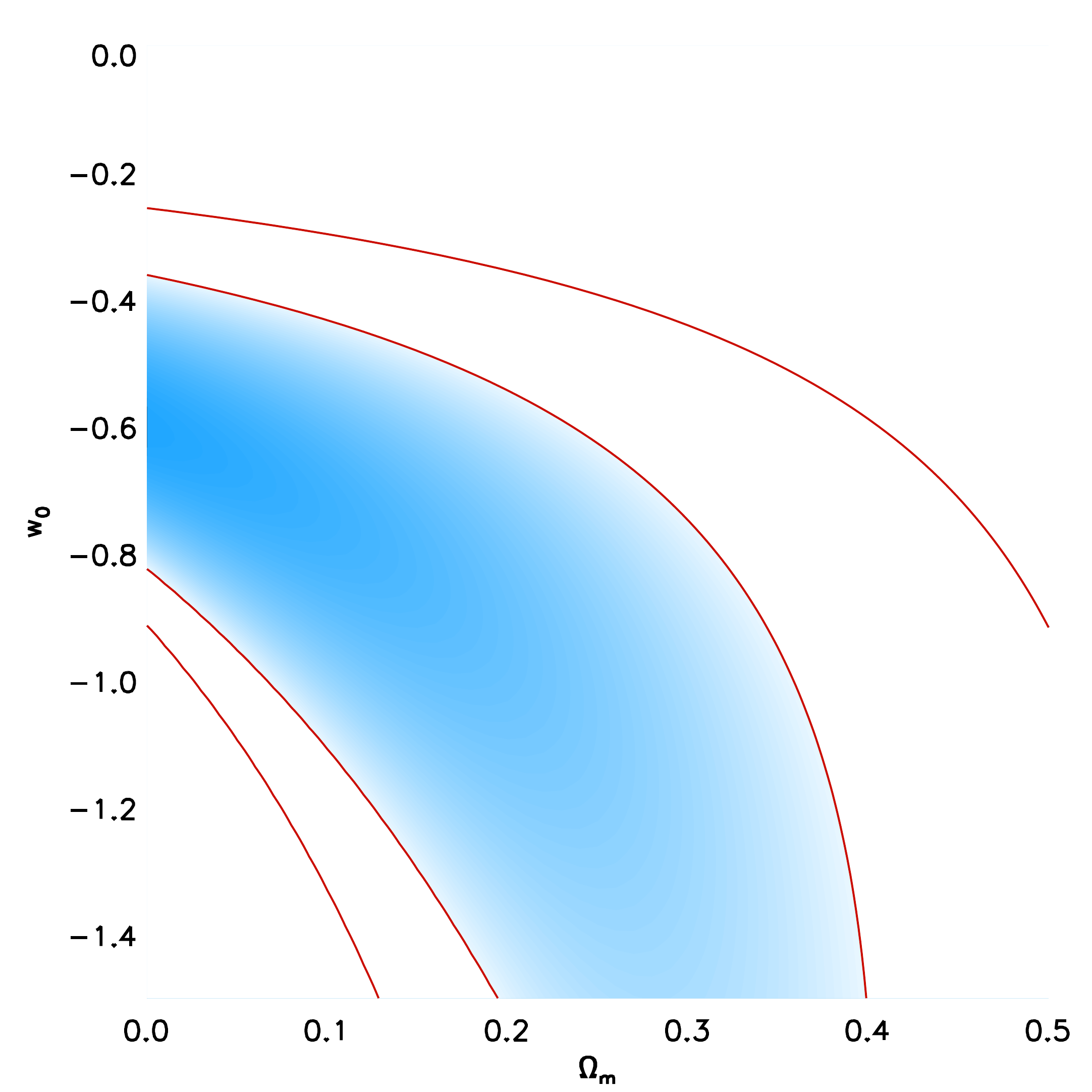}}
  \subfloat[]{\label{test:2}\includegraphics[width=80mm]{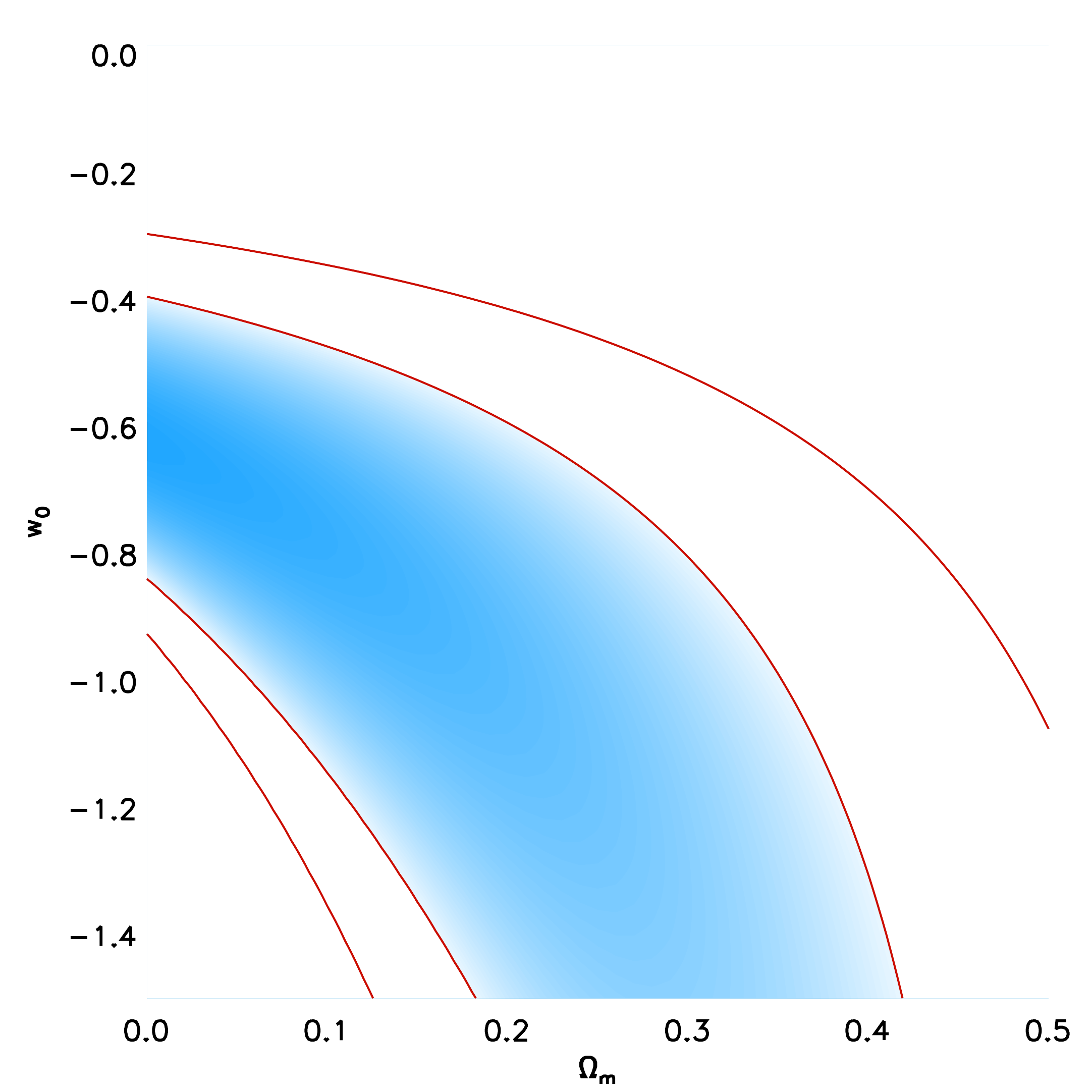}}
  \\
  \subfloat[]{\label{test:3}\includegraphics[width=80mm]{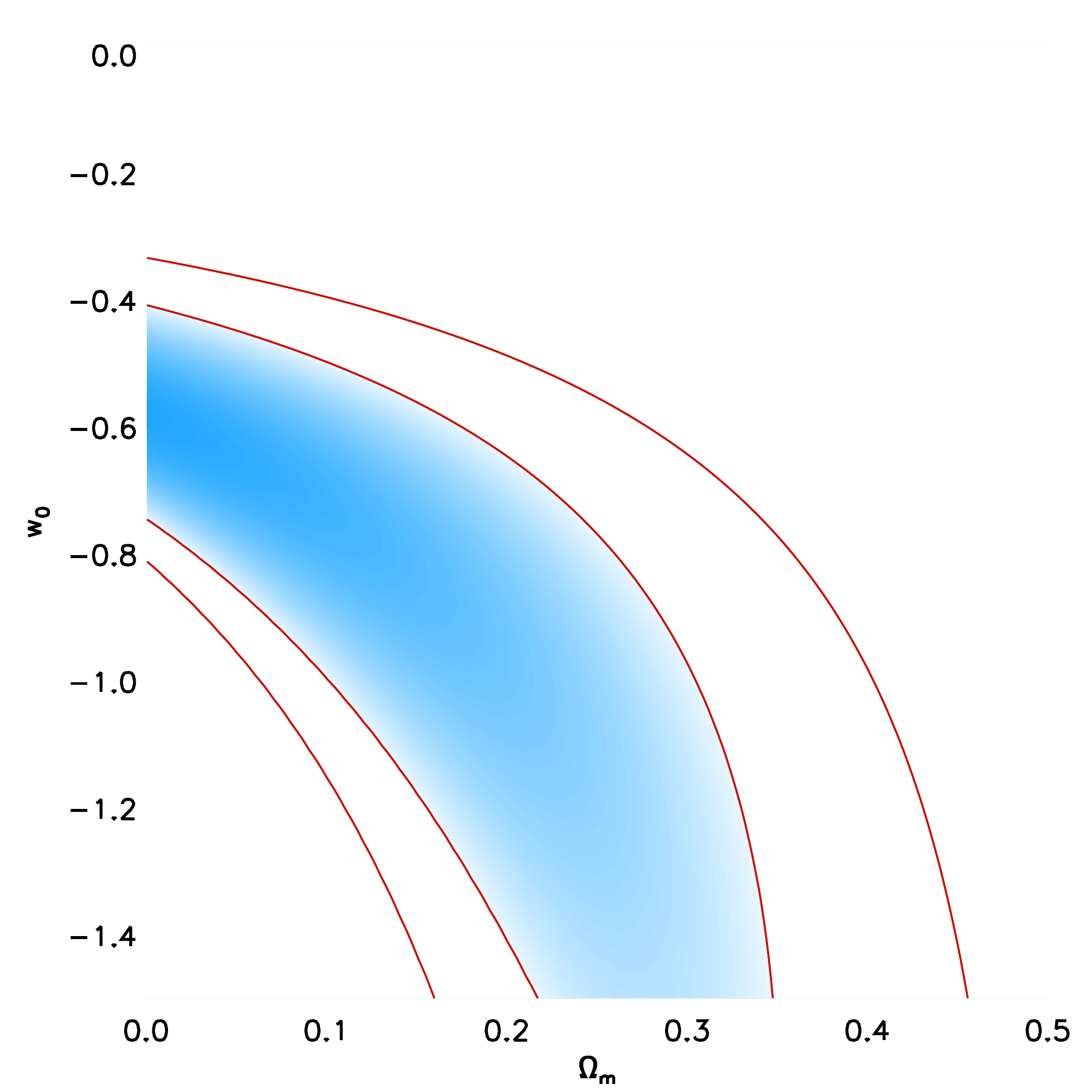}}
  \subfloat[]{\label{test:4}\includegraphics[width=80mm]{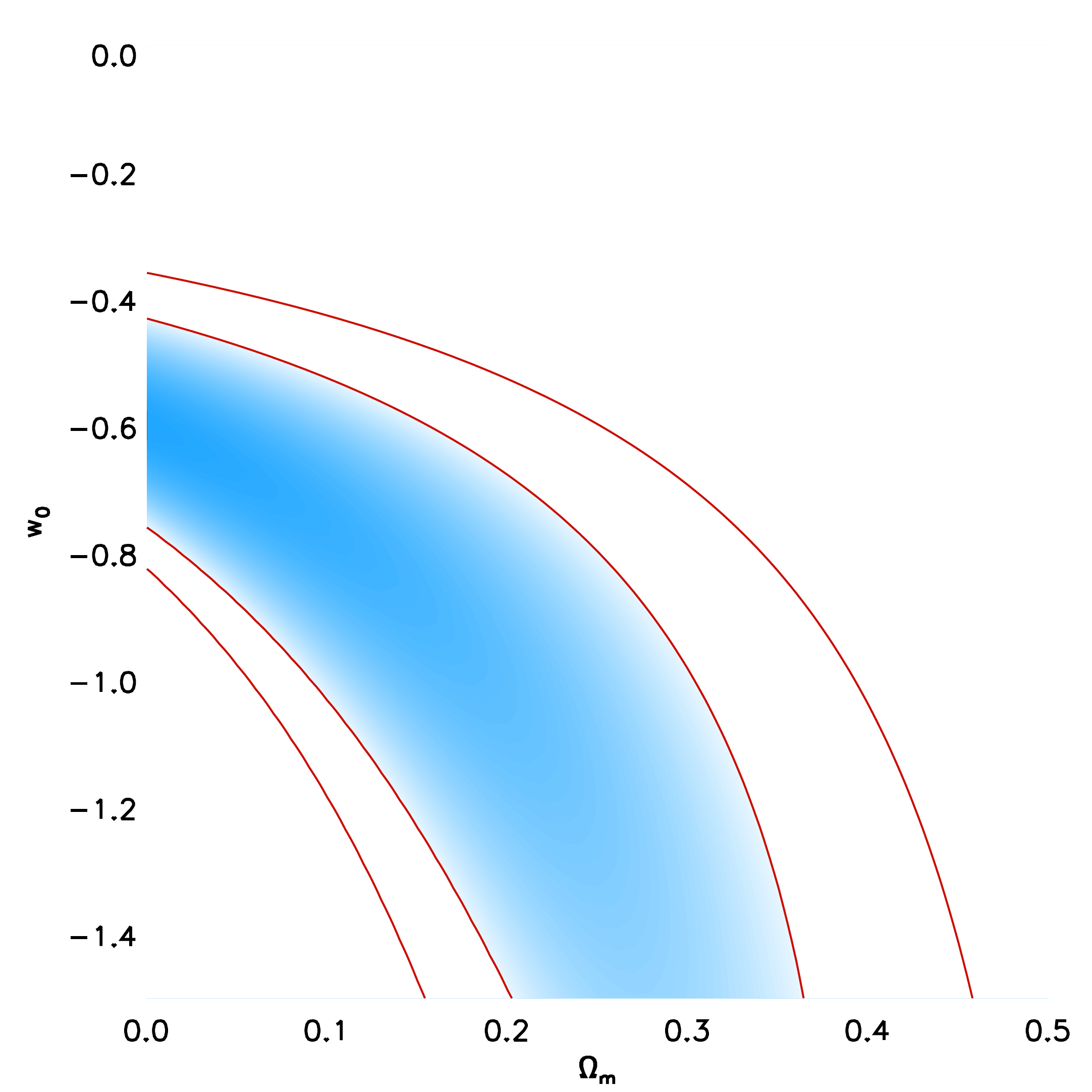}}
  \caption[Solution spaces for $\mathbf{p} = \{\Omega_m, w_0\}$]{\small Solution spaces for $\mathbf{p} = \{\Omega_m, w_0\}$. Panel (a) shows the solution space when we are using our observed sample of  6 high-$z$ \hii\ galaxies and are assuming a cosmological constant and asymmetrical errors in the logarithm of luminosity.  Panel (b) shows the same but joining the 6 high-$z$ objects with the local sample (S3) of 107 objects. Panel (c) shows the same solution space but using the combined 21 objects high-$z$ sample. Panel (d) is as (c) but joining the high-$z$ sample with the local (S3) sample. In all panels we show the 1 and 2-$\sigma$ contours.}
  \label{fig:Omw0}
\end{figure}

Assuming the value of $H_0$ obtained in the previous chapter, $w_1 = 0$ and again using different combinations of high-$z$ and local  \hii\  galaxies samples, we obtain the results shown in Figure \ref{fig:Omw0} for $\mathbf{p} = \{\Omega_m, w_0\}$, again we have used asymmetrical errors in the logarithm of luminosity as explained above. %If additionally we add (S3) and following the same procedure, we obtain the results shown in Figure \ref{fig:Omw0LS}.
From the figure it is clear that we have obtained only weak restrictions in the $\{\Omega_m, w_0\}$ plane, but we must remember that we are using only 21 high-$z$ objects and that we can combine the obtained results with other determinations e.g. SNe Ia results.

%%%%%%%%%%%%%%%%%%%%%%%%%% Figure 9 %%%%%%%%%%%%%%

%%%%%%%%%%%%%%%%%%%%%%%%%%%%%%%%%%%%%%%%%%%%

\begin{figure}
  \centering
  \subfloat[]{\label{test:1}\includegraphics[width=80mm]{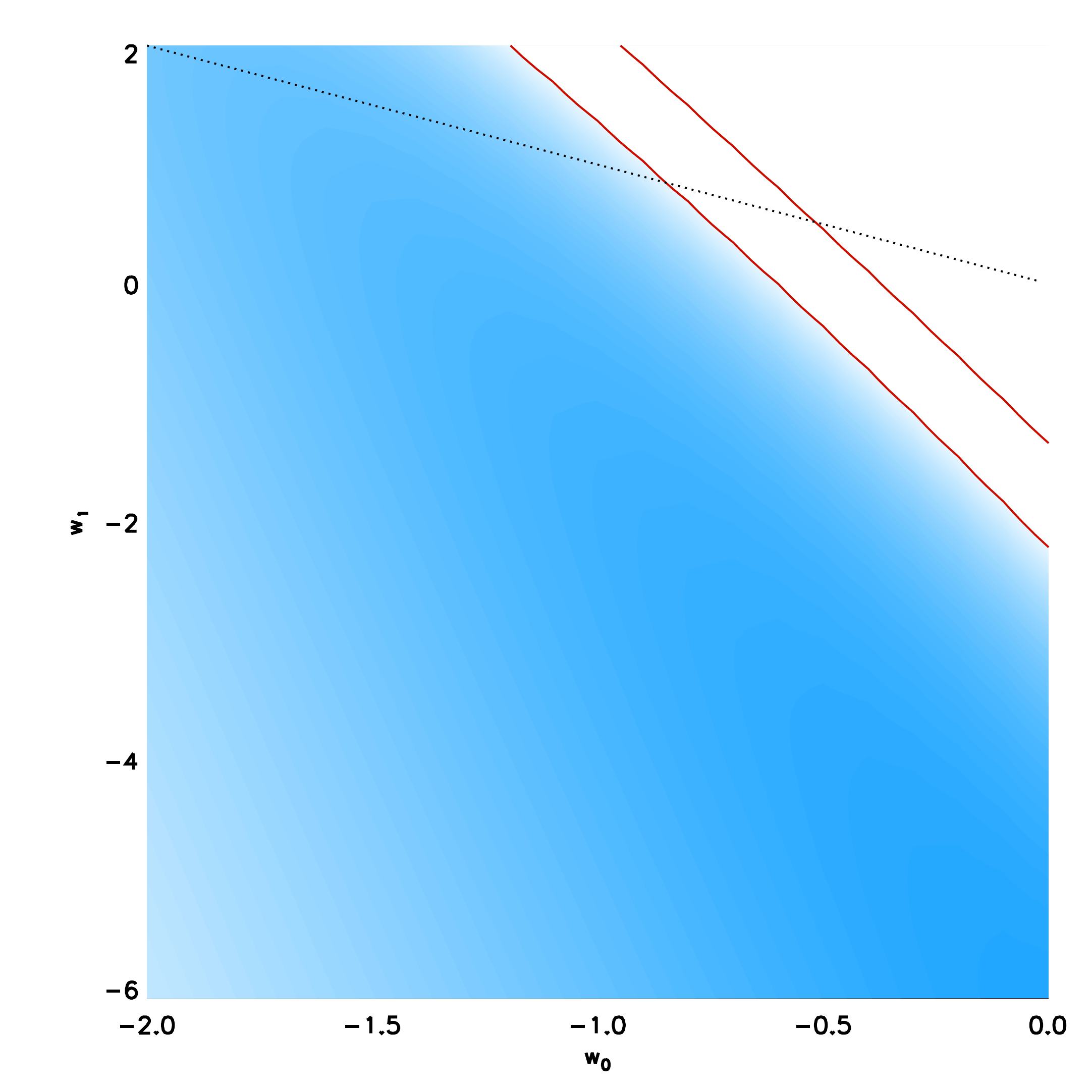}}
  \subfloat[]{\label{test:2}\includegraphics[width=80mm]{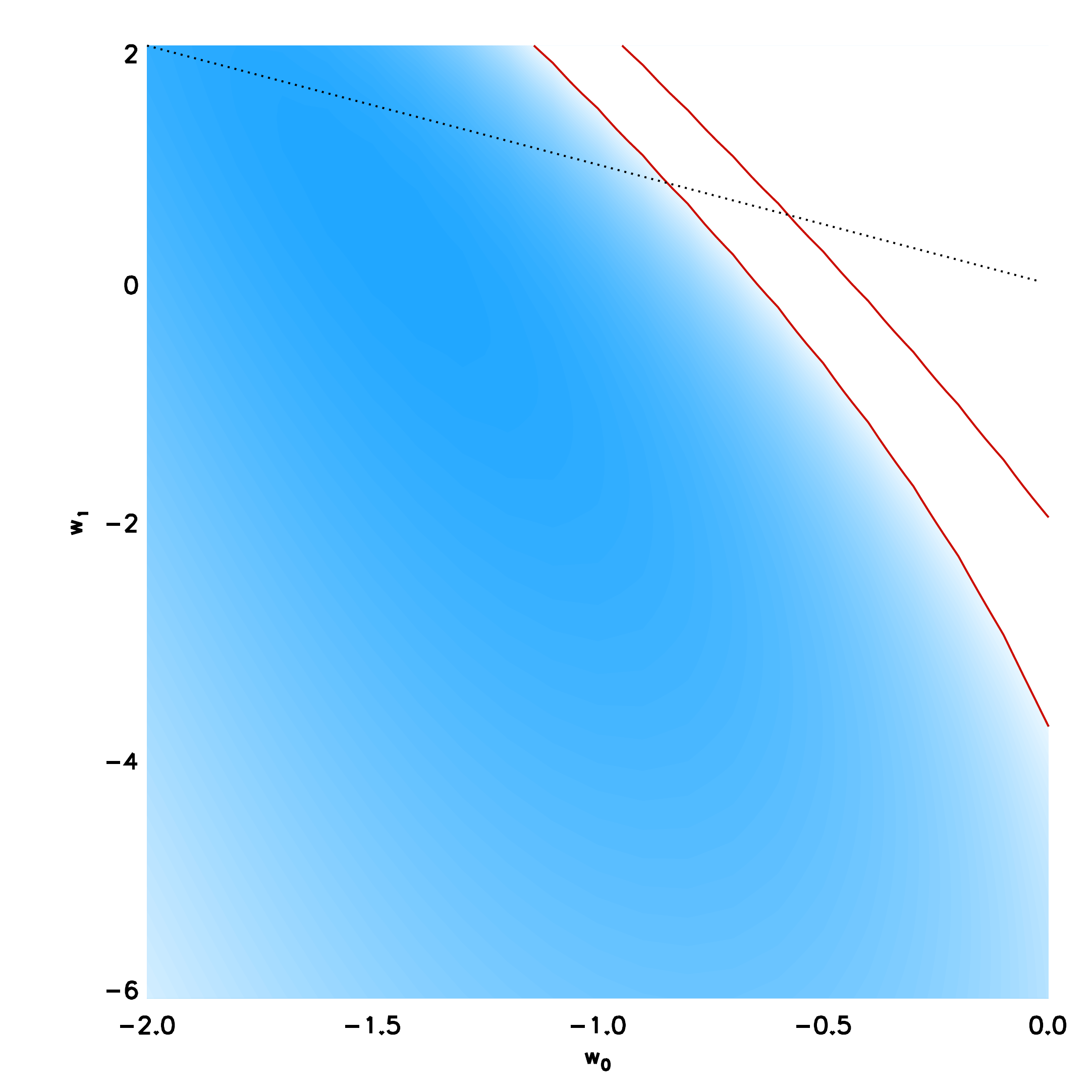}}
  \\
  \subfloat[]{\label{test:3}\includegraphics[width=80mm]{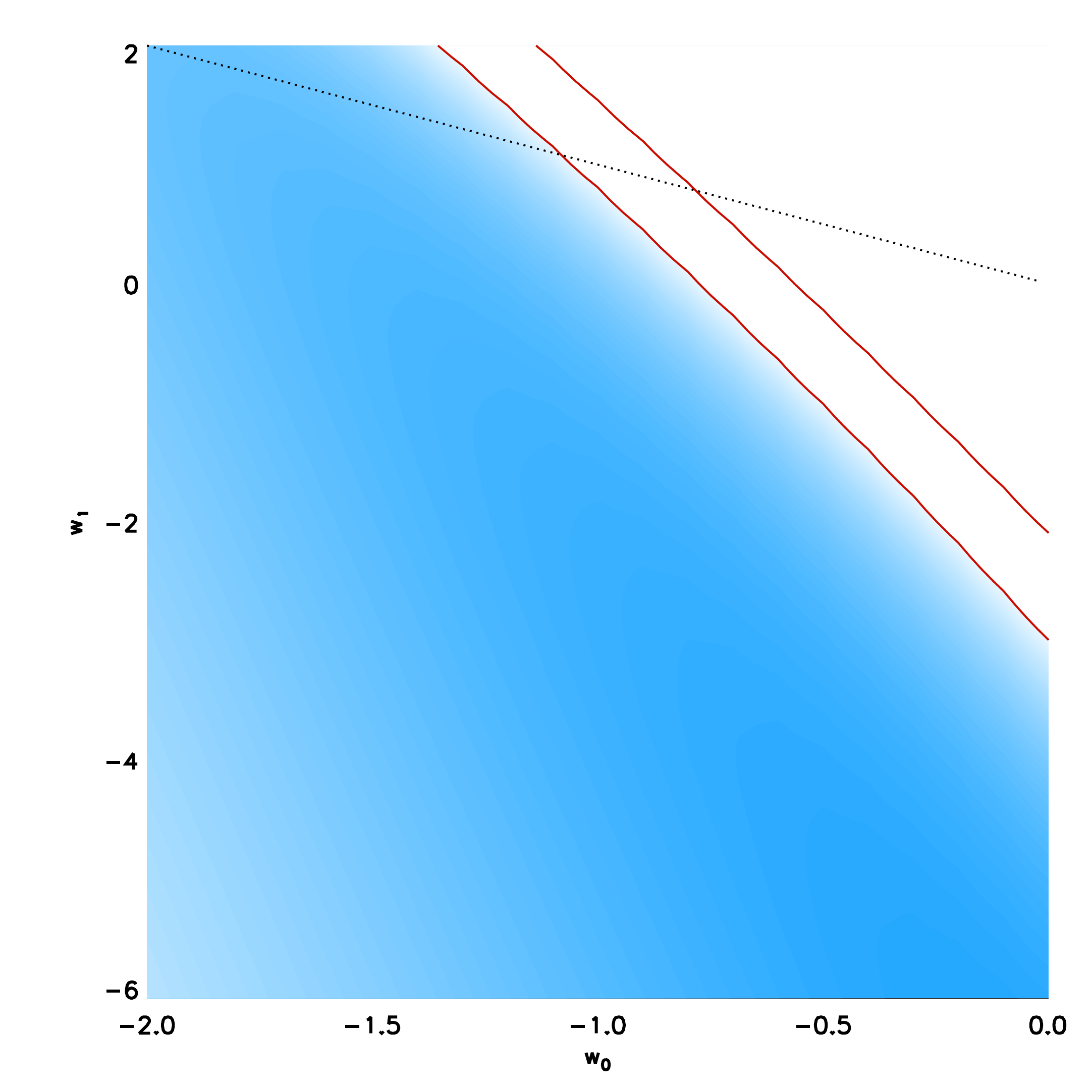}}
  \subfloat[]{\label{test:4}\includegraphics[width=80mm]{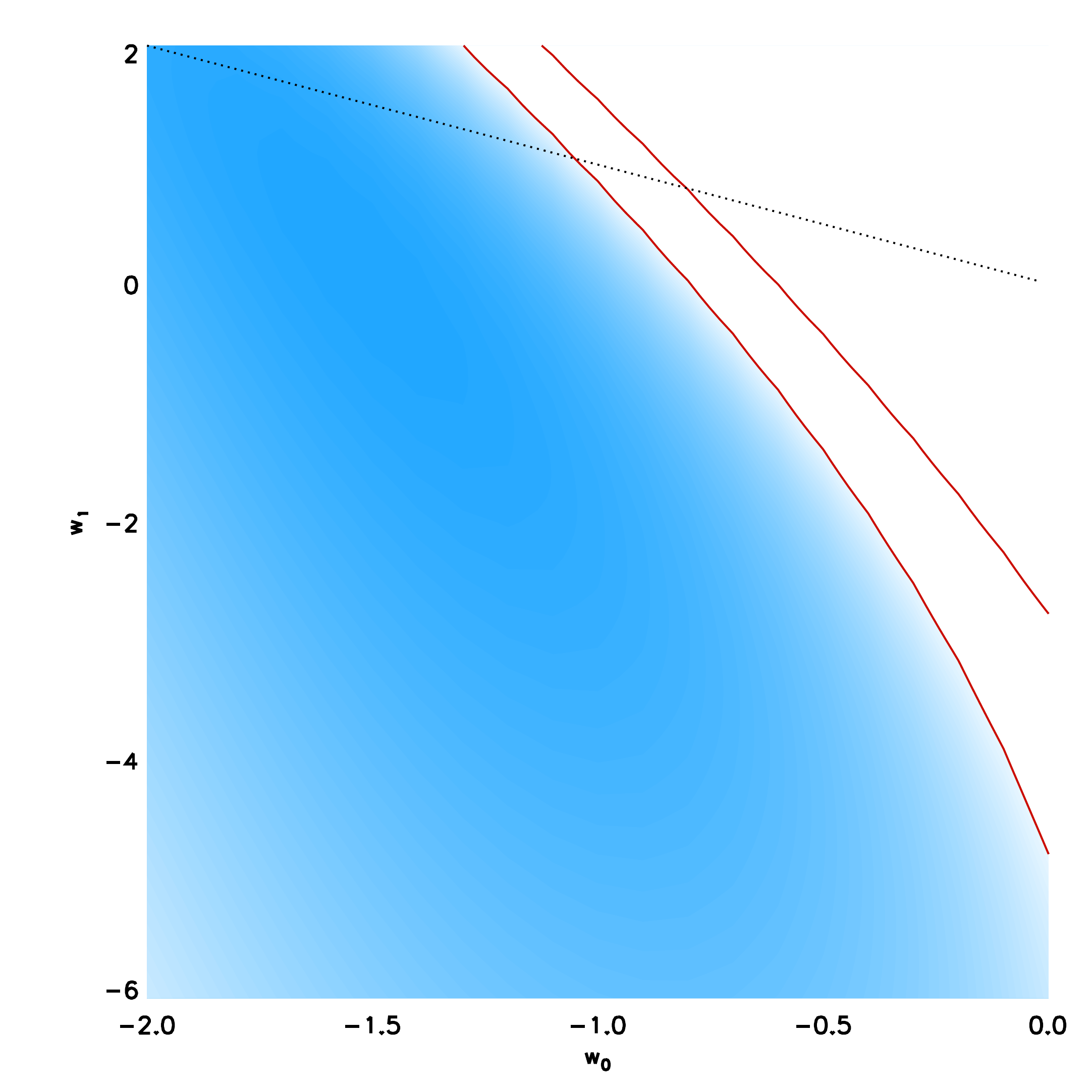}}
  \caption[Solution spaces for $\mathbf{p} = \{w_0, w_1\}$]{\small Solution spaces for $\mathbf{p} = \{w_0, w_1\}$. Panel (a) shows the solution space where we are using our observed sample of  6 high-$z$ \hii\ galaxies and are assuming a cosmological constant and asymmetrical errors in the logarithm of luminosity.  Panel (b) shows the same but joining with the local sample (S3). Panel (c) shows the same solution space but using the combined  21 objects  high-$z$ sample and asymmetrical errors. Panel (d) is as (c) but joining with the local (S3) sample. In all panels we show the 1 and 2-$\sigma$ contours and the line $w_0  = w_1$.}
  \label{fig:w0w1}
\end{figure}

Following the same procedure, assuming the concordance value for $\Omega_m$, and using combinations of our high-$z$ and local \hii\ galaxies samples, we analyze the restrictions in the plane  
$\mathbf{p} = \{w_0, w_1\}$, the results are shown in Figure \ref{fig:w0w1}, together with the line $w_0  = w_1$, points above this line violate early
matter domination and are disfavoured by the data. From the figure, it is clear that we need a larger sample for obtaining meaningful restrictions on this plane, the obtained restrictions are extremely weak.

%%%%%%%%%%%%%%%%%%%%%%%%%% Figure 9 %%%%%%%%%%%%%%

%%%%%%%%%%%%%%%%%%%%%%%%%%%%%%%%%%%%%%%%%%%%

\begin{figure}
  \centering
  \subfloat[]{\label{test:1}\includegraphics[width=68mm]{ch06/fig/OmW0ZXSH+LS.pdf}}
  \subfloat[]{\label{test:2}\includegraphics[width=70mm]{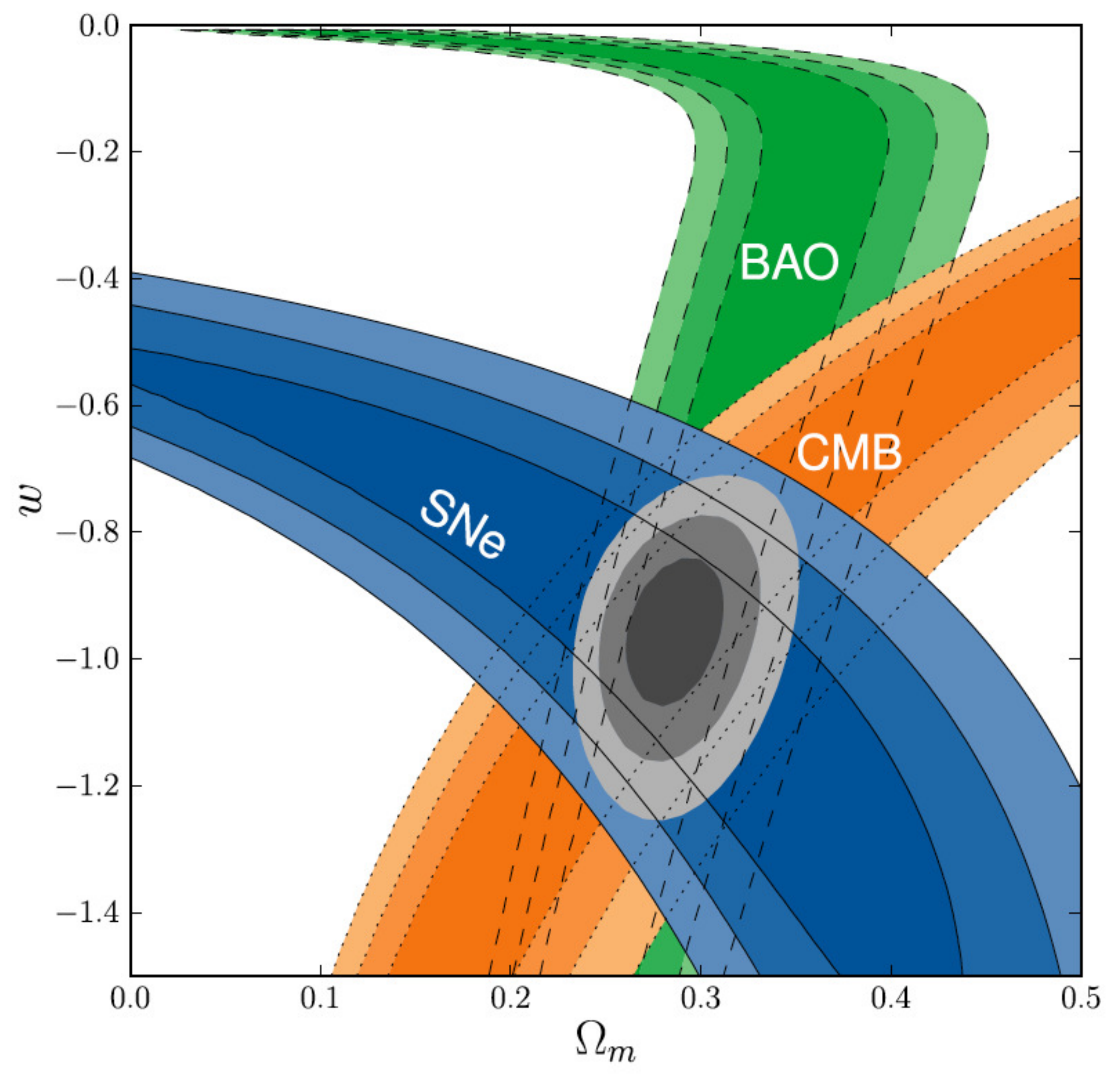}}
  \caption[Comparison on restrictions of the plane $\mathbf{p} = \{\Omega_m, w_0\}$]{\small Comparison on restrictions of the plane $\mathbf{p} = \{\Omega_m, w_0\}$. Panel (a) shows our results, obtained as described in the text for the combined high-$z$ sample of 21 \hii\ galaxies and local (S3) sample, we show the 1 and 2-$\sigma$ contours (random). Panel (b) results for SNe Ia, CMB and BAOs, the 1, 2 and 3-$\sigma$ contours are shown (random plus systematic). Taken from \citet{Suzuki2012}. }
  \label{fig:omwoco}
\end{figure}

%\subsection{Comparison with other results}
In Figure \ref{fig:omwoco}, we compare our results for the space  $\mathbf{p} = \{\Omega_m, w_0\}$ with recent results from SNe Ia, CMB and BAO. Form the figure it is clear that our results are weaker than the ones for SNe Ia, but this is not surprising since in our case we have only 128 objects vs 580 SNe Ia, altough we must remember that we have 17 high-$z$ objects ($z > 1.5$) whereas the maximum redshift for SNe Ia is $\sim 1.5$.

\begin{figure}
  \centering
  \subfloat[]{\label{test:1}\includegraphics[width=68mm]{ch06/fig/W0W1ZXSH+LSR.pdf}}
  \subfloat[]{\label{test:2}\includegraphics[width=68mm]{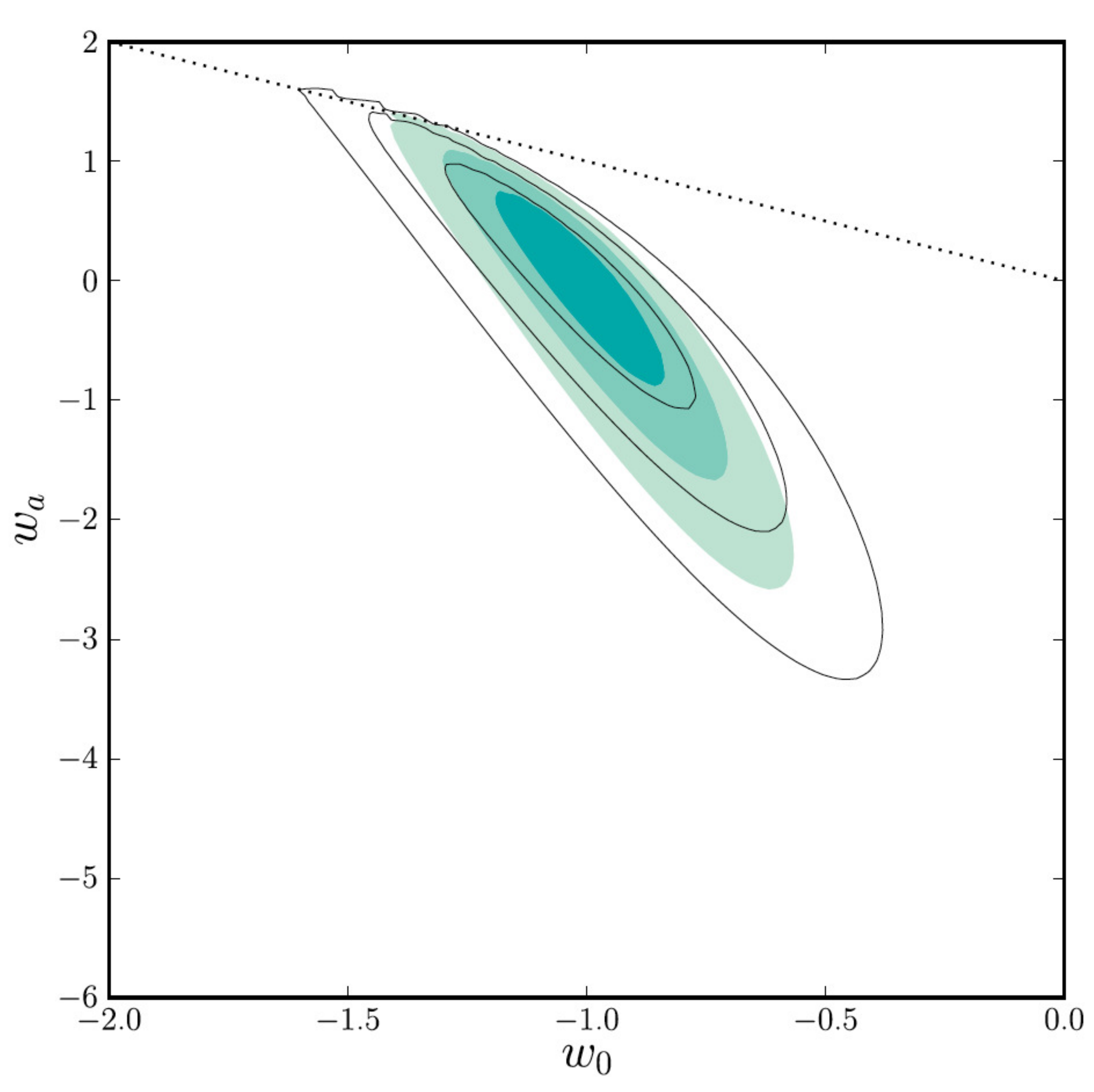}}
  \caption[Comparison on restrictions of the plane $\mathbf{p} = \{\Omega_m, w_0\}$]{\small Comparison on restrictions of the plane $\mathbf{p} = \{\Omega_m, w_0\}$. Panel (a) shows our results as described in the text for the combined high-$z$ sample of 21 \hii\ galaxies and local (S3) sample, we show the 1 and 2-$\sigma$ contours (random). Panel (b) shows the same plane but with recent combined results from SNe Ia, CMB, BAOs, and $H_0$; the 1, 2 and 3-$\sigma$ contours are shown both with (solid contours) and without (shaded contours) systematic errors. Points above the dotted line ($w_0 + w_a > 0$) violate early matter domination and are disfavoured by the data.  Taken from \citet{Suzuki2012}. }
  \label{fig:w0w1co}
\end{figure}

In Figure \ref{fig:w0w1co}, we compare our results for the space  $\mathbf{p} = \{w_0, w_1\}$ with recent combined results from SNe Ia, CMB and BAO. Form the figure it is clear that our results are much weaker that the combined ones for the other methods, but we must consider that we are comparing results form only one method versus the combined from several ones. In the other hand, one compelling result is that most of our solution space is below the region $w_0 + w_1 > 0$, and then is compatible with other results.

%%%%%%%%%%%%%%%%%%%%%%%%%% 

%%%%%%%%%%%%%%%%%%%%%%%%%%%%%%%%%%%%%%%%%%%%
%%%%%%%%%%%%%%%%%%%%%%%%%% 

%%%%%%%%%%%%%%%%%%%%%%%%%%%%%%%%%%%%%%%%%%%%

In general the obtained results are in line with other recent determinations, the general weakness of our determination can be attributed to the small number of intermediate to high-$z$ data (only 21 objects), of which only 6 have high quality observations, whereas the other 15 objects are taken from the literature, and have observations conducted with other purposes in mind and hence are of lower quality for our requirements. 

As discussed in \citet{Plionis2011} (and references therein), we can estimate the number of high-$z$ tracers required  to reduce by a certain factor the cosmological parameters solution space using the figure of merit (FoM), defined as the reciprocal area of the $2\sigma$ contour in the parameter space of any two degenerate cosmological parameters. In this way a larger FoM indicates better restrictions to the cosmological parameters. \citet{Plionis2011} use the parameter $S$ or `reduction factor' to compare the ratio of FoM of SNe Ia  + high-$z$ tracers to that of only SNe Ia. They found that the number of high-$z$ tracers needed to obtain a given S in the QDE model ( in which $w_1 = 0$ but $w_0$ changes in time)  and when combined with the intermediate and low-$z$ SNa Ia, can be expressed as:
\begin{equation}
	N_{Hz} \simeq 187 S/S_{100} - 88\, 
\end{equation} 
where $S_{100} = 1.87 \logd (\mean{\sigma_{\mu}}^{-1} + 0.74) + 1.28$, and $\mean{\sigma_{\mu}}$ is the mean distance modulus error for the tracer. In the case of \hii\ galaxies, ${\sigma_{\mu}} \simeq 0.6$, then $S_{100} \simeq 1.99$ and consequently in order to obtain a reduction to $2$ in $S$, we need around  100 high-$z$ \hii\ galaxies. We can visualize that in Figure \ref{fig:omwoco} panel (b) for the $\{w_0, w_1\}$ plane, where a factor 2 reduction means that $2\sigma$ contours would be similar to the actual $1\sigma$ SNe Ia contours. For the case of the CPL model, where $w_1$ is variable then  
the same result  can be expressed as:
\begin{equation}
	N_{Hz} \simeq 404 S/S_{100} - 300\, 
\end{equation} 
where $S_{100} = 0.49 \logd (\mean{\sigma_{\mu}}^{-1} + 0.65) + 1.09$. In the case of \hii\ galaxies  $S_{100} \simeq 1.27$ and consequently in order to obtain a reduction of  2 in $S$, we need around 340 high-$z$ \hii\ galaxies. 

As already said, a substantial improvement to the restrictions of cosmological parameters can be obtained combining a few hundreds of \hii\ galaxies with the existing SNe Ia data. The observation of this few hundreds of objects, is technically not so difficult, given the relative abundance of \hii\ galaxies, many of them appear in the field of view of many of the currently operational medium spectral resolution integrated field spectrographs (IFUs) in 8 m class telescopes, e.g. using KMOS at VLT we would require only a few nights of observation to obtain the required data.

\section{Summary}
We have used a sample of 6 \hii\ galaxies that were observed using the XShooter medium spectral resolution spectrograph at the ESO-VLT, in a redshift range of $0.6 \leq z \leq 2.3$ combined with a sample of 15 \hii\ galaxies taken from the literature and our sample of 121 local \hii galaxies to obtain constraints on the $\{H_0, \Omega_m\}$, $\{\Omega_m, w_0\}$ and$ \{w_0, w_1\}$ planes. 

The obtained constraints are, in general, in line with other determinations, although weaker. The weakness of our results can be attributed to the small size of the intermediate to high-$z$ sample, and the considerable size of the uncertainties in the data taken from the literature. 

The best way to obtain better constraints on the diverse cosmological parameters would be to obtain medium spectral resolution data for a sample of a few hundreds of objects and combine them with the local \hii\ galaxies and SNe Ia data. 

%The obtained constraints are in general weak but given the small size of the high quality high-$z$ sample this fact is not surprising. We expect that observing a larger sample of high-$z$ \hii\ galaxies we could obtain much better results. 

%% file: ch06/tab/tab01.tex
%Table  01 --- H II galaxies high-z sample FWHM measurements. 

\onecolumn
%\begin{landscape}
\scriptsize
\begin{center}
\begin{longtable}{@{} c l c c c c c @{}}
 
\caption[High-$z$ sample FWHM]{\small FWHM of $[\Oiii]\ \lambda 5007$ and $\ha$ from the XShooter high dispersion spectra. } 
\label{tab:tab601} \\

%This is the header for the first page of the table...
\hline \hline \\[-2ex]
\multicolumn{1}{c}{(1)} &
\multicolumn{1}{c}{(2)} &
\multicolumn{1}{c}{(3)} &
\multicolumn{1}{c}{(4)} &
\multicolumn{1}{c}{(5)} &
\multicolumn{1}{c}{(6)} & 
\multicolumn{1}{c}{(7)} \\ 

\multicolumn{1}{c}{\bf{Index}} &
\multicolumn{1}{c}{\bf{Name}} &
\multicolumn{1}{c}{$\mathbf{z_{hel}}$} &
\multicolumn{1}{c}{ \textbf{$\mathbf{\lambda_{c}}$ ($\mathbf{[\Oiii]\ \lambda  5007}$)}} &
\multicolumn{1}{c}{ \textbf{FWHM($\mathbf{[\Oiii]\ \lambda  5007}$)}} &
\multicolumn{1}{c}{ \textbf{$\mathbf{\lambda_{c}}$ ($\mathbf{\ha}$)}} &
\multicolumn{1}{c}{ \textbf{FWHM ($\mathbf{\ha}$)}} \\

\multicolumn{1}{c}{} &
\multicolumn{1}{c}{} &
\multicolumn{1}{c}{ } &
\multicolumn{1}{c}{(nm)} & 
\multicolumn{1}{c}{($\mathrm{km\ s^{-1}}$)} & 
\multicolumn{1}{c}{(nm)} & 
\multicolumn{1}{c}{($\mathrm{km\ s^{-1}}$)}  \\[0.5ex] \hline \\[-1.8ex]

\endfirsthead

%This is the header for the remaining page(s) of the table...
\multicolumn{1}{c}{(1)} &
\multicolumn{1}{c}{(2)} &
\multicolumn{1}{c}{(3)} &
\multicolumn{1}{c}{(4)} &
\multicolumn{1}{c}{(5)} &
\multicolumn{1}{c}{(6)} & 
\multicolumn{1}{c}{(7)} \\ 

\multicolumn{1}{c}{\bf{Index}} &
\multicolumn{1}{c}{\bf{Name}} &
\multicolumn{1}{c}{$\mathbf{z_{hel}}$} &
\multicolumn{1}{c}{ \textbf{$\mathbf{\lambda_{c}}$ ($\mathbf{[\Oiii]\ \lambda  5007}$)}} &
\multicolumn{1}{c}{ \textbf{FWHM($\mathbf{[\Oiii]\ \lambda  5007}$)}} &
\multicolumn{1}{c}{ \textbf{$\mathbf{\lambda_{c}}$ ($\mathbf{\ha}$)}} &
\multicolumn{1}{c}{ \textbf{FWHM ($\mathbf{\ha}$)}} \\

\multicolumn{1}{c}{} &
\multicolumn{1}{c}{} &
\multicolumn{1}{c}{ } &
\multicolumn{1}{c}{(nm)} & 
\multicolumn{1}{c}{($\mathrm{km\ s^{-1}}$)} & 
\multicolumn{1}{c}{(nm)} & 
\multicolumn{1}{c}{($\mathrm{km\ s^{-1}}$)}  \\[0.5ex] \hline \\[-1.8ex]
\endhead

\multicolumn{2}{l}{{Continued on Next Page\ldots}} \\
\endfoot

\\[-1.8ex] \hline 

\endlastfoot			
401	&	HoyosD2 5	&	0.6364	&	819.359	$\pm$	0.002	&	96.6	$\pm$	1.3	&	1073.861	$\pm$	0.012	&	113.3	$\pm$	7.8	\\
273	&	Q2343-BM133 	&	1.4774	&	1240.440	$\pm$	0.003	&	134.1	$\pm$	1.5	&	1625.881	$\pm$	0.005	&	142.6	$\pm$	2.1	\\
301	&	Q2343-BX660 	&	2.1735	&	1589.181	$\pm$	0.006	&	150.9	$\pm$	2.5	&	2082.558	$\pm$	0.012	&	---	\\
404	&	HoyosD2 1	&	0.8510	&	926.740	$\pm$	0.015	&	118.2	$\pm$	11.4	&	---	&	---	\\
287	&	Q2343-BX435 	&	2.1119	&	1557.924	$\pm$	0.034	&	171.3	$\pm$	15.6	&	2042.258	$\pm$	0.021	&	178.6	$\pm$	7.2	\\
285	&	Q2343-BX418 	&	2.3052	&	1654.870	$\pm$	0.002	&	135.3	$\pm$	1.0	&	---	&	---	\\
7	&	Q2343-BX436 	&	2.3277	&	1666.438	$\pm$	0.025	&	180.0	$\pm$	10.8	&	---	&	---	\\
8	&	MatsudaC3HAE3	&	2.2397	&	1622.102	$\pm$	0.054	&	292.7	$\pm$	23.4	&	---	&	---	\\
9	&	HoyosD2 12	&	0.6800	&	841.988	$\pm$	0.002	&	84.4	$\pm$	1.2	&	---	&	---	\\

\end{longtable}
%\end{ThreePartTable}
\end{center}
\normalsize 
%\end{landscape}
%\twocolumn

%% file: ch06/tab/tab02.tex
%Table  01 --- H II galaxies high-z sample FWHM measurements. 

\onecolumn
%\begin{landscape}
\scriptsize
\begin{center}
\begin{longtable}{@{} c l c c c c @{}}
 
\caption[High-$z$ \hii\ galaxies parameters adopted values]{\small Adopted values for the $L - \sigma$ relation of high-$z$ \hii\ Galaxies.} 
\label{tab:tab602} \\

%This is the header for the first page of the table...
\hline \hline \\[-2ex]
\multicolumn{1}{c}{(1)} &
\multicolumn{1}{c}{(2)} &
\multicolumn{1}{c}{(3)} &
\multicolumn{1}{c}{(4)} &
\multicolumn{1}{c}{(5)} &
\multicolumn{1}{c}{(6)} \\ 

\multicolumn{1}{c}{\bf{Index}} &
\multicolumn{1}{c}{\bf{Name}} &
\multicolumn{1}{c}{$\mathbf{z_{hel}}$} &
\multicolumn{1}{c}{ \textbf{$\mathbf{\log \sigma}$ ($\mathbf{\ha}$)}} &
\multicolumn{1}{c}{ \textbf{$\mathbf{\log L}$ ($\mathbf{\hb}$)}} &
\multicolumn{1}{c}{ \textbf{$\mathbf{f}$ ($\mathbf{\hb}$)}} \\

\multicolumn{1}{c}{} &
\multicolumn{1}{c}{} &
\multicolumn{1}{c}{} &
\multicolumn{1}{c}{($\mathrm{km\ s^{-1}}$)} & 
\multicolumn{1}{c}{($\mathrm{erg\ s^{-1}}$)} & 
\multicolumn{1}{c}{($10^{-17}\ \mathrm{erg\ s^{-1}\ cm^{-2}}$)}   \\[0.5ex] \hline \\[-1.8ex]

\endfirsthead

%This is the header for the remaining page(s) of the table...
\multicolumn{1}{c}{(1)} &
\multicolumn{1}{c}{(2)} &
\multicolumn{1}{c}{(3)} &
\multicolumn{1}{c}{(4)} &
\multicolumn{1}{c}{(5)} &
\multicolumn{1}{c}{(6)} \\ 

\multicolumn{1}{c}{\bf{Index}} &
\multicolumn{1}{c}{\bf{Name}} &
\multicolumn{1}{c}{$\mathbf{z_{hel}}$} &
\multicolumn{1}{c}{ \textbf{$\mathbf{\log \sigma}$ ($\mathbf{\ha}$)}} &
\multicolumn{1}{c}{ \textbf{$\mathbf{\log L}$ ($\mathbf{\hb}$)}} &
\multicolumn{1}{c}{ \textbf{$\mathbf{f}$ ($\mathbf{\hb}$)}} \\

\multicolumn{1}{c}{} &
\multicolumn{1}{c}{} &
\multicolumn{1}{c}{} &
\multicolumn{1}{c}{($\mathrm{km\ s^{-1}}$)} & 
\multicolumn{1}{c}{($\mathrm{erg\ s^{-1}}$)} & 
\multicolumn{1}{c}{($10^{-17}\ \mathrm{erg\ s^{-1}\ cm^{-2}}$)}   \\[0.5ex] \hline \\[-1.8ex] 

\endhead

\multicolumn{2}{l}{{Continued on Next Page\ldots}} \\
\endfoot

\\[-1.8ex] \hline 

\endlastfoot	
6	&	HDF-BX1277	&	2.2713	&	1.799	$\pm$	0.062	&	$41.812	^{+0.038}	_{-0.042}$	&	1.853	$\pm$	0.070	\\
10	&	HDF-BX1332	&	2.2136	&	1.732	$\pm$	0.129	&	$41.704	^{+0.044}	_{-0.050}$	&	1.538	$\pm$	0.105	\\
17	&	HDF-BX1479	&	2.3745	&	1.663	$\pm$	0.170	&	$41.532	^{+0.048}	_{-0.053}$	&	0.874	$\pm$	0.070	\\
22	&	Q0201-B13	&	2.1663	&	1.792	$\pm$	0.070	&	$41.418	^{+0.039}	_{-0.043}$	&	0.839	$\pm$	0.035	\\
25	&	Q1623-BX214	&	2.4700	&	1.740	$\pm$	0.110	&	$41.900	^{+0.046}	_{-0.052}$	&	1.853	$\pm$	0.140	\\
26	&	Q1623-BX215	&	2.1814	&	1.845	$\pm$	0.093	&	$41.726	^{+0.043}	_{-0.048}$	&	1.678	$\pm$	0.105	\\
33	&	Q1623-BX429	&	2.0160	&	1.756	$\pm$	0.168	&	$41.669	^{+0.053}	_{-0.060}$	&	1.783	$\pm$	0.175	\\
34	&	Q1623-BX432	&	2.1817	&	1.732	$\pm$	0.121	&	$41.777	^{+0.041}	_{-0.046}$	&	1.888	$\pm$	0.105	\\
38	&	Q1623-BX453	&	2.1816	&	1.785	$\pm$	0.028	&	$42.185	^{+0.035}	_{-0.038}$	&	4.825	$\pm$	0.070	\\
71	&	Q1700-MD154	&	2.6291	&	1.756	$\pm$	0.259	&	$41.855	^{+0.060}	_{-0.069}$	&	1.434	$\pm$	0.175	\\
273	&	Q2343-BM133	&	1.4774	&	1.756	$\pm$	0.007	&	$42.087	^{+0.037}	_{-0.040}$	&	10.035	$\pm$	0.280	\\
74	&	Q2343-BM181	&	1.4951	&	1.591	$\pm$	0.189	&	$41.173	^{+0.068}	_{-0.080}$	&	1.189	$\pm$	0.175	\\
285	&	Q2343-BX418	&	2.3052	&	1.758	$\pm$	0.004	&	$42.006	^{+0.036}	_{-0.040}$	&	2.797	$\pm$	0.070	\\
86	&	Q2343-BX429	&	2.1751	&	1.708	$\pm$	0.136	&	$41.723	^{+0.043}	_{-0.048}$	&	1.678	$\pm$	0.105	\\
287	&	Q2343-BX435	&	2.1119	&	1.863	$\pm$	0.019	&	$41.919	^{+0.040}	_{-0.044}$	&	2.832	$\pm$	0.140	\\
301	&	Q2343-BX660	&	2.1735	&	1.808	$\pm$	0.008	&	$42.014	^{+0.039}	_{-0.043}$	&	3.287	$\pm$	0.140	\\
105	&	Q2346-BX120	&	2.2664	&	1.792	$\pm$	0.084	&	$41.810	^{+0.042}	_{-0.046}$	&	1.853	$\pm$	0.105	\\
107	&	Q2346-BX244	&	1.6465	&	1.623	$\pm$	0.279	&	$41.478	^{+0.068}	_{-0.081}$	&	1.888	$\pm$	0.280	\\
109	&	Q2346-BX405	&	2.0300	&	1.699	$\pm$	0.035	&	$42.115	^{+0.035}	_{-0.038}$	&	4.895	$\pm$	0.070	\\
401	&	HoyosD2 5	&	0.6364	&	1.597	$\pm$	0.008	&	$41.438	^{+0.022}	_{-0.023}$	&	17.960	$\pm$	1.490	\\
404	&	HoyosD2 1	&	0.8510	&	1.695	$\pm$	0.045	&	$41.711	^{+0.022}	_{-0.023}$	&	16.525	$\pm$	1.360	\\

\end{longtable}
%\end{ThreePartTable}
\end{center}
\normalsize 
%\end{landscape}
%\twocolumn

%% file: ch07/ch07.tex
%chapter ch06.tex
\chapter{General Conclusions}

\epigraph{\emph{We are such stuff\\
As dreams are made on; and our little life\\
Is rounded with a sleep.}}{--- \textup{William Shakespeare}, The Tempest (1611), IV, i}

\lettrine{W}{e have} established a workable methodology to exploit \hii\ galaxies as cosmic tracers through their \lsig\ relation. Using them as tracers, we have obtained constraints in the value of $H_0$ and in the $\{H_0, \Omega_m\}$, $\{\Omega_m, w_0\}$ and$ \{w_0, w_1\}$ spaces. The obtained constraints are in line with the results obtained to date from other methodologies.

The relationship between the integrated H$\beta$ line luminosity and the velocity dispersion of the ionised gas of \hii\ galaxies and giant extragalactic \hii\ regions has been known for some time \citep{Terlevich1981, Melnick1987}. It represents an extremely interesting standard candle that, in principle, can be used up to $z \sim 4$ \citep{Plionis2011}. Locally  we have used the \hii\ galaxies \lsig\ relation to obtain high precision measurements of the local Hubble parameter and at intermediate and high redshift to constrain the dark energy equation of state parameters space. The use of the \hii\ galaxies \lsig\  relation as a cosmic tracer has been recognised as a potential research case in the E-ELT MOS white paper \citep{Evans2013}. 

The cosmic acceleration problem is one of the most important open issues in the whole of physics. Its solution could shed light over many other important problems in physics and astronomy. The observational evidence for the cosmic acceleration is strong and originates in many distinct probes. All the current probes are limited by systematic errors, that in many cases require a better comprehension.  

The physical mechanism responsible for the cosmic acceleration is unclear. The current evidence is consistent with a cosmological constant but without ruling out dark energy theories or some other dynamical models. %Form the theoretical side, the vacuum energy theories face the cosmological constant problem.

Large observational efforts are necessary in order to better constrain the dark energy equation of state parameters. In \citet{Plionis2011}, we show that to use a sample of high redshift probes is the best strategy to obtain better constraints in the parameters of the dark energy equation of state.

We have proposed the use of \hii\ galaxies $L(\hb)-\sigma$ relation, to determine the Hubble function to intermediate and high redshifts and to obtain stringent constraints for the dark energy equation of state. We have shown that   a good determination of the zero point of the  $L(\hb)-\sigma$ relation are necessary; in addition we must take care of all possible systematics that could affect this correlation, in order to succeed in our aims.

In \citet{Chavez2014}, with a sample of 107 \hii\ galaxies, we find that the \lsig\ relation is stable to changes in the sample definition. Additionally, we tested the behaviour or the \lsig\ relation when additional parameters are considered, our results point clearly to the existence of  a Fundamental Plane in \hii\ galaxies suggesting that the main mechanism of line broadening is linked to the gravitational potential of the young massive cluster, although affected by age and metallicity. In \citet{Koulouridis2013} we find that the \lsig\ relation is independent of the \hii\ galaxies environment. 

In \citet{Chavez2012} we used a sample of 69 \hii\ galaxies ($z < 0.16$), observed with high dispersion spectroscopy at Subaru-HDS and the VLT-UVES, and low dispersion spectrophotometry at 2m class telescopes, together with a local calibration of the \lsig\ relation based on 23 GEHR in 9 nearby galaxies, whose distances are known from primary distance indicators, as an alternative `geometric' approach for estimating the Hubble constant. We obtained $H_0 = 74.3 \pm 3.1$ (random) $\pm 2.9$ (systematic) $\mathrm{km\ s^{-1}\ Mpc^{-1}}$, in excellent agreement with, and independently confirming, the most recent SNe Ia-based results \citep{Riess2011, Freedman2012}. It is important to note that by using mutually independent methods for the $H_0$ determination, it is possible to achieve a better comprehension of the systematic uncertainties' effects over every one of those methods, hence the importance of having alternative determinations of the Hubble constant.
 
We have applied the \hii\ galaxies \lsig\ relation as a cosmological probe at $0.6 < z < 3$. We have used a sample, selected from emission line galaxies surveys \citet[e.g.][]{Hoyos2005, Erb2006, Matsuda2011}, of 6 objects observed using X-Shooter at VLT combined with a sample of 15 objects taken from the literature to obtain constraints in the $\{H_0, \Omega_m\}$, $\{\Omega_m, w_0\}$ and$ \{w_0, w_1\}$ planes. The obtained constraints are in line with other determinations although weaker but we have shown that using a larger sample at high-$z$ we will be able to put much more stringent constraint on the dark energy equation of state parameters. 

\section{Future Work}

The work to be done can be thought basically in terms of two partial objectives that must be accomplished: 

\begin{enumerate}
 \item To improve the $L(\hb) - \sigma$ relation zero point through new high quality observations of  GEHR. 
 \item To obtain a large sample (about 300) of  high quality velocity dispersion and flux data for high-$z$ \hii\ galaxies and using the improved $L(\hb) - \sigma$ relation zero-point to construct a better \hii\ galaxies Hubble diagram to high-$z$.
\end{enumerate}

Having accomplished the above two objectives, we will use the Hubble diagram of \hii\ galaxies (possibly combined with SNe Ia data) to put stringent constraints to the dark energy equation of state (i.e. the expansion history of the Universe). 

%Through the first section of this chapter we explore the work to do in order to achieve the above mentioned first objective, whereas the second section explores the work to do in order to arrive at the second objective. Finally, conclusions to this work are given in the third section.

\section{Improving the $L(\hb) - \sigma$ Relation Zero Point}
In \citet{Chavez2012} the zero-point value, provided by the local calibrator of the GEHRs,  represents the weak link towards a more
accurate value of $H_0$ from the \lsig\ relation. This is due to the fact that for the local GEHRs, the high-dispersion spectroscopic data, that provided
the $\sigma$ measurements, and the spectrophotometry, that provided the $L(\rm{H}\beta)$ values, 
date back mostly to the 1980's, and better measurements can be carried out with modern instrumentation. 

In order to improve the $L(\hb) - \sigma$ relation zero point we will need accurate integrated flux and high spectral resolution velocity dispersion measurements for a local GEHR sample. 

We have assembled a sample of $\sim 50$ GEHRs in $\sim 20$ galaxies having accurate Cepheid distances for which we have measured their integrated $\hb$ fluxes at OAN-SPM and OAGH-Cananea. We are now measuring the $\sigma$ of their nebular emission lines from high-dispersion spectroscopy using  CanHiS at OAGH and MEzCal at OAN.
This larger sample,  observed with modern instrumentation, will lead to a factor of two reduction in
the uncertainty in the $H_0$ value measured via the L(H$_{Balmer}) - \sigma$ relation. It is important to note that our sample includes three GEHRs in NGC 4258, the so called `maser galaxy' for which very precise `geometric' distance measurements are available. 

\section{Cosmological Constraints from \hii\ Galaxies} 
The second main task to do, is to select a large sample of \hii\  galaxies to intermediate and high-$z$ from emission line sources catalogs. This large sample will be observed using middle resolution spectrographs at 8 m class telescopes.

There are many emission line galaxies catalogs already published \citep[e.g.][]{Erb2006, Erb2006b, Kakazu2007, Matsuda2011, Hoyos2005, Atek2010}, from which we can select the candidates to be observed. The selection criteria will be roughly the same as those described in section \ref{sec:4_1}.

In oder to observe the selected sample, one of the best options is to use IFUs, at large telescopes, e.g. VLT-KMOS. The reason for this is that several objects in the same field can be observed simultaneously, thus increasing notably the observation efficiency, and in this way exploiting one of the main advantages of \hii\ galaxies over other e.g. SNe Ia, i.e. their relatively high number density.

%The construction of the Hubble diagram of  is the fundamental mean to fulfill our main objective which is to obtain stringent constraints for the dark energy equation of state. 
%
%Among the important tasks to perform is the selection and observation of larger intermediate and high-$z$ samples.

%% file: apa/apa.tex
%chapter apa.tex ------- Cosmological Field Equations --- last modification: 27 May 2009 %ET 15 Jun 10s
\chapter{Cosmological Field Equations}
\label{ApA}
The purpose of this appendix is to derive the Cosmological Field Equations from the General Relativity (GR) Field Equations; the approach followed for the derivation is variational since this method is intuitive, easy to follow and, not the least, very powerful.

\section{The General Relativity Field Equations}
The GR Field Equations can be written as 
\begin{equation}
 R_{\mu \nu} - \frac{1}{2} g_{\mu \nu} R + \Lambda g_{\mu \nu} = -\kappa T_{\mu \nu},
\end{equation}
\nomenclature{$\kappa$}{The Einstein's gravitational constant.}%
or alternatively as
\begin{equation}
 R_{\mu \nu} = - \kappa \left(T_{\mu \nu} - \frac{1}{2} T g_{\mu \nu} \right) + \Lambda g_{\mu \nu},
\label{eq:grfe}
\end{equation}
where $R_{\mu \nu}$ is the Ricci Tensor, $T_{\mu \nu}$ is the Energy-Momentum Tensor, $g_{\mu \nu}$ is the Metric Tensor, $\Lambda$ is the cosmological constant and $\kappa$ is a constant given by
\begin{equation}
 \kappa = 8 \pi G,
\end{equation}
note that we are using units in which $c = 1$.

Our general approach to obtain the GR field equations for the FRW metric will be simply to obtain variationaly the Ricci Tensor and then to use the value of the Energy-Momentum Tensor for a perfect fluid to obtain the right-hand side of the GR field equations.

\section{The Euler-Lagrange Equations}
From the calculus of variations we know that if we want to find a function that makes an integral dependent on that function stationary, on a certain interval, we can proceed as follows; first we have the integral that we want to make stationary
\begin{equation}
 S = \int_{a}^{b}{L(q^a, \dot{q}^a, t) dt},
\end{equation}
 where we define $S$ as the action, $L$ is the Lagrangian which is dependent on $q^a$, a set of generalized coordinates ($a$ is an index running over all the elements of the set), $\dot{q}^a$, the set of the generalized coordinates time derivatives, $\dot{q}^a \equiv dq^a / dt$ and $t$, the time, a parameter.

The variation of the action can be written as 
\begin{eqnarray}
 \delta S &=& \int_{a}^{b}{\left(\frac{\partial L}{\partial q^a} \delta q^a + \frac{\partial L}{\partial \dot{q}^a} \delta \dot{q}^a \right) dt}\\
          &=& \int_{a}^{b}{\frac{\partial L}{\partial q^a} \delta q^a dt} +\int_a^b{\frac{\partial L}{\partial \dot{q}^a} \delta \dot{q}^a dt},
\end{eqnarray}
integrating the last term by parts and requiring the variation $\delta S$ to be zero (the condition for  $S$ to be stationary), we have
\begin{eqnarray}
  \int_{a}^{b}{\frac{\partial L}{\partial q^a} \delta q^a dt} + \left[\frac{\partial L}{\partial \dot{q}^a} \delta q^a \right]_a^b - \int_a^b{\frac{d}{dt}\left(\frac{\partial L}{\partial \dot{q}^a} \right) \delta q^a dt} &=& 0\\
  \left[\frac{\partial L}{\partial \dot{q}^a} \delta q^a \right]_a^b + \int_{a}^{b}{\left[\frac{\partial L}{\partial q^a} - \frac{d}{dt}\left(\frac{\partial L}{\partial \dot{q}^a} \right) \right] \delta q^a dt} &=& 0,
\end{eqnarray}
since $a$ and $b$ are fixed then the first term vanishes and in order for the integral to be zero, since $\delta q^a$ is arbitrary, then 
\begin{equation}
 \frac{d}{dt}\left(\frac{\partial L}{\partial \dot{q}^a} \right) - \frac{\partial L}{\partial q^a} = 0
\end{equation}

 These are the Euler-Lagrange equations that must be satisfied in order to make the action stationary.

\section{Variational Method for Geodesics}
In order to obtain the equations for the geodesics, and from them read out the metric connection coefficients, we must solve the Euler-Lagrange equations for the Lagrangian
\begin{equation}
 L = \frac{1}{2} g_{ab} \dot{x}^a \dot{x}^b,
\end{equation}
where $g_{ab}$ are the metric elements and $\dot{x}^a$ are the coordinates time derivatives. Applying the Euler-Lagrange equations over the Lagrangian we obtain
\begin{eqnarray}
 \frac{d}{dt}(g_{ac} \dot{x}^a) - \frac{1}{2} (\partial_c g_{ab}) \dot{x}^a \dot{x}^b = 0\\
 \dot{g}_{ac} \dot{x}^a + g_{ac} \ddot{x}^a - \frac{1}{2} (\partial_c g_{ab}) \dot{x}^a \dot{x}^b = 0\\
 (\partial_b g_{ac}) \dot{x}^a \dot{x}^b + g_{ac} \ddot{x}^a - \frac{1}{2} (\partial_c g_{ab}) \dot{x}^a \dot{x}^b = 0\\
 g_{ac} \ddot{x}^a + (\partial_b g_{ac}) \dot{x}^a \dot{x}^b - \frac{1}{2} (\partial_c g_{ab}) \dot{x}^a \dot{x}^b = 0 \label{eqAF},
\end{eqnarray}
since $\dot{x}^a$ and $\dot{x}^b$ commutes, then we have
\begin{eqnarray}
 g_{ac} \ddot{x}^a + \frac{1}{2}(\partial_b g_{ac} + \partial_a g_{bc} - \partial_c g_{ab}) \dot{x}^a \dot{x}^b = 0\\
 g^{dc} [\ddot{x}^a + \frac{1}{2}(\partial_b g_{ac} + \partial_a g_{bc} - \partial_c g_{ab}) \dot{x}^a \dot{x}^b] = 0\\
 \ddot{x}^d + \frac{1}{2} g^{dc}(\partial_b g_{ac} + \partial_a g_{bc} - \partial_c g_{ab})\dot{x}^a \dot{x}^b = 0 \label{eqAI}\\
 \ddot{x}^d + \Gamma^d_{ab} \dot{x}^a \dot{x}^b = 0\\
 \ddot{x}^a + \Gamma^a_{bc} \dot{x}^b \dot{x}^c = 0,
\end{eqnarray}
\nomenclature{$\Gamma^a_{bc}$}{The metric connection coefficients or Christoffel symbols.}%
where $\Gamma^a_{bc}$ are the metric connection coefficients and were clearly defined as
\begin{equation}
 \Gamma^a_{bc} = \frac{1}{2} g^{dc}(\partial_b g_{ac} + \partial_a g_{bc} - \partial_c g_{ab})
\end{equation}
and  from (\ref{eqAI}) we can read without effort the metric connection coefficients.

\section{Application to the FRW Metric}
Using the FRW metric a distance element can be written as
\begin{equation}
 ds^2 = dt^2 - a^2(t)\left[\frac{dr^2}{1- k r^2} + r^2 (d \theta^2 + \sin^2 \theta d\phi^2) \right]
\end{equation}
then the metric is given by
\begin{equation}
 [g_{ab}] = \left( \begin{array}{cccc}
                    1 & 0 & 0 & 0\\
                    0 & -\frac{a^2(t)}{1 - k r^2} & 0 & 0\\
                    0 & 0 & - a^2(t) r^2 & 0\\
                    0 & 0 & 0& -a^2(t) r^2 \sin^2 \theta\\
                   \end{array} \right).
\end{equation}

From the previous section, equation (\ref{eqAF}) is the easiest to use; then we will apply this equation successively for values of the index $c$ running from $0$ to $3$. In the case in which $c = 0$ we have
\begin{equation}
 g_{00} \ddot{x}^0 - \frac{1}{2}[(\partial_0 g_{11}) \dot{x}^1 \dot{x}^1 + (\partial_0 g_{22}) \dot{x}^2 \dot{x}^2 + (\partial_0 g_{33}) \dot{x}^3 \dot{x}^3] = 0,
\end{equation}
then substituting and solving we obtain
\begin{equation}
 \ddot{t} + \frac{a \dot{a}}{1 - k r^2} (\dot{r})^2 +  a \dot{a} r^2 (\dot{\theta})^2 + a \dot{a} r^2 \sin^2{\theta} (\dot{\phi})^2 = 0,
\end{equation}
from here we can read the metric connection coefficients
\begin{eqnarray}
 \Gamma^0_{11} &=& \frac{a \dot{a}}{1 - k r^2}\\
 \Gamma^0_{22} &=& a \dot{a} r^2\\
 \Gamma^0_{33} &=& a \dot{a} r^2 \sin^2{\theta}.
\end{eqnarray}

For the case when $c = 1$ we have
\begin{equation}
 g_{11} \ddot{x}^1 +(\partial_0 g_{11})\dot{x}^1 \dot{x}^0 + (\partial_1 g_{11}) \dot{x}^1 \dot{x}^1 - \frac{1}{2}[(\partial_1 g_{11}) \dot{x}^1 \dot{x}^1 + (\partial_1 g_{22}) \dot{x}^2 \dot{x}^2 + (\partial_1 g_{33}) \dot{x}^3 \dot{x}^3] = 0,
\end{equation}
then substituting and solving we obtain
\begin{equation}
 \ddot{r} +2\frac{\dot{a}}{a} \dot{t} \dot{r} +\frac{ k r}{1 - k r^2} (\dot{r})^2 -  r (1 - k r^2) (\dot{\theta})^2 - r (1 - k r^2) \sin^2 {\theta} (\dot{\phi})^2 = 0,
\end{equation}
from here we can read the metric connection coefficients
\begin{eqnarray}
 \Gamma^1_{01} &=& \frac{\dot{a}}{a}\\
 \Gamma^1_{11} &=& \frac{k r}{1-kr^2}\\
 \Gamma^1_{22} &=& -  r (1 - k r^2)\\
 \Gamma^1_{33} &=& - r (1 - k r^2) \sin^2 {\theta}.
\end{eqnarray}

For the case when $c = 2$ we have
\begin{equation}
 g_{22} \ddot{x}^2 +(\partial_0 g_{22})\dot{x}^2 \dot{x}^0 + (\partial_1 g_{22}) \dot{x}^2 \dot{x}^1 - \frac{1}{2}(\partial_2 g_{33}) \dot{x}^3 \dot{x}^3  = 0,
\end{equation}
then substituting and solving we obtain
\begin{equation}
 \ddot{\theta} +2\frac{\dot{a}}{a} \dot{\theta} \dot{t} + 2 \frac{1}{r} \dot{\theta} \dot{r} - \sin{\theta} \cos{\theta} (\dot{\phi})^2 = 0,
\end{equation}
from here we can read the metric connection coefficients
\begin{eqnarray}
 \Gamma^2_{02} &=& \frac{\dot{a}}{a}\\
 \Gamma^2_{12} &=& \frac{1}{r} \\
 \Gamma^2_{33} &=& - \sin{\theta} \cos{\theta}.
\end{eqnarray}

For the case when $c = 3$ we have
\begin{equation}
 g_{33} \ddot{x}^3 +(\partial_0 g_{33})\dot{x}^3 \dot{x}^0 + (\partial_1 g_{33}) \dot{x}^3 \dot{x}^1 + (\partial_2 g_{33}) \dot{x}^3 \dot{x}^2  = 0,
\end{equation}
then substituting and solving we obtain
\begin{equation}
 \ddot{\phi} +2\frac{\dot{a}}{a} \dot{\phi} \dot{t} + 2 \frac{1}{r} \dot{\phi} \dot{r} + 2 \frac{\cos{\theta}}{\sin{\theta}} \dot{\phi} \dot{\theta} = 0,
\end{equation}
from here we can read the metric connection coefficients
\begin{eqnarray}
 \Gamma^3_{03} &=& \frac{\dot{a}}{a}\\
 \Gamma^3_{12} &=& \frac{1}{r} \\
 \Gamma^3_{23} &=& \frac{\cos{\theta}}{\sin{\theta}} = \cot{\theta}.
\end{eqnarray}

\section{Obtaining the Ricci Tensor}
Having the metric connection coefficients, the next step is to obtain the independent values of the Ricci tensor which is given by
\begin{equation}
 R_{\mu \nu} = \partial_{\nu} \Gamma^{\sigma}_{\mu \sigma} - \partial_{\sigma} \Gamma^{\sigma}_{\mu \nu} + \Gamma^{\rho}_{\mu \sigma} \Gamma^{\sigma}_{\rho \nu} - \Gamma^{\rho}_{\mu \nu} \Gamma^{\sigma}_{\rho \sigma}.
\end{equation}

From the metric connection coefficients we obtain that
\begin{eqnarray}
 R_{00} &=& 3 \partial_0 \Gamma^{1}_{01} + 3 (\Gamma^{1}_{01})^2\\
        &=& 3 \left[\left(\frac{\ddot{a}}{a} - \left(\frac{\dot{a}}{a} \right)^2 \right) + \left(\frac{\dot{a}}{a}\right)^2 \right]\\
        &=& 3 \frac{\ddot{a}}{a},
\end{eqnarray}
for $R_{11}$ we obtain
\begin{eqnarray}
 R_{11} &=& 2 \partial_1 \Gamma^{2}_{12} - \partial_0 \Gamma^{0}_{11} - \Gamma^{0}_{11} \Gamma^{1}_{01} - 2 \Gamma^{1}_{11} \Gamma^{2}_{12} + 2(\Gamma^{2}_{12})^2\\
        &=& -\frac{2}{r^2} - \frac{(\dot{a}^2 + a \ddot{a})}{1 - k r^2} - \frac{\dot{a}^2}{1- kr^2} - \frac{2k}{1 - kr^2} + \frac{2}{r^2}\\
        &=& - \frac{a \ddot{a} + 2 \dot{a}^2 + 2 k}{1 - kr^2},
\end{eqnarray}
for $R_{22}$ we have
\begin{eqnarray}
 R_{22} &=& \partial_2 \Gamma^{3}_{23} - \partial_0 \Gamma^{0}_{22} - \partial_1 \Gamma^{1}_{22} + 2 \Gamma^{0}_{22} \Gamma^{2}_{02} + 2 \Gamma^{1}_{22} \Gamma^2_{12} + (\Gamma^3_{23})^2\\
\nonumber & & - 3 \Gamma^0_{22} \Gamma^1_{01} - \Gamma^1_{22} \Gamma^{1}_{11} - 2 \Gamma^{1}_{22} \Gamma^2_{12}\\
        &=& -\csc^2{\theta} - r^2 (\dot{a}^2 + \ddot{a}a)  + (1 - kr^2) - 2 kr^2 + 2 r^2 \dot{a}^2\\
\nonumber& & - 2(1 - k r^2) + \cot^2{\theta} - 3 r^2 \dot{a}^2 + kr^2 + 2(1-kr^2)\\
        &=& -r^2(a \ddot{a} + 2 \dot{a}^2 + 2 k),
\end{eqnarray}
finally for $R_{33}$ we have
\begin{eqnarray}
R_{33} &=& -\partial_0 \Gamma^0_{33} - \partial_1 \Gamma^1_{33} - \partial_2 \Gamma^2_{33}+ 2 \Gamma^0_{33} \Gamma^3_{03} + 2 \Gamma^{1}_{33} \Gamma^3_{13} + 2 \Gamma^2_{33} \Gamma^3_{23}\\
\nonumber & & - 3 \Gamma^0_{33} \Gamma^1_{01} - \Gamma^{1}_{33} \Gamma^1_{11} - 2 \Gamma^1_{33} \Gamma^2_{12} - \Gamma^2_{33} \Gamma^3_{23}\\
       &=& -r^2 \sin^2 \theta (\dot{a}^2 + a \ddot{a}) - 3kr^2 \sin^2 \theta + \sin^2 \theta + \cos^2 \theta - \sin^2 \theta \\
\nonumber & & + 2 r^2 \sin^2 \theta \dot{a}^2 - 2 \sin^2 \theta (1 - kr^2) - 2 \cos^2 \theta -3 r^2 \sin^2 \theta \dot{a}^2 \\
\nonumber & & + kr^2 \sin^2 \theta + 2 \sin^2 \theta (1 - kr^2) + \cos \theta \\
       &=& - r^2 \sin^2 \theta (a \ddot{a} + 2 \dot{a}^2 + 2 k)
\end{eqnarray}

\section{The Energy-Momentum Tensor}
In order to simplify we will assume that the matter that fills the Universe can be characterized as a perfect fluid, this assumption implies that we are neglecting any shear-viscous, bulk-viscous and heat-conductive properties of the matter \citep{Hobson2005}. The energy-momentum tensor is given by
\begin{equation}
 T^{\mu \nu} = (\rho + p)u^{\mu} u^{\nu} - p g^{\mu \nu}.
\label{eq:emt}
\end{equation}
Since in a comoving coordinate system the 4-velocity is given simply by $u^{\mu} = \delta_{0}^{\mu}$ and $u_{\mu} = \delta_{\mu}^{0}$, then we have
\begin{eqnarray}
 T_{\mu \nu} = (\rho + p) \delta_{\mu}^{0} \delta_{\nu}^{0} - p g_{\mu \nu}.
\end{eqnarray}
For the contracted energy-momentum tensor we have
\begin{eqnarray}
 T &=& T_{\mu}^{\mu}\\
   &=& (\rho + p) - p\delta_{\mu}^{\mu}\\
   &=& \rho + p - 4p\\
   &=& \rho - 3p,
\end{eqnarray}
then, we have that
\begin{eqnarray}
 T_{\mu \nu} - \frac{1}{2} T g_{\mu \nu} &=& (\rho + p) \delta_{\mu}^0 \delta_{\nu}^0 - p g_{\mu \nu} - \frac{1}{2} (\rho - 3 p)g_{\mu \nu}\\
 &=& (\rho + p) \delta_{\mu}^0 \delta_{\nu}^0 - \frac{1}{2}(\rho + p)g_{\mu \nu},
\end{eqnarray}
from here we can substitute in the right hand side of (\ref{eq:grfe}) to obtain
\begin{eqnarray}
 -\kappa(T_{00} - \frac{1}{2} T g_{00}) + \Lambda g_{00} &=& -\frac{1}{2} \kappa (\rho + 3 p) + \Lambda\\
 -\kappa(T_{11} - \frac{1}{2} T g_{11}) + \Lambda g_{11} &=& -\left[\frac{1}{2} (\rho - p) + \Lambda \right] \frac{a^2}{1 - k r^2}\\
 -\kappa(T_{22} - \frac{1}{2} T g_{22}) + \Lambda g_{22} &=& -\left[\frac{1}{2} (\rho - p) + \Lambda \right] a^2 r^2\\
 -\kappa(T_{33} - \frac{1}{2} T g_{33}) + \Lambda g_{33} &=& -\left[\frac{1}{2} (\rho - p) + \Lambda \right] a^2 r^2 \sin^2 \theta.
\end{eqnarray}

\section{The Cosmological Field Equations}
In the two previous sections we have derived both sides of the GR Field Equations, then at this point the reamaining step is to combine these results to obtain the Cosmological Field Equations. For $R_{00}$ we have
\begin{eqnarray}
 3\frac{\ddot{a}}{a} &=& -\frac{1}{2} \kappa (\rho + 3 p) + \Lambda\\
 3\frac{\ddot{a}}{a} &=& -\frac{8 \pi G}{2} (\rho + 3 p) + \Lambda\\
 \frac{\ddot{a}}{a} &=& -\frac{4 \pi G}{3} (\rho + 3 p) + \frac{1}{3} \Lambda; \label{eq:aa}
\end{eqnarray}
for $R_{11}$ we have
\begin{eqnarray}
 -\frac{a \ddot{a} + 2 \dot{a}^2 + 2 k }{1 - kr^2} &=& -\left[\frac{1}{2} \kappa (\rho - p) + \Lambda \right] \frac{a^2}{1- kr^2}\\
 a \ddot{a} + 2 \dot{a}^2 + 2 k &=& ( 4 \pi G  (\rho - p) + \Lambda) a^2, \label{eq:base}
\end{eqnarray}
substituting the value for $\ddot{a}$ from (\ref{eq:aa}) we have
\begin{eqnarray}
 -\frac{4 \pi G}{3} (\rho + 3 p) a^2 + \frac{1}{3} \Lambda a^2 + 2 \dot{a}^2 + 2 k &=& 4 \pi G (\rho - p) a^2 + \Lambda a^2\\
 2 \dot{a}^2 &=& \frac{4 \pi G}{3} (4 \rho) a^2 + \frac{2}{3} \Lambda a^2 - 2 k\\
 \left(\frac{\dot{a}}{a}\right)^2 &=& \frac{8 \pi G \rho}{3} - \frac{k}{a^2} + \frac{\Lambda}{3}; \label{eq:a}
\end{eqnarray}
for $R_{22}$ we have
\begin{eqnarray}
 -r^2(a \ddot{a} + 2 \dot{a}^2 + 2 k ) &=& - \left[\frac{1}{2} \kappa (\rho - p) + \Lambda \right] a^2 r^2\\
 a \ddot{a} + 2 \dot{a}^2 + 2 k &=& ( 4 \pi G  (\rho - p) + \Lambda) a^2,
\end{eqnarray}
we have obtained (\ref{eq:base}), then the equation given by $R_{22}$ is not independent. For $R_{33}$ we have
\begin{eqnarray}
 -r^2 \sin^2 \theta (a \ddot{a} + 2 \dot{a}^2 + 2 k ) &=& - \left[\frac{1}{2} \kappa (\rho - p) + \Lambda \right] a^2 r^2 \sin^2 \theta\\
 a \ddot{a} + 2 \dot{a}^2 + 2 k &=& ( 4 \pi G  (\rho - p) + \Lambda) a^2,
\end{eqnarray}
anew, we have obtained (\ref{eq:base}) and then the equation $R_{33}$ is redundant.

From the previous discusion, only two of the four equations are independent (equation (\ref{eq:aa}) and equation (\ref{eq:a}) ):
\begin{eqnarray}
 \frac{\ddot{a}}{a} &=& -\frac{4 \pi G}{3} (\rho + 3 p) + \frac{1}{3} \Lambda\\
 \left(\frac{\dot{a}}{a}\right)^2 &=& \frac{8 \pi G \rho}{3} - \frac{k}{a^2} + \frac{\Lambda}{3},
\end{eqnarray}
these are the Cosmological Field Equations.

%% file: apc/apc.tex
%chapter apa.tex ------- The Cosmic Distance Ladder --- last modification: 24 May 2010 %ET1606
\chapter{The Cosmic Distance Ladder }
\label{ApC}

From the relation (\ref{eqDL}) we can see that knowing the values for the absolute luminosity $L$ and the flux $f$ for an object we can obtain immediately the value of the luminosity distance $D_L$; if we obtain $D_L$ and $z$ for a great number of objects we can determine an approximate value for $H_0$, as can be seen from (\ref{eqHR}), or constrain the cosmological model by means of the relation (\ref{eqLD}); then the knowledge of $D_L$ is of great importance, although, the difficult problem is to determine the value of the absolute luminosity. 

Conventionally, the objects used to measure distances in cosmology, are classified as primary and secondary distance indicators. The primary distance indicators are those whose absolute luminosities are measured either directly, by kinematic methods, or indirectly, by means of the association of these objects with others  whose distance was measured by kinematic methods. The primary distance indicators are not bright enough to be studied at distances farther than the corresponding to values of $z$ around $0.01$. The secondary distance indicators are bright enough to be studied at larger distances and their absolute luminosities are known through their association  with primary distance indicators \citep{Weinberg2008}; is by means of these last objects that we can constrain a cosmological model since, aside of other considerations, their value of $z$ is large enough to make negligible the contribution of the peculiar velocities to the redshift determination.

%%Kinematic Methods to Distance Determinations
\section{Kinematic Methods to Distance Determinations}
As already said, in cosmology the primary distance indicators are of importance as calibrators of the secondary distance indicators which can be used to constrain a cosmological model, but these primary distance indicators must be calibrated by means of distance determinations carried out by kinematic methods. Below we will briefly discuss the kinematic methods used to measure the distance to the primary distance indicators.

\subsection{Trigonometric parallax}
%Figure ---- Parallax scheme 
\begin{figure}[ht]
\centering
\resizebox{0.5\textwidth}{!}{\includegraphics{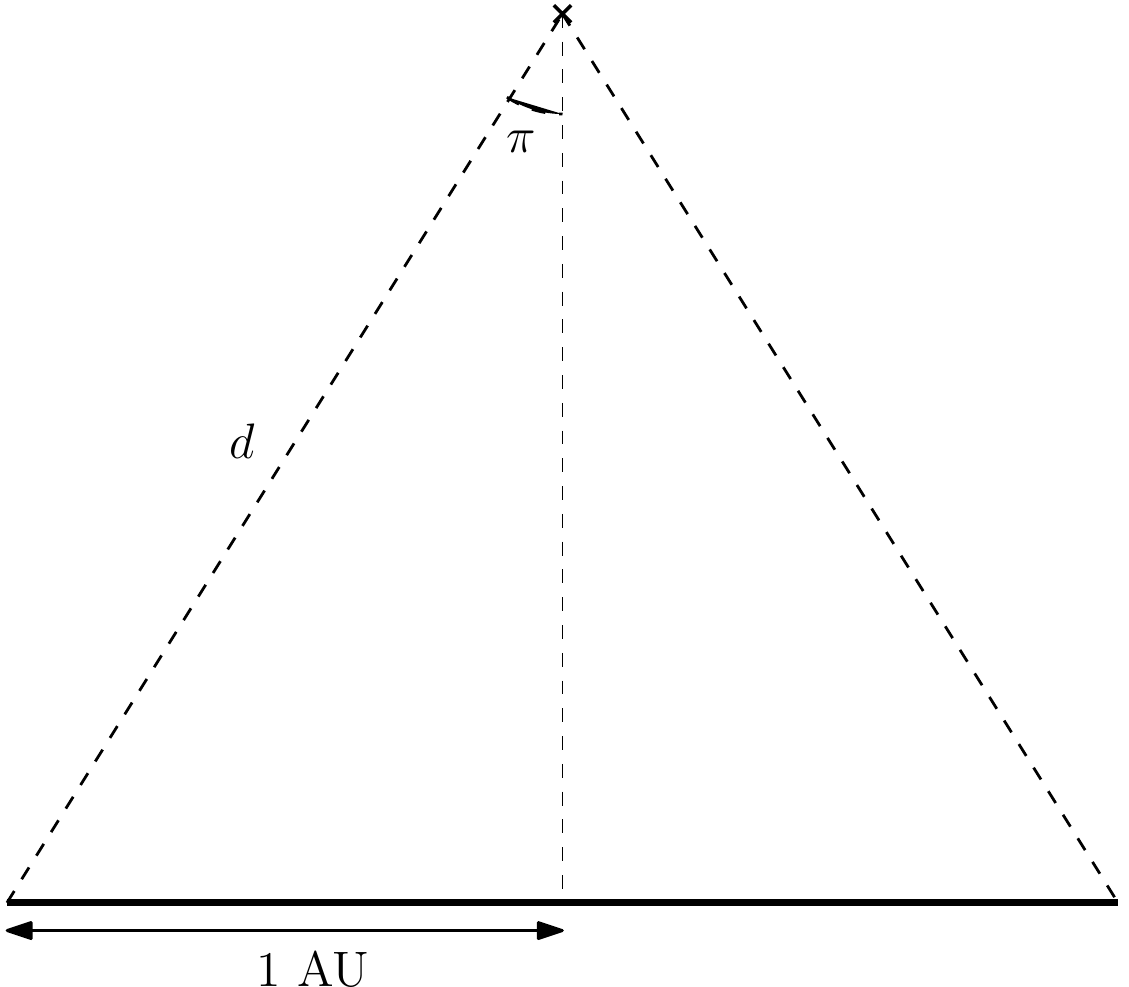}}
\caption[Trigonometric Parallax scheme] {\small Scheme illustrating the Trigonometric Parallax}
\label{fig:par}
\end{figure}

While the earth's annual motion around the sun takes place, the stars appear to have an elliptical motion due to the true movement of our planet, the maximum angular radius of this motion is called parallax, $\pi$; this situation is shown schematically  in Figure \ref{fig:par}. We can see that it is possible to calculate the actual distance to a star by means of an accurate measure of its parallax and knowing the mean distance between the sun and the earth, which is called an astronomical unit (AU). The distance to the star is given by
\begin{equation}
 d = \frac{1\ \rm{AU}}{\sin {\pi}},
\end{equation}
if we assume that $\pi \ll 1\ \rm{rad}$, which is the case for all the stars, then $\sin {\pi} \simeq \pi$, with enough approximation; even more, if we give $\pi$ in arcseconds, we obtain the relation
\begin{equation}
 \frac{d}{\rm{pc}} = \left(\frac{\pi}{\rm{arcsec}} \right)^{-1},
\label{eqDPa}
\end{equation}
where 1 parsec (pc) has been defined as the distance of an object when $\pi = 1''$ and the measure baseline is 1 AU, since $1\ \rm{rad} = 206264.8''$ and $1\ \rm{AU} = 1.49 \times 10^{13}\ \rm{cm}$, then
$$
 1\ \rm{pc} = 206264.8\ \rm{AU} = 3.09 \times 10^{18}\ \rm{cm}.
$$
This simple trigonometric method can not be applied accurately from the earth surface for stars with $\pi < 0.03''$ due to atmospheric turbulence effects (seeing) which blurs the star's image; then using ground-based telescopes this method can only be used to measure distances to stars that are about $30$ pc from us \citep{Weinberg2008}. 

From 1989 to 1993 the Hipparcos satellite, launched by the European Space Agency (ESA), measured parallaxes for more than 100 000 stars in the solar neighbourhood with a median accuracy of $\sigma = 0.97\ \rm{mas}$ \citep{Perryman1997}; this remarkable accuracy can be obtained since the observations were carried out from space and the usual problems related with the terrestrial atmosphere and gravitational field were not present.

\subsection{The moving-cluster method}

%Figure ---- Moving-cluster method scheme 
\begin{figure}[tb]
\centering
\resizebox{0.5\textwidth}{!}{\includegraphics{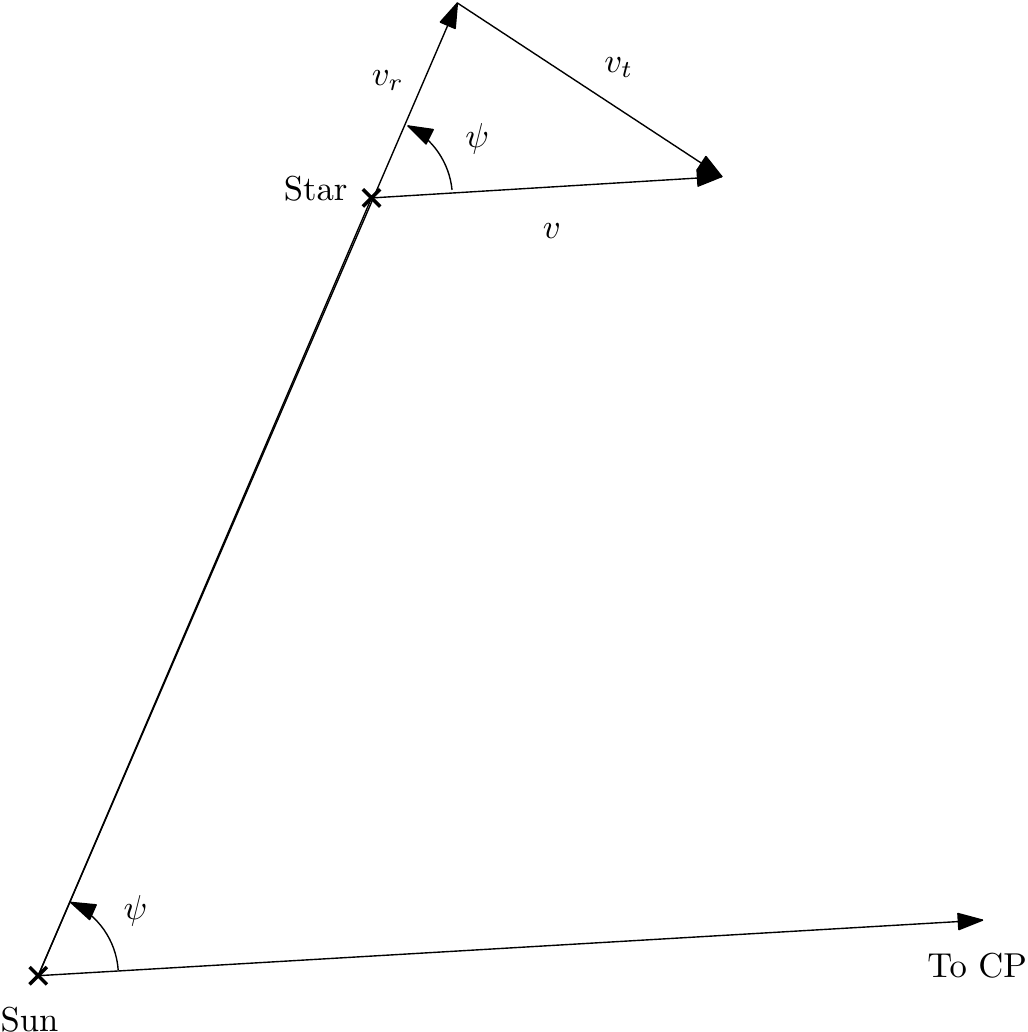}}
\caption[Moving-cluster method scheme] {\small Scheme that shows the geometric construction for the moving-cluster method; adapted from \citet{Binney1998}.}
\label{fig:movclus}
\end{figure}

The fundamental assumption over which this method is constructed is that of the parallelism in the space motion of the member stars of an open cluster; i.e, the space velocity vectors of the  members of the cluster, must point in the same direction. The implications of the previous assumption are that the random motions, the expansion or contraction velocities and the space velocities due to rotation, for the individual members, must be negligible \citep{Hanson1975}.

Since the space velocity vectors of all the stars in the cluster are parallel, then for an observer for whom the cluster is receding (or approaching), all the stars appear to be moving to (from) a convergent point (CP), the geometry for this situation is depicted in Figure \ref{fig:movclus}. From the figure we can see that the angle between the positions of the stars and the CP on the sky $\psi$\footnote{Note that this angle is seen by an outside fixed observer, from the point of view of an observer on one of the stars there is no such CP at all.}, and the angle between the star's space velocity vector and the Sun-star line of sight are the same, then we have that
\begin{equation}
 v_t = v_r \tan{\psi},
\end{equation}
where $v_r$ is the radial velocity, i.e the space velocity vector component in the direction of the line of sight, and $v_t$ is the tangent velocity defined as 
\begin{equation}
v_t = \mu d, 
\end{equation}
where $\mu$ is the proper motion of the star, i.e. its angular apparent motion on the sky plane, and $d$ is the distance from the sun to the star; then from the two previous definitions we have that 
\begin{equation}
 d = \frac{v_r \tan {\psi}}{\mu},
\end{equation}
or using the definition (\ref{eqDPa})
\begin{equation}
 \frac{\pi}{\rm{mas}} = \frac{4.74}{\tan {\psi}} \left(\frac{v_r}{\rm{km\ s^{-1}}} \right)^{-1} \frac{\mu}{\rm{mas\ yr^{-1}}}.
\end{equation}

From the above relation we can determine the parallax or the distance to every star member of the cluster under consideration, using its observed proper motion, radial velocity (easily obtained measuring the shift of spectral lines) and its value of $\psi$ \citep{Binney1998}.

%\subsubsection{Statistical Parallaxes}

%\subsubsection{Pulsation Parallaxes}

\section{Primary Distance Indicators}
As  previously pointed out, the primary distance indicators are of importance in the calibration of the secondary distance indicators.

\subsection{Cepheids}

The Cepheids are one of the best known primary distance indicators. These variable stars are very bright and since they exhibit a regular variation of their luminosity with time, they are useful to measure distances outside our galaxy. In 1912 Henrietta Swan Leavitt \citep{Leavitt1912} observed that the Cepheid variables that she was studying in the Small Magellanic Cloud (SMC) have fluxes that vary as a function of the period of the variation in luminosity (Leavitt law). The Cepheids pulsation periods are from 2 to over 100 days whereas their brightness variations go from $-2  <  M_{V} < -6\  \mathrm{mag}$ \citep{Freedman2010}.

The basic physics behind the Leavitt law is well understood, the Stephan-Boltzmann law can be written as 
\begin{equation}
 L = 4 \pi R^{2} \sigma T_{e}^{4},
\end{equation}
where, $L$, in this case, is the bolometric luminosity, $R$ is the star radius and $T_{e}$ is the star effective temperature. Expressing the above relation in therms of magnitudes, we have
\begin{equation}
 M_{\rm{BOL}} = - 5 \logd R - 10 \logd T_{e} + C;
\end{equation}
thereafter we can map $\logd T_{e}$ into an observable intrinsic color like $(B - V)_{o}$ or $(V - I)_{o}$ and map the radius into an observable period using a period-mean-density relation \footnote{A relation of the type $\omega_{\rm{dyn}} = 2 \pi / P = (G M /R^{3})^{1/2} \approx (G \bar{\rho})^{1/2}$, where $\omega_{\rm{dyn}}$ is the dynamical frequency and is proportional to the inverse of a free fall over the distance of a stellar radius.}, then we obtain the period-luminosity-color (PLC) relation for Cepheids as \citep{Freedman2010} 
\begin{equation}
 M_{V} = \alpha \logd P + \beta (B - V)_{o} + \gamma.
\end{equation} 

Today the slope of the Period-Luminosity (PL) relation is generally taken from the Cepheids in the Large Magellanic Cloud (LMC). The values of the PL relation given by the Hubble Space Telescope (HST) key project \citep{Freedman2001}, assuming that the LMC distance modulus is $\mu(LMC) = 18.50\ \rm{mag}$, are
\begin{eqnarray}
 M_V &=& -2.760 [\pm 0.03](\log_{10}P -1) -4.218[\pm0.02] \\
 M_I &=& -2.962 [\pm 0.02](\log_{10}P -1) -4.904[\pm0.01],
\end{eqnarray}
where $P$ is the period in days; but these results have been under discussion due to considerations of metallicity effects in the determinations of the LMC distance modulus \citep{Cole1998, Girardi1998, Salaris2003}.

The calibration of the PL relation can be done by observations of galactic Cepheids, in which case trigonometric parallax determinations are generally used. Using data from Hipparcos, the PL relation has been given as \citep{Feast1997}
\begin{equation}
 M_V = -2.81 \log_{10} P -1.43 [\pm0.10].
\end{equation}
Assuming the slope given by the last equation, \citet{Feast2005} has parametrized the PL relation as
\begin{equation}
 M_V = -2.81 \log_{10} P + \gamma,
\end{equation}
where $\gamma$ is the PL relation zero-point, and using four distinct methods he has obtained a mean value of $\gamma = -1.40$.

Finally, recent work points out that no significant difference exists in the slopes of the PL relation between our Galaxy and the LMC \citep{Fouque2007}, and gives for our Galaxy
\begin{eqnarray}
 M_V &=& -2.678 [\pm 0.076] \log_{10} P  -1.275 [\pm 0.023]\\
 M_I &=& -2.980 [\pm 0.074] \log_{10} P  -1.726 [\pm 0.022];
\end{eqnarray}
and for the LMC
\begin{eqnarray}
 M_V &=& -2.734 [\pm 0.029] \log_{10} P -1.348 [\pm 0.007]\\
 M_I &=& -2.957 [\pm 0.020] \log_{10} P -1.811 [\pm 0.005];
\end{eqnarray}
where it has been assumed that the LMC distance modulus is $\mu(LMC) = 18.40\ \rm{mag}$, which is consistent with recent results \citep{Benedict2007}.

% 29 April 2010
\subsection{Tip of the red giant branch method}

The tip of the red giant branch (TRGB) is a technique for determining distances to nearby galaxies. This method uses the well understood \citep{Salaris2002} discontinuity in the luminosity function (LF) of stars evolving up the red giant branch (RGB) in old, low metallicity stellar populations that has been calibrated using Galactic globular clusters; necessary condition for its application beeing that the observed RGB LF is well populated ($\sim 100$ stars within 1 mag form the TRGB) \citep{Madore1995}. 

The empirical calibration of the TRGB is typically given as:
\begin{equation}
 M_{I}^{TRGB} = f([\rm{Fe}/\rm{H}]) + ZP
\end{equation}
where, $M_{I}^{TRGB}$ is the absolute magnitude for the TRGB\footnote{In this case for the Cousins' I passband, but the model is similar in other bands.}, $f([\rm{Fe}/\rm{H}]) $ is a function of the metallicity (typically a polynomial), and $ZP$ is the calibration Zero Point. This kind of models neglect the impact of other parameters on the calibration and then induce uncertainties of order $\pm0.1$ mag in the determination of $M_{I}^{TRGB}$ \citep{Bellazzini2008}.

\section{Secondary Distance Indicators}
The primary distance indicators are not sufficiently  bright to be observed at $z > 0.01$, brighter objects are needed as tracers to constrain a cosmological model, these brighter objects can be galaxies or supernovae, which are as bright as galaxies. We need methods to obtain the luminosity of these objects in order to determine their distances.

%\subsubsection{Maser Galaxies}

%\subsubsection{Surface Brightness Fluctuations} 

% 29 April 2010
\subsection{Type Ia supernovae}

Type Ia Supernovae (SNe Ia) are the result of the thermonuclear destruction of an accreting carbon-oxygen white dwarf star approaching the Chandrasekhar mass limit. Observationally, the defining characteristic of SNe Ia is the absence of $\rm{H}$ and $\rm{He}$ lines and the presence of strong $\rm{Si}$ absorption lines in their spectra. 

In spite of the fact that the details of the nature of the SNe Ia explosion are still obscure, the origin of the observed light curve is relatively well understood. It is powered by the radioactive decay of $\indices{^{56}}\rm{Ni}$ into $\indices{^{56}}\rm{Co}$, and then into $\indices{^{56}}\rm{Fe}$. The SN ejecta is heated by energetic gamma rays, produced by the radioactive decay, and then radiates thermally to produce the observed light curve. Photometrically, SN Ia rises to maximum light in a period of 20 days, followed by a decline of $\sim 3$ mag in the following month and $\sim 1 $ mag per month subsequently \citep{Freedman2010, Wolschin2010}. 

SNe Ia are not intrinsically standard candles, but can be standardized by means of simple empirical correspondences. The first of these relations is the light-curve width--luminosity relationship (WLR) or `Phillips relation' \citep{Phillips1993}; essentially SN Ia peak luminosities are strongly correlated with the width of their light curve. Furthermore, SN Ia light curves can be parametrized using a `stretch' parameter, which stretches or contracts a template light curve to match an observed one \citep{Perlmutter1997}. As an aside, the physical origin of the Phillips relation is yet not completely clear \citep{Kasen2007, Wolschin2010}.

Another --though poorly understood-- relation is between the SNe Ia luminosity and their color B - V \citep{Tripp1998, Wolschin2010}. The two previous relationships can be applied to observed peak magnitudes $m$:
\begin{equation}
 m_{\rm{corr}} = m + \alpha (s - 1) - \beta C,
\end{equation}
where the stretch-luminosity is parametrized by $\alpha$, and the color-luminosity relation by $\beta$. After applying the calibration to SNe Ia measurements, precise distance estimates (to $0.12 - 0.14$ mag) can be obtained.

\subsection{Tully-Fisher relation}

\citet{Tully1977} proposed the existence of a correlation between the global \hi\ line (21 cm) profile width and the absolute blue magnitude of spiral galaxies; later, after the study of the correlation of the \hi\ width and infrared luminosity, the physical basis for this relation was understood, i.e that the 21 cm line is widened by Doppler effect, caused by the rotation of the galaxy; therefore the  \hi\ line width is an indicator of the maximum speed of rotation of the galaxy $V_{rot}$, which by gravity is related to the mass of the galaxy, which in turn is related to the luminosity $L$ by the mass-luminosity ratio \citep{Aaronson1979}. Roughly, we have
\begin{equation}
 L \sim V_{rot}^4.
\end{equation}

The Tully-Fisher relation, calibrated with Cepheids distances and metallicity-corrected, has been given as \citep{Sakai2000}
\begin{eqnarray}
 B_{T,Z}^c &=& -(8.07 \pm 0.72) (\log_{10} W_{20}^c - 2.5) - (19.88 \pm 0.11)\\
 I_{T,Z}^c &=& -(9.46 \pm 0.76) (\log_{10} W_{20}^c - 2.5) - (21.19 \pm 0.12),
\end{eqnarray}
where $X_{T,Z}^c$ are aperture magnitudes corrected for metallicity and Galactic and internal extinction, and $W_{20}^c$ are the 20\% line widths corrected for inclination and redshift.

\subsection{Faber-Jackson relation}

For elliptical galaxies a correlation exists that is  similar to the Tully-Fisher relation, only  in this case between the luminosity and the velocity dispersion. The theoretical basis for this is too the Virial theorem \citep{Faber1976}. The analytical form of this relation can be given roughly as 
\begin{equation}
 L_e \sim \sigma_0^4,
\end{equation}
where $L_e$ is the luminosity inside  the effective radius and $\sigma_0$ is the central velocity dispersion measured from spectral line broadening \citep{Binney1998}.

%% file: apb/apb.tex
\chapter{Statistical Techniques in Cosmology}
\label{ApB}
In order to analyse the large data sets that are now available for cosmological work it is absolutely necessary the use of more and more sophisticated statistical tools. Here we present a few basic statistical techniques that are used through this work and that in general can be applied  in cosmological data sets analysis. Through this appendix we closely follow the work of \citet{Verde2010}.

\section{Bayes Theorem and Statistical Inference}
The fundamental rules of probability are (hereafter $\mathcal{P}$ is the probability of an event):
\begin{enumerate}
\item $\mathcal{P} \geq 0$.
\item $\int_{-\infty}^{\infty}{dx \mathcal{P}(x)} = 1$.
\item For mutually exclusive events $\mathcal{P}( x \cup y)  = \mathcal{P}(x) + \mathcal{P}(y)$.
\item For dependent events $\mathcal{P}(x \cap y) = \mathcal{P}(x) \mathcal{P}(y | x)$, where $\mathcal{P}(y | x)$ is the conditional probability of $y$ given that $x$ has already occurred. 
\end{enumerate}
from the last relation we can derive the Bayes theorem (writing $\mathcal{P}(x, y) = \mathcal{P}(y, x)$):
\begin{equation}
 \mathcal{P}(H | D) = \frac{\mathcal{P}(H) \mathcal{P}(D | H)}{\mathcal{P}(D)}
\end{equation}
where $D$ stands for \emph{data}, $H$ for \emph{hypothesis} or model, $\mathcal{P}(H | D)$ is called the \emph{posterior}, $\mathcal{P}(D | H)$ is the \emph{likelihood} and $\mathcal{P}(H)$ is called the \emph{prior}.

Bayes theorem is at the base of statistical inference, let us assume that we have some already collected data set, then $\mathcal{P}(D) = 1$, and we have a model characterized by some set of parameters $\mathbf{p}$, in general we want to know the probability distribution for the model parameters given the data $\mathcal{P}(\mathbf{p} | D)$ (from a bayesian view point as opposed to a frequentist one). However, usually we can compute accurately the likelihood which, by Bayes theorem, is related to the posterior by the prior.

One fundamental problem with the above approach is that the use of distinct priors leads to different posteriors since e.g. if we have a prior in two distinct equally valid variables, then we have a distinct probability distributions for every prior, say $\mathcal{P}(x)$ and $\mathcal{G}(y)$, then in order to transform from one distribution to the other one we have
\begin{align}
 \mathcal{P}(x) dx &= \mathcal{G} (y) dy,\\
 \mathcal{P}(x) &= \mathcal{G}(y) \left | \frac{dy}{dx}\right |.
\end{align}

Another important concept is the marginalization procedure. If we have a multivariate distribution, say $\mathcal{P}(x, y)$ and we want to know the probability distribution $\mathcal{P}(x)$ regardless of the values of $y$, then we \emph{marginalize} with respect to $y$:
\begin{equation}
 \mathcal{P}(x) = \int{dy \mathcal{P}(x, y)}.
\end{equation}

\section{Chi-square and Goodness of Fit}
In order to find the model, characterized by a set of parameters $\mathbf{p}$, that better fit a given data set, we must define a merit function that quantifies the correspondence between the model and the data. 

The least squares fitting is given by 
\begin{equation}
 \chi^2 = \sum_{i}{w_i[D_i - y (x_i | \mathbf{p})]^2},
\end{equation}
where $D_i$ are the data points, $y(x_i | \mathbf{p})$ is the model and $w_i$ are suitably defined weights. The minimum variance weight is $w_i = 1 / \sigma_i^2$ where $\sigma_i$ denotes the error on data point $i$. With these weights the least squares is called chi-square. The best fit parameters are those that minimize the $\chi^2$.

If the data are correlated, the chi-square becomes 
\begin{equation}
 \chi^2 = \sum_{ij}{[D_i - y(x_i | \mathbf{p})] Q_{ij}[D_j - y(x_j | \mathbf{p})]},
\end{equation} 
where $Q$ denotes the inverse of the covariance matrix. 

The probability distribution for the values of $\chi^2$ around its minimum value, is given by a $\chi^2$ distribution for $\nu = n - m $ degrees of freedom, where $n$ is the number of independent data points and $m$ is the number of parameters. The probability that the value of $\chi^2$ obtained from the fit exceeds by chance the value $\hat{\chi}$ for the \emph{correct} model is $Q(\nu, \hat{\chi}) = 1 - \Gamma (\nu/2, \hat{\chi}/2)$ where $\Gamma$ is the incomplete Gamma function. $Q$ measures the goodness of the fit. 

\section{Likelihood}
If in the Bayes theorem we take $\mathcal{P}(D) = 1$ since we assume that we already have the data, and $\mathcal{P}(H) = 1$ since we ignore the prior, then estimating the likelihood we obtain the posterior. However, since we have ignored the prior then we can not give the goodness of fit or the absolute probability for a model in which case we can only obtain relative probabilities. Assuming that the data are gaussianly distributed the likelihood is given by a multi-variate Gaussian:
\begin{equation}
 \mathcal{L} = \frac{1}{(2 \pi)^{n/2} |det(C)|^{1/2}} \exp{\left[- \frac{1}{2} \sum_{ij}(D - y)_i C_{ij}^{-1}(D - y)_j \right]},
\end{equation}  
where $C_{ij} $ is the covariance matrix.

For Gaussian distributions we have $\mathcal{L} \propto \exp{[-1/2 \chi^2]}$ and minimizing the $\chi^2$ is equivalent to maximizing the likelihood. 

The likelihood ratio is used in order to obtain results independently of the prior, it is the comparison between the likelihood at a point and the maximum likelihood, $\mathcal{L}_{max}$. Then, a model is acceptable if the likelihood ratio, 
\begin{equation}
 \Lambda = - 2 \ln \left[\frac{\mathcal{L}(p)}{\mathcal{L}_{max}} \right],
\end{equation} 
is above a given threshold.

\section{Fisher Matrix}
The Fisher matrix allows to estimate the parameters error for a given model. It is defined as
\begin{equation}
  F_{ij} =  - \left \langle \frac{\partial^2 \ln \mathcal{L}}{\partial p_i \partial p_j} \right \rangle ,
\end{equation}
where the average is the ensemble average over observational data (those that would be gathered if the real Universe was given by the model).

For a parameter $i$ the marginalized error is given by
\begin{equation}
 \sigma_{p_i} \geq  (F^{-1})^{1/2}_{ii},
\end{equation}
this last equation is the Kramer-Rao inequality that implies that the Fisher matrix always gives an optimistic estimate of the errors. This inequality is an equality only if the likelihood is Gaussian, this happens when the data are gaussianly distributed and the model depends linearly on the parameters.

\section{Monte Carlo Methods}
The methodology of Monte Carlo methods for error analysis can be described as follows. Given a measured data set $D_0$, we can fit some model to it and obtain a set of parameters $p_0$ and their errors. With the intention of exploring the errors for $p_0$, we assume that the fitted parameters $p_0$ are the \emph{true} ones. Subsequently, we construct an ensemble of simulated sets of parameters $p_i^s$ taking care of the observational errors associated with the data set $D_0$. Finally, we can construct the distribution $p_i^s - p_0 $ from which we can explore the parameters error.

The Monte Carlo methods for error determinations are specially useful when complicated effects can be simulated but not described analytically by a model. 

%% file: apd/apd.tex
\chapter{Profile Fits to the High Resolution H$\beta$ Lines}
\label{ApD}
%In this appendix we describe the fitting procedure to the high resolution H$\beta$ spectral lines. 

We have used three independent fit procedures for each object.
\begin{enumerate}

\item A single gaussian fit to the line using the \verb|gaussfit| IDL routine.

\item Two different gaussians using the \verb|arm_asymgaussfit| routine in order to explore possible  asymmetries.

\item Three separate gaussians using the \verb|arm_multgaussfit| routine to investigate the role of the extended `non-gaussian' wings. 
For this  case we constructed a grid of parameters to use as seeds for the routine, as described in the main text.

\end{enumerate}

\begin{figure}
  \centering
  \caption[VLT-UVES instrumental profile]{\small VLT-UVES instrumental profile and its gaussian fit, as obtained from the OI 5577 \AA\ sky line. 
  The observed line is shown in black and the gaussian fit in red. This, as all the 
  following profiles, is shown in a 20 \AA\ wide window.}
    \includegraphics[width=69mm]{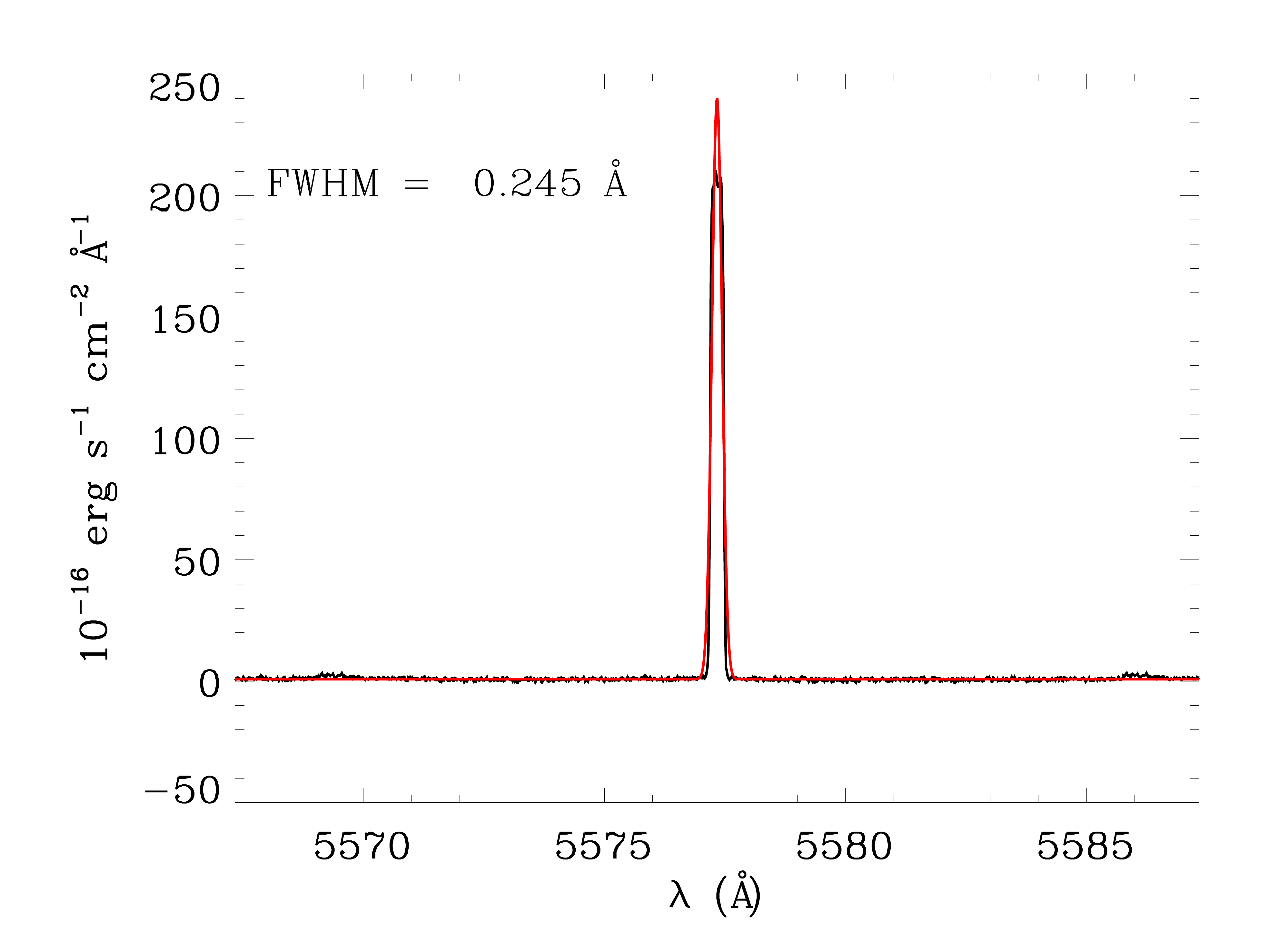}
  \label{Afig00}
\end{figure}

In Figure \ref{Afig00} we show the UVES instrumental profile and its gaussian fit obtained from the OI 5577 \AA\ sky line. 
Figures D.2 to D.11 show the best fits for the H$\beta$ lines. Each plot presents the fits to a different HIIGx. 
The upper panel shows the three independent fits while the lower panel shows their residuals. 
The insets indicate the results of the fits and the distribution resulting from the Montecarlo 
simulation used to estimate the errors in the FWHM (see main text).  

Figure D.12 shows the HDS instrumental profile and its gaussian fit, obtained from 
the OI 5577 \AA\ sky line. 
Figures D.13 to D.24 show the best fits corresponding to the HDS observations. The details 
are like those for the UVES spectra. 

\emph{Note: Due to file size limitations we have only included a sample of the appendix. I can provide the full file upon request to ricardoc@inaoep.mx}

\begin{figure}
  \centering
  \label{Afig01}\caption[H$\beta$ lines best fits for VLT UVES data]{\small H$\beta$ lines best fits for VLT UVES data. 
  The observed $\hb$ line and the three different fits are shown in a 20 \AA\ wide window for each object as labelled. 
  \emph{Upper panel:} The single gaussian fit 
  is indicated by a dashed line (thick black),the asymmetric gaussian fit is indicated by 
  a dash-dotted line (blue) and the three separate gaussians fit is indicated by long-dashed 
  lines (red) with its total fit  shown by a dash-double-dotted line (yellow); the parameters 
  of the fits are listed in the top left corner.  \emph{Lower panel:} Shows the residuals 
  from the fitting procedures following the same colour code with crosses for the 
  single gaussian fit and  continuous lines both for the asymmetric and three gaussian fits. 
  The inset shows the results  from the Montecarlo simulation  to estimate the errors in 
  the FWHM of the best fit. Details are described in the main text. }
  
  \subfloat[J051519-391741]{\label{Afig01:1}\includegraphics[width=69mm]{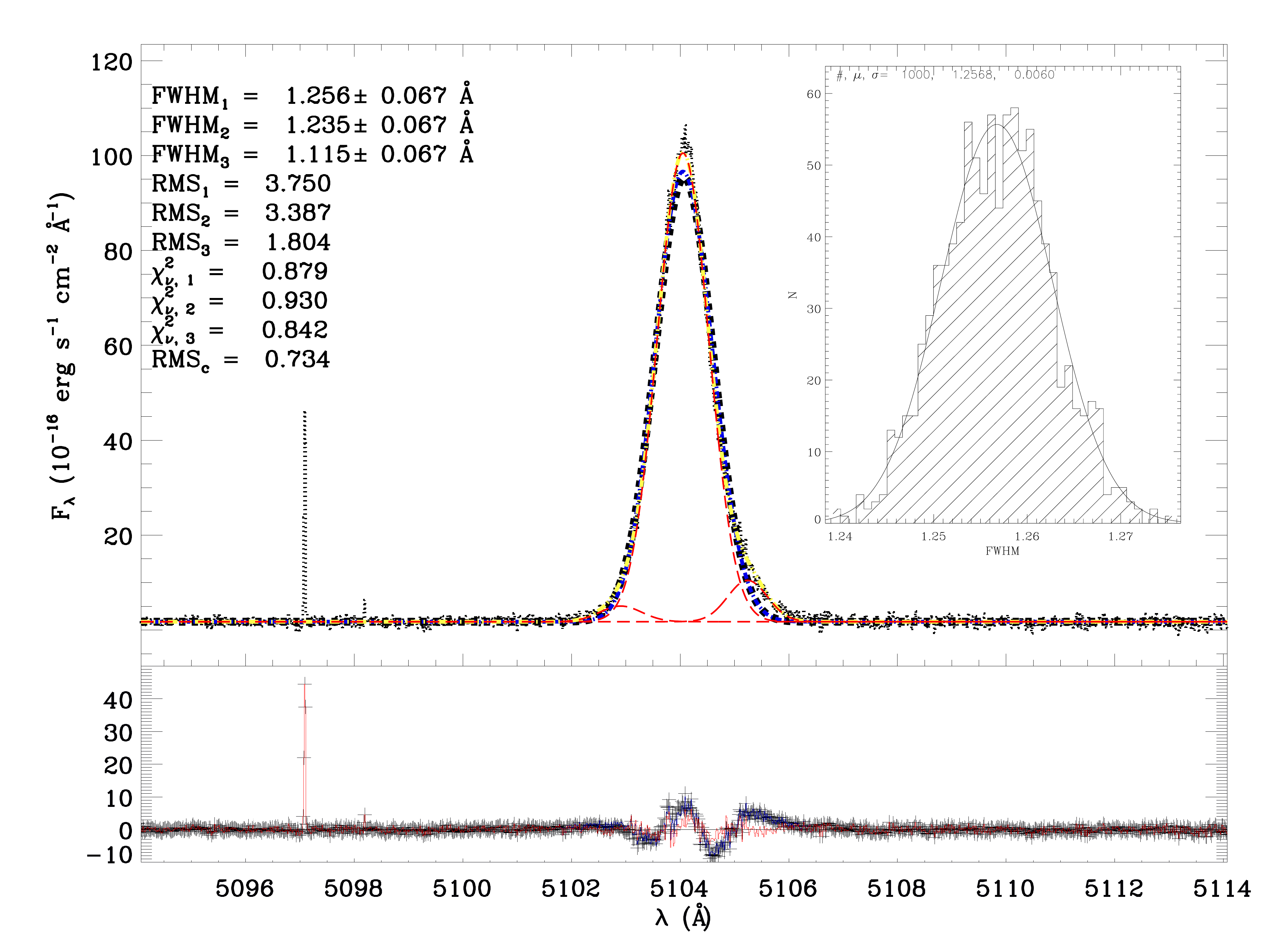}}
  \subfloat[J074806+193146]{\label{Afig01:2}\includegraphics[width=69mm]{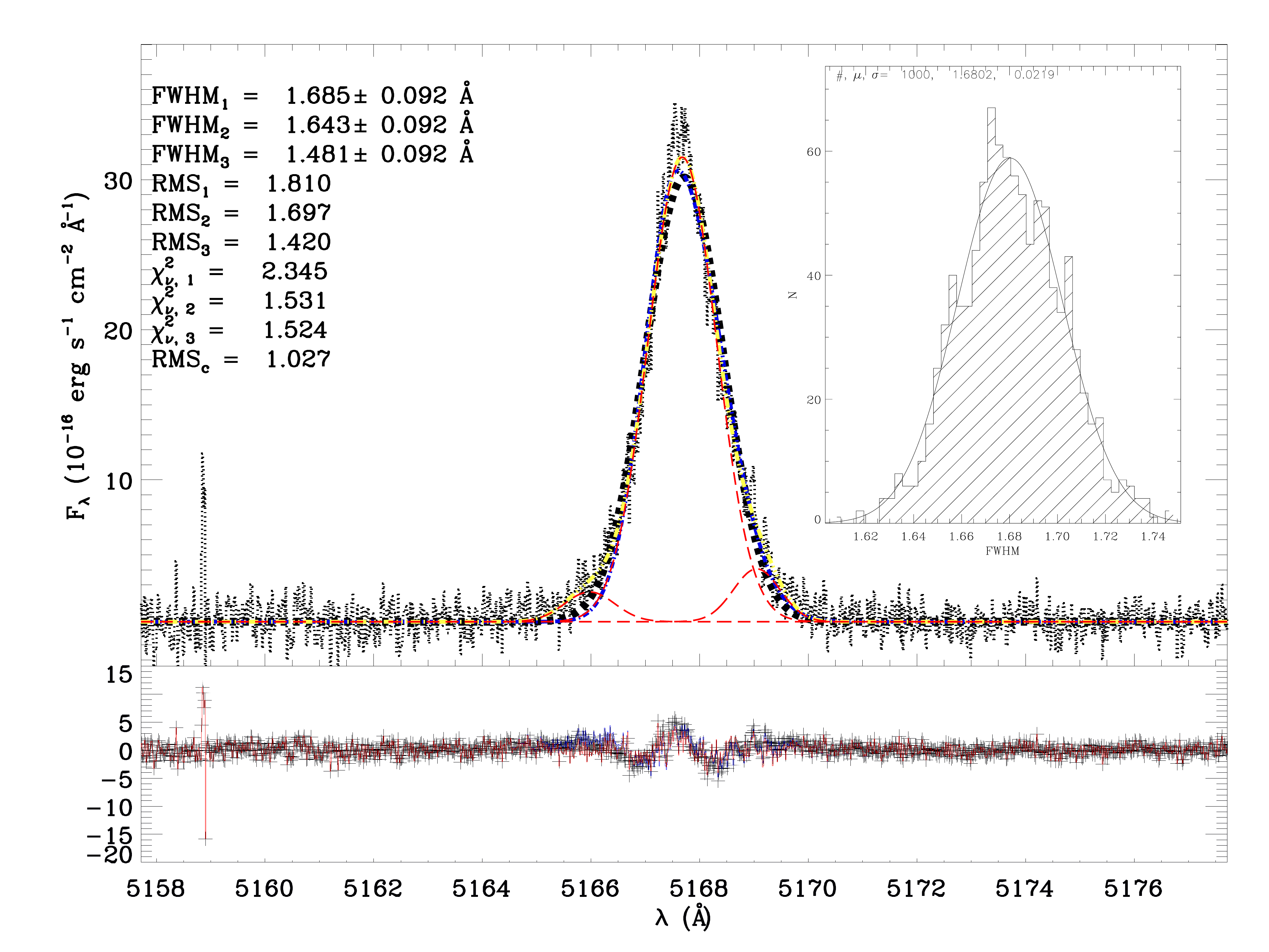}}
  \\
  \subfloat[J074947+154013]{\label{Afig01:3}\includegraphics[width=69mm]{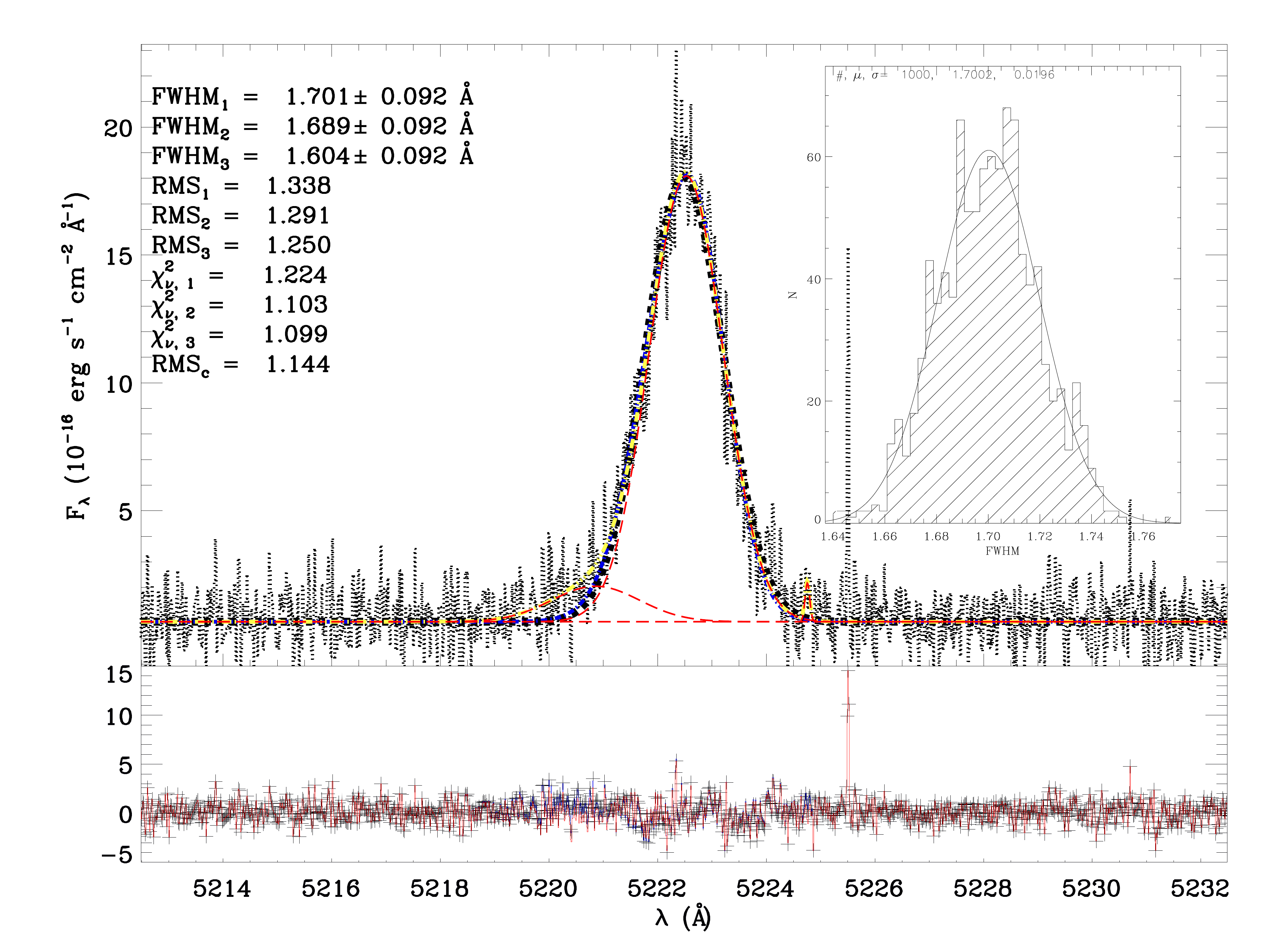}}
  \subfloat[J080000+274642]{\label{Afig01:4}\includegraphics[width=69mm]{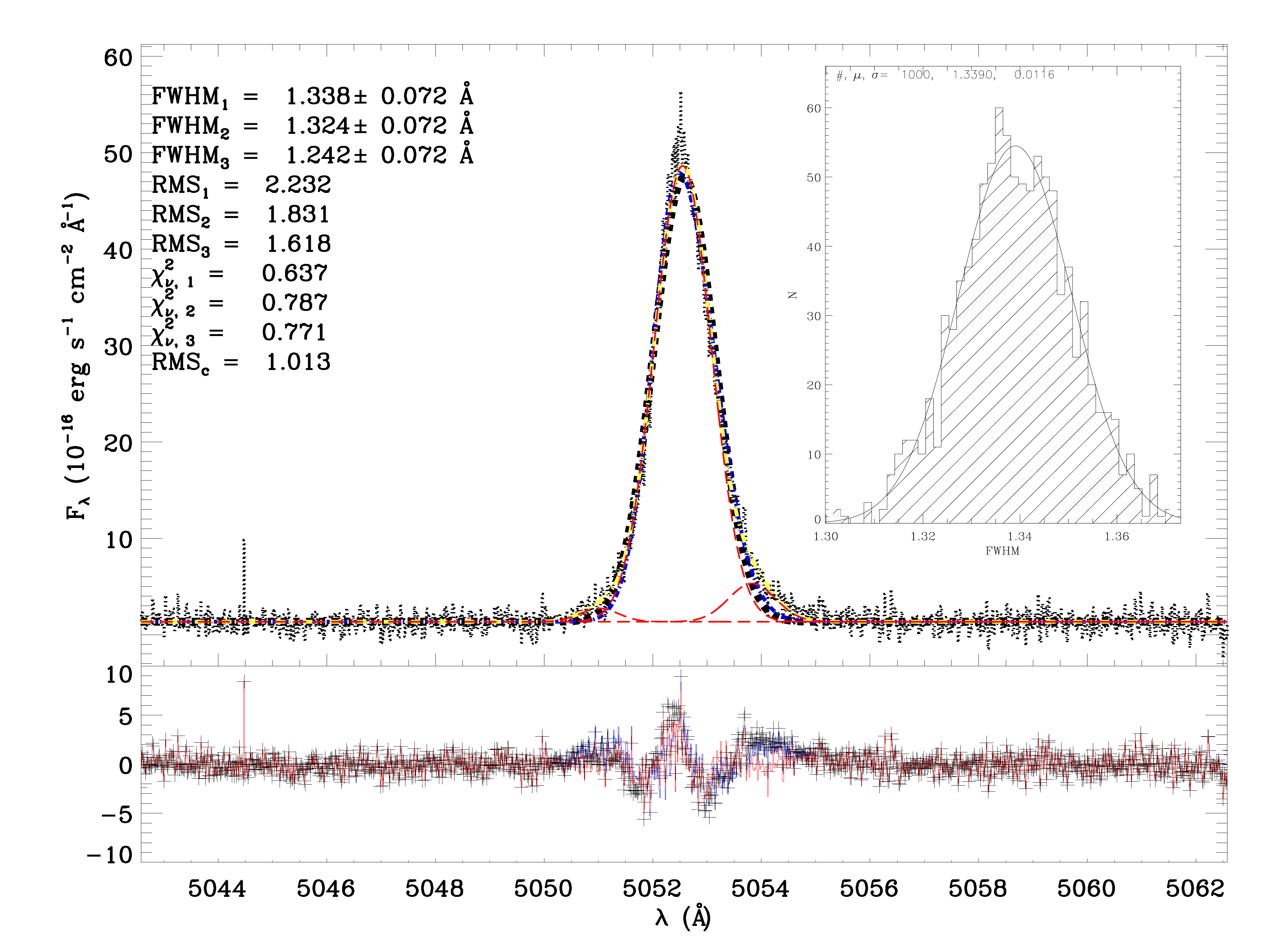}}
\end{figure}

\begin{figure*}
  \centering
  \label{Afig02} \caption{\small H$\beta$ lines best fits continued.}
  \subfloat[J081403+235328]{\label{Afig02:1}\includegraphics[width=80mm]{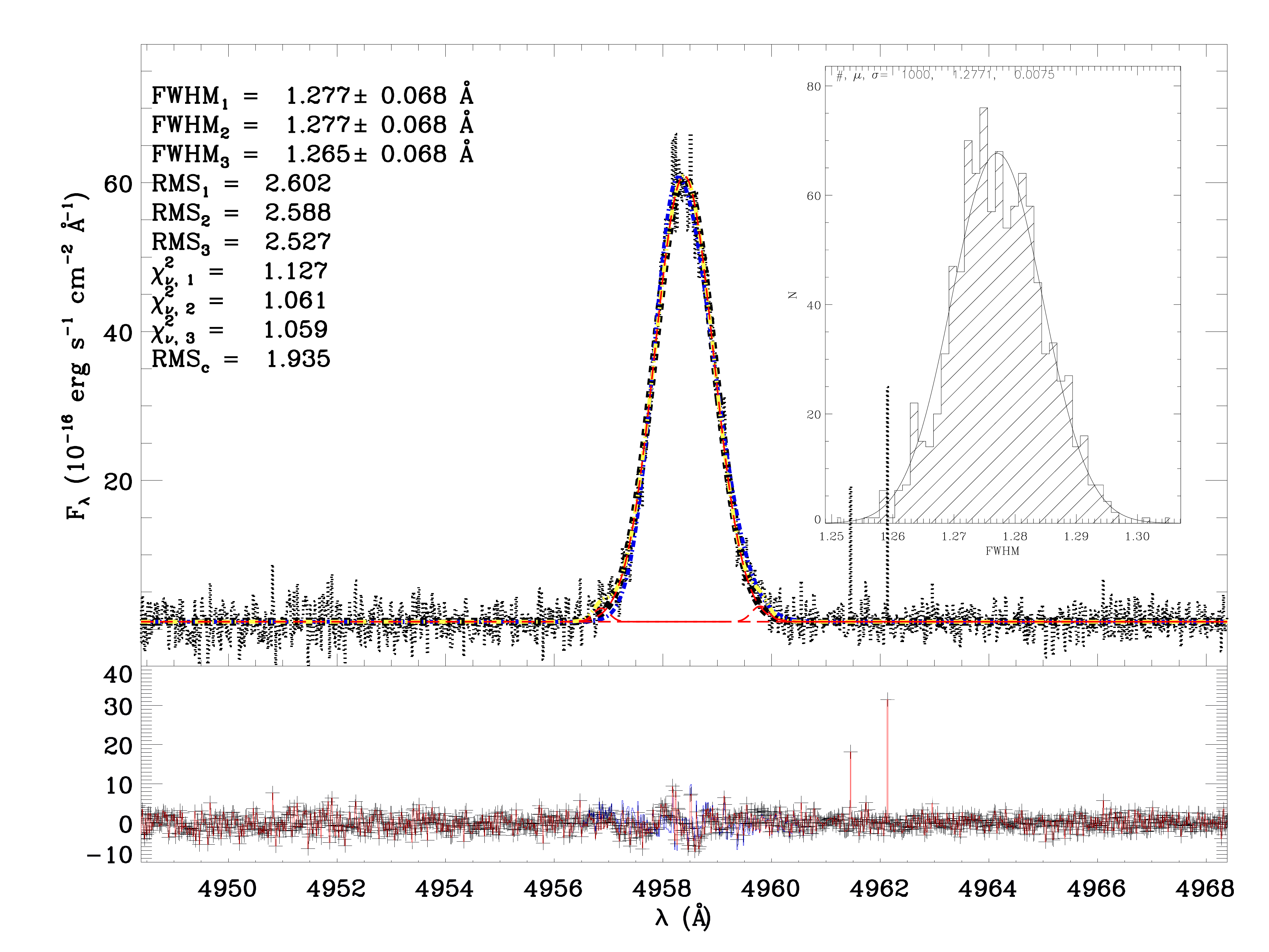}}
  \subfloat[J082520+082723]{\label{Afig02:2}\includegraphics[width=80mm]{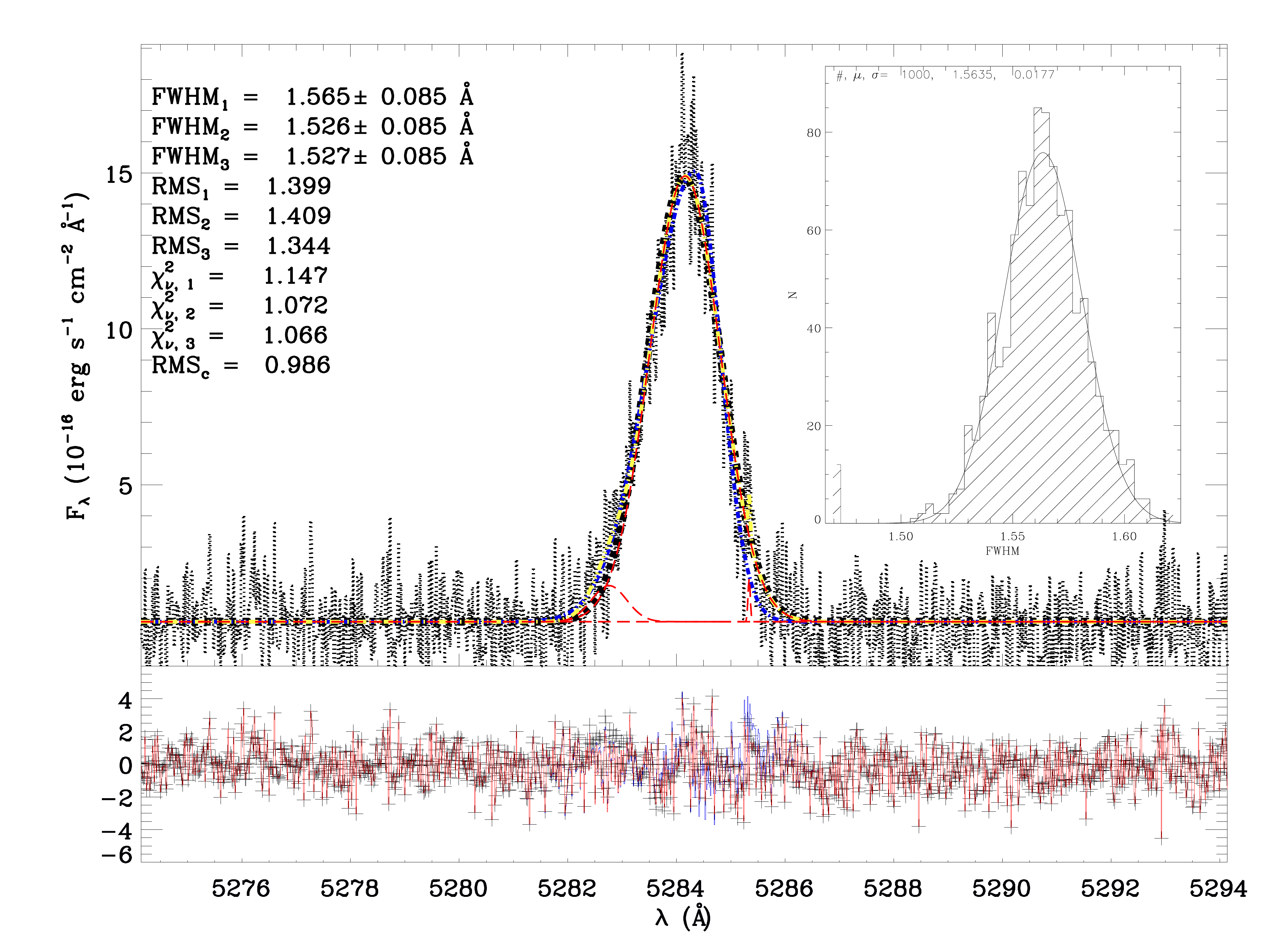}}
  \\
   \subfloat[J082520+082723]{\label{Afig02:3}\includegraphics[width=80mm]{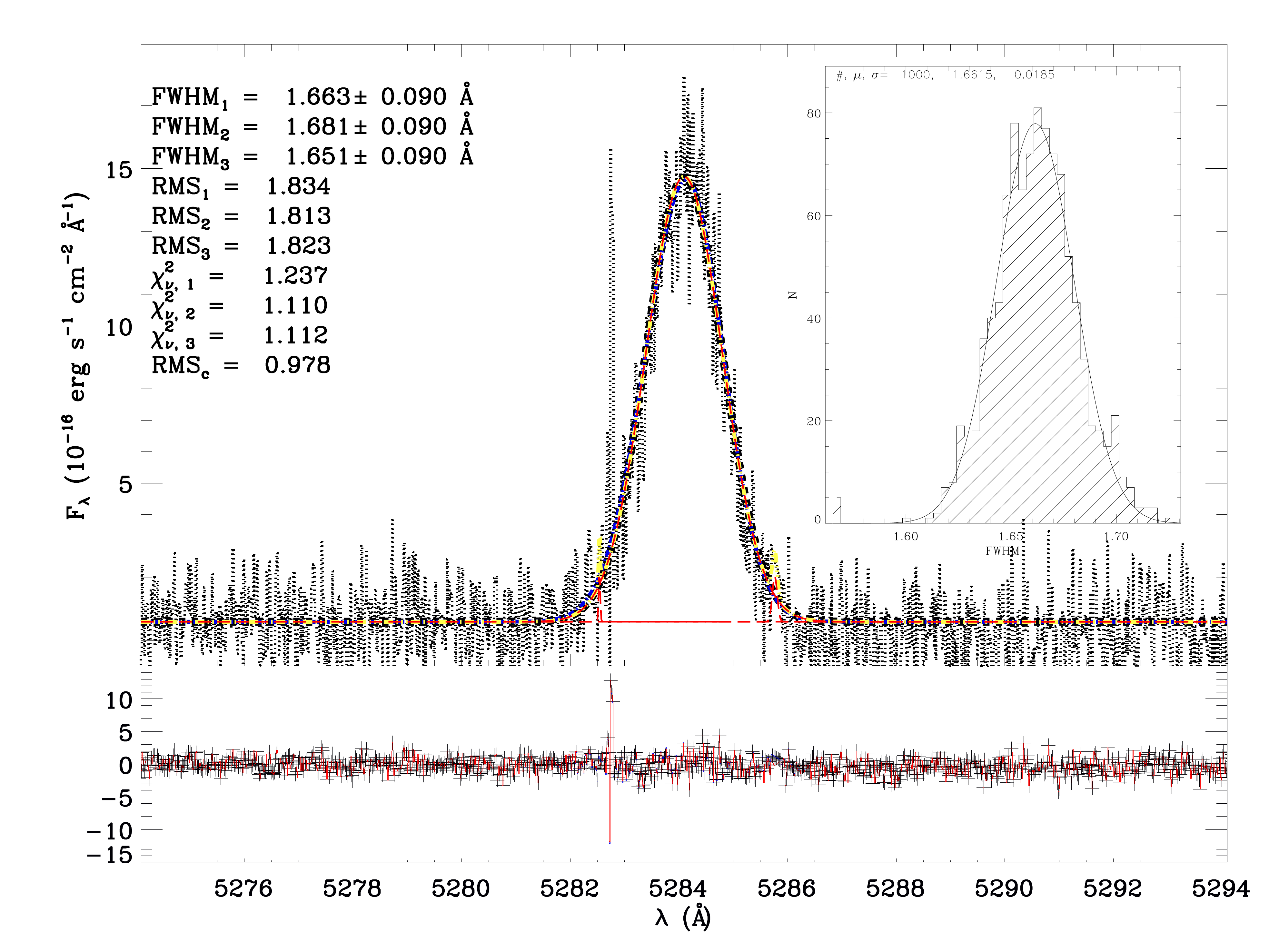}}
   \subfloat[J082722+202612]{\label{Afig02:4}\includegraphics[width=80mm]{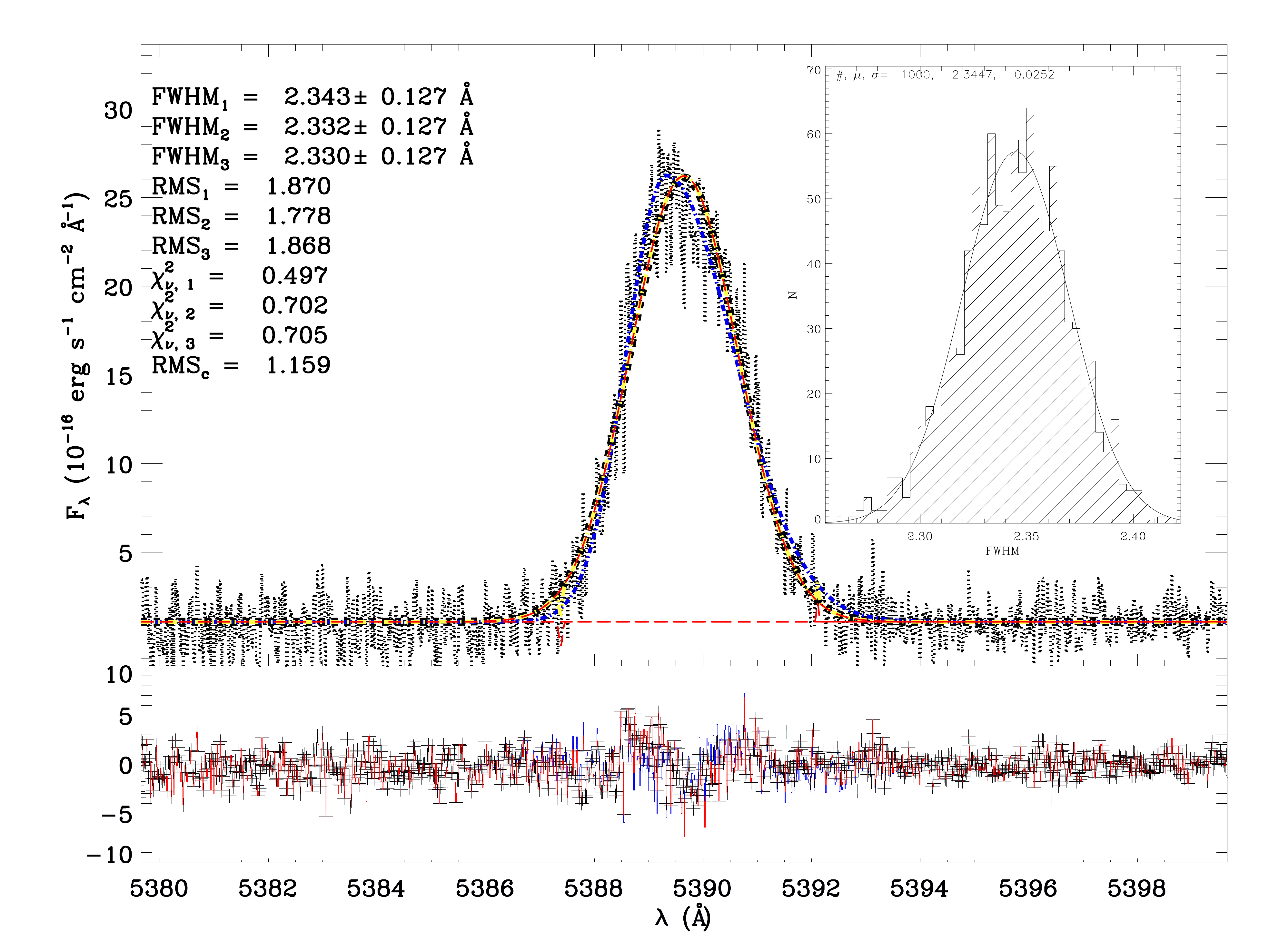}}
  \\
   \subfloat[J083946+140033]{\label{Afig02:5}\includegraphics[width=80mm]{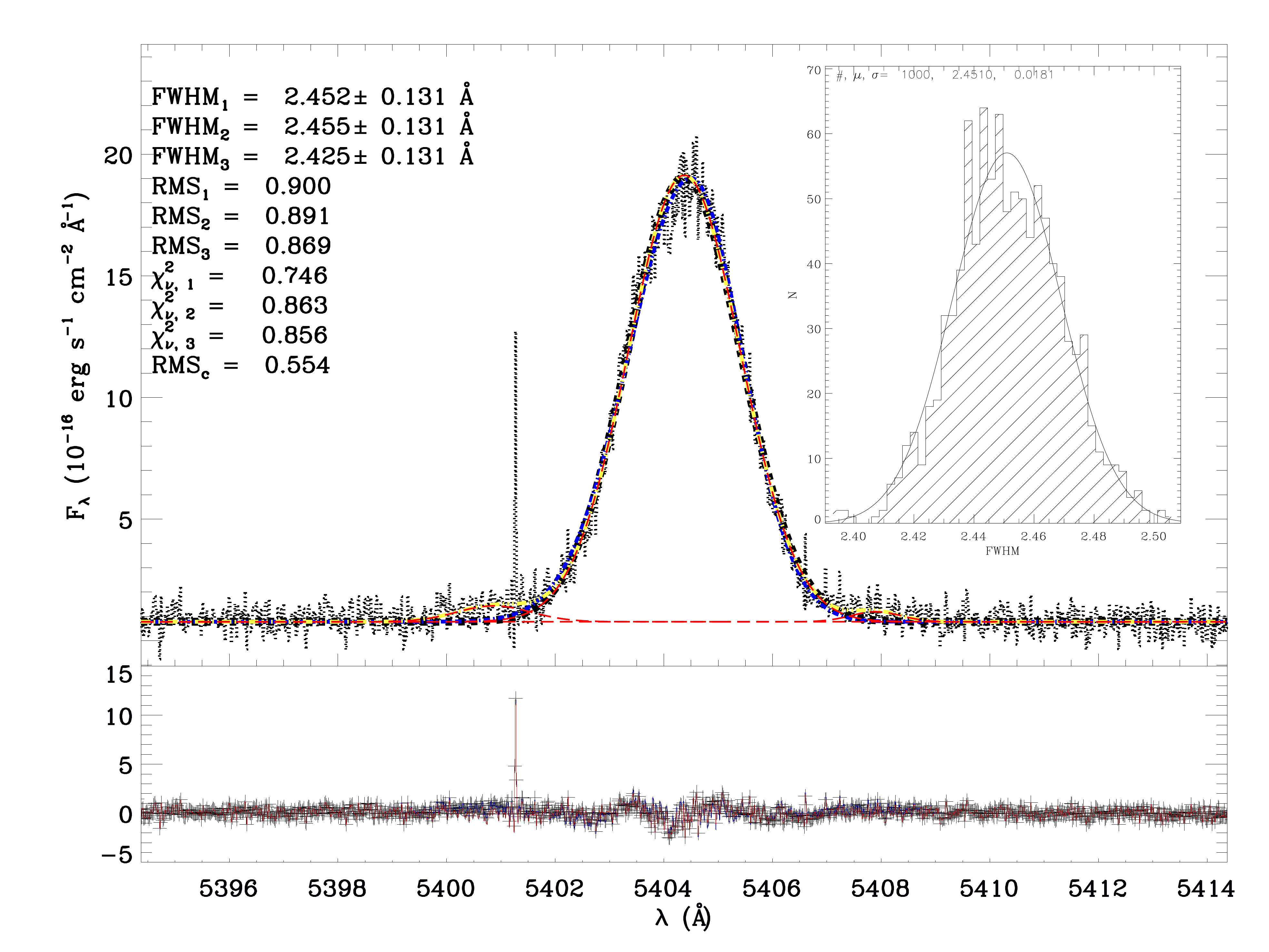}}
   \subfloat[J084000+180531]{\label{Afig02:6}\includegraphics[width=80mm]{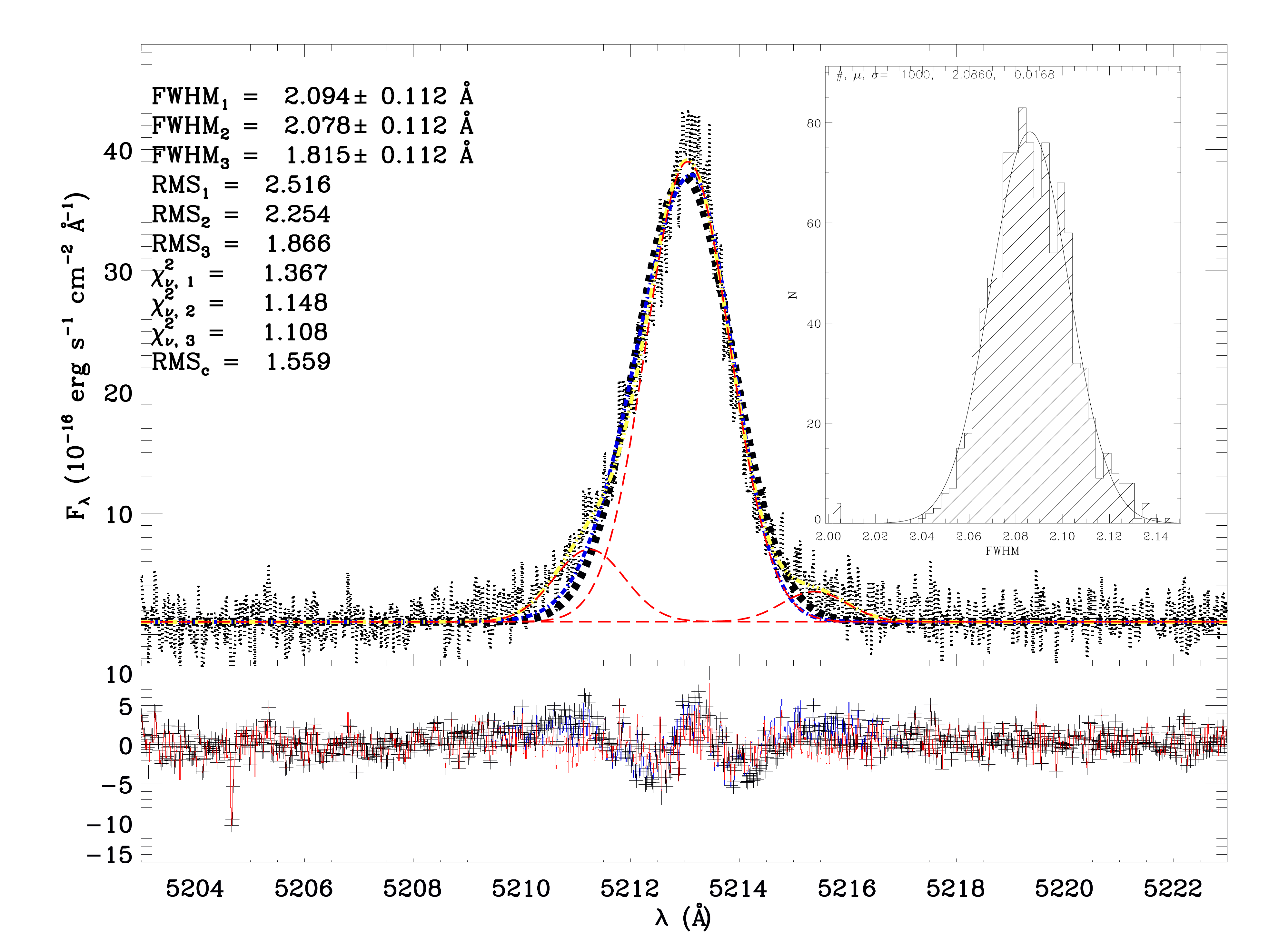}}
\end{figure*}